\documentclass{article}

\usepackage{amsmath}
\usepackage{amssymb}
\usepackage{authblk}
\usepackage{hyperref}
\usepackage{graphicx}
\usepackage{framed}
\usepackage{setspace}
\usepackage{fullpage}

\title{Computational General Relativity in the Wolfram Language using \textsc{Gravitas} I: Symbolic and Analytic Computation}
\author[1,2]{Jonathan Gorard}
\affil[1]{Wolfram Institute\footnote{\href{mailto:jonathang@wolfram.com}{jonathang@wolfram.com}}}
\affil[2]{University of Cambridge, Cambridge, UK\footnote{\href{mailto:jg865@cantab.ac.uk}{jg865@cantab.ac.uk}}}

\begin{document}

\maketitle

\begin{abstract}
We introduce a new, open-source computational general relativity framework for the Wolfram Language called \textsc{Gravitas}, which boasts a number of novel and distinctive features as compared to the many pre-existing computational and numerical relativity frameworks currently available within the open-source community. These include, but are not limited to: seamless integration of its powerful symbolic and numerical subsystems, and, by extension, seamless transition between analytic/continuous representations and numerical/discrete representations of arbitrary spacetime geometries; highly modular, general and extensible representations of spacetime geometries, spacetime topologies, gauge conditions, coordinate systems, matter fields, evolution equations and initial data; ability to set up and run complex numerical relativity simulations, and to perform 2D and 3D visualizations, symbolic computations and numerical analysis (including the extraction of gravitational wave signals) on the resulting data, all from within a single notebook environment; and a totally-unstructured adaptive refinement scheme based on hypergraph rewriting, allowing for exceedingly efficient discretization and numerical evolution of Cauchy initial data for a wide range of challenging computational problems involving strong relativistic field dynamics. In this first in a series of two articles covering the framework, we focus on the design and capabilities of \textsc{Gravitas}'s symbolic subsystem, including its general and flexible handling of arbitrary geometries parametrized by arbitrary curvilinear coordinate systems (along with an in-built library of standard metrics and coordinate conditions), as well as its various high-level tensor calculus and differential geometry features. We proceed to show how this subsystem can be used in parallel with \textsc{Gravitas}'s representation of arbitrary energy-matter distributions and its specialized relativistic electromagnetism functionality (along with in-built libraries of standard stress-energy tensors and energy conditions) in order to solve the Einstein and Einstein-Maxwell field equations, both analytically and numerically, in a highly automated and generalizable fashion. We conclude by providing motivation for the second article in the series (covering numerical relativity functionality), as well as discussing planned extensions to the framework and potential future research applications in mathematical relativity, astrophysics and quantum gravity.
\end{abstract}

\section{Introduction}

The Einstein field equations of general relativity, though they are often expressed in a deceptively simple form by means of abstract tensor notation and general covariance, remain notoriously difficult to solve, both analytically and numerically\cite{stephani}. Indeed, all known analytical solutions to the field equations derived to date have involved the application of one or more strong simplifying assumptions, such as a high degree of a priori symmetry\cite{krasinski}\cite{bicak} (as in the case of the Friedmann-Lema\^itre-Robertson-Walker/FLRW solution\cite{friedmann}\cite{lemaitre}\cite{robertson}\cite{walker} for a uniformly expanding or contracting universe, in which the homogeneity and isotropy assumptions on the spacetime ensure both translational and rotational symmetry of the spatial geometry, thus allowing one to reduce the ten independent components of the Einstein field equations down to just two) or some other kind of ``algebraic specialness''\cite{pravda} condition (such as in Petrov type II solutions\cite{petrov}, in which two of the four principal null directions of the Weyl tensor are assumed to coincide\cite{penrose}\cite{penrose2}, again reducing the total number of independent components of the metric tensor). This difficulty in obtaining analytic solutions derives in large part from the fact that the Einstein field equations, when represented in a fully explicit form within a chosen coordinate basis, constitute a system of ten tightly-coupled and highly-nonlinear second-order partial differential equations in the components of the metric tensor, of mixed hyperbolic-elliptic character (though they can generally be simplified down to six independent coupled equations via judicious application of the contracted Bianchi identities). This places the discovery of analytic solutions firmly out of the reach of perturbation theory and other standard mathematical approaches, and indeed the construction of such solutions remains an active area of research in differential geometry, geometric analysis and the analysis of partial differential equations more generally. To this end, a wide variety of different computational frameworks have been developed to facilitate the requisite tensor calculus and differential geometry operations, including notably the \href{http://www.xact.es/}{\texttt{xAct}} suite of packages developed by Mart\'in-Garc\'ia et al. in the Wolfram Language (designed primarily for dealing with abstract tensor calculus, such as the \href{http://www.xact.es/Invar/index.html}{\texttt{Invar}} package\cite{martingarcia}\cite{martingarcia2} for manipulating scalar invariants of Riemann tensors and the \href{http://www.xact.es/xPerm/index.html}{\texttt{xPerm}} package\cite{martingarcia3} for canonicalizing tensor indices with respect to permutation symmetries, though also with additional support for concrete component calculus through the \href{http://www.xact.es/xCoba/index.html}{\texttt{xCoba}} package), as well as frameworks such as the \href{https://einsteinpy.org}{\texttt{EinsteinPy}} package\cite{bapat} in Python (which, in addition to facilitating certain classes of elementary tensor manipulations of relevance for typical general relativity, also allows for visualizations and geodesic computations over a small handful of in-built spacetime geometries).

On the other hand, even attempting to solve the Einstein field equations using numerical approximations presents its own unique set of challenges. Schemes for solving systems of hyperbolic partial differential equations numerically typically involve defining the initial data for the system on some Cauchy surface and then evolving that data forwards in time using an explicit time-stepping method. However, with the Einstein field equations, the notion of time is intimately tied to the spacetime metric itself, whose components are also precisely the variables being evolved, so one has a rather non-trivial ``mixing'' of independent and dependent variables of the system as a consequence of covariance. This conceptual difficulty can be circumvented by imposing a gauge choice that ``foliates'' the spacetime into a time-ordered sequence of codimension-1 spacelike hypersurfaces, otherwise known as a ``${3 + 1}$ decomposition'' of the metric, using some variant of the ADM formalism of Arnowitt, Deser and Misner\cite{arnowitt}\cite{arnowitt2} (though the typical mathematical form of the ${3 + 1}$ decomposition used in practice was originally due to York\cite{york}). Such a decomposition splits the ten mixed hyperbolic-elliptic Einstein field equations into a system of six evolution equations for the components of the induced spatial metric tensor (or, equivalently, the components of the extrinsic curvature tensor on the spacelike hypersurfaces) which are generally purely hyperbolic in nature, and four constraint equations on the gauge variables (arising from projections of the contracted Bianchi identities) which are generally purely elliptic in nature\cite{alcubierre}. However, this, in turn, presents further difficulties, since iterative algorithms for solving the elliptic constraint equations are typically radically different in character and significantly more computationally expensive than the explicit time-stepping algorithms used for solving the hyperbolic evolution equations, and so one must either reformulate the constraint equations to be solvable using purely hyperbolic methods (e.g. using the constraint-violation damping approach of Gundlach et al.\cite{gundlach} based on the ${\lambda}$-system formalism of Brodbeck et al.\cite{brodbeck}), or otherwise develop novel numerical schemes for performing constrained evolution in a computationally efficient way. For these purposes, a range of different open-source numerical relativity codes have been released, including the \href{http://www.cactuscode.org}{\texttt{Cactus}} framework\cite{goodale}, deployed as part of the much larger open-source \href{https://einsteintoolkit.org/}{\texttt{Einstein}} toolkit\cite{zilhao} (with \href{http://www.cactuscode.org}{\texttt{Cactus}} itself containing implementations such as \href{https://einsteintoolkit.org/arrangementguide/McLachlan/documentation.html}{\texttt{McLachlan}}\cite{brown}\cite{reisswig}, designed for solving the Einstein field equations using a finite-difference discretization scheme with up to eighth-order accuracy, and based on the Baumgarte-Shapiro-Shibata-Nakamura/BSSN formulation of the field equations\cite{nakamura}\cite{shibata}\cite{baumgarte} with block-structured adaptive mesh refinement/AMR), and the \href{https://www.black-holes.org/code/SpEC.html}{SpEC} code\cite{pfeiffer} (which solves the Einstein field equations by using pseudospectral methods applied to the generalized harmonic formulation of the equations).

By and large, these analytical and numerical aspects of the field of computational general relativity interact only loosely: most tensor calculus and differential geometry packages do not implement any powerful numerical algorithms, and most purpose-built numerical relativity codes can perform only very crude symbolic tensor computations (if any at all). One standard pipeline within numerical relativity is to use a symbolic framework such as \href{http://kranccode.org}{\texttt{Kranc}}\cite{husa} in the Wolfram Language or \href{https://nrpyplus.net}{\texttt{SENR/NRPy+}}\cite{ruchlin} in Python, both of which allow one to input evolution equations, gauge conditions, etc. in a high-level, clean and abstract tensorial form, perform automated algebraic simplifications on the resulting partial differential equations, and then generate highly-optimized, parallelized and low-level C/Fortran code for solving the equations numerically (thus reducing the probability of coding errors), which then typically must be integrated into a pre-existing numerical framework such as \href{http://www.cactuscode.org}{\texttt{Cactus}}. However, even in these cases, the integration is far from seamless, often with different libraries, packages, frameworks, programming languages and software tools responsible for defining and simplifying the evolution equations, coordinate systems and gauge constraints, for constructing appropriate initial data, for performing the numerical evolutions, for visualizing the resulting evolutions, for extracting relevant simulation parameters (such as gravitational waveform data) and performing the requisite scientific analysis on those parameters. In this article, we introduce a new, open-source Wolfram Language framework called \textsc{Gravitas}, which allows one to perform all of the aforementioned operations (and more) using a single, clean, unified and high-level programming interface, thus allowing one, among other things, to configure, run, visualize and analyze a complex numerical relativity simulation from within a single notebook environment. One of the core design principles of \textsc{Gravitas} is that it does not make any distinction between continuous spacetimes represented using symbolic/analytical functions, and discrete spacetimes represented using numerical approximations; although the internal representations of these two cases are inevitably very different, the user can interact with, and convert between, the two in a fully coherent and unified way.

This article is the first in a series of two articles intended to introduce the core design and functionality of \textsc{Gravitas}, and will focus primarily on \textsc{Gravitas}'s analytical and symbolic capabilities, including its general handling of metrics and coordinate systems, its powerful tensor calculus and differential geometry subsystems, and its ability to solve the Einstein field equations (both analytically and numerically) in a wide range of scenarios, including in the presence of arbitrary energy-matter distributions and non-gravitational fields, with particular support for electromagnetic fields via the Einstein-Maxwell equations. The second (forthcoming) article in the series will focus more heavily on \textsc{Gravitas}'s capabilities for numerical relativity, including its handling of ADM formalism and ${3 + 1}$ decompositions of spacetime, definition and enforcement of gauge conditions, construction of Cauchy initial data, satisfaction of Hamiltonian and momentum constraints, totally unstructured adaptive refinement algorithms based on hypergraph rewriting/Wolfram model evolution\cite{gorard}\cite{gorard2}\cite{gorard3} (enabling the handling of arbitrary curvilinear coordinate systems and arbitrary spacetime topologies within its numerical algorithms), and its various analysis and visualization tools. Although the present article will include some cases in which the Einstein field equations are solved numerically/non-exactly, these will all correspond to cases in which no gauge choice has been imposed, no metric decomposition has been performed and full general covariance of the spacetime has been preserved. The core elements of the \textsc{Gravitas} framework, and especially the abstract hypergraph rewriting\cite{gorard4}\cite{gorard5}\cite{gorard6} and hypergraph canonicalization\cite{gorard7} algorithms that underlie its numerical subsystems, have already been applied to great effect in the study of binary black hole collisions\cite{gorard8}, quantum field theory in curved spacetime\cite{gorard9} and idealized gravitational collapse models/singularity theorems\cite{gorard10}, all taking place within highly general and topologically-unstructured discrete spacetime settings.

We begin in Section \ref{sec:Section1} by introducing the design of the \texttt{MetricTensor} object (the most fundamental object in the \textsc{Gravitas} framework) and illustrating how \textsc{Gravitas} is able to represent arbitrary Riemannian/pseudo-Riemannian/Lorentzian metrics and their corresponding manifold geometries, as well as demonstrating how many equivalent geometries may be represented using different coordinate systems. We also show excerpts from \textsc{Gravitas}'s in-built library of geometries, metrics and coordinate systems (including most standard black hole and cosmological metrics), and also highlight functionality for determining lengths of and angles between tangent vectors, ascertaining the timelike/lightlike/spacelike nature of tangent vectors, converting to and from line element and volume form representations of the metric, and performing automatic code generation for various key geometrical operations, all for any geometry and coordinate system that can be represented by \texttt{MetricTensor}. In Section \ref{sec:Section2}, we move on to demonstrating how \textsc{Gravitas} handles the definition of the Levi-Civita connection (a special case of the more general metric and affine connections) over an arbitrary (pseudo-)Riemannian manifold, using the \texttt{ChristoffelSymbols} object to represent the corresponding connection coefficients, and we show how imposing such a connection immediately facilitates the straightforward computation of a wide range of different curvature tensors over the manifold, including the \texttt{RiemannTensor}, \texttt{WeylTensor}, \texttt{RicciTensor}, \texttt{EinsteinTensor}, \texttt{SchoutenTensor} and \texttt{BachTensor} objects. We also demonstrate many of \textsc{Gravitas}'s symbolic tensor calculus and differential geometry capabilities, including automatic raising and lowering of indices, automatic handling of index contractions via Einstein summation convention, computation of covariant derivatives, computation of various (scalar) curvature invariants, verification of the (differential) Bianchi identities and their contractions in various forms, determination of necessary and sufficient conditions for Ricci-flatness and conformal-flatness of manifolds, computation of Taylor expansions of metric volume elements and volume forms, and several other things. In Section \ref{sec:Section3}, we illustrate how these tensor calculus and differential geometry features may be leveraged in order to solve the Einstein field equations automatically, both analytically and numerically, under a wide variety of conditions, starting with the vacuum Einstein field equations (with solutions represented by \texttt{VacuumSolution} objects produced via the \texttt{SolveVacuumEinsteinEquations} function) before proceeding to the case of the full Einstein field equations equipped with an arbitrary \texttt{StressEnergyTensor} object (with solutions represented by \texttt{EinsteinSolution} objects produced via the \texttt{SolveEinsteinEquations} function). We provide an overview of part of \textsc{Gravitas}'s in-built library of standard energy-matter distributions, explain how various quantities such as relativistic energy densities, momentum densities and stress tensors can be extracted, how relativistic energy conditions (including the standard null, weak, dominant and strong conditions) can be enforced, how relativistic continuity equations can be derived and solved, and how relativistic angular momentum tensors can be computed and manipulated. We also provide an in-depth summary of \textsc{Gravitas}'s capabilities for handling general relativistic electromagnetism through the \texttt{ElectromagneticTensor} object, including formulations of Maxwell's equations within arbitrary curved spacetimes, and how the (vacuum) Einstein-Maxwell equations can be solved, both analytically and numerically (with solutions represented by \texttt{ElectrovacuumSolution} objects produced via the \texttt{SolveElectrovacuumEinsteinEquations} function), for arbitrary electromagnetic potentials. Finally, we conclude in Section \ref{sec:Section4} with a summary of some planned future symbolic and analytic functionality (as well as of other functionality that is presently still in the implementation phase), and also of several planned and ongoing research applications of the \textsc{Gravitas} framework to mathematical relativity, quantum gravity, relativistic astrophysics and relativistic cosmology.

Note that many of the core \textsc{Gravitas} functions discussed within this article and its forthcoming companion are currently fully-documented and exposed via the \textit{Wolfram Function Repository}, including \href{https://resources.wolframcloud.com/FunctionRepository/resources/MetricTensor/}{\texttt{MetricTensor}}, \href{https://resources.wolframcloud.com/FunctionRepository/resources/ChristoffelSymbols/}{\texttt{ChristoffelSymbols}}, \href{https://resources.wolframcloud.com/FunctionRepository/resources/RiemannTensor/}{\texttt{RiemannTensor}}, \href{https://resources.wolframcloud.com/FunctionRepository/resources/RicciTensor}{\texttt{RicciTensor}}, \href{https://resources.wolframcloud.com/FunctionRepository/resources/EinsteinTensor/}{\texttt{EinsteinTensor}}, \href{https://resources.wolframcloud.com/FunctionRepository/resources/StressEnergyTensor/}{\texttt{StressEnergyTensor}}, \href{https://resources.wolframcloud.com/FunctionRepository/resources/SolveVacuumEinsteinEquations/}{\texttt{SolveVacuumEinsteinEquations}}, \href{https://resources.wolframcloud.com/FunctionRepository/resources/SolveEinsteinEquations/}{\texttt{SolveEinsteinEquations}}, \href{https://resources.wolframcloud.com/FunctionRepository/resources/ADMDecomposition/}{\texttt{ADMDecomposition}} and \href{https://resources.wolframcloud.com/FunctionRepository/resources/DiscreteHypersurfaceDecomposition/}{\texttt{DiscreteHypersurfaceDecomposition}}. However, many other functions are yet to be documented and exposed in this way, and many of the documented versions may be out-of-date; an up-to-date version of the \textsc{Gravitas} codebase, included all experimental and research functionality, can instead always be obtained from its official \href{https://github.com/JonathanGorard/Gravitas/}{GitHub Repository}. The codebase is fairly large, at least by Wolfram Language package standards, currently standing at over 27,000 lines at time of writing (and rapidly growing, as the framework remains in very active development), yet is organized into several smaller and largely independent package files, for the purposes of retaining modularity, maintainability and interoperability of its various components. Within the framework, and throughout the remainder of this article, we assume a purely geometric unit system in which ${c = G = \hbar = 1}$; the only ``dimensionful'' constant that will appear within this article is the vacuum magnetic permeability constant ${\mu_0}$, which occurs in the context of general relativistic electromagnetism, since this cannot be fully eliminated when using geometric units. Henceforth, we also adopt the Einstein summation convention (in which all repeated tensor indices are implicitly summed over), and we assume a metric signature of ${\left( -, +, +, + \right)}$ in all relevant cases.

\section{Metrics, Geometries and Coordinate Systems}
\label{sec:Section1}

\begin{figure}[ht]
\centering
\begin{framed}
\includegraphics[width=0.495\textwidth]{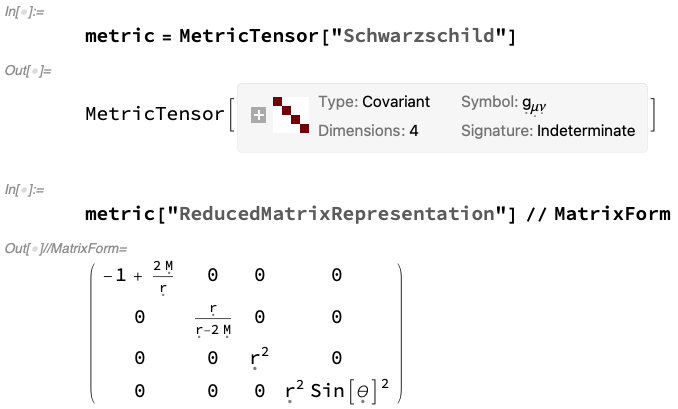}
\vrule
\includegraphics[width=0.495\textwidth]{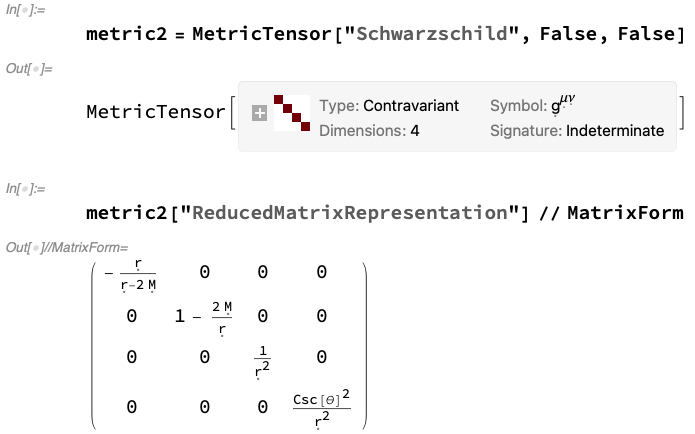}
\end{framed}
\caption{On the left, the \texttt{MetricTensor} object for a Schwarzschild geometry (representing e.g. an uncharged, non-rotating black hole of mass $M$ in Schwarzschild/spherical polar coordinates ${\left( t, r, \theta, \phi \right)}$) in covariant matrix form, with both indices lowered/covariant. On the right, the \texttt{MetricTensor} object for a Schwarzschild geometry (representing e.g. an uncharged, non-rotating black hole of mass $M$ in Schwarzschild/spherical polar coordinates ${\left( t, r, \theta, \phi \right)}$) in contravariant matrix form, with both indices raised/contravariant.}
\label{fig:Figure1}
\end{figure}

\begin{figure}[ht]
\centering
\begin{framed}
\includegraphics[width=0.495\textwidth]{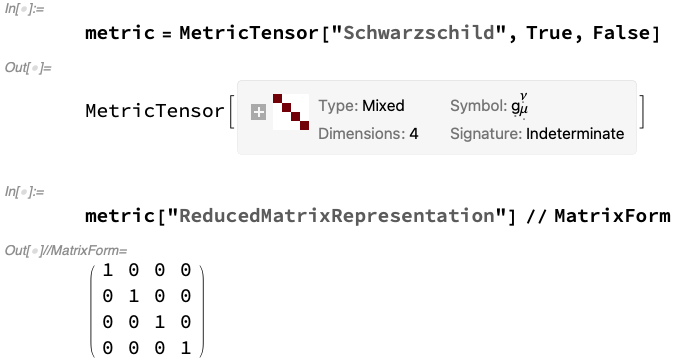}
\vrule
\includegraphics[width=0.495\textwidth]{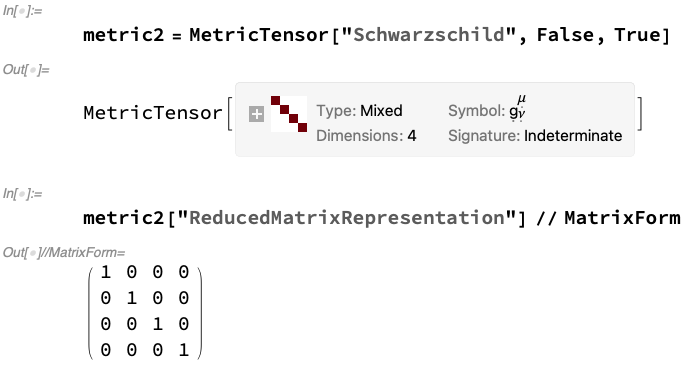}
\end{framed}
\caption{The \texttt{MetricTensor} objects for a Schwarzschild geometry (representing e.g. an uncharged, non-rotating black hole of mass $M$ in Schwarzschild/spherical polar coordinates ${\left( t, r, \theta, \phi \right)}$) in mixed-index matrix form, with one index lowered/covariant and one index raised/contravariant, showing that they are both equal to the identity tensor ${\delta_{\mu}^{\nu} = \delta_{\nu}^{\mu}}$ (i.e. the Kronecker delta function).}
\label{fig:Figure2}
\end{figure}

The most fundamental symbolic object in the \textsc{Gravitas} framework, from which all other objects inherit many of their properties and capabilities, is the \texttt{MetricTensor} object, which is consequently afforded special treatment within the framework's core design. A \texttt{MetricTensor} object represents a symmetric bilinear form ${g_{\mathbf{x}}}$, defined over the tangent space ${T_{\mathbf{x}} \mathcal{M}}$ of a differentiable manifold ${\mathcal{M}}$ at a point ${\mathbf{x} \in \mathcal{M}}$, which can be evaluated to give a real number ${g_{\mathbf{x}} \left( \mathbf{u}, \mathbf{v} \right)}$ for any pair of tangent vectors ${\mathbf{u}, \mathbf{v} \in T_{\mathbf{x}} \mathcal{M}}$ at ${\mathbf{x} \in \mathcal{M}}$\cite{lee}:

\begin{equation}
\forall \mathbf{u}, \mathbf{v} \in T_{\mathbf{x}} \mathcal{M}, \qquad g_{\mathbf{x}} \left( \mathbf{u}, \mathbf{v} \right) = g_{\mathbf{x}} \left( \mathbf{v}, \mathbf{u} \right) \in \mathbb{R}.
\end{equation}
\texttt{MetricTensor} generally does not make any distinction between an individual metric tensor ${g_{\mathbf{x}}}$ (defined at a specific point ${\mathbf{x} \in \mathcal{M}}$) and the overall metric tensor \textit{field} $g$ (defined over all such points in ${\mathcal{M}}$), and we shall henceforth use the terms ``tensor'' and ``tensor field'' somewhat interchangeably. Concretely, \texttt{MetricTensor} represents the abstract metric tensor ${g_{\mathbf{x}}}$ by a symmetric matrix defined within a given local coordinate basis ${\left\lbrace x^{\mu} \right\rbrace}$ (i.e. a coordinate basis defined over a local patch of ${\mathcal{M}}$ in the neighborhood of ${\mathbf{x} \in \mathcal{M}}$), in which all components of the matrix transform either covariantly (in which case the matrix is denoted ${g_{\mu \nu} \left( \mathbf{x} \right)}$) or contravariantly (in which case the matrix is denoted ${g^{\mu \nu} \left( \mathbf{x} \right)}$):

\begin{equation}
g_{\mathbf{x}} = g_{\mu \nu} \left( \mathbf{x} \right) d x^{\mu} \otimes d x^{\nu}, \qquad \text{ or } \qquad g_{\mathbf{x}} = g^{\mu \nu} \left( \mathbf{x} \right) d x_{\mu} \otimes d x_{\nu},
\end{equation}
where ${d x^{\mu}}$ and ${d x^{\nu}}$ are vector gradients,  and ${d x_{\mu}}$ and ${d x_{\nu}}$ the corresponding covector/1-form gradients, of the coordinates ${x^{\mu}}$. Henceforth, for the sake of notational cleanness, we shall suppress the explicit tensor product symbol ${\otimes}$ in all tensorial expressions. The covariant ${g_{\mu \nu}}$ and contravariant ${g^{\mu \nu}}$ forms of the metric tensor are then related by means of the ordinary matrix inverse, i.e. ${g^{\mu \nu} = \left( g_{\mu \nu} \right)^{-1}}$. These two representations are shown in Figure \ref{fig:Figure1}, for the case of the Schwarzschild metric\cite{schwarzschild}, independently discovered by Droste\cite{droste}, describing the exterior spacetime geometry surrounding an uncharged, spherically-symmetric and non-rotating mass distribution (e.g. an uncharged, non-rotating black hole) of mass $M$:

\begin{equation}
d s^2 = g_{\mu \nu} d x^{\mu} d x^{\nu} = - \left( 1 - \frac{2 M}{r} \right) d t^2 + \left( 1 - \frac{2 M}{r} \right)^{-1} d r^2 + r^2 \left( d \theta^2 + \sin^2 \left( \theta \right) d \phi^2 \right),
\end{equation}
in Schwarzschild/spherical polar coordinates ${\left( t, r, \theta, \phi \right)}$. This, in turn, implies that the mixed-index forms of the metric tensor ${g_{\mu}^{\nu} = g_{\nu}^{\mu}}$ will simply be equal to the identity tensor ${\delta_{\mu}^{\nu} = \delta_{\nu}^{\mu}}$ (i.e. the Kronecker delta function), as illustrated in Figure \ref{fig:Figure2}. Note that here, and henceforth, the keyword \textit{``Reduced''} within any property names such as \textit{``ReducedMatrixRepresentation''} is used to indicate to \textsc{Gravitas} that it should attempt to apply all known algebraic equivalences and to present the result in its canonical/simplest form (\textsc{Gravitas} automatically keeps track of all non-trivial equivalences between tensor expressions behind-the-scenes, for the purposes of facilitating such algebraic simplifications). There is also a closely related keyword \textit{``Symbolic''}, which is used to indicate to \textsc{Gravitas} that it should not attempt to evaluate any of the (partial) derivatives appearing within a given tensor expression, and should instead leave them as purely symbolic derivatives; though we shall not make use of this feature here, it will become of great relevance to the numerical relativity capabilities outlined within the second article of this series.

\begin{figure}[ht]
\centering
\begin{framed}
\includegraphics[width=0.495\textwidth]{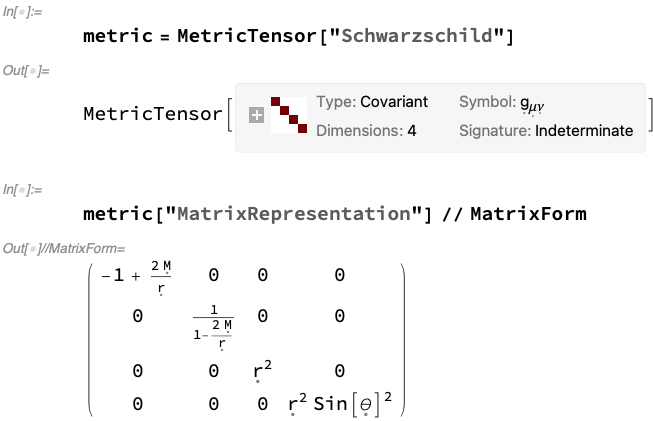}
\vrule
\includegraphics[width=0.495\textwidth]{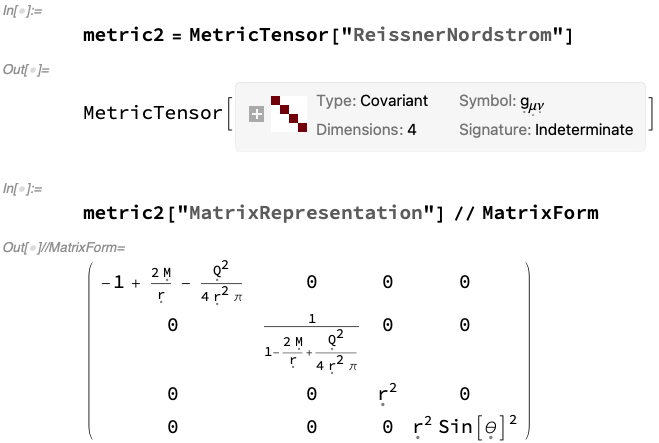}
\end{framed}
\caption{On the left, the \texttt{MetricTensor} object for a Schwarzschild geometry (representing e.g. an uncharged, non-rotating black hole of mass $M$) in Schwarzschild/spherical polar coordinates ${\left( t, r, \theta, \phi \right)}$. On the right, the \texttt{MetricTensor} object for a Reissner-Nordstr\"om geometry (representing e.g. a charged, non-rotating black hole of mass $M$ and electric charge $Q$) in Schwarzschild/spherical polar coordinates ${\left( t, r, \theta, \phi \right)}$.}
\label{fig:Figure3}
\end{figure}

\begin{figure}[ht]
\centering
\begin{framed}
\includegraphics[width=0.495\textwidth]{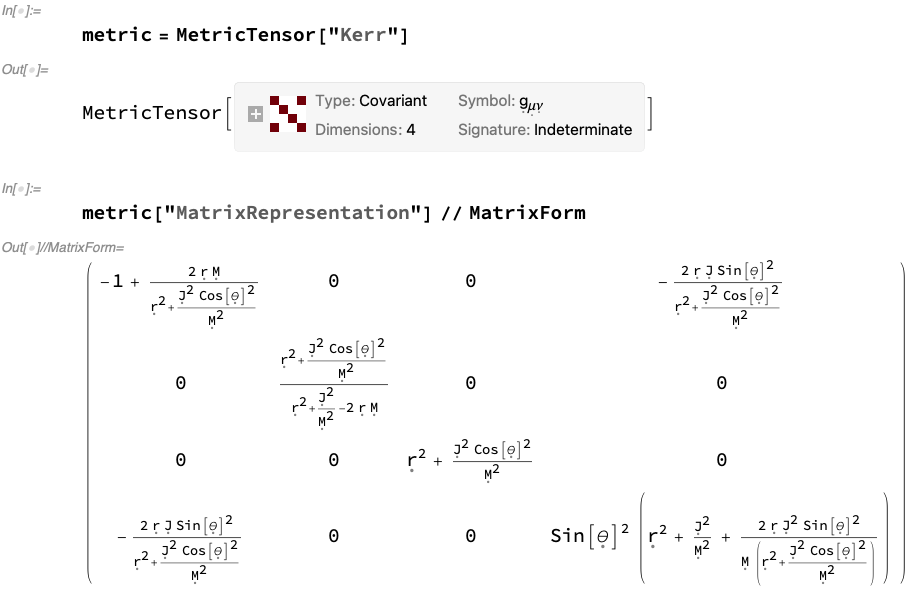}
\vrule
\includegraphics[width=0.495\textwidth]{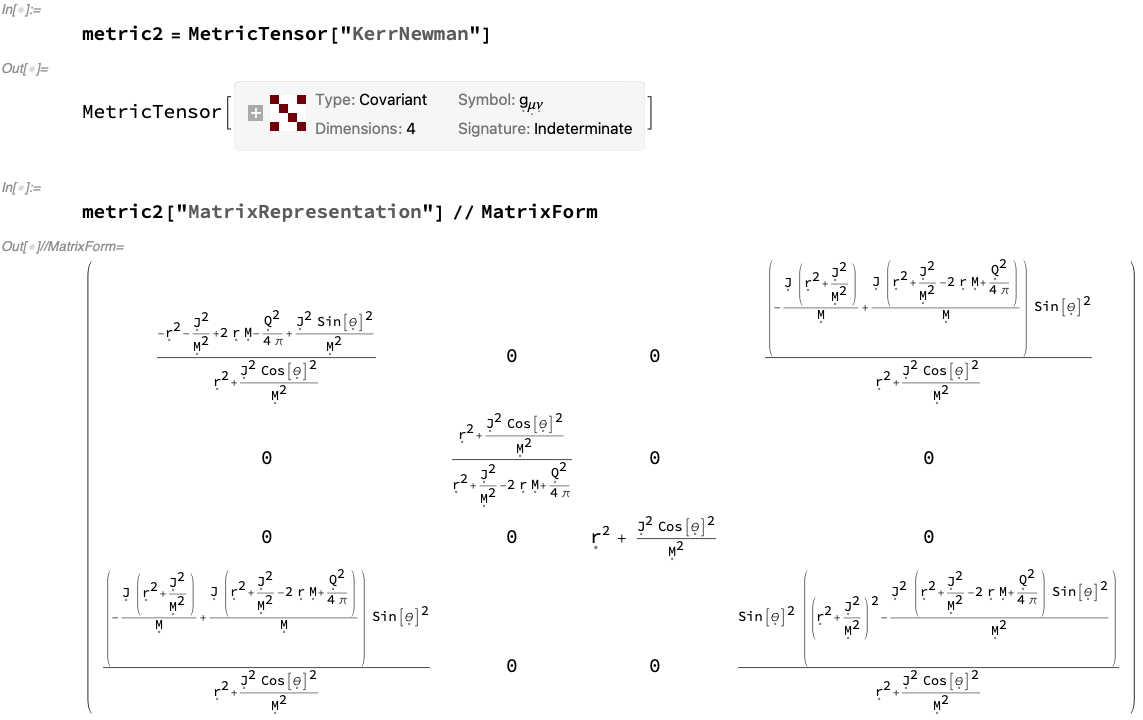}
\end{framed}
\caption{On the left, the \texttt{MetricTensor} object for a Kerr geometry (representing e.g. an uncharged, spinning black hole of mass $M$ and angular momentum $J$) in Boyer-Lindquist/oblate spheroidal coordinates ${\left( t, r, \theta, \phi \right)}$. On the right, the \texttt{MetricTensor} object for a Kerr-Newman geometry (representing e.g. a charged, spinning black hole of mass $M$, angular momentum $J$ and electric charge $Q$) in Boyer-Lindquist/oblate spheroidal coordinates ${\left( t, r, \theta, \phi \right)}$.}
\label{fig:Figure4}
\end{figure}

\begin{figure}[ht]
\centering
\begin{framed}
\includegraphics[width=0.495\textwidth]{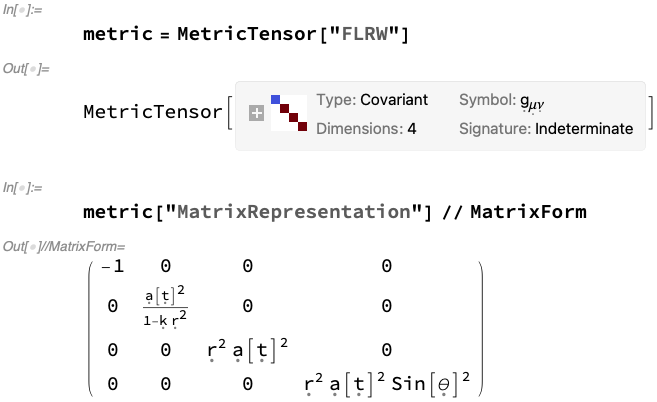}
\vrule
\includegraphics[width=0.495\textwidth]{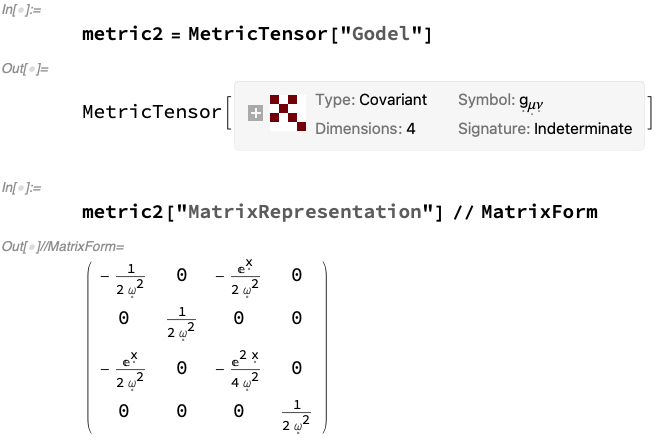}
\end{framed}
\caption{On the left, the \texttt{MetricTensor} object for a Friedmann-Lema\^itre-Robertson-Walker/FLRW geometry (representing e.g. a homogeneous, isotropic and uniformly expanding/contracting universe with global curvature $k$ and scale factor ${a \left( t \right)}$) in spherical polar coordinates ${\left( t, r, \theta, \phi \right)}$. On the right, the \texttt{MetricTensor} object for a G\"odel geometry (representing e.g. a rotating, dust-filled universe with global angular velocity ${\omega}$) in G\"odel's Cartesian-like coordinates ${\left( t, x, y, z \right)}$.}
\label{fig:Figure5}
\end{figure}

\begin{figure}[ht]
\centering
\begin{framed}
\includegraphics[width=0.495\textwidth]{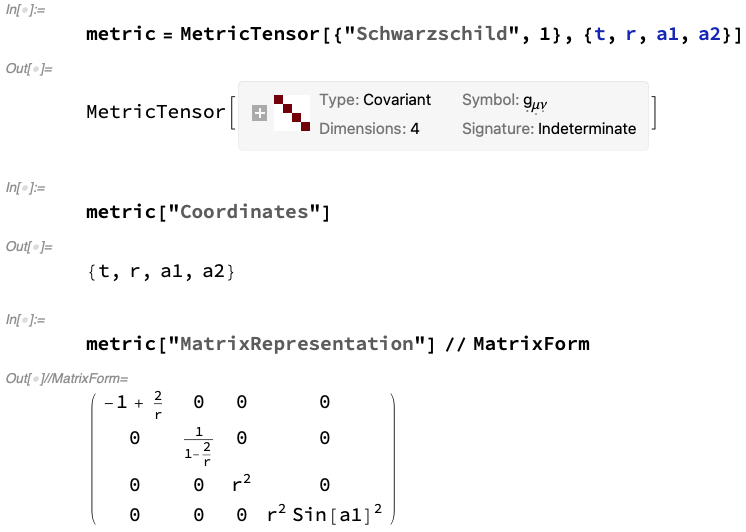}
\vrule
\includegraphics[width=0.495\textwidth]{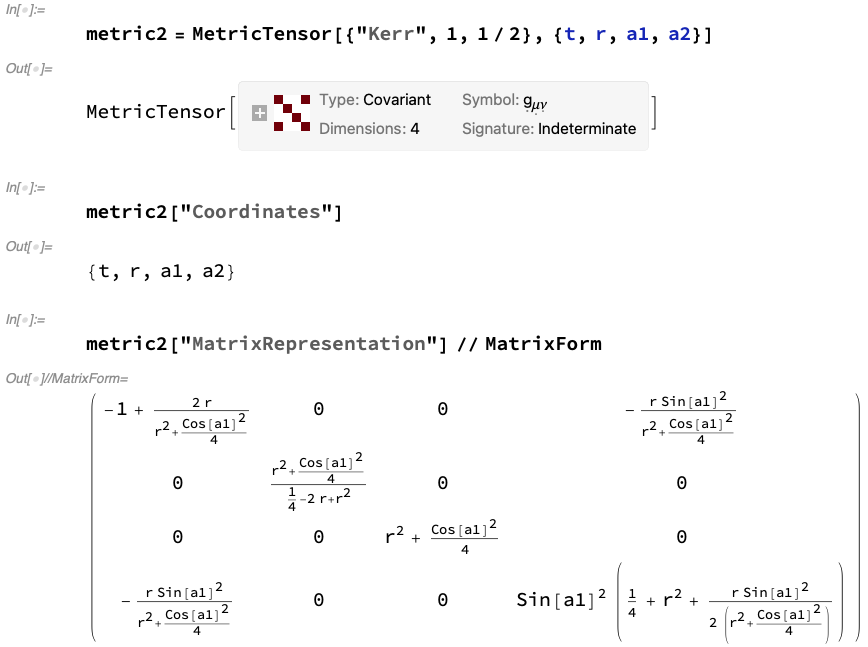}
\end{framed}
\caption{On the left, the \texttt{MetricTensor} object for a Schwarzschild geometry (representing e.g. an uncharged non-rotating black hole of numerical mass 1) in modified Schwarzschild/spherical polar coordinates ${\left( t, r, a1, a2 \right)}$. On the right, the \texttt{MetricTensor} object for a Kerr geometry (representing e.g. an uncharged, spinning black hole of numerical mass 1 and numerical angular momentum ${\frac{1}{2}}$) in modified Boyer-Lindquist/oblate spheroidal coordinates ${\left( t, r, a1, a2 \right)}$.}
\label{fig:Figure6}
\end{figure}

In addition to the Schwarzschild metric, \texttt{MetricTensor} also includes a small library of in-built metrics and spacetime geometries (with many more planned for future inclusion), including, but not limited to: the Reissner-Nordstr\"om metric\cite{reissner}\cite{nordstrom}, independently discovered by Weyl\cite{weyl} and Jeffery\cite{jeffery}, describing the exterior spacetime geometry surrounding a charged, spherically-symmetric and non-rotating mass distribution (e.g. a charged, non-rotating black hole) of mass $M$ and electric charge $Q$:

\begin{equation}
d s^2 = g_{\mu \nu} d x^{\mu} d x^{\nu} = - \left( 1 - \frac{2 M}{r} + \frac{Q^2}{4 \pi r^2} \right) d t^2 + \left( 1 - \frac{2 M}{r} + \frac{Q^2}{4 \pi r^2} \right)^{-1} d r^2 + r^2 \left( d \theta^2 + \sin^2 \left( \theta \right) d \phi^2 \right),
\end{equation}
in Schwarzschild/spherical polar coordinates ${\left( t, r, \theta, \phi \right)}$, as shown in Figure \ref{fig:Figure3}; the Kerr\cite{kerr} and Kerr-Newman\cite{newman} metrics describing the exterior spacetime geometries surrounding an (un)charged, axially-symmetric, and rotating compact mass distribution (e.g. an (un)charged, spinning black hole) of mass $M$ and angular momentum $J$ for the case of the Kerr metric:

\begin{multline}
d s^2 = g_{\mu \nu} d x^{\mu} d x^{\nu} = - \left( 1 - \frac{2 M}{\left( r^2 + \left( \frac{J}{M} \right)^2 \cos^2 \left( \theta \right) \right)} \right) d t^2 + \left( \frac{r^2 + \left( \frac{J}{M} \right)^2 \cos^2 \left( \theta \right)}{r^2 - 2 M + \left( \frac{J}{M} \right)^2} \right) d r^2\\
+ \left( r^2 + \left( \frac{J}{M} \right)^2 \cos^2 \left( \theta \right) \right) d \theta^2 + \left( r^2 + \left( \frac{J}{M} \right)^2 + \frac{2 J^2 \sin^2 \left( \theta \right)}{M \left( r^2 + \left( \frac{J}{M} \right)^2 \cos^2 \left( \theta \right) \right)} \right) \sin^2 \left( \theta \right) d \phi^2\\
- \left( \frac{4 J \sin^2 \left( \theta \right)}{r^2 + \left( \frac{J}{M} \right)^2 \cos^2 \left( \theta \right)} \right) d t d \phi,
\end{multline}
and with an additional electric charge parameter $Q$ for the case of the Kerr-Newman metric:

\begin{multline}
d s^2 = g_{\mu \nu} d x^{\mu} d x^{\nu} = - \left( \frac{d r^2}{r^2 - 2 M + \left( \frac{J}{M} \right)^2 + \frac{Q^2}{4 \pi}} + d \theta^2 \right) \left( r^2 + \left( \frac{J}{M} \right)^2 \cos^2 \left( \theta \right) \right)\\
+ \left( d t - \left( \frac{J}{M} \right) \sin^2 \left( \theta \right) d \phi \right)^2 \left( \frac{r^2 - 2 M + \left( \frac{J}{M} \right)^2 + \frac{Q^2}{4 \pi}}{r^2 + \left( \frac{J}{M} \right)^2 \cos^2 \left( \theta \right)} \right)\\
- \left( \left( r^2 + \left( \frac{J}{M} \right)^2 \right) d \phi - \left( \frac{J}{M} \right) d t \right)^2 \left( \frac{\sin^2 \left( \theta \right)}{r^2 + \left( \frac{J}{M} \right)^2 \cos^2 \left( \theta \right)} \right),
\end{multline}
in Boyer-Lindquist/oblate spheroidal coordinates\cite{boyer} ${\left( t, r, \theta, \phi \right)}$, as shown in Figure \ref{fig:Figure4}; and the Friedmann-Lema\^itre-Robertson-Walker/FLRW metric\cite{friedmann}\cite{lemaitre}\cite{robertson}\cite{walker} describing the overall spacetime geometry of a homogeneous and isotropic universe which is either uniformly expanding or uniformly contracting with global curvature $k$ and scale factor ${a \left( t \right)}$:

\begin{equation}
d s^2 = g_{\mu \nu} d x^{\mu} d x^{\nu} = - d t^2 + a^2 \left( t \right) \left( \frac{d r^2}{1 - k r^2} + r^2 \left( d \theta^2 + \sin^2 \left( \theta \right) d \phi^2 \right) \right),
\end{equation}
in spherical polar coordinates ${\left( t, r, \theta, \phi \right)}$, or the G\"odel metric\cite{godel} describing the overall spacetime geometry of a rotating, dust-filled universe with global angular velocity ${\omega}$:

\begin{equation}
d s^2 = g_{\mu \nu} d x^{\mu} d x^{\nu} = \frac{1}{2 \omega^2} \left( - \left( d t^2 + e^x d y \right)^2 + \frac{1}{2} e^{2 x} d y^2 + d z^2 \right),
\end{equation}
in G\"odel's Cartesian-like coordinates ${\left( t, x, y, z \right)}$, as shown in Figure \ref{fig:Figure5}. By default, appropriate formal symbols are assigned to the various parameters of the metrics (e.g. mass $M$ or angular momentum $J$), as well as to the coordinates (e.g. time coordinate $t$ and radial coordinate $r$), although these defaults can easily be overridden using additional arguments, as shown in Figure \ref{fig:Figure6}.

\begin{figure}[ht]
\centering
\begin{framed}
\includegraphics[width=0.495\textwidth]{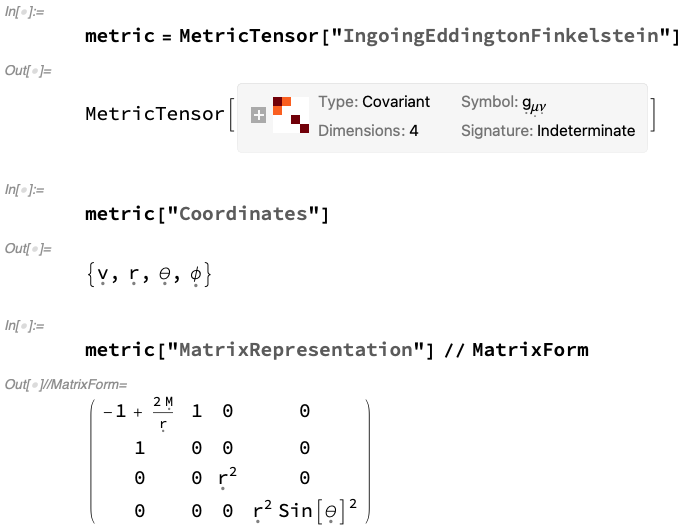}
\vrule
\includegraphics[width=0.495\textwidth]{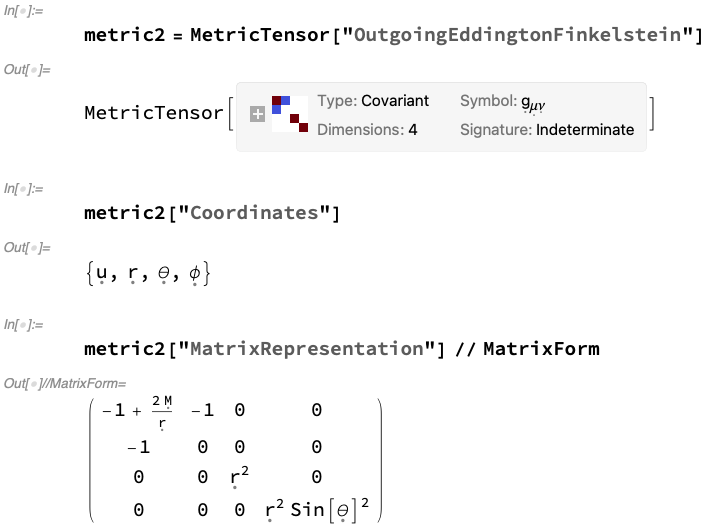}
\end{framed}
\caption{On the left, the \texttt{MetricTensor} object for a Schwarzschild geometry (representing e.g. an uncharged, non-rotating black hole of mass $M$) in ingoing Eddington-Finkelstein coordinates ${\left( v, r, \theta, \phi \right)}$, i.e. the coordinate system adapted to inward-traveling radial lightlike geodesics. On the right, the \texttt{MetricTensor} object for a Schwarzschild geometry (representing e.g. an uncharged, non-rotating black hole of mass $M$) in outgoing Eddington-Finkelstein coordinates ${\left( u, r, \theta, \phi \right)}$, i.e. the coordinate system adapted to outward-traveling radial lightlike geodesics.}
\label{fig:Figure7}
\end{figure}

\begin{figure}[ht]
\centering
\begin{framed}
\includegraphics[width=0.495\textwidth]{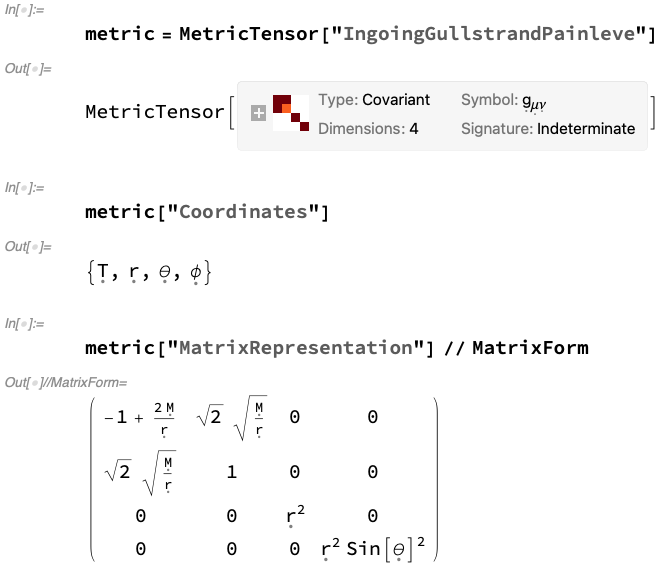}
\vrule
\includegraphics[width=0.495\textwidth]{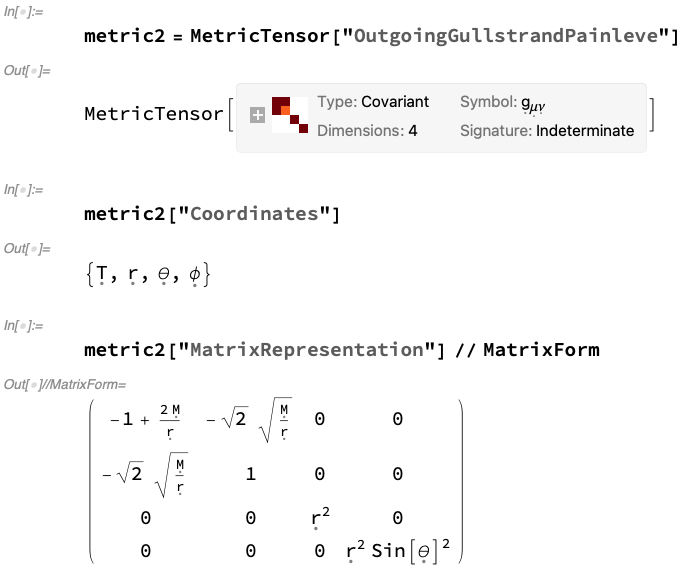}
\end{framed}
\caption{On the left, the \texttt{MetricTensor} object for a Schwarzschild geometry (representing e.g. an uncharged, non-rotating black hole of mass $M$) in ingoing Gullstrand-Painlev\'e coordinates ${\left( T, r, \theta, \phi \right)}$, i.e. the coordinate system whose time coordinate follows the proper time of a free-falling observer, falling inwards from infinity. On the right, the \texttt{MetricTensor} object for a Schwarzschild geometry (representing e.g. an uncharged, non-rotating black hole of mass $M$) in outgoing Gullstrand-Painlev\'e coordinates ${\left( T, r, \theta, \phi \right)}$, i.e. the coordinate system whose time coordinate follows the proper time of a free-falling observer, falling outwards to infinity.}
\label{fig:Figure8}
\end{figure}

\begin{figure}[ht]
\centering
\begin{framed}
\includegraphics[width=0.395\textwidth]{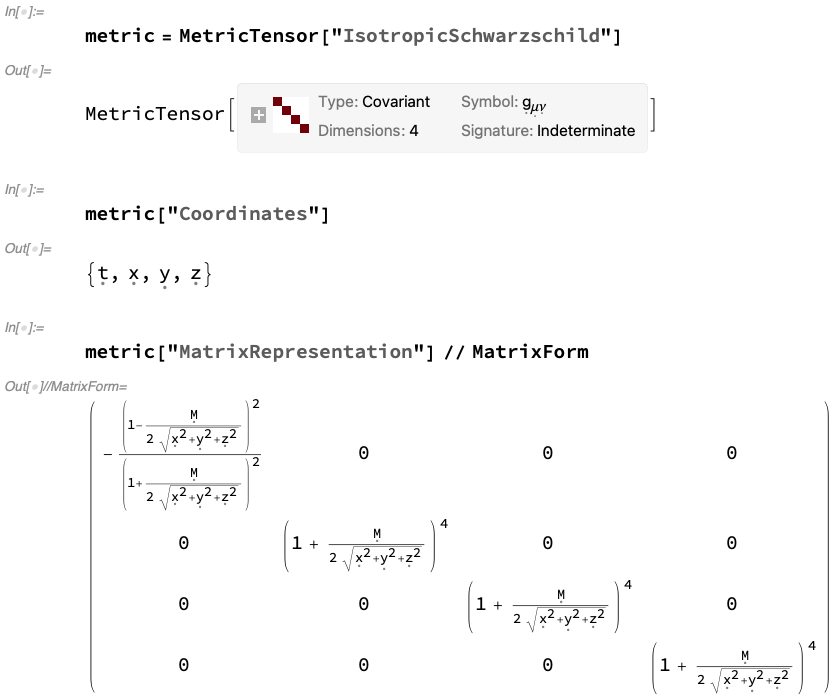}
\vrule
\includegraphics[width=0.595\textwidth]{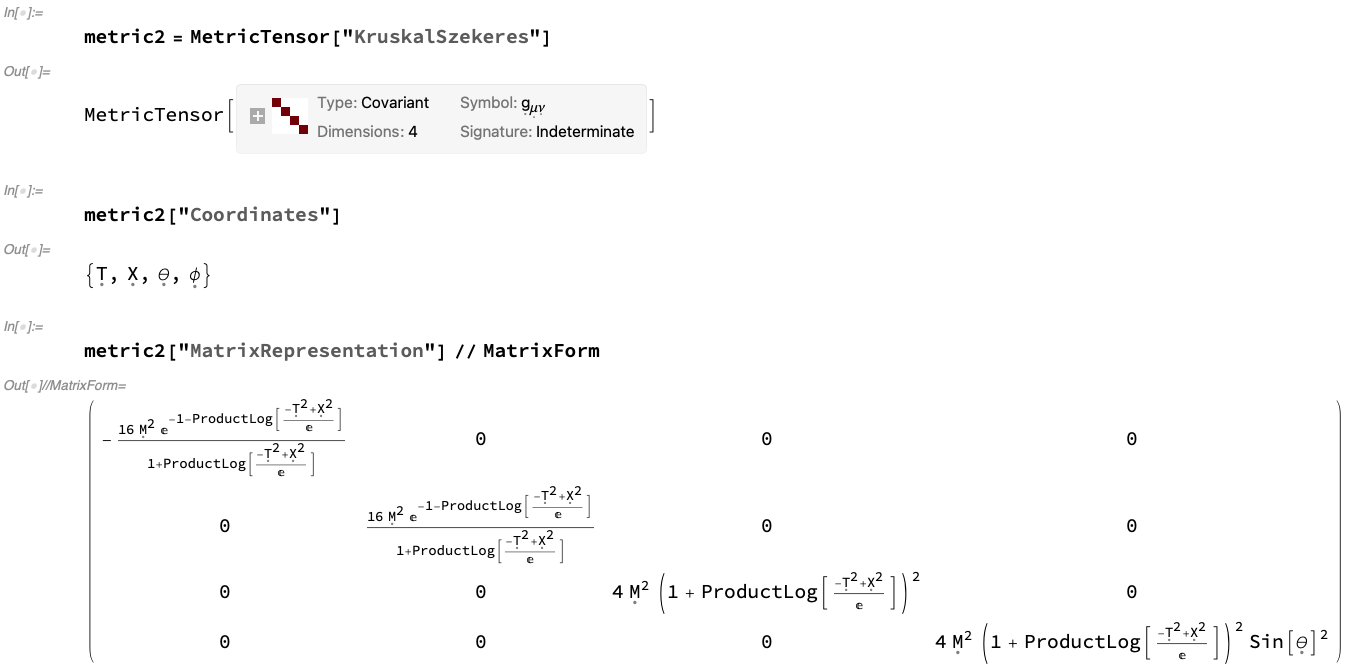}
\end{framed}
\caption{On the left, the \texttt{MetricTensor} object for a Schwarzschild geometry (representing e.g. an uncharged, non-rotating black hole of mass $M$) in isotropic coordinates ${\left( t, x, y, z \right)}$, i.e. the Cartesian-like coordinate system in which radial distances are defined in such a way that all light cones appear round on constant-time slices. On the right, the \texttt{MetricTensor} object for a Schwarzschild geometry (representing e.g. an uncharged, non-rotating black hole of mass $M$) in Kruskal-Szekeres coordinates ${\left( T, X, \theta, \phi \right)}$, i.e. the coordinate system that covers the entire spacetime manifold of the maximal analytic extension of the Schwarzschild solution.}
\label{fig:Figure9}
\end{figure}

Although all of the above geometries have used default choices of coordinate systems (e.g. Schwarzschild/spherical polar coordinates for the Schwarzschild geometry or Boyer-Lindquist/oblate spheroidal coordinates for the Kerr geometry), \texttt{MetricTensor} also allows for the handling of, and transformation between, arbitrary curvilinear coordinate systems. Recall that, in general, when applying a generic coordinate transformation ${x^{\mu} \to \widetilde{x^{\mu}}}$ to a (pseudo-)Riemannian manifold, the components of the metric tensor ${g_{\mu \nu}}$ (when represented in covariant matrix form) transform as:

\begin{equation}
\widetilde{g_{\mu \nu}} = \left( \frac{\partial x^{\rho}}{\partial \widetilde{x^{\mu}}} \right) \left( \frac{\partial x^{\sigma}}{\partial \widetilde{x^{\nu}}} \right) g_{\rho \sigma}.
\end{equation}
For instance, we can choose to represent the Schwarzschild geometry (e.g. for an uncharged, non-rotating black hole of mass $M$) shown above in terms of the ingoing Eddington-Finkelstein coordinates\cite{eddington}\cite{finkelstein}\cite{penrose3} ${\left( v, r, \theta, \phi \right)}$:

\begin{equation}
d s^2 = g_{\mu \nu} d x^{\mu} d x^{\nu} = - \left( 1 - \frac{2 M}{r} \right) d v^2 + 2 d v d r + r^2 \left( d \theta^2 + \sin^2 \left( \theta \right) d \phi^2 \right),
\end{equation}
or the outgoing Eddington-Finkelstein coordinates ${\left( u, r, \theta, \phi \right)}$:

\begin{equation}
d s ^2 = g_{\mu \nu} d x^{\mu} d x^{\nu} = - \left( 1 - \frac{2 M}{r} \right) d u ^2 - 2 d u d r + r^2 \left( d \theta^2 + \sin^2 \left( \theta \right) d \phi^2 \right),
\end{equation}
which are adapted to inward-traveling and outward-traveling radial lightlike geodesics, respectively, where we have introduced the new time coordinates ${v = t + r^{*}}$ and ${u = t - r^{*}}$, and where the so-called \textit{tortoise coordinate} ${r^{*}}$ is defined so as to satisfy the differential equation:

\begin{equation}
\frac{d r^{*}}{d r} = \left( 1 - \frac{2 M}{r} \right)^{-1}, \qquad \text{ i.e. } \qquad r^{*} = r + 2 M \log \left( \left\lvert \frac{r}{2 M} - 1 \right\rvert \right),
\end{equation}
as shown in Figure \ref{fig:Figure7}. We could equally choose the ingoing:

\begin{equation}
d s^2 = g_{\mu \nu} d x^{\mu} d x^{\nu} = - \left( 1 - \frac{2 M}{r} \right) d T^2 + 2 \sqrt{\frac{2 M}{r}} d T d r + d r^2 + r^2 \left( d \theta^2 + \sin^2 \left( \theta \right) d \phi^2 \right),
\end{equation}
or outgoing:

\begin{equation}
d s^2 = g_{\mu \nu} d x^{\mu} d x^{\nu} = - \left( 1 - \frac{2 M}{r} \right) d T^2 - 2 \sqrt{\frac{2 M}{r}} d T d r + d r^2 + r^2 \left( d \theta^2 + \sin^2 \left( \theta \right) d \phi^2 \right),
\end{equation}
Gullstrand-Painlev\'e coordinates\cite{painleve}\cite{gullstrand} ${\left( T, r, \theta, \phi \right)}$, whose time coordinates $T$ follow the proper times of free-falling observers falling inwards from infinity, and falling outwards to infinity, respectively, and where we have introduced these new time coordinates by means of the following integral:

\begin{equation}
T = t \pm \int \frac{\sqrt{\frac{2 M}{r}}}{1 - \frac{2 M}{r}} d r = t \pm 2 M \left( 2 \left( \sqrt{\frac{r}{2 M}} \right) - \log \left( \frac{\sqrt{\frac{r}{2 M}} + 1}{\sqrt{\frac{r}{2 M}} - 1} \right) \right),
\end{equation}
as shown in Figure \ref{fig:Figure8}. We could also choose to use the isotropic coordinate system ${\left( t, x, y, z \right)}$:

\begin{equation}
d s^2 = g_{\mu \nu} d x^{\mu} d x^{\nu} = - \frac{\left( 1 - \frac{M}{2 \sqrt{x^2 + y^2 + z^2}} \right)^2}{\left( 1 + \frac{M}{2 \sqrt{x^2 + y^2 + z^2}} \right)^2} d t^2 + \left( 1 + \frac{M}{2 \sqrt{x^2 + y^2 + z^2}} \right)^4 \left( d x^2 + d y^2 + d z^2 \right),
\end{equation}
namely the Cartesian-like coordinate system in which radial distances ${\sqrt{x^2 + y^2 + z^2}}$ are defined in such a way that all light cones appear round (at least on all constant-time slices); alternatively, we could choose to use the Kruskal-Szekeres coordinates\cite{kruskal}\cite{szekeres} ${\left( T, X, \theta, \phi \right)}$:

\begin{equation}
d s ^2 = g_{\mu \nu} d x^{\mu} d x^{\nu} = \frac{32 M^3}{r} \exp \left(- \frac{r}{2 M} \right) \left( - d T^2 + d X^2 \right) + r^2 \left( d \theta^2 \sin^2 \left( \theta \right) d \phi^2 \right),
\end{equation}
namely the coordinate system that naturally parametrizes the entire spacetime manifold of the maximal analytic extension of the Schwarzschild solution, where we have introduced a new timelike coordinate $T$ and a new spacelike coordinate $X$, defined by:

\begin{equation}
T = \left( \sqrt{\frac{r}{2 M} - 1} \right) \exp \left( \frac{r}{4 M} \right) \sinh \left( \frac{t}{4 M} \right), \qquad \text{ and } \qquad X = \left( \sqrt{\frac{r}{2 M} - 1} \right) \exp \left( \frac{r}{4 M} \right) \cosh \left( \frac{t}{4 M} \right),
\end{equation}
respectively, in the region ${r > 2 M}$ outside the event horizon, and:

\begin{equation}
T = \left( \sqrt{\frac{r}{2 M} - 1} \right) \exp \left( \frac{r}{4 M} \right) \cosh \left( \frac{t}{4 M} \right), \qquad \text{ and } \qquad X = \left( \sqrt{\frac{r}{2 M} - 1} \right) \exp \left( \frac{r}{4 M} \right) \sinh \left( \frac{t}{4 M} \right),
\end{equation}
respectively, in the region ${0 < r \leq 2 M}$ inside the event horizon, and where the radial coordinate $r$ is defined so as to be the unique solution to the exponential equation:

\begin{equation}
T^2 - X^2 = \left( 1 - \frac{r}{2 M} \right) \exp \left( \frac{r}{2 M} \right), \qquad \text{ where } \qquad T^2 - X^2 < 1,
\end{equation}
namely:

\begin{equation}
r = 2 M \left( 1 + W_0 \left( \frac{X^2 - T^2}{e} \right) \right),
\end{equation}
where ${W_0}$ denotes the Lambert $W$ function, as shown in Figure \ref{fig:Figure9}.

\begin{figure}[ht]
\centering
\begin{framed}
\includegraphics[width=0.395\textwidth]{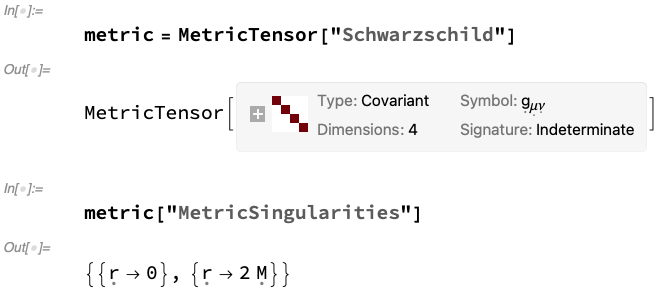}
\vrule
\includegraphics[width=0.595\textwidth]{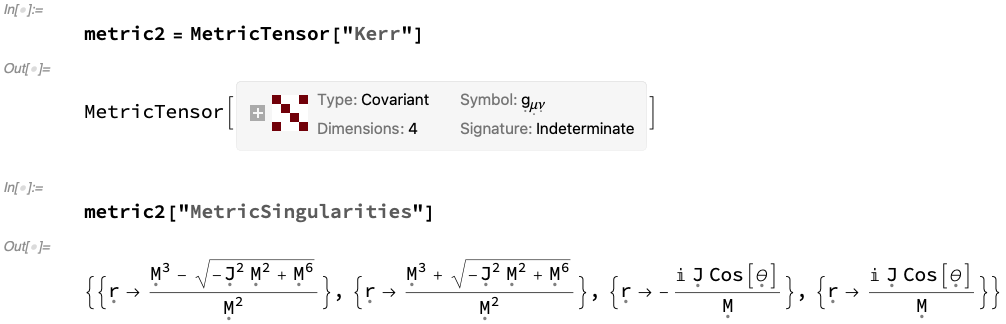}
\end{framed}
\caption{On the left, the list of coordinate values that cause the \texttt{MetricTensor} object for a Schwarzschild geometry (representing e.g. an uncharged, non-rotating black hole of mass $M$ in Schwarzschild/spherical polar coordinates ${\left( t, r, \theta, \phi \right)}$) to become singular. On the right, the list of coordinate values that cause the \texttt{MetricTensor} object for a Kerr geometry (representing e.g. an uncharged, spinning black hole of mass $M$ and angular momentum $J$ in Boyer-Lindquist/oblate spheroidal coordinates ${\left( t, r, \theta, \phi \right)}$) to become singular.}
\label{fig:Figure10}
\end{figure}

\begin{figure}[ht]
\centering
\begin{framed}
\includegraphics[width=0.495\textwidth]{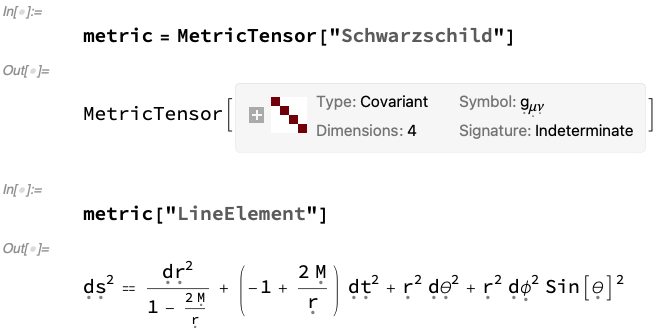}
\vrule
\includegraphics[width=0.495\textwidth]{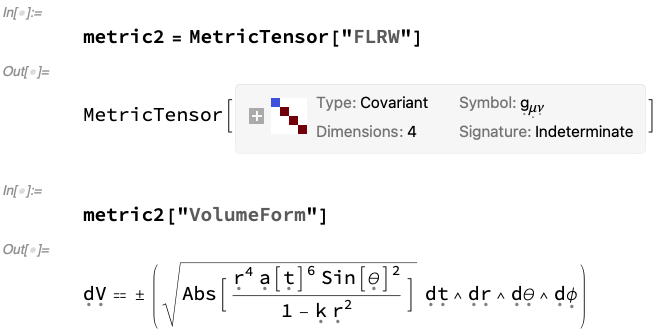}
\end{framed}
\caption{On the left, the line element representation, i.e. the algebraic relationship between the proper time/first fundamental form ${d s^2}$ and the differential 1-form symbols for the coordinates, of the \texttt{MetricTensor} object for a Schwarzschild geometry (representing e.g. an uncharged, non-rotating black hole of mass $M$ in Schwarzschild/spherical polar coordinates ${\left( t, r, \theta, \phi \right)}$). On the right, the volume form representation, i.e. the algebraic relationship between the infinitesimal volume element/metric volume form ${d V}$ and the determinant of the metric, of the \texttt{MetricTensor} object for an FLRW geometry (representing e.g. a homogeneous, isotropic and uniformly expanding/contracting universe with global curvature $k$ and scale factor ${a \left( t \right)}$ in spherical polar coordinates ${\left( t, r, \theta, \phi \right)}$).}
\label{fig:Figure11}
\end{figure}

\begin{figure}[ht]
\centering
\begin{framed}
\includegraphics[width=0.545\textwidth]{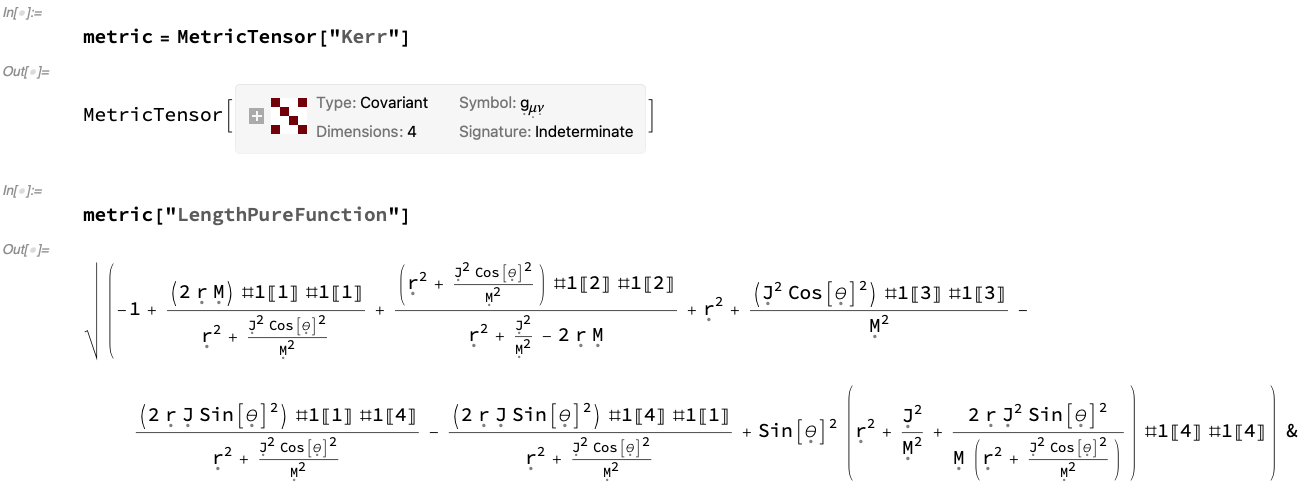}
\vrule
\includegraphics[width=0.445\textwidth]{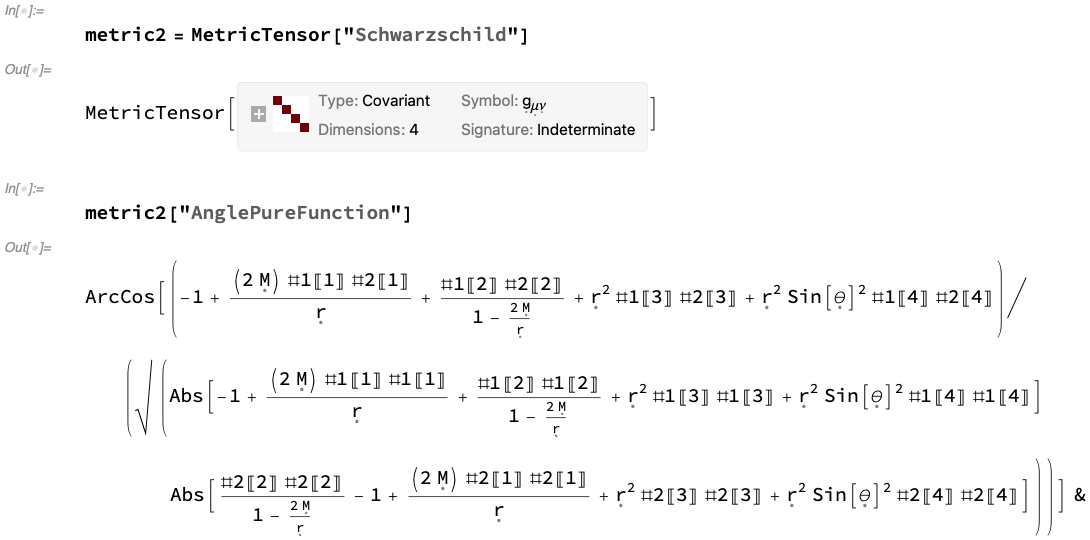}
\end{framed}
\caption{On the left, a pure function for determining the length of a given tangent vector using the \texttt{MetricTensor} object for a Kerr geometry (representing e.g. an uncharged, spinning black hole of mass $M$ and angular momentum $J$ in Boyer-Lindquist/oblate spheroidal coordinates ${\left( t, r, \theta, \phi \right)}$). On the right, a pure function for determining the angle between two given tangent vectors using the \texttt{MetricTensor} object for a Schwarzschild geometry (representing e.g. an uncharged, non-rotating black hole of mass $M$ in Schwarzschild/spherical polar coordinates ${\left( t, r, \theta, \phi \right)}$).}
\label{fig:Figure12}
\end{figure}

\begin{figure}[ht]
\centering
\begin{framed}
\includegraphics[width=0.495\textwidth]{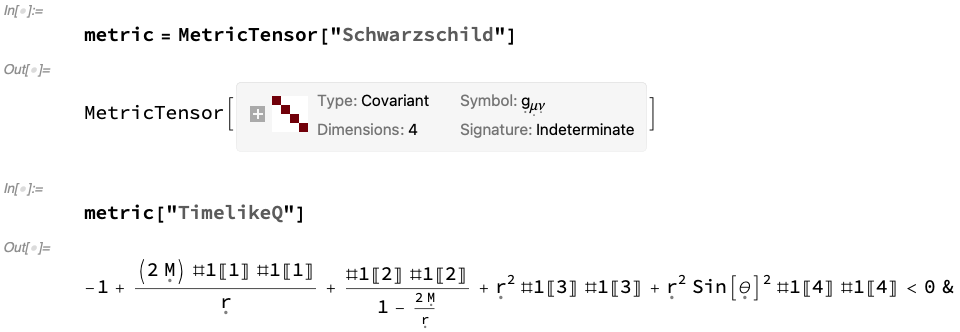}
\vrule
\includegraphics[width=0.495\textwidth]{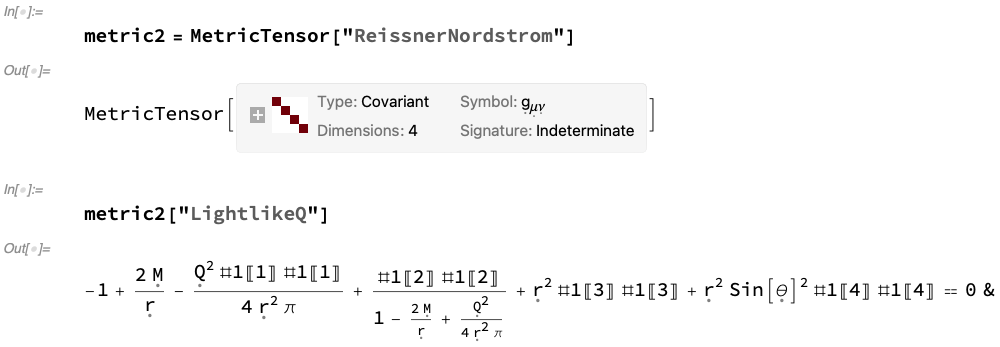}
\end{framed}
\caption{On the left, a pure function for determining whether a given tangent vector is timelike, i.e. whether it lies strictly on the interior of the light cone at ${\mathbf{x} \in \mathcal{M}}$, using the \texttt{MetricTensor} object for a Schwarzschild geometry (representing e.g. an uncharged, non-rotating black hole of mass $M$ in Schwarzschild/spherical polar coordinates ${\left( t, r, \theta, \phi \right)}$).  On the right, a pure function for determining whether a given tangent vector is lightlike, i.e. whether it lies on the boundary of the light cone at ${\mathbf{x} \in \mathcal{M}}$, using the \texttt{MetricTensor} object for a Reissner-Nordstr\"om geometry (representing e.g. a charged, non-rotating black hole of mass $M$ and electric charge $Q$ in Schwarzschild/spherical polar coordinates ${\left( t, r, \theta, \phi \right)}$).}
\label{fig:Figure13}
\end{figure}

\begin{figure}[ht]
\centering
\begin{framed}
\includegraphics[width=0.545\textwidth]{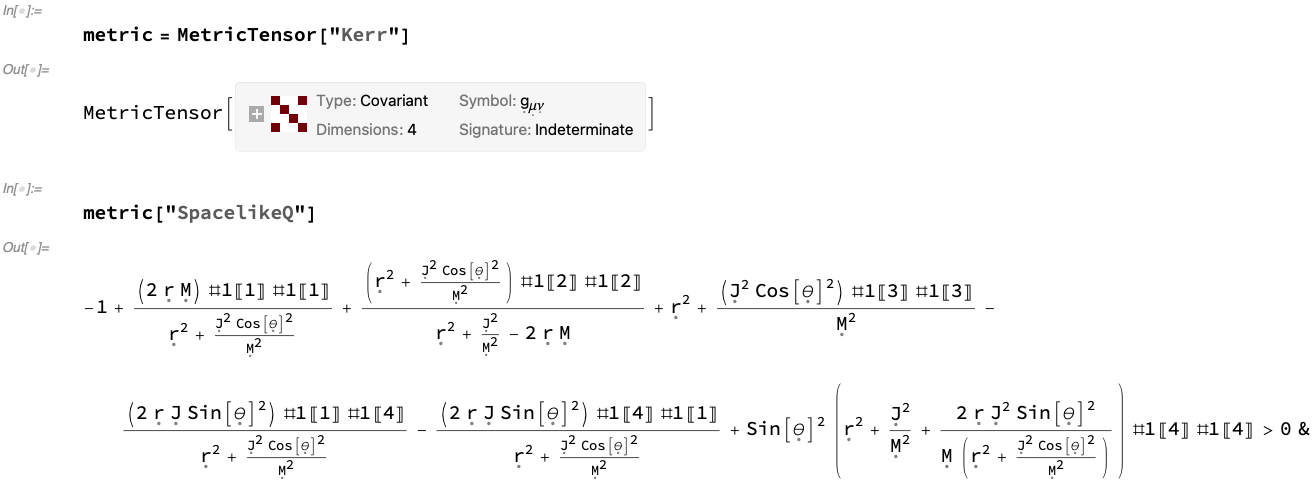}
\vrule
\includegraphics[width=0.445\textwidth]{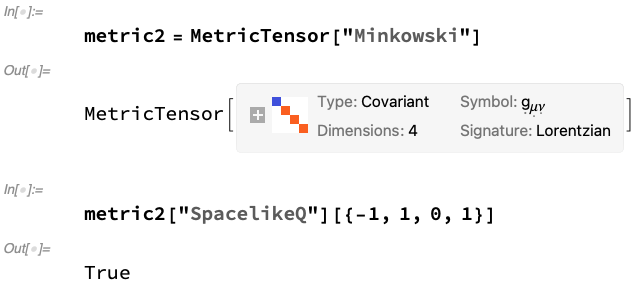}
\end{framed}
\caption{On the left, a pure function for determining whether a given tangent vector is spacelike, i.e. whether it lies strictly on the exterior of the light cone at ${\mathbf{x} \in \mathcal{M}}$, using the \texttt{MetricTensor} object for a Kerr geometry (representing e.g. an uncharged, spinning black hole of mass $M$ and angular momentum $J$ in Boyer-Lindquist/oblate spheroidal coordinates ${\left( t, r, \theta, \phi \right)}$). On the right, an illustration that the tangent vector ${\left( -1, 1, 0, 1 \right)}$ is spacelike within a Minkowski geometry (representing a flat spacetime in Cartesian coordinates ${\left( t, x^1, x^2, x^3 \right)}$).}
\label{fig:Figure14}
\end{figure}

\begin{figure}[ht]
\centering
\begin{framed}
\includegraphics[width=0.595\textwidth]{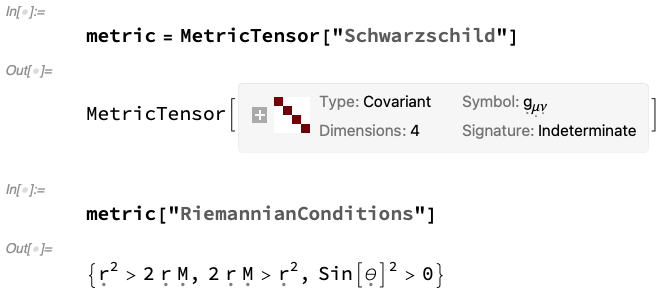}
\vrule
\includegraphics[width=0.395\textwidth]{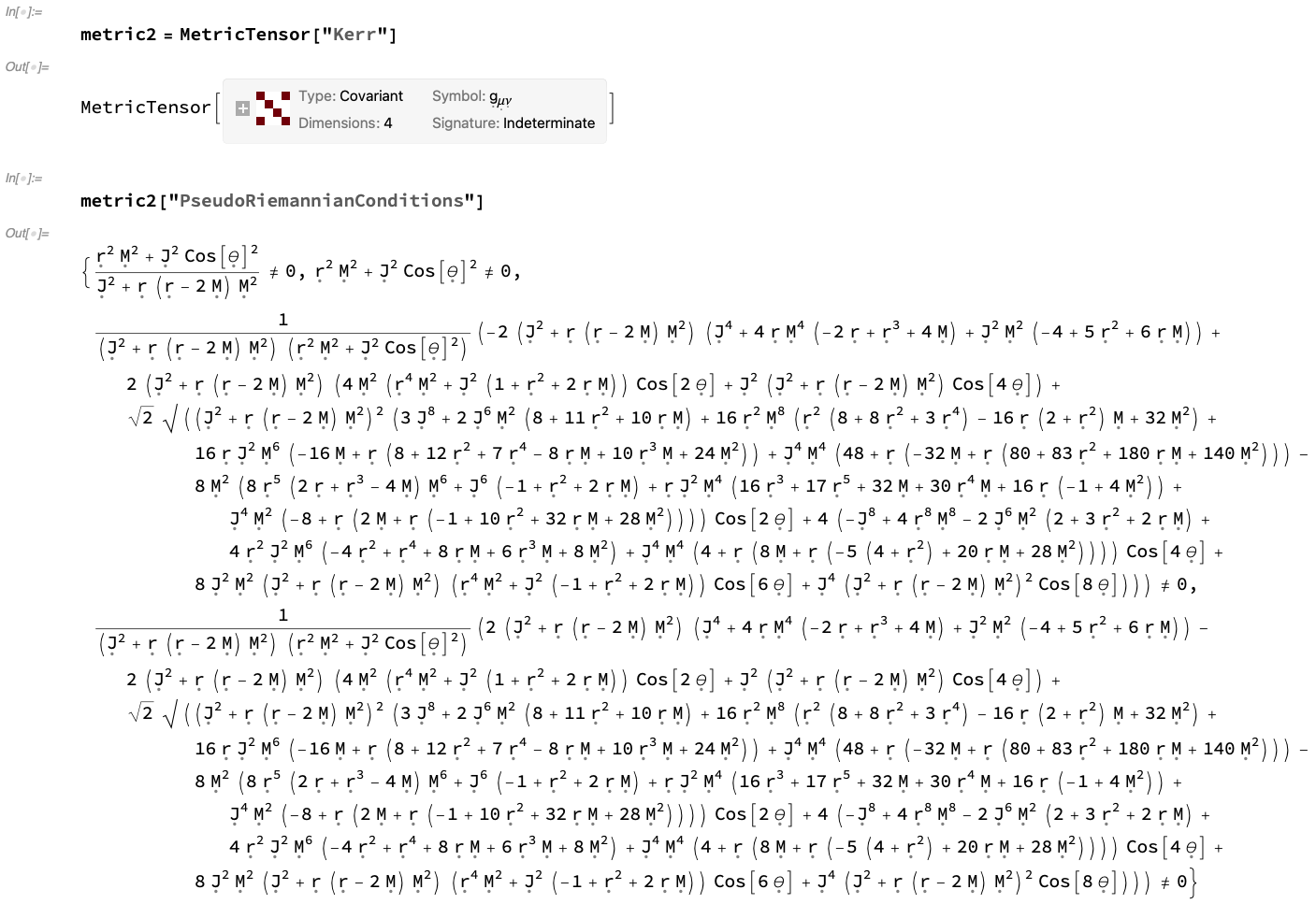}
\end{framed}
\caption{On the left, the list of coordinate conditions required to guarantee that the manifold described by the \texttt{MetricTensor} object for a Schwarzschild geometry (representing e.g. an uncharged, non-rotating black hole of mass $M$ in Schwarzschild/spherical polar coordinates ${\left( t, r, \theta, \phi \right)}$) is Riemannian, i.e. all eigenvalues are positive. On the right, the list of coordinate conditions required to guarantee that the manifold described by the \texttt{MetricTensor} object for a Kerr geometry (representing e.g. an uncharged, spinning black hole of mass $M$ and angular momentum $J$ in Boyer-Lindquist/oblate spheroidal coordinates ${\left( t, r, \theta, \phi \right)}$) is pseudo-Riemannian, i.e. all eigenvalues are non-zero.}
\label{fig:Figure15}
\end{figure}

\begin{figure}[ht]
\centering
\begin{framed}
\includegraphics[width=0.495\textwidth]{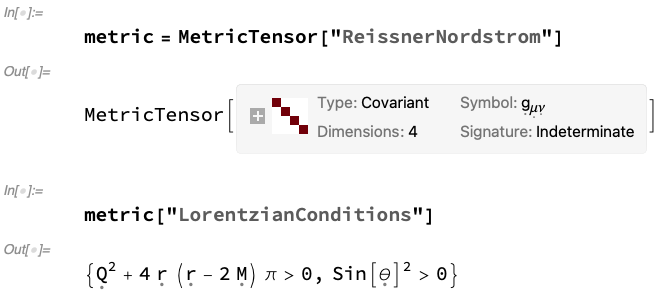}
\vrule
\includegraphics[width=0.495\textwidth]{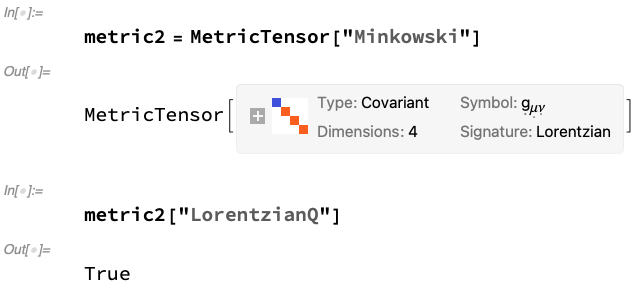}
\end{framed}
\caption{On the left, the list of coordinate conditions required to guarantee that the manifold described by the \texttt{MetricTensor} object for a Reissner-Nordstr\"om geometry (representing e.g. a charged, non-rotating black hole of mass $M$ and electric charge $Q$ in Schwarzschild/spherical polar coordinates ${\left( t, r, \theta, \phi \right)}$) is Lorentzian, i.e. the ``time'' eigenvalue is negative and all other eigenvalues are positive. On the right, an illustration that the Minkowski geometry (representing a flat spacetime in Cartesian coordinates ${\left( t, x^1, x^2, x^3 \right)}$) is Lorentzian.}
\label{fig:Figure16}
\end{figure}

A wide range of different geometrical properties of a manifold, including its singularity structure, can be computed directly from its symbolic representation as a \texttt{MetricTensor} object. For instance, the Schwarzschild metric in Schwarzschild/spherical polar coordinates is known to become singular at coordinate values ${r = 0}$ (i.e. the ``physical'' singularity at the origin of the spherical geometry) and ${r = 2 M}$ (i.e. the ``coordinate'' singularity at the event horizon), whereas the Kerr metric in Boyer-Lindquist/oblate spheroidal coordinates admits coordinate singularities at both its interior and exterior horizons ${r_{H}^{\pm}}$:

\begin{equation}
r_{H}^{\pm} = M \pm \sqrt{M^2 - \left( \frac{J}{M} \right)^2},
\end{equation}
and at the interior and exterior boundaries of its ergosphere ${r_{E}^{\pm}}$:

\begin{equation}
r_{E}^{\pm} = M \pm \sqrt{M^2 - \left( \frac{J}{M} \right)^2 \cos^2 \left( \theta \right)},
\end{equation}
as shown in Figure \ref{fig:Figure10}. The \texttt{MetricTensor} object in the local coordinate basis ${\left\lbrace x^{\mu} \right\rbrace}$ can also be represented purely in terms of its line element/first fundamental form (typically denoted ${d s^2}$), or in terms of its volume element/metric volume form (typically denoted ${d V}$ or ${\omega}$), namely:

\begin{equation}
d s^2 = g_{\mu \nu} d x^{\mu} d x^{\nu}, \qquad \text{ or } \qquad d V = \omega = \sqrt{\det \left( g_{\mu \nu} \right)} d x^{1} \wedge d x^{2} \wedge \cdots \wedge d x^{3},
\end{equation}
where ${\wedge}$ here denotes the exterior/wedge product in the corresponding Grassmann algebra of differential forms, as shown in Figure \ref{fig:Figure11}. Note that, in the above (and henceforth), we use the notation ${\det \left( g_{\mu \nu} \right)}$ to indicate that we are taking a determinant of the metric tensor, represented as an explicit covariant matrix: the indices ${\mu}$ and ${\nu}$ are to be thought of as being purely structural. Recalling that, if ${g_{\mathbf{x}}}$ denotes the value of the metric tensor field at a point ${\mathbf{x} \in \mathcal{M}}$ in the manifold, and ${\mathbf{u}, \mathbf{v} \in T_{\mathbf{x}} \mathcal{M}}$ denote tangent vectors at that point, then the length ${\left\lVert \mathbf{u} \right\rVert}$ of any such tangent vector, and the angle ${\theta \left( \mathbf{u}, \mathbf{v} \right)}$ between any such pair of tangent vectors, can be computed directly from the metric tensor as:

\begin{equation}
\forall \mathbf{u} \in T_{\mathbf{x}} \mathcal{M}, \qquad \left\lVert \mathbf{u} \right\rVert = \sqrt{g_{\mathbf{x}} \left( \mathbf{u}, \mathbf{u} \right)} = \sqrt{g_{\mu \nu} u^{\mu} u^{\nu}},
\end{equation}
and:

\begin{equation}
\forall \mathbf{u}, \mathbf{v} \in T_{\mathbf{x}} \mathcal{M}, \qquad \theta \left( \mathbf{u}, \mathbf{v} \right) = \arccos \left( \frac{g_{\mathbf{x}} \left( \mathbf{u}, \mathbf{v} \right)}{\sqrt{g_{\mathbf{x}} \left( \mathbf{u}, \mathbf{u} \right)} \sqrt{g_{\mathbf{x}} \left( \mathbf{v}, \mathbf{v} \right)}} \right) = \arccos \left( \frac{g_{\mu \nu} u^{\mu} v^{\nu}}{\sqrt{g_{\mu \nu} u^{\mu} u^{\nu}} \sqrt{g_{\mu \nu} v^{\mu} v^{\nu}}} \right),
\end{equation}
respectively, we note that \textsc{Gravitas} is able to synthesize symbolic Wolfram Language code automatically (in the form of bespoke pure functions) for computing such quantities directly from any symbolic \texttt{MetricTensor} object, as demonstrated in Figure \ref{fig:Figure12}. Similar pure functions can also be automatically synthesized for determining whether a given tangent vector ${\mathbf{u} \in T_{\mathbf{x}} \mathcal{M}}$ is \textit{timelike}, \textit{lightlike} or \textit{spacelike}, i.e. whether the quantity ${g_{\mathbf{x}} \left( \mathbf{u}, \mathbf{u} \right) = g_{\mu \nu} u^{\mu} u^{\nu}}$ is negative, identically zero, or positive, and therefore whether the tangent vector ${\mathbf{u} \in T_{\mathbf{x}} \mathcal{M}}$ lies on the interior, the boundary or the exterior of the corresponding light cone at ${\mathbf{x} \in \mathcal{M}}$, as illustrated in Figures \ref{fig:Figure13} and \ref{fig:Figure14}, along with an example application of one of these pure functions to the tangent vector ${\left( -1, 1, 0, 1 \right)}$ within a Minkowski geometry (representing a flat spacetime in Cartesian coordinates ${\left( t, x^1, x^2, x^3 \right)}$), demonstrating that it is indeed spacelike. Finally, based on the eigenvalues of the \texttt{MetricTensor} object (when represented as an explicit matrix in covariant form), the underlying manifold ${\mathcal{M}}$ may be classified as either Riemannian (i.e. all eigenvalues are positive, or more generally have the same sign), pseudo-Riemannian (i.e. all eigenvalues are non-zero, but not all necessarily have the same sign) or Lorentzian (i.e. all eigenvalues are positive/have the same sign, except for one eigenvalue corresponding to the ``time'' coordinate, which is negative/has the opposite sign), and appropriate conditions on the coordinates can be imposed so as to guarantee that a given local patch of the manifold satisfies any of these conditions, as shown in Figures \ref{fig:Figure15} and \ref{fig:Figure16}, along with an illustration that the Minkowski geometry is, indeed, Lorentzian.

Although, for the purposes of the present article, the majority of our attention will be focused upon named/in-built \texttt{MetricTensor} objects and standard coordinate systems, it is important to note that the \texttt{MetricTensor} function itself can accept any arbitrary symbolic matrix (in any number of dimensions) as input, defined with respect to any arbitrary symbolic coordinate system. Relatedly, all of the examples presented within this article will be defined with respect to geometries in four dimensions; however, the vast majority of the \textsc{Gravitas} functionality also works for geometries in arbitrary numbers of dimensions, and wherever this is not the case (e.g. wherever, for instance, the algebraic properties of the Hodge star operator ${\star}$ or the totally-antisymmetric Levi-Civita symbol ${\varepsilon_{\rho \sigma \mu \nu}}$ constrain certain operations to be definable only in four dimensions) we have been clear to indicate such limitations within the text.

\clearpage

\section{Tensor Calculus and Differential Geometry}
\label{sec:Section2}

Since our manifold ${\mathcal{M}}$ is, by hypothesis, differentiable, we are now in a position to be able to introduce an affine connection ${\nabla}$, i.e. a bilinear map of the general form\cite{jost}:

\begin{equation}
\nabla : \Gamma \left( \bigsqcup\limits_{\mathbf{x} \in \mathcal{M}} T_{\mathbf{x}} \mathcal{M} \right) \times \Gamma \left( \bigsqcup\limits_{\mathbf{x} \in \mathcal{M}} T_{\mathbf{x}} \mathcal{M} \right) \to \Gamma \left( \bigsqcup\limits_{\mathbf{x} \in \mathcal{M}} T_{\mathbf{x}} \mathcal{M} \right),
\end{equation}
connecting neighboring tangent spaces together, and more specifically sending:

\begin{equation}
\forall \mathbf{X}, \mathbf{Y} \in \Gamma \left( \bigsqcup\limits_{\mathbf{x} \in \mathcal{M}} T_{\mathbf{x}} \mathcal{M} \right), \qquad \left( \mathbf{X}, \mathbf{Y} \right) \mapsto \nabla_{\mathbf{X}} \mathbf{Y},
\end{equation}
where the disjoint union of tangent spaces ${\bigsqcup\limits_{\mathbf{x} \in \mathcal{M}} T_{\mathbf{x}} \mathcal{M}}$ simply denotes the tangent bundle on ${\mathcal{M}}$, and ${\Gamma \left( \bigsqcup\limits_{\mathbf{x} \in \mathcal{M}} T_{\mathbf{x}} \mathcal{M} \right)}$ denotes the space of smooth sections of that tangent bundle, i.e. the space of vector fields definable on ${\mathcal{M}}$ (thus making ${\mathbf{X}}$ and ${\mathbf{Y}}$ arbitrary vector fields on ${\mathcal{M}}$ within the above map). Introducing an affine connection ${\nabla}$ over the tangent bundle ${\bigsqcup\limits_{\mathbf{x} \in \mathcal{M}} T_{\mathbf{x}} \mathcal{M}}$ on ${\mathcal{M}}$ is equivalent to defining a notion of parallel transport, or a notion of covariant differentiation (i.e. a notion of differentiation along tangent vectors), over the underlying manifold ${\mathcal{M}}$. This affine connection specializes to become a (Riemannian) metric connection in the particular case where the covariant derivative of the metric tensor $g$ vanishes for every vector field ${\mathbf{X} \in \Gamma \left( \bigsqcup\limits_{\mathbf{x} \in \mathcal{M}} T_{\mathbf{x}} \mathcal{M} \right)}$:

\begin{equation}
\forall \mathbf{X} \in \Gamma \left( \bigsqcup\limits_{\mathbf{x} \in \mathcal{M}} T_{\mathbf{x}} \mathcal{M} \right), \qquad \nabla_{\mathbf{X}} g = 0,
\end{equation}
or, in a more explicit (component-based) form:

\begin{equation}
\forall \mathbf{X}, \mathbf{Y}, \mathbf{Z} \in \Gamma \left( \bigsqcup\limits_{\mathbf{x} \in \mathcal{M}} T_{\mathbf{x}} \mathcal{M} \right), \qquad \frac{\partial}{\partial X^{\rho}} \left( g_{\mu \nu} Y^{\mu} Z^{\nu} \right) = g_{\mu \nu} \left( \nabla_{\rho} Y^{\mu} \right) Z^{\nu} + g_{\mu \nu} Y^{\mu} \left( \nabla_{\rho} Z^{\nu} \right),
\end{equation}
where ${\nabla_{\rho}}$ designates covariant differentiation with respect to the component ${X^{\rho}}$, and therefore the metric is preserved under parallel transport (i.e. one has \textit{metric compatibility}). If the (Riemannian) metric connection is, moreover, torsion-free, i.e. if the (Cartan) torsion tensor\cite{cartan}\cite{cartan2} vanishes for all pairs of vector fields ${\mathbf{X}, \mathbf{Y} \in \Gamma \left( \bigsqcup\limits_{\mathbf{x} \in \mathcal{M}} T_{\mathbf{x}} \mathcal{M} \right)}$:

\begin{equation}
\forall \mathbf{X}, \mathbf{Y} \in \Gamma \left( \bigsqcup\limits_{\mathbf{x} \in \mathcal{M}} T_{\mathbf{x}} \mathcal{M} \right), \qquad \nabla_{\mathbf{X}} \mathbf{Y} - \nabla_{\mathbf{Y}} \mathbf{X} - \left[ \mathbf{X}, \mathbf{Y} \right] = \mathbf{0},
\end{equation}
where ${\left[ \mathbf{X}, \mathbf{Y} \right]}$ denotes the usual Lie bracket of vector fields, i.e. in explicit (component-based) form, using the local coordinate basis ${\left\lbrace x^{\mu} \right\rbrace}$:

\begin{equation}
\forall \mathbf{X}, \mathbf{Y} \in \Gamma \left( \bigsqcup\limits_{\mathbf{x} \in \mathcal{M}} T_{\mathbf{x}} \mathcal{M} \right), \qquad \left[ \mathbf{X}, \mathbf{Y} \right]^{\mu} = X^{\nu} \frac{\partial}{\partial x^{\nu}} Y^{\mu} - Y^{\nu} \frac{\partial}{\partial x^{\nu}} X^{\mu},
\end{equation}
then one obtains the Levi-Civita connection\cite{levicivita} (whose existence and uniqueness are both guaranteed by virtue of the fundamental theorem of Riemannian geometry\cite{helgason}).

\begin{figure}[ht]
\centering
\begin{framed}
\includegraphics[width=0.445\textwidth]{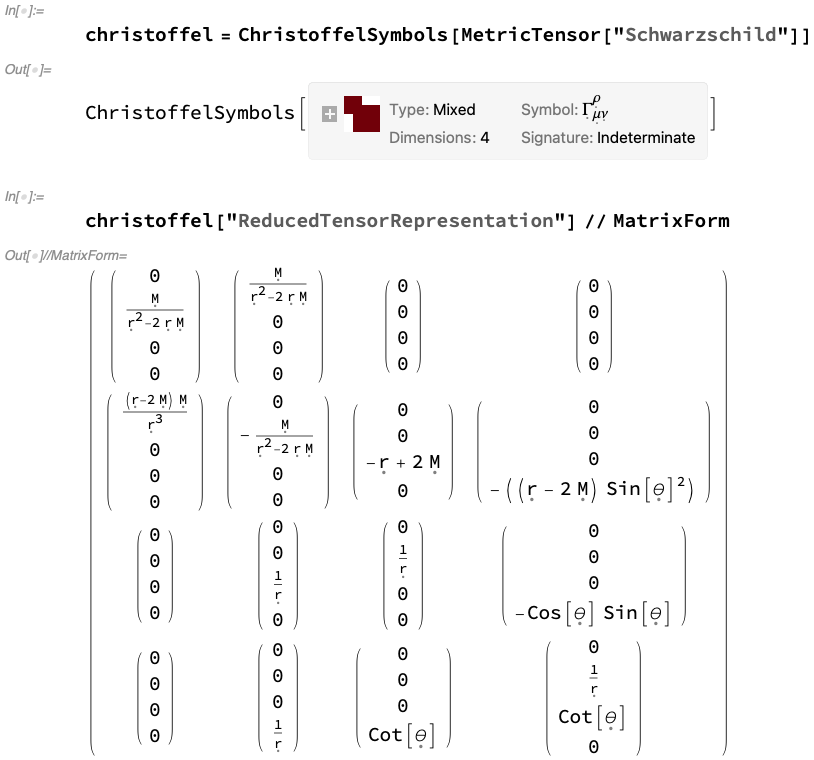}
\vrule
\includegraphics[width=0.545\textwidth]{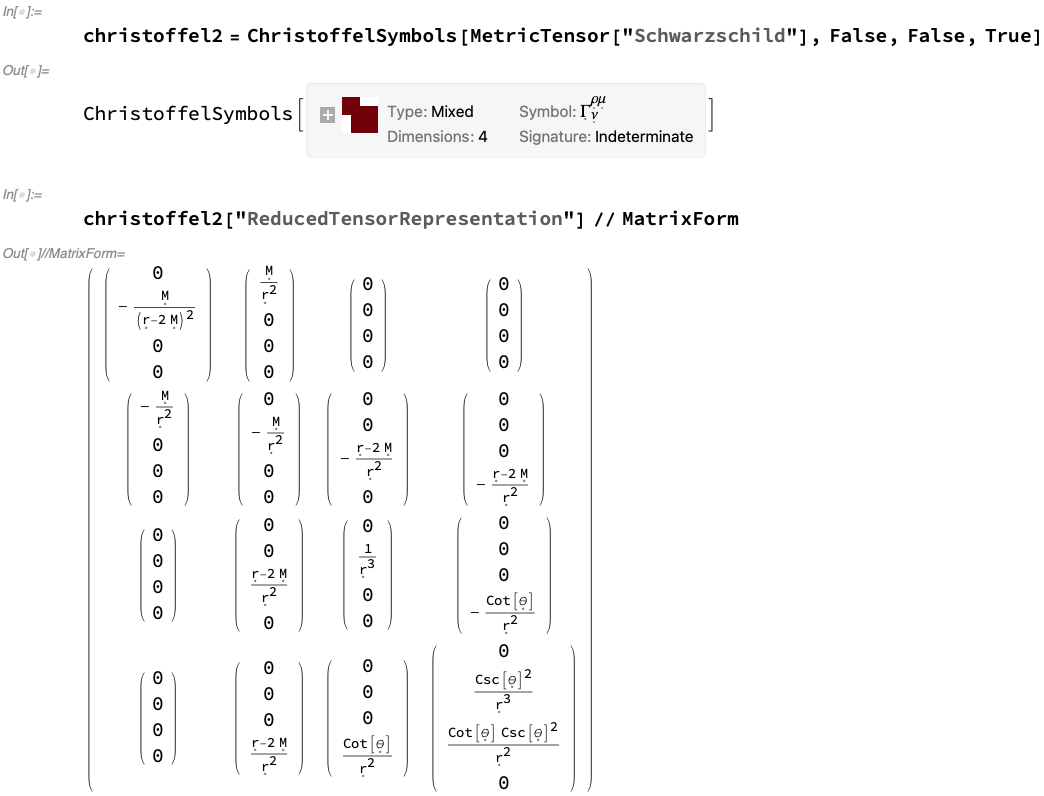}
\end{framed}
\caption{On the left, the \texttt{ChristoffelSymbols} object for a Schwarzschild geometry (representing e.g. an uncharged, non-rotating black hole of mass $M$ in Schwarzschild/spherical polar coordinates ${\left( t, r, \theta, \phi \right)}$) in explicit mixed-index array form, with the first index raised/contravariant and the latter two indices lowered/covariant (default). On the right, the \texttt{ChristoffelSymbols} object for a Schwarzschild geometry (representing e.g. an uncharged, non-rotating black hole of mass $M$ in Schwarzschild/spherical polar coordinates ${\left( t, r, \theta, \phi \right)}$) in explicit mixed-index array form, with the first two indices raised/contravariant and the latter index lowered/covariant.}
\label{fig:Figure17}
\end{figure}

\begin{figure}[ht]
\centering
\begin{framed}
\includegraphics[width=0.445\textwidth]{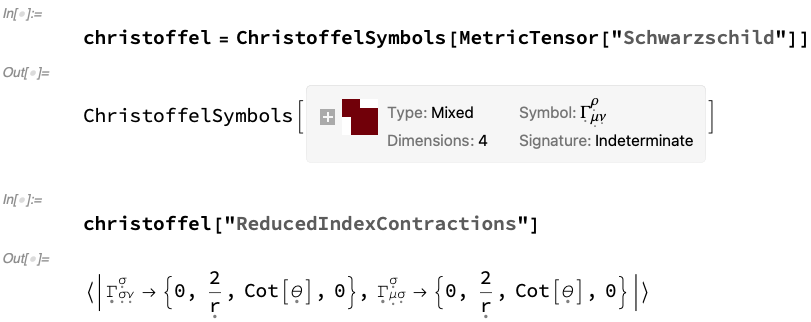}
\vrule
\includegraphics[width=0.545\textwidth]{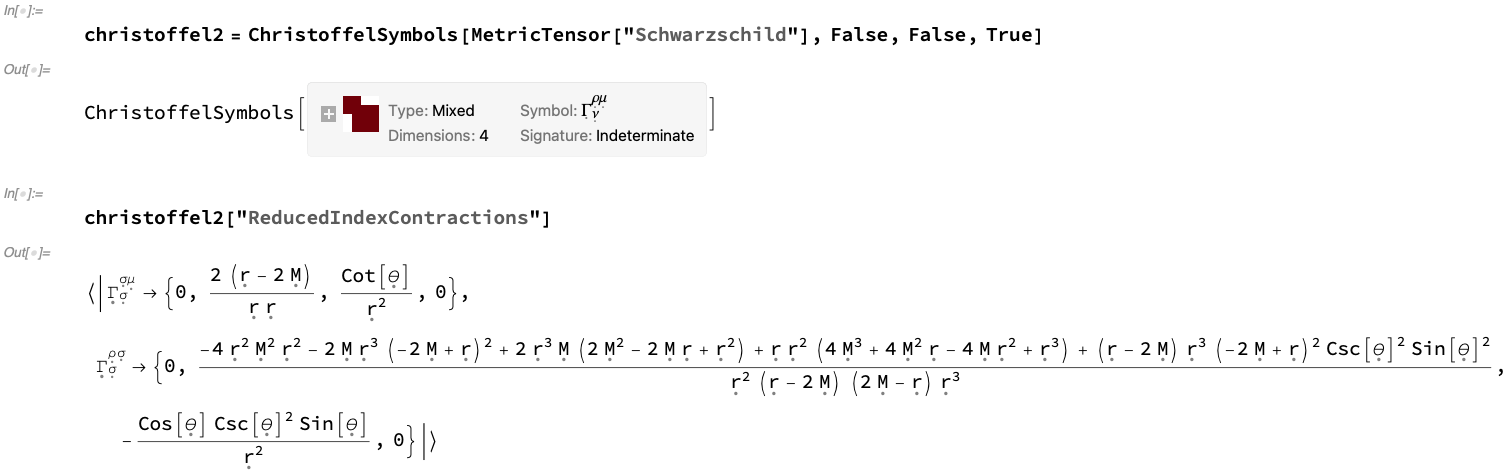}
\end{framed}
\caption{On the left, the association of all possible index contractions of the \texttt{ChristoffelSymbols} object for a Schwarzschild geometry (representing e.g. an uncharged, non-rotating black hole of mass $M$ in Schwarzschild/spherical polar coordinates ${\left( t, r, \theta, \phi \right)}$) with the first index raised/contravariant and the latter two indices lowered/covariant (default). On the right, the association of all possible index contractions of the \texttt{ChristoffelSymbols} object for a Schwarzschild geometry (representing e.g. an uncharged, non-rotating black hole of mass $M$ in Schwarzschild/spherical polar coordinates ${\left( t, r, \theta, \phi \right)}$) with the first two indices raised/contravariant and the latter index lowered/covariant.}
\label{fig:Figure18}
\end{figure}

\begin{figure}[ht]
\centering
\begin{framed}
\includegraphics[width=0.495\textwidth]{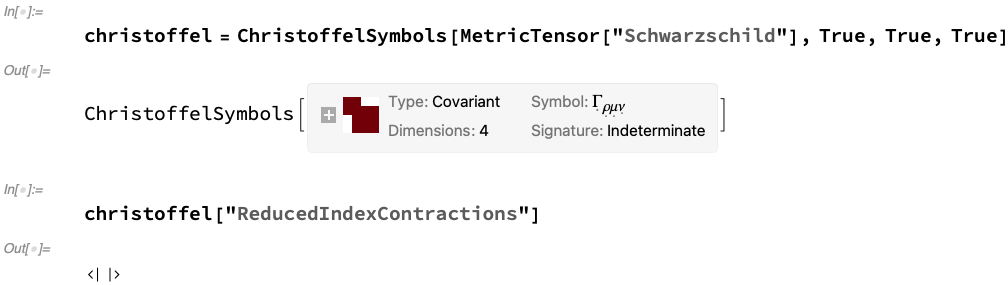}
\vrule
\includegraphics[width=0.495\textwidth]{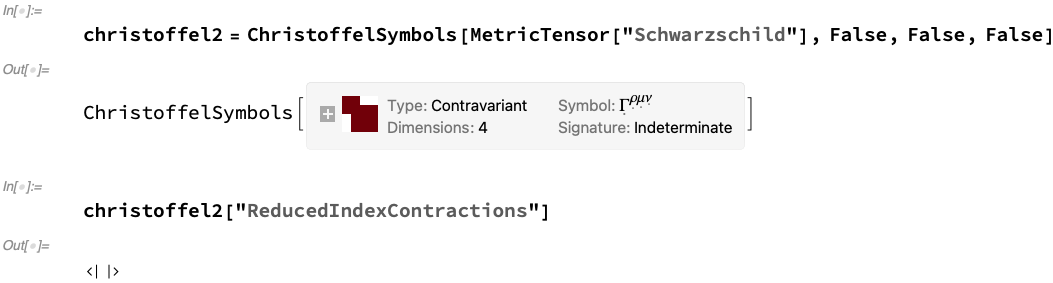}
\end{framed}
\caption{On the left, the association of all possible index contractions, i.e. none, of the \texttt{ChristoffelSymbols} object for a Schwarzschild geometry (representing e.g. an uncharged, non-rotating black hole of mass $M$ in Schwarzschild/spherical polar coordinates ${\left( t, r, \theta, \phi \right)}$) with all indices lowered/covariant. On the right, the association of all possible index contractions, i.e. none, of the \texttt{ChristoffelSymbols} object for a Schwarzschild geometry (representing e.g. an uncharged, non-rotating black hole of mass $M$ in Schwarzschild/spherical polar coordinates ${\left( t, r, \theta, \phi \right)}$) with all indices raised/contravariant.}
\label{fig:Figure19}
\end{figure}

The coefficients of the Levi-Civita connection, namely the Christoffel symbols\cite{christoffel} ${\Gamma_{\mu \nu}^{\rho}}$, may then be defined abstractly in terms of the relationship between the covariant derivative operator ${\nabla}$ and the partial derivative operator ${\partial}$ (assuming a local coordinate basis ${\left\lbrace x^{\mu} \right\rbrace}$):

\begin{equation}
\nabla_{\mu} \frac{\partial}{\partial x^{\nu}} = \Gamma_{\mu \nu}^{\rho} \frac{\partial}{\partial x^{\rho}},
\end{equation}
or, represented more concretely in terms of partial derivatives of the metric tensor ${g_{\mu \nu}}$ (in explicit covariant matrix form):

\begin{equation}
\Gamma_{\mu \nu}^{\rho} = \frac{1}{2} g^{\rho \sigma} \left( \frac{\partial}{\partial x^{\mu}} \left( g_{\sigma \nu} \right) + \frac{\partial}{\partial x^{\nu}} \left( g_{\mu \sigma} \right) - \frac{\partial}{\partial x^{\sigma}} \left( g_{\mu \nu} \right) \right).
\end{equation}
Building upon the general geometric interpretation of the coefficients for the affine and (Riemannian) metric connections, we can interpret the Christoffel symbols ${\Gamma_{\mu \nu}^{\rho}}$ as defining how the basis vectors change as one moves from point to point along tangent vectors ${\mathbf{X} \in \Gamma \left( \bigsqcup\limits_{\mathbf{x} \in \mathcal{M}} T_{\mathbf{x}} \mathcal{M} \right)}$ on the manifold ${\mathcal{M}}$. Under a generic coordinate transformation of the form ${x^{\mu} \to \widetilde{x^{\mu}}}$, the components of the Christoffel symbols transform as:

\begin{equation}
\widetilde{\Gamma_{\mu \nu}^{\rho}} = \left( \frac{\partial \widetilde{x^{\rho}}}{\partial x^{\alpha}} \right) \left( \frac{\partial x^{\beta}}{\partial \widetilde{x^{\mu}}} \right) \left( \frac{\partial x^{\gamma}}{\partial \widetilde{x^{\nu}}} \right) \Gamma_{\beta \gamma}^{\alpha} + \left( \frac{\partial^2 x^{\alpha}}{\partial \widetilde{x^{\mu}} \partial \widetilde{x^{\nu}}} \right) \left( \frac{\partial \widetilde{x^{\rho}}}{\partial x^{\alpha}} \right),
\end{equation}
the first term of which is simply the standard tensor transformation law for rank-3 tensors with mixed indices, with the second (inhomogeneous) term representing the fact that Christoffel symbols only transform as tensors under \textit{linear} coordinate transformations (in which case the inhomogeneous term vanishes identically): more generally, Christoffel symbols transform as functions on the jet bundle of the frame bundle over the manifold ${\mathcal{M}}$. Thus, although they are not tensors (merely rank-3 arrays of numbers), they inherit many of the core algebraic properties of tensors, in particular the rules for raising and lowering (and hence contracting) indices with respect to the metric tensor ${g_{\mu \nu}}$, i.e:

\begin{equation}
V^{\mu} = g^{\mu \sigma} V_{\sigma}, \qquad \text{ and } \qquad V_{\mu} = g_{\mu \sigma} V^{\sigma},
\end{equation}
for the case of rank-1 tensor fields $V$ (i.e. vector and covector fields), or:

\begin{equation}
T^{\mu \nu} = g^{\mu \sigma} T_{\sigma}^{\nu} = g^{\sigma \nu} T_{\sigma}^{\mu} = g^{\mu \sigma} g^{\lambda \nu} T_{\sigma \lambda}, \qquad T_{\mu}^{\nu} = g_{\mu \sigma} T^{\sigma \nu} = g_{\mu \sigma} g^{\lambda \nu} T_{\lambda}^{\sigma} = g^{\sigma \nu} T_{\mu \sigma},
\end{equation}
\begin{equation}
T_{\nu}^{\mu} = g_{\sigma \nu} T^{\mu \sigma} = g^{\mu \sigma} g_{\lambda \nu} T_{\sigma}^{\lambda} = g^{\mu \sigma} T_{\sigma \nu}, \qquad \text{ and } \qquad T_{\mu \nu} = g_{\mu \sigma} g_{\lambda \nu} T^{\sigma \lambda} = g_{\sigma \nu} T_{\mu}^{\sigma} = g_{\mu \sigma} T_{\nu}^{\sigma},
\end{equation}
for the case of rank-2 tensor fields $T$, and so on for objects of higher rank. A representation of the Christoffel symbols for the Schwarzschild metric (representing e.g. an uncharged, non-rotating black hole of mass $M$ in Schwarzschild/spherical polar coordinates ${\left( t, r, \theta, \phi \right)}$) using the \texttt{ChristoffelSymbols} function is shown in Figure \ref{fig:Figure17}, including both the default case with the first index raised/contravariant and the latter two indices lowered/covariant (i.e. ${\Gamma_{\mu \nu}^{\rho}}$), and the case with the first two indices raised/contravariant and the latter index lowered/covariant (i.e. ${\Gamma_{\nu}^{\rho \mu}}$). These two cases each give rise to two possible index contractions, namely ${\Gamma_{\sigma \nu}^{\sigma}}$ and ${\Gamma_{\mu \sigma}^{\sigma}}$ in the first instance, and ${\Gamma_{\sigma}^{\sigma \mu}}$ and ${\Gamma_{\sigma}^{\rho \sigma}}$ in the second instance, as illustrated in Figure \ref{fig:Figure18}. Since covariant indices may only ever be contracted with contravariant ones, no index contractions are possible in cases where the Christoffel symbols are either fully covariant (i.e. ${\Gamma_{\rho \mu \nu}}$) or fully contravariant (i.e. ${\Gamma^{\rho \mu \nu}}$), as demonstrated in Figure \ref{fig:Figure19}. Note that, much like \texttt{MetricTensor} objects, \texttt{ChristoffelSymbols} objects therefore lack some of the core functionality and properties that all other objects of abstract type \texttt{Tensor} within the \textsc{Gravitas} framework possess, such as the ability to evaluate covariant derivatives.

\begin{figure}[ht]
\centering
\begin{framed}
\includegraphics[width=0.495\textwidth]{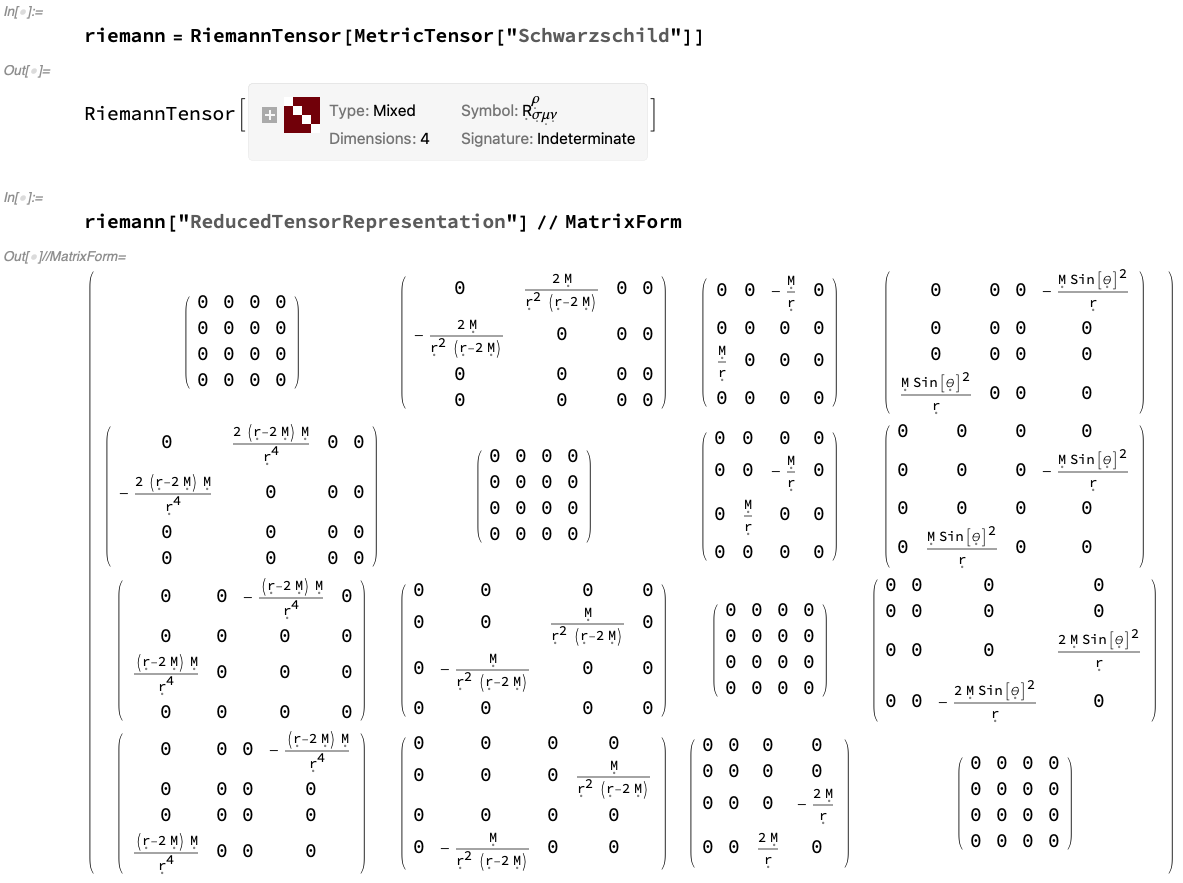}
\vrule
\includegraphics[width=0.495\textwidth]{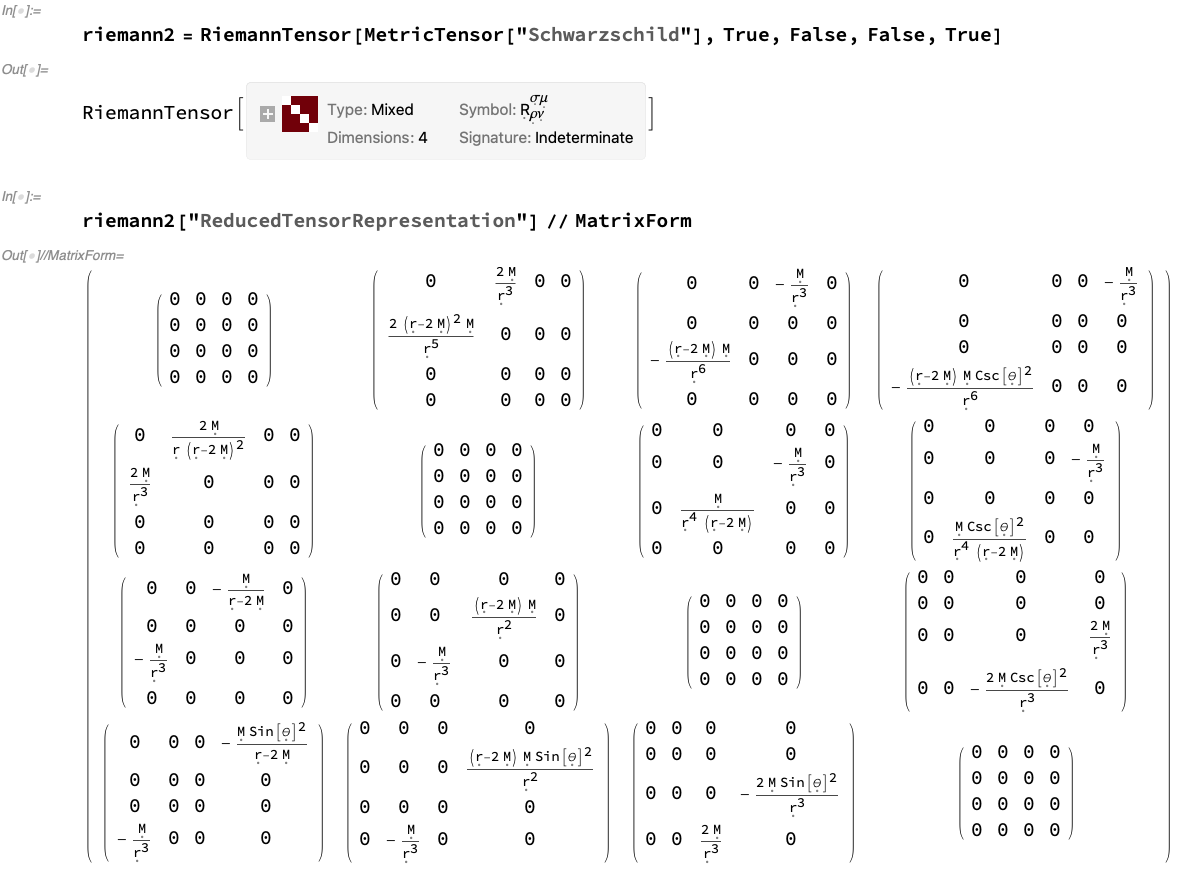}
\end{framed}
\caption{On the left, the \texttt{RiemannTensor} object for a Schwarzschild geometry (representing e.g. an uncharged, non-rotating black hole of mass $M$ in Schwarzschild/spherical polar coordinates ${\left( t, r, \theta, \phi \right)}$) in explicit mixed-index array form, with the first index raised/contravariant and the latter three indices lowered/covariant (default). On the right, the \texttt{RiemannTensor} object for a Schwarzschild geometry (representing e.g. an uncharged, non-rotating black hole of mass $M$ in Schwarzschild/spherical polar coordinates ${\left( t, r, \theta, \phi \right)}$) in explicit mixed-index array form, with the first and last indices lowered/covariant and all other indices raised/contravariant.}
\label{fig:Figure20}
\end{figure}

\begin{figure}[ht]
\centering
\begin{framed}
\includegraphics[width=0.495\textwidth]{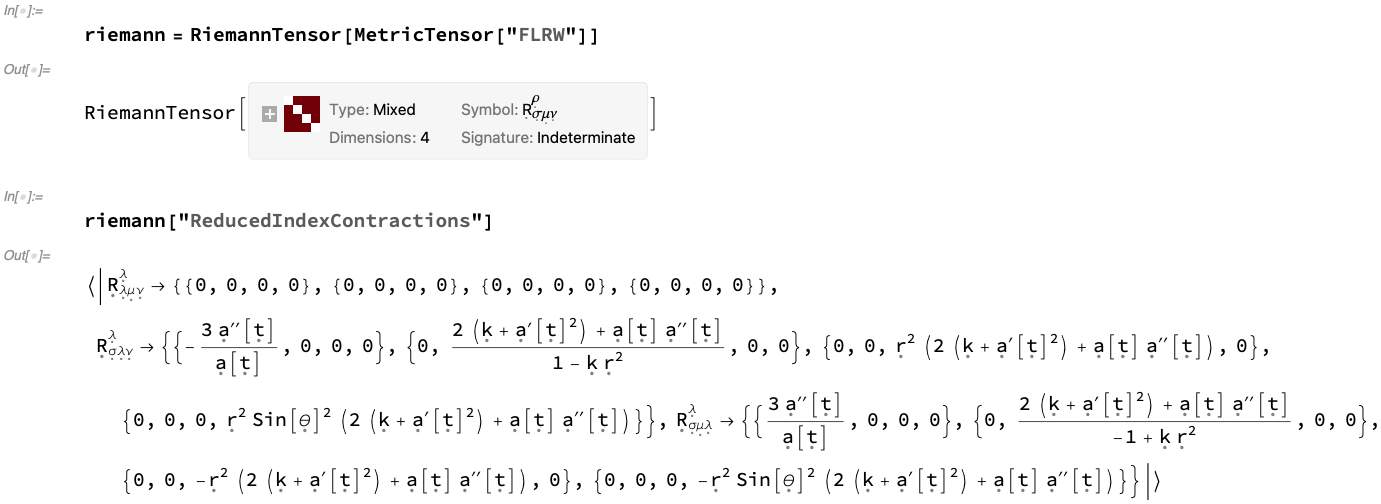}
\vrule
\includegraphics[width=0.495\textwidth]{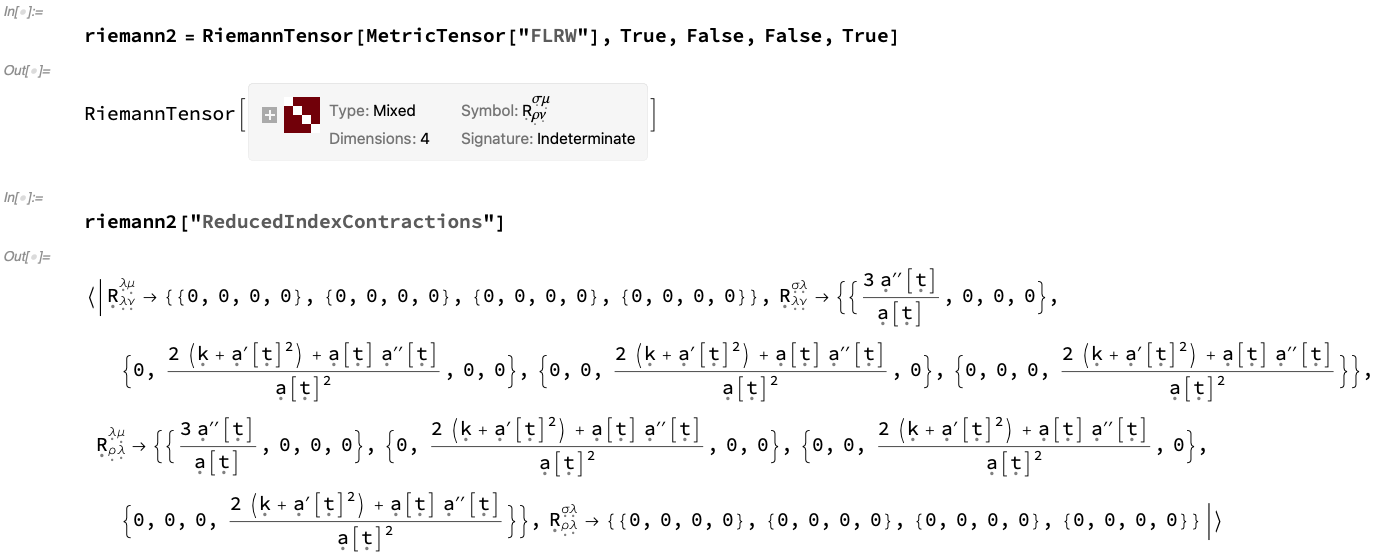}
\end{framed}
\caption{On the left, the association of all possible index contractions, i.e. 3, of the \texttt{RiemannTensor} object for an FLRW geometry (representing e.g. a homogeneous, isotropic and uniformly expanding/contracting universe with global curvature $k$ and scale factor ${a \left( t \right)}$ in spherical polar coordinates ${\left( t, r, \theta, \phi \right)}$) with the first index raised/contravariant and the latter three indices lowered/covariant (default). On the right, the association of all possible index contractions, i.e. 4, of the \texttt{RiemannTensor} object for an FLRW geometry (representing e.g. a homogeneous, isotropic and uniformly expanding/contracting universe with global curvature $k$ and scale factor ${a \left( t \right)}$ in spherical polar coordinates ${\left( t, r, \theta, \phi \right)}$) with the first and last indices lowered/covariant and all other indices raised/contravariant.}
\label{fig:Figure21}
\end{figure}

\begin{figure}[ht]
\centering
\begin{framed}
\includegraphics[width=0.495\textwidth]{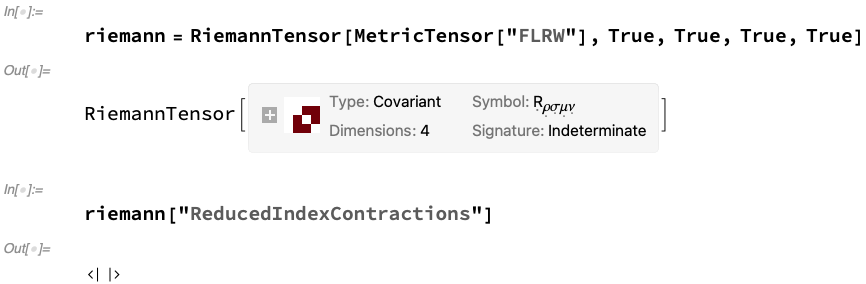}
\vrule
\includegraphics[width=0.495\textwidth]{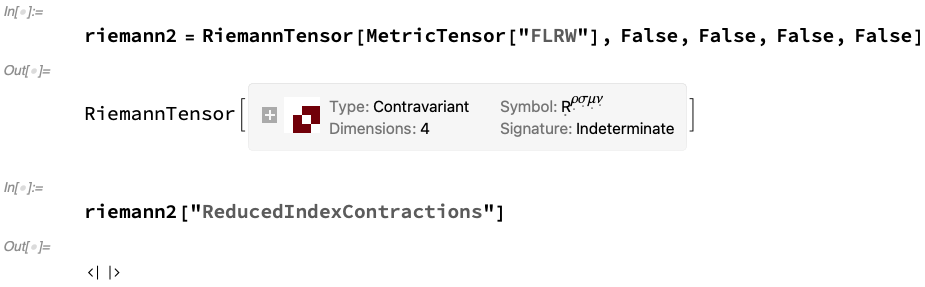}
\end{framed}
\caption{On the left, the association of all possible index contractions, i.e. none, of the \texttt{RiemannTensor} object for an FLRW geometry (representing e.g. a homogeneous, isotropic and uniformly expanding/contracting universe with global curvature $k$ and scale factor ${a \left( t \right)}$ in spherical polar coordinates ${\left( t, r, \theta, \phi \right)}$) with all indices lowered/covariant. On the right, the association of all possible index contractions, i.e. none, of the \texttt{RiemannTensor} object for an FLRW geometry (representing e.g. a homogeneous, isotropic and uniformly expanding/contracting universe with global curvature $k$ and scale factor ${a \left( t \right)}$ in spherical polar coordinates ${\left( t, r, \theta, \phi \right)}$) with all indices raised/contravariant.}
\label{fig:Figure22}
\end{figure}

\begin{figure}[ht]
\centering
\begin{framed}
\includegraphics[width=0.495\textwidth]{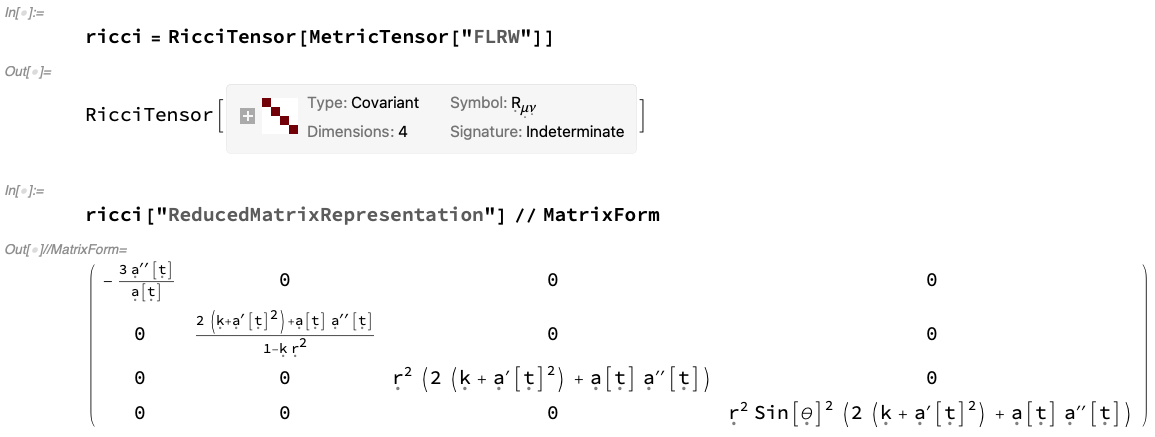}
\vrule
\includegraphics[width=0.495\textwidth]{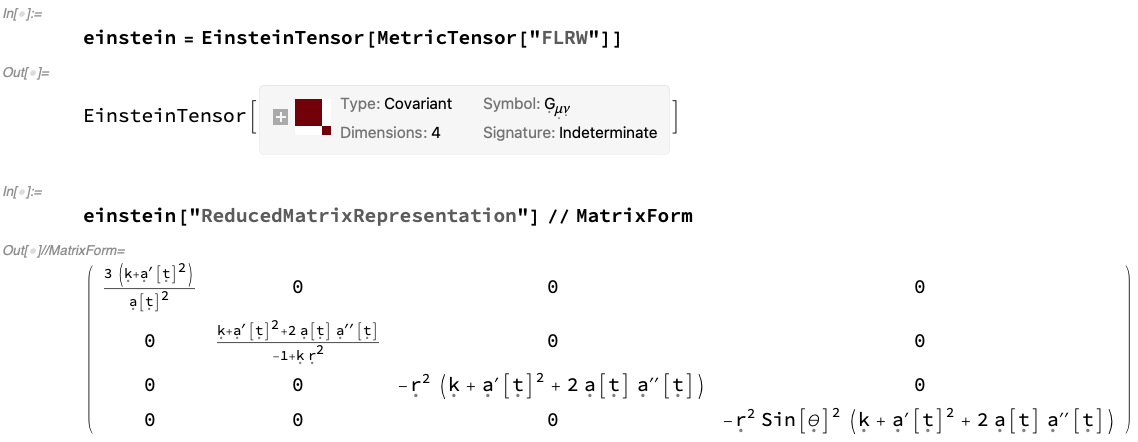}
\end{framed}
\caption{On the left, the \texttt{RicciTensor} object for an FLRW geometry (representing e.g. a homogeneous, isotropic and uniformly expanding/contracting universe with global curvature $k$ and scale factor ${a \left( t \right)}$ in spherical polar coordinates ${\left( t, r, \theta, \phi \right)}$) in explicit covariant matrix form (default). On the right, the \texttt{EinsteinTensor} object for an FLRW geometry (representing e.g. a homogeneous, isotropic and uniformly expanding/contracting universe with global curvature $k$ and scale factor ${a \left( t \right)}$ in spherical polar coordinates ${\left( t, r, \theta, \phi \right)}$) in explicit covariant matrix form (default).}
\label{fig:Figure23}
\end{figure}

\begin{figure}[ht]
\centering
\begin{framed}
\includegraphics[width=0.495\textwidth]{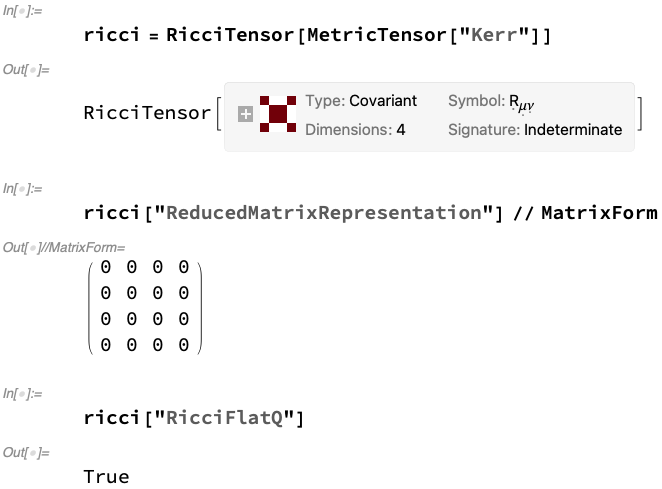}
\vrule
\includegraphics[width=0.495\textwidth]{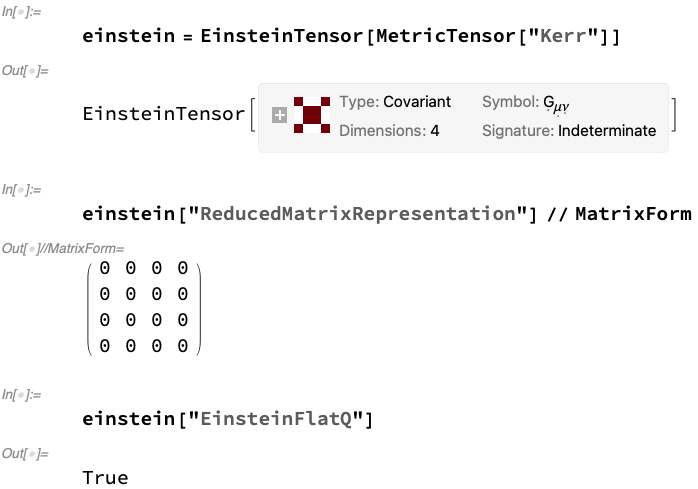}
\end{framed}
\caption{On the left, the \texttt{RicciTensor} object for a Kerr geometry (representing e.g. an uncharged, spinning black hole of mass $M$ and angular momentum $J$ in Boyer-Lindquist/oblate spheroidal coordinates ${\left( t, r, \theta, \phi \right)}$) in explicit covariant matrix form (default), illustrating that the Kerr metric is Ricci-flat. On the right, the \texttt{EinsteinTensor} object for a Kerr geometry (representing e.g. an uncharged, spinning black hole of mass $M$ and angular momentum $J$ in Boyer-Lindquist/oblate spheroidal coordinates ${\left( t, r, \theta, \phi \right)}$) in explicit covariant matrix form (default), illustrating that the Kerr metric is Einstein-flat.}
\label{fig:Figure24}
\end{figure}

We are now able to quantify the degree to which our connection fails to be exact, and hence the extent to which our covariant derivative operator ${\nabla}$ fails to be commutative, abstract by means of the rank-4 Riemann (curvature) tensor $R$\cite{kobayashi}\cite{kobayashi2}:

\begin{equation}
\forall \mathbf{X}, \mathbf{Y} \in \Gamma \left( \bigsqcup\limits_{\mathbf{x} \in \mathcal{M}} T_{\mathbf{x}} \mathcal{M} \right), \qquad R \left( \mathbf{X}, \mathbf{Y} \right) = \left[ \nabla_{\mathbf{X}}, \nabla_{\mathbf{Y}} \right] - \nabla_{\left[ \mathbf{X}, \mathbf{Y} \right]},
\end{equation}
where ${\left[ \nabla_{\mathbf{X}}, \nabla_{\mathbf{Y}} \right]}$ designates a commutator of differential operators and, as above, ${\left[ \mathbf{X}, \mathbf{Y} \right]}$ denotes the usual Lie bracket of vector fields; in other words, we can characterize the Riemann tensor $R$ abstractly in terms of its action on an arbitrary vector field ${\mathbf{Z} \in \Gamma \left( \bigsqcup\limits_{\mathbf{x} \in \mathcal{M}} T_{\mathbf{x}} \mathcal{M} \right)}$, as:

\begin{equation}
\forall \mathbf{X}, \mathbf{Y}, \mathbf{Z} \in \Gamma \left( \bigsqcup\limits_{\mathbf{x} \in \mathcal{M}} T_{\mathbf{x}} \mathcal{M} \right), \qquad R \left( \mathbf{X}, \mathbf{Y} \right) \mathbf{Z} = \nabla_{\mathbf{X}} \left( \nabla_{\mathbf{Y}} \mathbf{Z} \right) - \nabla_{\mathbf{Y}} \left( \nabla_{\mathbf{X}} \mathbf{Z} \right) - \nabla_{\left[ \mathbf{X}, \mathbf{Y} \right]} \mathbf{Z},
\end{equation}
or, in explicit (component-based) form, exploiting the fact that the Levi-Civita connection is torsion-free and therefore that the Lie bracket term ${\nabla_{\left[ \mathbf{X}, \mathbf{Y} \right]} \mathbf{Z}}$ in the above must vanish identically:

\begin{equation}
\forall \mathbf{Z} \in \Gamma \left( \bigsqcup\limits_{\mathbf{x} \in \mathcal{M}} T_{\mathbf{x}} \mathcal{M} \right), \qquad R_{\sigma \mu \nu}^{\rho} Z^{\sigma} = \nabla_{\mu} \left( \nabla_{\nu} Z^{\rho} \right) - \nabla_{\nu} \left( \nabla_{\mu} Z^{\rho} \right).
\end{equation}
Thus, at least within the local coordinate basis ${\left\lbrace x^{\mu} \right\rbrace}$, we can express the Riemann tensor ${R_{\sigma \mu \nu}^{\rho}}$ directly in terms of (partial derivatives of) the Christoffel symbols ${\Gamma_{\mu \nu}^{\rho}}$ (which are themselves expressed directly in terms of partial derivatives of the metric tensor ${g_{\mu \nu}}$):

\begin{equation}
R_{\sigma \mu \nu}^{\rho} = \frac{\partial}{\partial x^{\mu}} \left( \Gamma_{\sigma \mu}^{\rho} \right) - \frac{\partial}{\partial x^{\nu}} \left( \Gamma_{\mu \sigma}^{\rho} \right) + \Gamma_{\mu \lambda}^{\rho} \Gamma_{\sigma \nu}^{\lambda} - \Gamma_{\lambda \nu}^{\rho} \Gamma_{\mu \sigma}^{\lambda}.
\end{equation}
A representation of the Riemann tensor for the Schwarzschild metric (representing e.g. an uncharged, non-rotating black hole of mass $M$ in Schwarzschild/spherical polar coordinates ${\left( t, r, \theta, \phi \right)}$) using the \texttt{RiemannTensor} function is shown in Figure \ref{fig:Figure20}, including both the default case with the first index raised/contravariant and the latter three indices lowered/covariant (i.e. ${R_{\sigma \mu \nu}^{\rho}}$), and the case with the first and last indices lowered/covariant and all other indices raised/contravariant (i.e. ${R_{\rho \nu}^{\sigma \mu}}$). The first of these cases gives rise to three possible index contractions, namely ${R_{\lambda \mu \nu}^{\lambda}}$, ${R_{\sigma \lambda \nu}^{\lambda}}$ and ${R_{\sigma \mu \lambda}^{\lambda}}$, and the second case gives rise to four possible index contractions, namely ${R_{\lambda \nu}^{\lambda \mu}}$, ${R_{\rho \lambda}^{\lambda \mu}}$, ${R_{\lambda \nu}^{\sigma \lambda}}$ and ${R_{\rho \lambda}^{\sigma \lambda}}$, as illustrated in Figure \ref{fig:Figure21}. As ever, no index contractions are possible whenever the Riemann tensor is in either of its fully covariant (i.e. ${R_{\rho \sigma \mu \nu}}$) or fully contravariant (i.e. ${R^{\rho \sigma \mu \nu}}$) forms, as demonstrated in Figure \ref{fig:Figure22}. Due to the various index/slot permutation symmetries of the Riemann tensor, there exists (at least modulo a sign) a single, unique non-zero contraction of ${R_{\sigma \mu \nu}^{\rho}}$ down to a rank-2 tensor, namely the Ricci tensor\cite{besse} ${R_{\mu \nu}}$, from which one can directly compute its trace-reversed form ${G_{\mu \nu}}$, i.e. the Einstein tensor\cite{lovelock}\cite{lovelock2}:

\begin{equation}
R_{\mu \nu} = R_{\sigma \lambda \nu}^{\lambda}, \qquad \text{ and } \qquad G_{\mu \nu} = R_{\mu \nu} - \frac{1}{2} R g_{\mu \nu},
\end{equation}
respectively, where ${R = R_{\sigma}^{\sigma}}$ denotes the Ricci scalar, i.e. the trace of the Ricci tensor, otherwise known as the scalar curvature. A representation of the Ricci and Einstein tensors for the FLRW metric (representing e.g. a homogeneous, isotropic and uniformly expanding/contracting universe with global curvature $k$ and scale factor ${a \left( t \right)}$ in spherical polar coordinates ${\left( t, r, \theta, \phi \right)}$) using the \texttt{RicciTensor} and \texttt{EinsteinTensor} functions is shown in Figure \ref{fig:Figure23}. An illustration that the Kerr metric (representing e.g. an uncharged, spinning black hole of mass $M$ and angular momentum $J$ in Boyer-Lindquist/oblate spheroidal coordinates ${\left( t, r, \theta, \phi \right)}$) is Ricci-flat and therefore Einstein-flat (i.e. that all components of the Ricci and Einstein tensors ${R_{\mu \nu}}$ and ${G_{\mu \nu}}$ vanish identically) can be found in Figure \ref{fig:Figure24}; due to the trace-reversed nature of the relationship between ${R_{\mu \nu}}$ and ${G_{\mu \nu}}$, a manifold is guaranteed to be Ricci-flat if and only if it is Einstein-flat, at least in dimensions ${n \neq 2}$.

\begin{figure}[ht]
\centering
\begin{framed}
\includegraphics[width=0.495\textwidth]{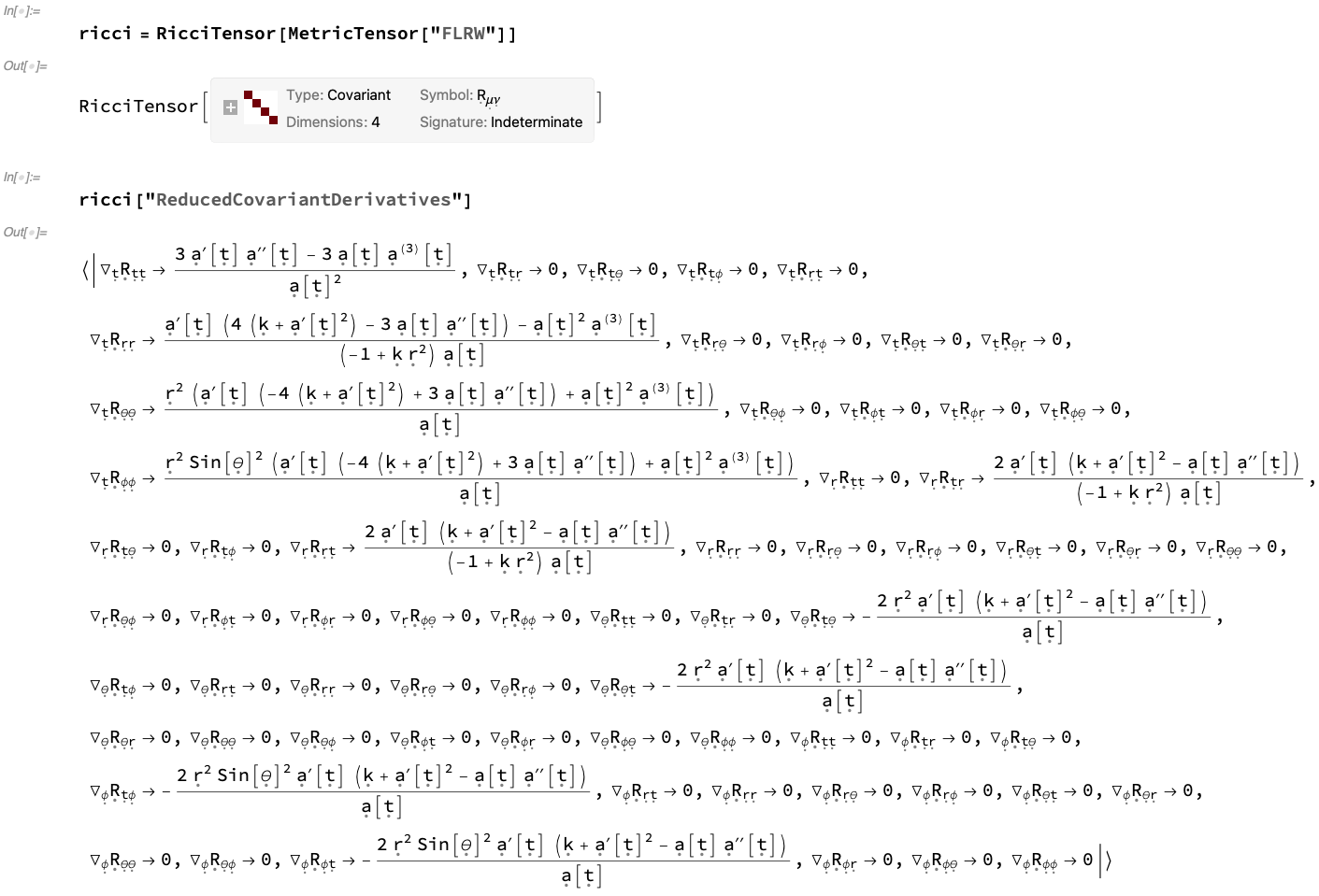}
\vrule
\includegraphics[width=0.495\textwidth]{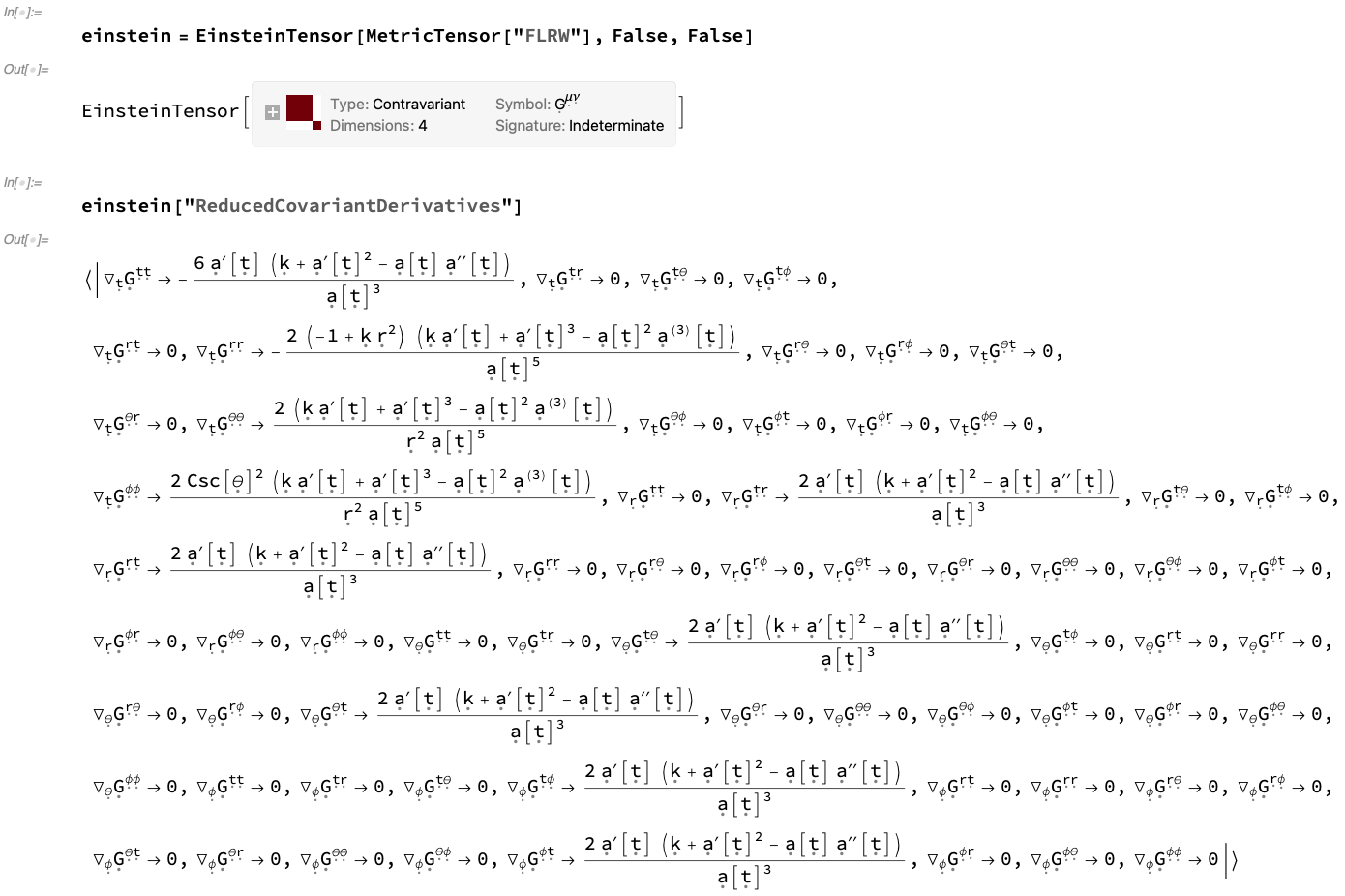}
\end{framed}
\caption{On the left, the association of all covariant derivatives of the \texttt{RicciTensor} object for an FLRW geometry (representing e.g. a homogeneous, isotropic and uniformly expanding/contracting universe with global curvature $k$ and scale factor ${a \left( t \right)}$ in spherical polar coordinates ${\left( t, r, \theta, \phi \right)}$) with both indices lowered/covariant (default). On the right, the association of all covariant derivatives of the \texttt{EinsteinTensor} object for an FLRW geometry (representing e.g. a homogeneous, isotropic and uniformly expanding/contracting universe with global curvature $k$ and scale factor ${a \left( t \right)}$ in spherical polar coordinates ${\left( t, r, \theta, \phi \right)}$) with both indices raised/contravariant.}
\label{fig:Figure25}
\end{figure}

\begin{figure}[ht]
\centering
\begin{framed}
\includegraphics[width=0.495\textwidth]{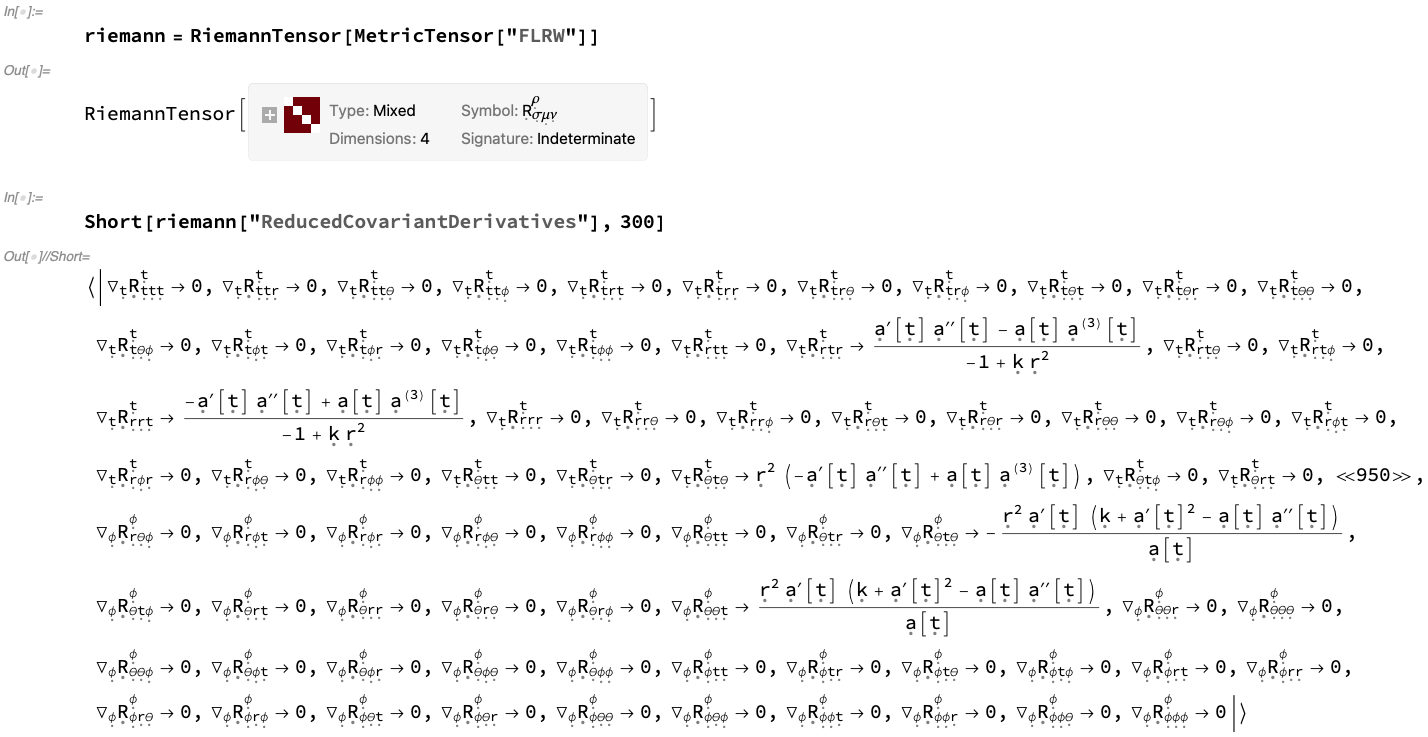}
\vrule
\includegraphics[width=0.495\textwidth]{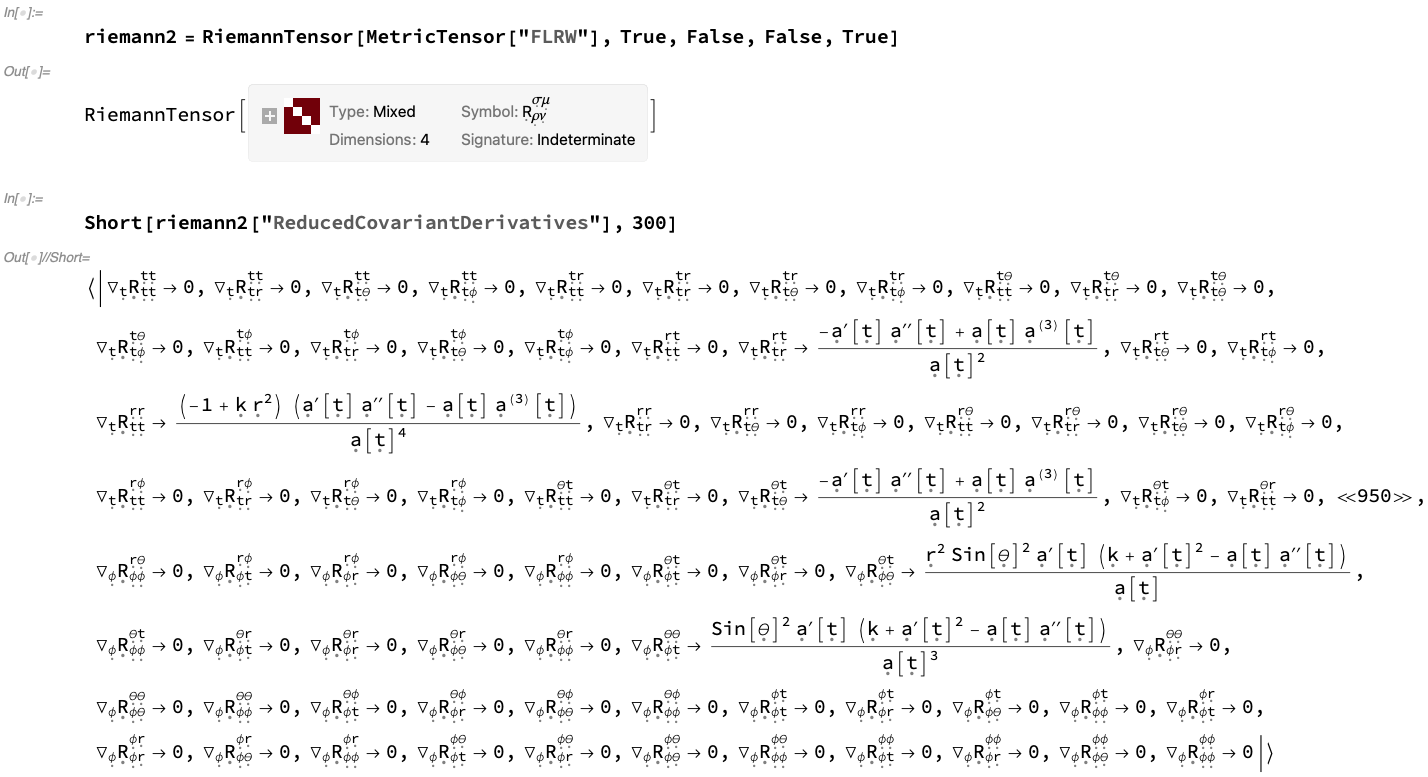}
\end{framed}
\caption{On the left, the association of all covariant derivatives of the \texttt{RiemannTensor} object for an FLRW geometry (representing e.g. a homogeneous, isotropic and uniformly expanding/contracting universe with global curvature $k$ and scale factor ${a \left( t \right)}$ in spherical polar coordinates ${\left( t, r, \theta, \phi \right)}$) with the first index raised/contravariant and the latter three indices lowered/covariant (default). On the right, the association of all covariant derivatives of the \texttt{RiemannTensor} object for an FLRW geometry (representing e.g. a homogeneous, isotropic and uniformly expanding/contracting universe with global curvature $k$ and scale factor ${a \left( t \right)}$ in spherical polar coordinates ${\left( t, r, \theta, \phi \right)}$) with the first and last indices lowered/covariant and all other indices raised/contravariant.}
\label{fig:Figure26}
\end{figure}

\begin{figure}[ht]
\centering
\begin{framed}
\includegraphics[width=0.545\textwidth]{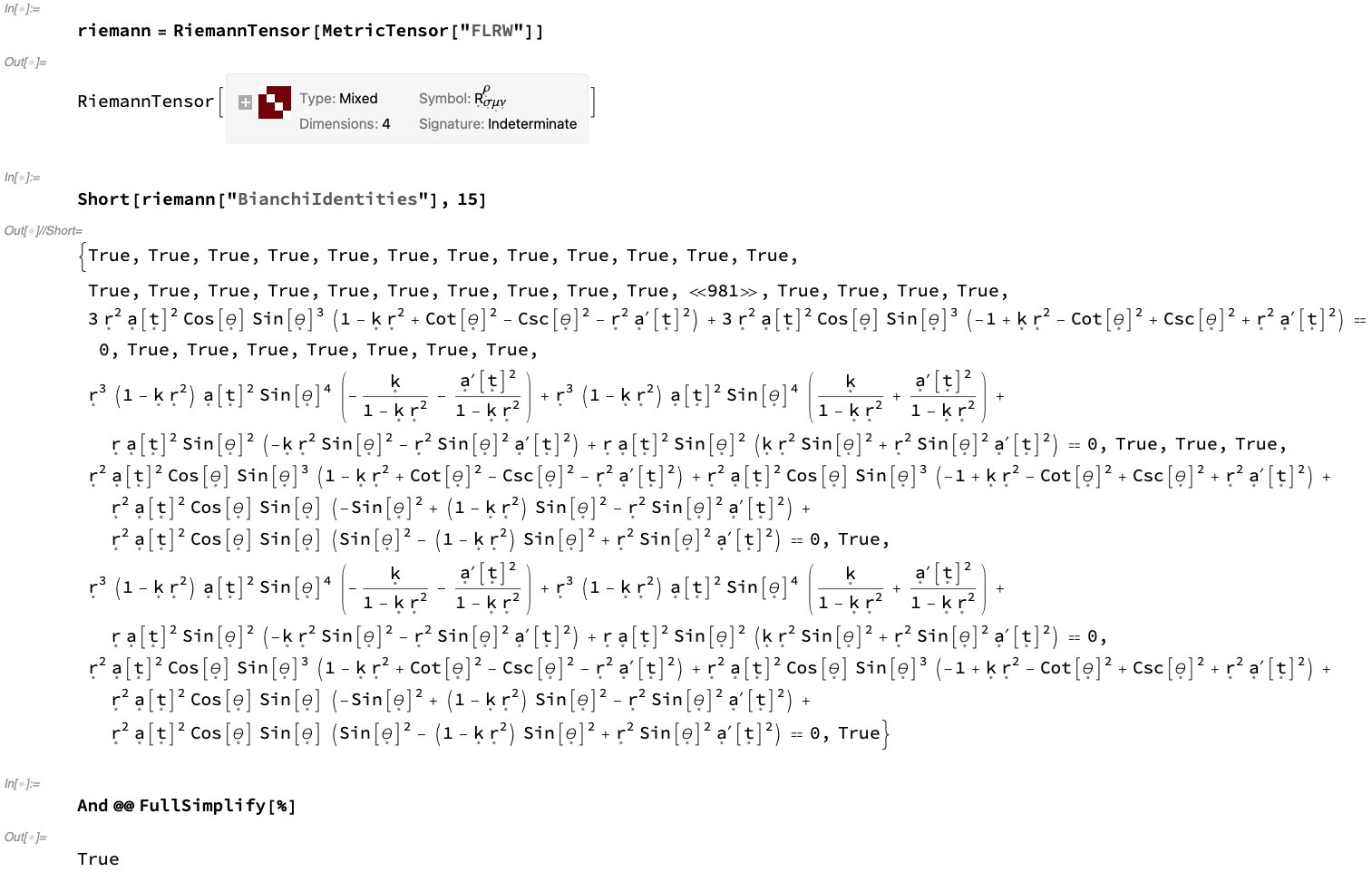}
\vrule
\includegraphics[width=0.445\textwidth]{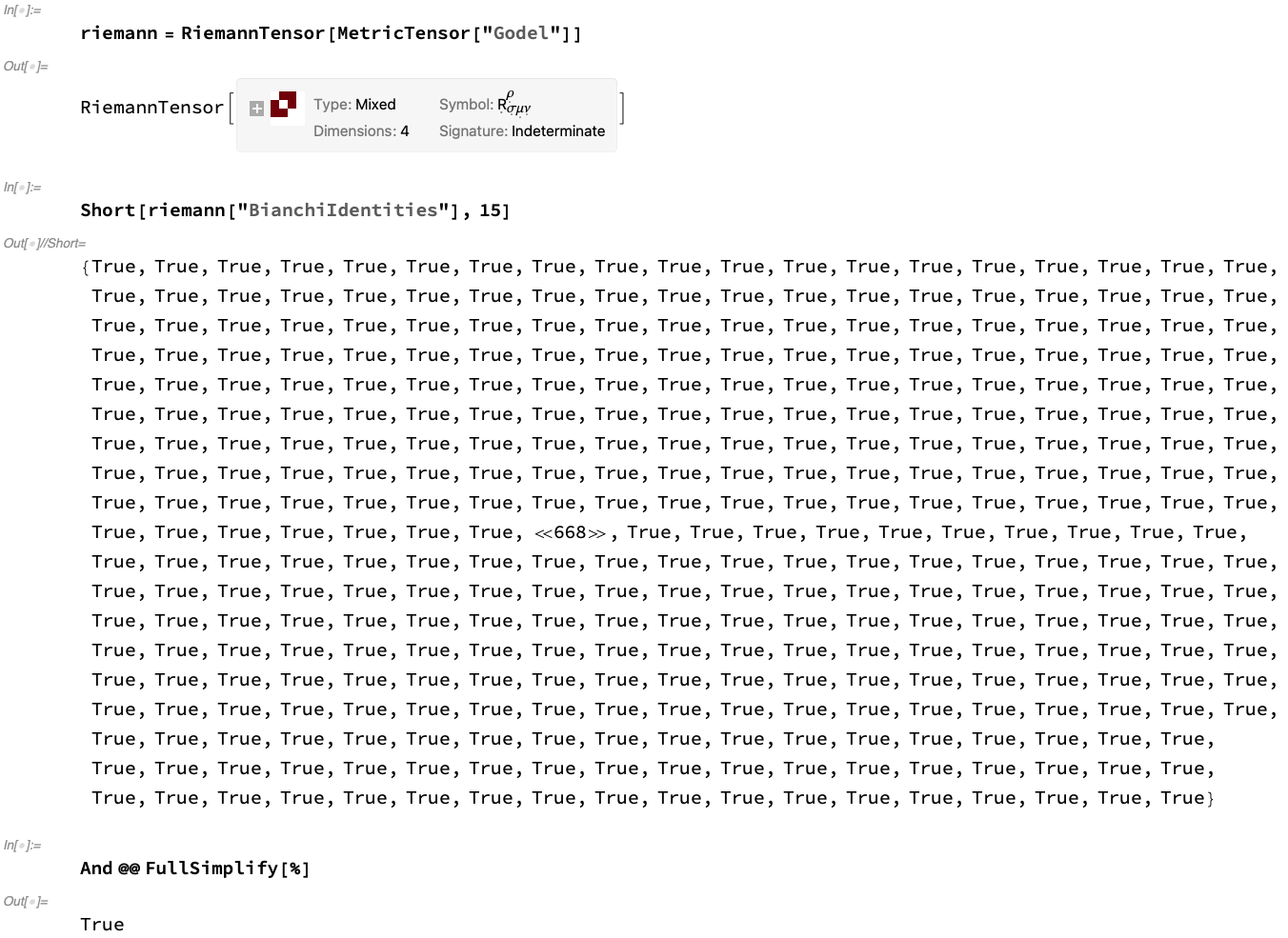}
\end{framed}
\caption{On the left, the list of Bianchi identities asserting the symmetries of the covariant derivatives of the \texttt{RiemannTensor} object for an FLRW geometry (representing e.g. a homogeneous, isotropic and uniformly expanding/contracting universe with global curvature $k$ and scale factor ${a \left( t \right)}$ in spherical polar coordinates ${\left( t, r, \theta, \phi \right)}$), together with a verification that they all hold identically. On the right, the list of Bianchi identities asserting the symmetries of the covariant derivatives of the \texttt{RiemannTensor} object for a G\"odel geometry (representing e.g. a rotating, dust-filled universe with global angular velocity ${\omega}$ in G\"odel's Cartesian-like coordinates ${\left( t, x, y, z \right)}$), together with a verification that they all hold identically.}
\label{fig:Figure27}
\end{figure}

\begin{figure}[ht]
\centering
\begin{framed}
\includegraphics[width=0.545\textwidth]{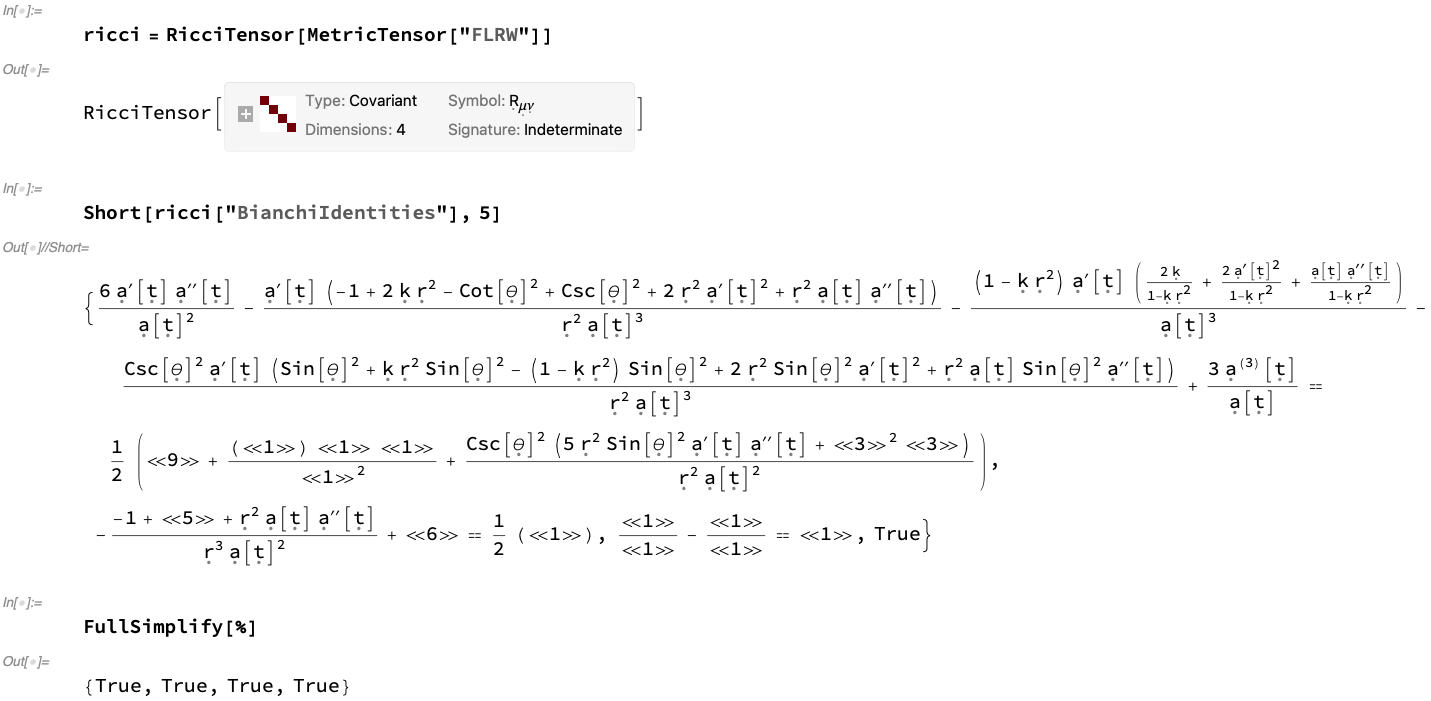}
\vrule
\includegraphics[width=0.445\textwidth]{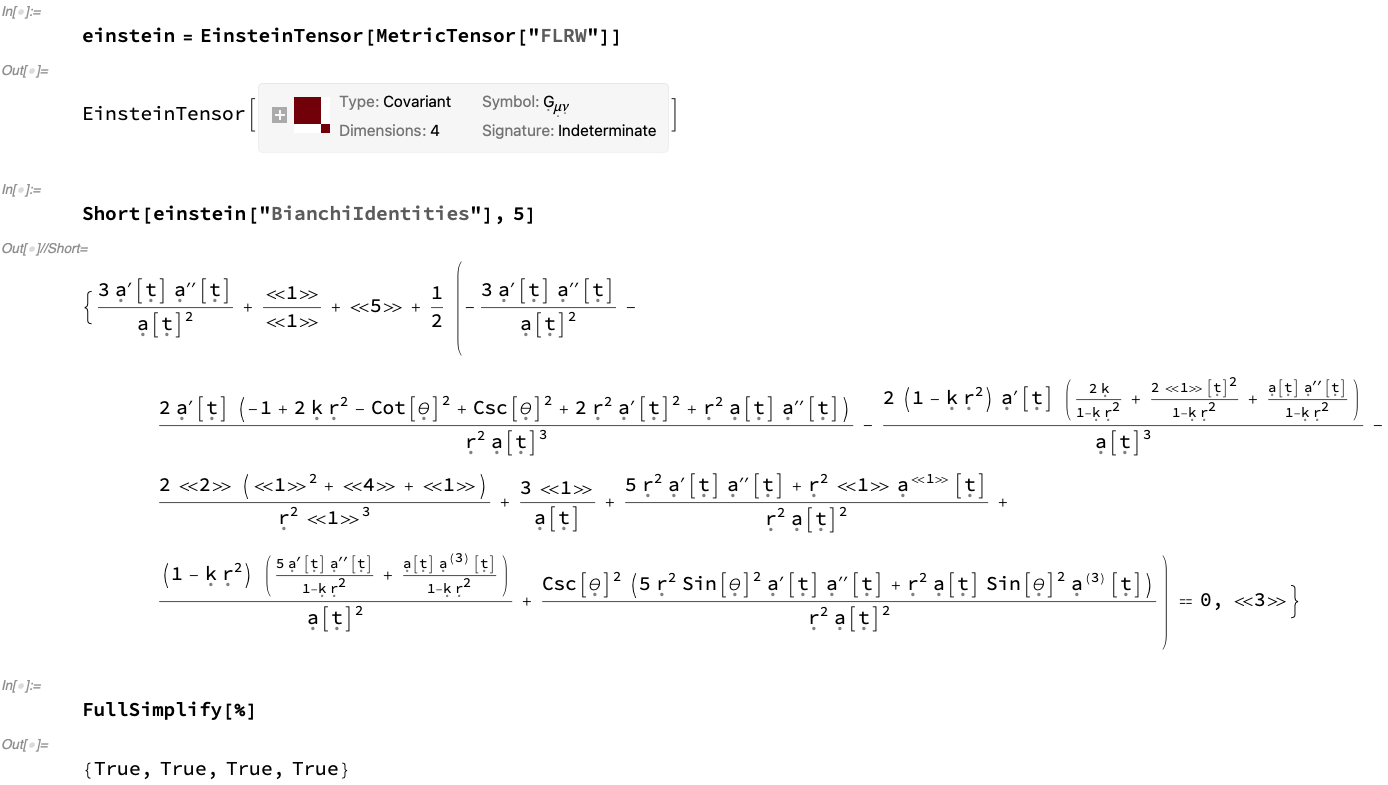}
\end{framed}
\caption{On the left, the list of contracted Bianchi identities relating the covariant divergence of the \texttt{RicciTensor} object for an FLRW geometry (representing e.g. a homogeneous, isotropic and uniformly expanding/contracting universe with global curvature $k$ and scale factor ${a \left( t \right)}$ in spherical polar coordinates ${\left( t, r, \theta, \phi \right)}$) and the covariant derivative of the corresponding Ricci scalar, together with a verification that they all hold identically. On the right, the list of contracted Bianchi identities asserting that the covariant divergence of the \texttt{EinsteinTensor} object for an FLRW geometry (representing e.g. a homogeneous, isotropic and uniformly expanding/contracting universe with global curvature $k$ and scale factor ${a \left( t \right)}$ in spherical polar coordinates ${\left( t, r, \theta, \phi \right)}$) vanishes, together with a verification that they all hold identically.}
\label{fig:Figure28}
\end{figure}

Since \texttt{RiemannTensor}, \texttt{RicciTensor} and \texttt{EinsteinTensor} are all represented internally as objects of abstract type \texttt{Tensor} within the \textsc{Gravitas} framework, one can perform standard tensor calculus operations such covariant differentiation on them, using all of the usual algebraic rules for differentiating along tangent vectors\cite{ricci}, i.e (assuming local coordinate basis ${\left\lbrace x^{\mu} \right\rbrace}$):

\begin{equation}
\nabla_{\rho} V^{\mu} = \frac{\partial}{\partial x^{\rho}} \left( V^{\mu} \right) + \Gamma_{\rho \sigma}^{\mu} V^{\sigma}, \qquad \text{ and } \qquad \nabla_{\rho} V_{\mu} = \frac{\partial}{\partial x^{\rho}} \left( V_{\mu} \right) - \Gamma_{\rho \mu}^{\sigma} V_{\sigma},
\end{equation}
for the case of rank-1 tensor fields $V$ (i.e. vector and covector fields), or:

\begin{equation}
\nabla_{\rho} T^{\mu \nu} = \frac{\partial}{\partial x^{\rho}} \left( T^{\mu \nu} \right) + \Gamma_{\rho \sigma}^{\mu} T^{\sigma \nu} + \Gamma_{\rho \sigma}^{\nu} T^{\mu \sigma}, \qquad \nabla_{\rho} T_{\mu}^{\nu} = \frac{\partial}{\partial x^{\rho}} \left( T_{\mu}^{\nu} \right) + \Gamma_{\rho \sigma}^{\nu} T_{\mu}^{\sigma} - \Gamma_{\rho \mu}^{\sigma} T_{\sigma}^{\nu},
\end{equation}
\begin{equation}
\nabla_{\rho} T_{\nu}^{\mu} = \frac{\partial}{\partial x^{\rho}} \left( T_{\nu}^{\mu} \right) + \Gamma_{\rho \sigma}^{\mu} T_{\nu}^{\sigma} - \Gamma_{\rho \nu}^{\sigma} T_{\sigma}^{\mu}, \qquad \text{ and } \qquad \nabla_{\rho} T_{\mu \nu} = \frac{\partial}{\partial x^{\rho}} \left( T_{\mu \nu} \right) - \Gamma_{\rho \mu}^{\sigma} T_{\sigma \nu} - \Gamma_{\rho \nu}^{\sigma} T_{\mu \sigma},
\end{equation}
for the case of rank-2 tensor fields $T$, and so on for objects of higher rank. This is illustrated in Figure \ref{fig:Figure25}, in which all covariant derivatives of the \texttt{RicciTensor} object in lowered-index/covariant form (i.e. ${R_{\mu \nu}}$, the default case), as well as all covariant derivatives of the \texttt{EinsteinTensor} object in raised-index/contravariant form (i.e. ${G^{\mu \nu}}$), are computed for the FLRW metric (representing e.g. a homogeneous, isotropic and uniformly expanding/contracting universe with global curvature $k$ and scale factor ${a \left( t \right)}$ in spherical polar coordinates ${\left( t, r, \theta, \phi \right)}$). The number of covariant derivative terms very quickly grows to become unmanageable in the case of higher-rank tensors, as demonstrated in Figure \ref{fig:Figure26}, in which all 1,024 covariant derivatives components of the \texttt{RiemannTensor} object for the FLRW metric are computed, both with the first index raised/contravariant and the latter three indices lowered/covariant (i.e. ${R_{\sigma \mu \nu}^{\rho}}$, the default case), and with the first and last indices lowered/covariant and all other indices raised/contravariant (i.e. ${R_{\sigma \nu}^{\rho \mu}}$); note that in both cases the output must be truncated in order to display reasonably. The (differential) Bianchi identities\cite{bianchi}, asserting the symmetries of the covariant derivatives of the Riemann tensor ${R_{\sigma \mu \nu}^{\rho}}$, namely:

\begin{equation}
\nabla_{\lambda} R_{\rho \sigma \mu \nu} + \nabla_{\mu} R_{\rho \sigma \nu \lambda} + \nabla_{\nu} R_{\rho \sigma \lambda \mu} = 0,
\end{equation}
i.e., in expanded form:

\begin{multline}
\left( \frac{\partial}{\partial x^{\lambda}} \left( R_{\rho \sigma \mu \nu} \right) - \Gamma_{\lambda \rho}^{\alpha} R_{\alpha \sigma \mu \nu} - \Gamma_{\lambda \sigma}^{\alpha} R_{\rho \alpha \mu \nu} - \Gamma_{\lambda \mu}^{\alpha} R_{\rho \sigma \alpha \nu} - \Gamma_{\lambda \nu}^{\alpha} R_{\rho \sigma \mu \alpha} \right)\\
+ \left( \frac{\partial}{\partial x^{\mu}} \left( R_{\rho \sigma \nu \lambda} \right) - \Gamma_{\mu \rho}^{\alpha} R_{\alpha \sigma \nu \lambda} - \Gamma_{\mu \sigma}^{\alpha} R_{\rho \alpha \nu \lambda} - \Gamma_{\mu \nu}^{\alpha} R_{\rho \sigma \alpha \lambda} - \Gamma_{\mu \lambda}^{\alpha} R_{\rho \sigma \nu \alpha} \right)\\
+ \left( \frac{\partial}{\partial x^{\nu}} \left( R_{\rho \sigma \lambda \mu} \right) - \Gamma_{\nu \rho}^{\alpha} R_{\alpha \sigma \lambda \mu} - \Gamma_{\nu \sigma}^{\alpha} R_{\rho \alpha \lambda \mu} - \Gamma_{\nu \lambda}^{\alpha} R_{\rho \sigma \alpha \mu} - \Gamma_{\nu \mu}^{\alpha} R_{\rho \sigma \lambda \alpha} \right) = 0,
\end{multline}
necessarily hold identically for any \texttt{RiemannTensor} object derived directly from a \texttt{MetricTensor} object in the manner described above, as illustrated in Figure \ref{fig:Figure27} for the cases of the FLRW metric (representing e.g. a homogeneous, isotropic and uniformly expanding/contracting universe with global curvature $k$ and scale factor ${a \left( t \right)}$ in spherical polar coordinates ${\left( t, r, \theta, \phi \right)}$) and the G\"odel metric (representing e.g. a rotating, dust-filled universe with global angular velocity ${\omega}$ in G\"odel's Cartesian-like coordinates ${\left( t, x, y, z \right)}$). Upon contracting both sides of the equation above with a pair of (inverse) metric tensors and rearranging, one obtains the contracted form of the Bianchi identities, asserting a relationship between the covariant divergence of the Ricci tensor ${R_{\mu \nu}}$ and the covariant derivative of the Ricci scalar ${R = R_{\sigma}^{\sigma}}$, or, equivalently, asserting that the covariant divergence of the Einstein tensor ${G_{\mu \nu}}$ vanishes identically:

\begin{equation}
\nabla_{\rho} R_{\mu}^{\rho} = \frac{1}{2} \nabla_{\mu} R, \qquad \text{ or } \qquad \nabla_{\nu} G^{\mu \nu} = 0,
\end{equation}
i.e., in expanded form:

\begin{equation}
\frac{\partial}{\partial x^{\rho}} \left( R_{\mu}^{\rho} \right) + \Gamma_{\rho \sigma}^{\rho} R_{\mu}^{\sigma} - \Gamma_{\rho \mu}^{\sigma} R_{\sigma}^{\rho} = \frac{1}{2} \left( \frac{\partial}{\partial x^{\mu}} \left( R \right) \right), \qquad \text{ or } \qquad \frac{\partial}{\partial x^{\nu}} \left( G^{\mu \nu} \right) + \Gamma_{\nu \sigma}^{\mu} G^{\sigma \nu} + \Gamma_{\nu \sigma}^{\nu} G^{\mu \sigma} = 0,
\end{equation}
which is illustrated in Figure \ref{fig:Figure28} for the case of \texttt{RicciTensor} and \texttt{EinsteinTensor} objects obtained from the FLRW metric.

\begin{figure}[ht]
\centering
\begin{framed}
\includegraphics[width=0.445\textwidth]{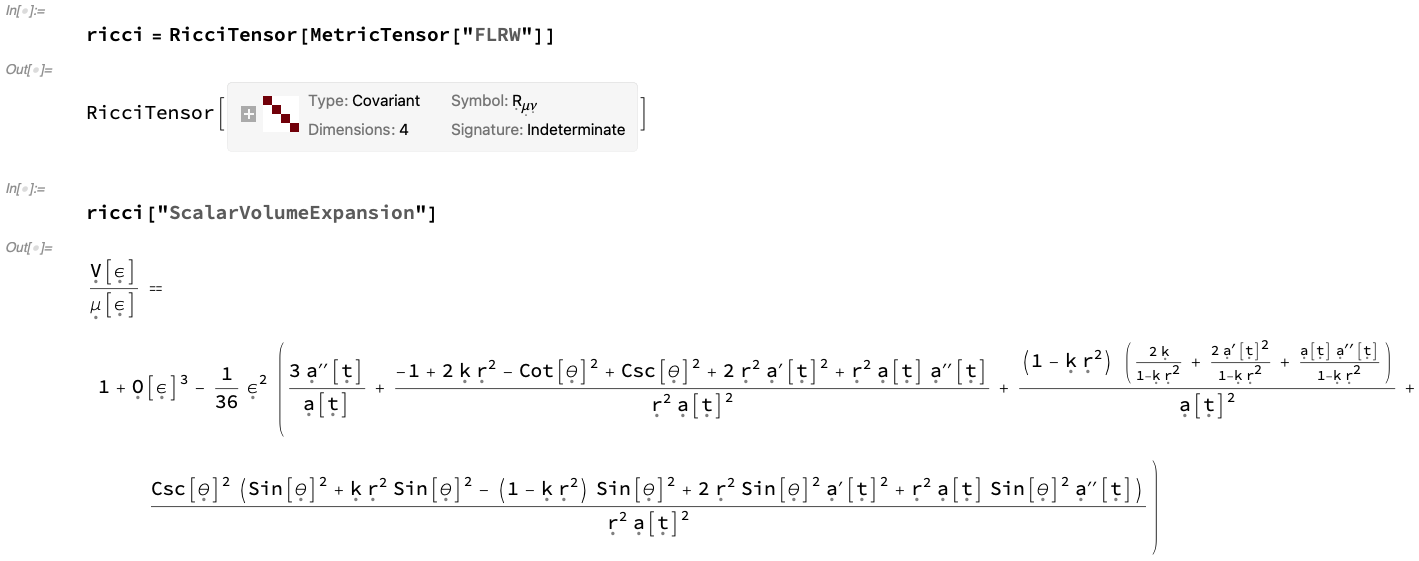}
\vrule
\includegraphics[width=0.545\textwidth]{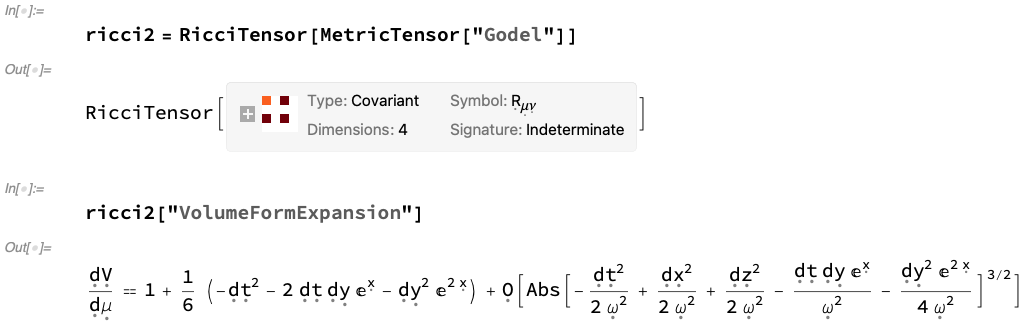}
\end{framed}
\caption{On the left, the Taylor expansion of the ratio between the volume of a small geodesic ball of radius ${\varepsilon}$ in the manifold to the volume of the corresponding ball of radius ${\varepsilon}$ in flat/Euclidean space, computed using the \texttt{RicciTensor} object for an FLRW geometry (representing e.g. a homogeneous, isotropic and uniformly expanding/contracting universe with global curvature $k$ and scale factor ${a \left( t \right)}$ in spherical polar coordinates ${\left( t, r, \theta, \phi \right)}$). On the right, the Taylor expansion along a Jacobi field of the ratio of the infinitesimal volume element/metric volume form ${d V}$ to the corresponding volume element/volume form in flat/Euclidean space ${d \mu}$, computed using the \texttt{RicciTensor} object for a G\"odel geometry (representing e.g. a rotating, dust-filled universe with global angular velocity ${\omega}$ in G\"odel's Cartesian-like coordinates ${\left( t, x, y, z \right)}$).}
\label{fig:Figure29}
\end{figure}

\begin{figure}[ht]
\centering
\begin{framed}
\includegraphics[width=0.695\textwidth]{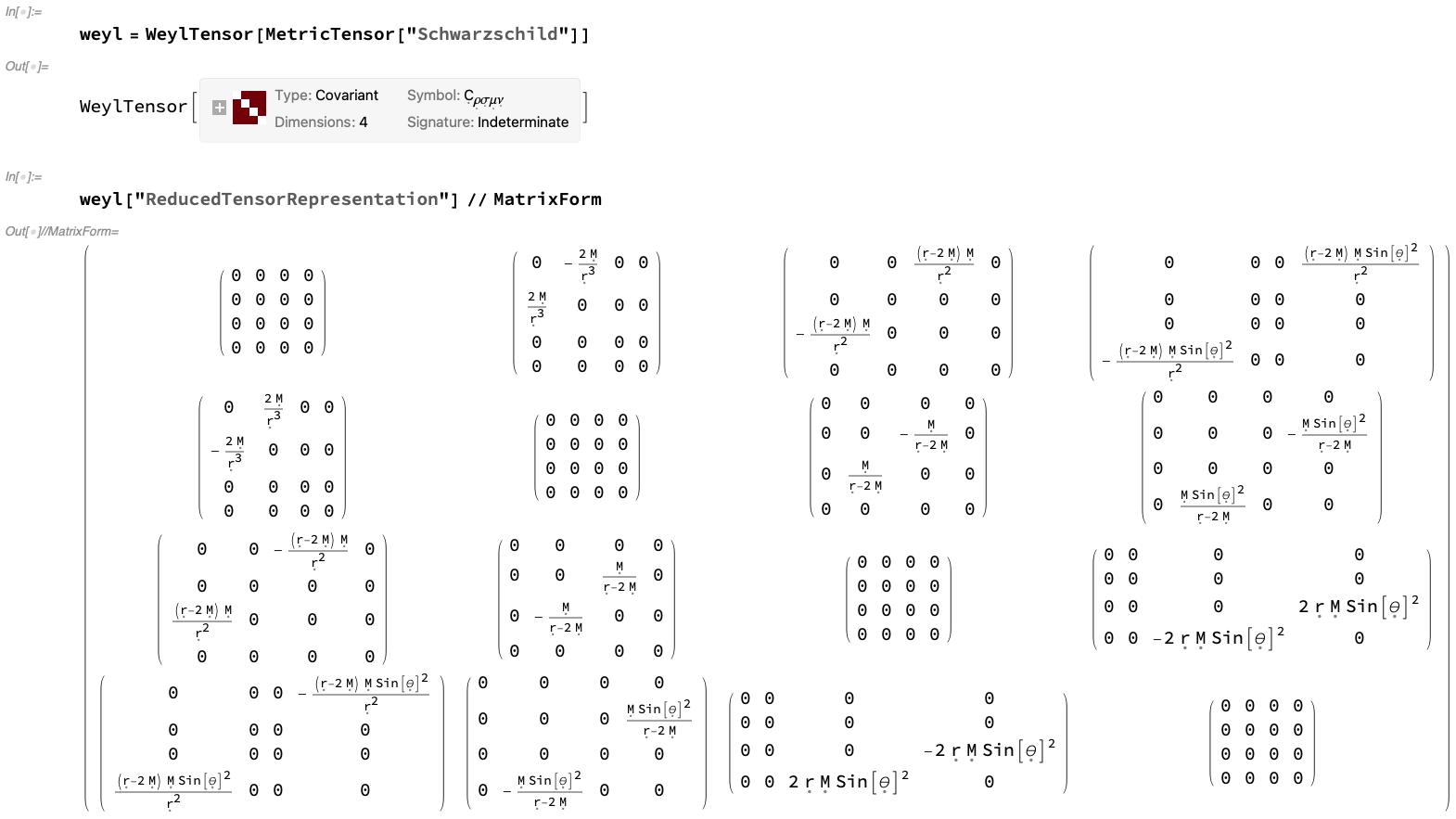}
\vrule
\includegraphics[width=0.295\textwidth]{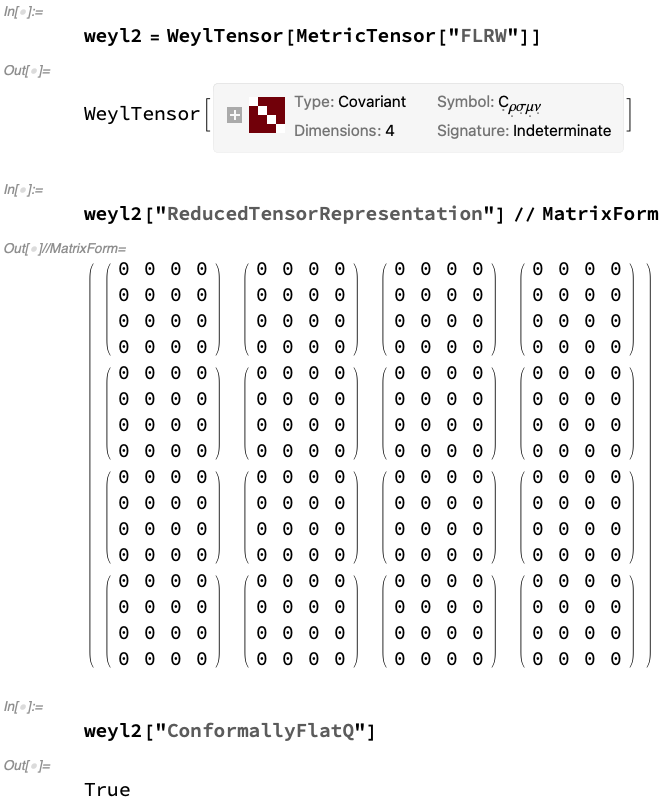}
\end{framed}
\caption{On the left, the \texttt{WeylTensor} object for a Schwarzschild geometry (representing e.g. an uncharged, non-rotating black hole of mass $M$ in Schwarzschild/spherical polar coordinates ${\left( t, r, \theta, \phi \right)}$) in explicit covariant array form, with all indices lowered/covariant (default). On the right, the \texttt{WeylTensor} object for an FLRW geometry (representing e.g. a homogeneous, isotropic and uniformly expanding/contracting universe with global curvature $k$ and scale factor ${a \left( t \right)}$ in spherical polar coordinates ${\left( t, r, \theta, \phi \right)}$) in explicit covariant array form, with all indices lowered/covariant (default), illustrating that the FLRW metric is conformally-flat.}
\label{fig:Figure30}
\end{figure}

\begin{figure}[ht]
\centering
\begin{framed}
\includegraphics[width=0.495\textwidth]{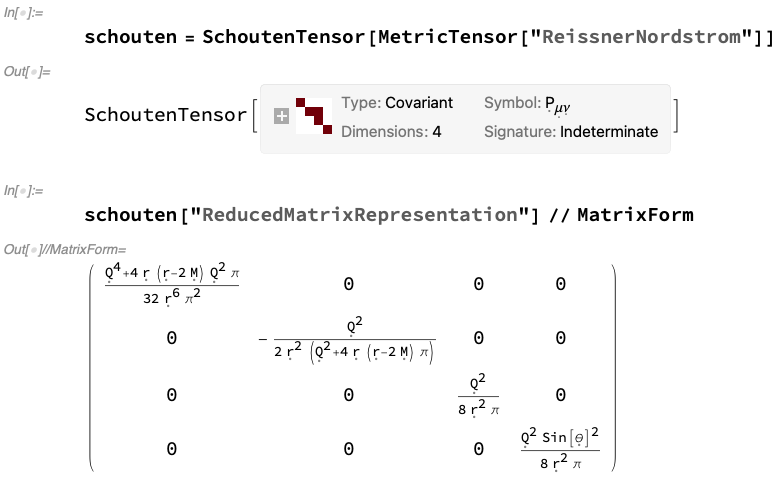}
\vrule
\includegraphics[width=0.495\textwidth]{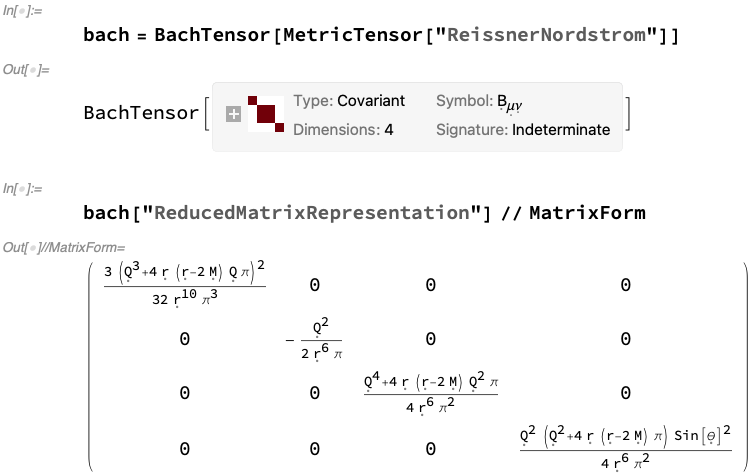}
\end{framed}
\caption{On the left, the \texttt{SchoutenTensor} object for a Reissner-Nordstr\"om geometry (representing e.g. a charged, non-rotating black hole of mass $M$ and electric charge $Q$ in Schwarzschild/spherical polar coordinates ${\left( t, r, \theta, \phi \right)}$) in explicit covariant matrix form (default). On the right, the \texttt{BachTensor} object for a Reissner-Nordstr\"om geometry (representing e.g. a charged, non-rotating black hole of mass $M$ and electric charge $Q$ in Schwarzschild/spherical polar coordinates ${\left( t, r, \theta, \phi \right)}$) in explicit covariant matrix form (default).}
\label{fig:Figure31}
\end{figure}

Both the Ricci tensor ${R_{\mu \nu}}$ and the Ricci scalar ${R = R_{\sigma}^{\sigma}}$ also admit very direct geometrical interpretations in terms of the distortions in the volumes of certain manifold regions due to the presence of curvature. More precisely, the value of the Ricci scalar ${R = R_{\sigma}^{\sigma}}$ at a point ${\mathbf{x} \in \mathcal{M}}$ in the manifold fully determines the second-order term in the Taylor expansion for the volume ${\mathrm{Vol} \left( B_{\varepsilon} \left( \mathbf{x} \right) \subset \mathcal{M} \right)}$, sometimes denoted ${V \left( \varepsilon \right)}$, of a small geodesic ball of radius ${\varepsilon}$ centered at that point, as compared to the volume ${\mathrm{Vol} \left( B_{\varepsilon} \left( \mathbf{0} \right) \subset \mathbb{R}^n \right)}$, sometimes denoted ${\mu \left( \varepsilon \right)}$, of the corresponding ball of radius ${\varepsilon}$ in ordinary flat/Euclidean space ${\mathbb{R}^n}$ (where $n$ here designates the dimension of the manifold ${\mathcal{M}}$)\cite{chavel}:

\begin{equation}
\frac{V \left( \varepsilon \right)}{\mu \left( \varepsilon \right)} = \frac{\mathrm{Vol} \left( B_{\varepsilon} \left( \mathbf{x} \right) \subset \mathcal{M} \right)}{\mathrm{Vol} \left( B_{\varepsilon} \left( \mathbf{0} \right) \subset \mathbb{R}^n \right)} = 1 - \frac{R}{6 \left( n + 2 \right)} \varepsilon^2 + O \left( \varepsilon^3 \right),
\end{equation}
in the limit as ${\varepsilon \to 0}$; likewise, the projections of the Ricci tensor ${R_{\mu \nu}}$ at a point ${\mathbf{x} \in \mathcal{M}}$ in the manifold fully determine the second-order term in the Taylor expansion along a Jacobi field (for a radial geodesic in geodesic normal coordinates) for the infinitesimal volume element/metric volume form ${d V}$, as compared to the corresponding volume element/volume form in ordinary flat/Euclidean space ${d \mu}$:

\begin{equation}
\frac{d V}{d \mu} = 1 - \frac{1}{6} R_{\mu \nu} x^{\mu} x^{\nu} + O \left( \left( x_{\sigma} x^{\sigma} \right)^{3/2} \right),
\end{equation}
in the limit as ${x_{\sigma} x^{\sigma} \to 0}$, both of which can be computed directly from the corresponding \texttt{RicciTensor} object, as demonstrated in Figure \ref{fig:Figure29}. The definition of the Ricci tensor ${R_{\mu \nu}}$, when combined with the Ricci decomposition theorem\cite{sharpe}, then allows us to construct a unique trace-free tensor of rank-4 (namely the Weyl tensor\cite{weyl2} ${C_{\rho \sigma \mu \nu}}$) by subtracting out all trace components (represented in terms of components of ${R_{\mu \nu}}$, interpreted here as the ``trace part'' of the Riemann curvature) from the full Riemann tensor ${R_{\sigma \mu \nu}^{\rho}}$:

\begin{multline}
C_{\rho \sigma \mu \nu} = R_{\rho \sigma \mu \nu} + \frac{1}{n - 2} \left( R_{\rho \nu} g_{\sigma \mu} - R_{\rho \mu} g_{\sigma \nu} + R_{\sigma \mu} g_{\rho \nu} - R_{\sigma \nu} g_{\rho \mu} \right)\\
+ \frac{R}{\left( n - 1 \right) \left( n - 2 \right)} \left( g_{\rho \mu} g_{\sigma \nu} - g_{\rho \nu} g_{\sigma \mu} \right).
\end{multline}
The Weyl tensor ${C_{\sigma \mu \nu}^{\rho}}$ is distinctive in that it remains invariant under conformal transformations of the manifold ${\mathcal{M}}$, i.e. under metric transformations of the form ${\widetilde{g_{\mu \nu}} = \Omega^2 g_{\mu \nu}}$ for some real conformal factor ${\Omega \in \mathbb{R}}$; indeed, at least in dimensions ${n \geq 4}$, the vanishing of all components of the Weyl tensor ${C_{\sigma \mu \nu}^{\rho}}$ is both a necessary and sufficient condition for the underlying manifold ${\mathcal{M}}$ to be conformally-flat, by the Weyl-Schouten theorem\cite{eisenhart} (in dimension ${n = 3}$, the Weyl tensor vanishes identically, and one must instead consider the vanishing of the rank-3 Cotton tensor\cite{cotton} ${C_{\rho \mu \nu}}$ as the appropriate necessary and sufficient condition for conformal-flatness). Intuitively, if the Ricci tensor ${R_{\mu \nu}}$ encodes the deformation in the volumes of manifold regions enclosed by tangent vectors due to the presence of curvature, the Weyl tensor ${C_{\sigma \mu \nu}^{\rho}}$ encodes the deformation in the \textit{angles} between tangent vectors due to the presence of curvature. Representations of the Weyl tensor for the Schwarzschild metric (representing e.g. an uncharged, non-rotating black hole of mass $M$ in Schwarzschild/spherical polar coordinates ${\left( t, r, \theta, \phi \right)}$) and the FLRW metric (representing e.g. a homogeneous, isotropic and uniformly expanding/contracting universe with global curvature $k$ and scale factor ${a \left( t \right)}$ in spherical polar coordinates ${\left( t, r, \theta, \phi \right)}$) using the \texttt{WeylTensor} function are shown in Figure \ref{fig:Figure30}, illustrating in particular that the FLRW metric is conformally-flat (i.e. that all components of the Weyl tensor ${C_{\rho \sigma \mu \nu}}$ vanish identically). One can also proceed to construct various related curvature tensors, such as the Schouten tensor\cite{alexakis} ${P_{\mu \nu}}$ (which, much like the Einstein tensor ${G_{\mu \nu}}$, is really nothing more than a ``trace-adjusted'' form of the Ricci tensor ${R_{\mu \nu}}$):

\begin{equation}
P_{\mu \nu} = \frac{1}{n - 2} \left( R_{\mu \nu} - \frac{R}{2 \left( n - 1 \right)} g_{\mu \nu} \right),
\end{equation}
and, from it, the Bach tensor\cite{bach} ${B_{\mu \nu}}$:

\begin{multline}
B_{\mu \nu} = P_{\rho \sigma} C_{\mu \nu}^{\rho \sigma} + \nabla^{\sigma} \left( \nabla_{\sigma} P_{\mu \nu} \right) - \nabla^{\sigma} \left( \nabla_{\mu} P_{\nu \sigma} \right) = P_{\rho \sigma} C_{\mu \nu}^{\rho \sigma} + g^{\sigma \rho} \nabla_{\rho} \left( \nabla_{\sigma} P_{\mu \nu} \right) - g^{\sigma \rho} \nabla_{\rho} \left( \nabla_{\mu} P_{\nu \sigma} \right),
\end{multline}
i.e., in expanded form:

\begin{multline}
B_{\mu \nu} = P_{\rho \sigma} C_{\mu \nu}^{\rho \sigma} + g^{\sigma \rho} \left( \frac{\partial}{\partial x^{\rho}} \left( D_{\sigma \mu \nu} \right) - \Gamma_{\rho \sigma}^{\lambda} D_{\lambda \mu \nu} - \Gamma_{\rho \mu}^{\lambda} D_{\sigma \lambda \nu} - \Gamma_{\rho \nu}^{\lambda} D_{\sigma \mu \lambda} \right)\\
- g^{\sigma \rho} \left( \frac{\partial}{\partial x^{\rho}} \left( D_{\mu \nu \sigma} \right) - \Gamma_{\rho \mu}^{\lambda} D_{\lambda \nu \sigma} - \Gamma_{\rho \nu}^{\lambda} D_{\mu \lambda \sigma} - \Gamma_{\rho \sigma}^{\lambda} D_{\mu \nu \lambda} \right),
\end{multline}
where we have introduced, for the sake of notational convenience, the rank-3 tensor ${D_{\sigma \mu \nu}}$ consisting of covariant derivatives of the Schouten tensor ${P_{\mu \nu}}$:

\begin{equation}
D_{\sigma \mu \nu} = \nabla_{\sigma} P_{\mu \nu} = \frac{\partial}{\partial x^{\sigma}} \left( P_{\mu \nu} \right) - \Gamma_{\sigma \mu}^{\lambda} P_{\lambda \nu} - \Gamma_{\sigma \nu}^{\lambda} P_{\mu \lambda}.
\end{equation}
The geometrical significance of the Schouten tensor ${P_{\mu \nu}}$ is that, under conformal transformations of the form ${\widetilde{g_{\mu \nu}} = \Omega^2 g_{\mu \nu}}$, the Schouten tensor transforms as:

\begin{equation}
\widetilde{P_{\mu \nu}} = P_{\mu \nu} - \nabla_{\mu} \left( \Omega^{-1} \frac{\partial \Omega}{\partial x^{\nu}} \right) + \left( \Omega^{-1} \frac{\partial \Omega}{\partial x^{\mu}} \right) \left( \Omega^{-1} \frac{\partial \Omega}{\partial x^{\nu}} \right) - \frac{1}{2} g^{\rho \sigma} \left( \Omega^{-1} \frac{\partial \Omega}{\partial x^{\rho}} \right) \left( \Omega^{-1} \frac{\partial \Omega}{\partial x^{\sigma}} \right) g_{\mu \nu},
\end{equation}
or, in slightly expanded form:

\begin{multline}
\widetilde{P_{\mu \nu}} = P_{\mu \nu} - \frac{\partial}{\partial x^{\mu}} \left( \Omega^{-1} \frac{\partial \Omega}{\partial x^{\nu}} \right) - \Gamma_{\mu \nu}^{\sigma} \left( \Omega^{-1} \frac{\partial \Omega}{\partial x^{\sigma}} \right) + \left( \Omega^{-1} \frac{\partial \Omega}{\partial x^{\mu}} \right) \left( \Omega^{-1} \frac{\partial \Omega}{\partial x^{\nu}} \right)\\
- \frac{1}{2} g^{\rho \sigma} \left( \Omega^{-1} \frac{\partial \Omega}{\partial x^{\rho}} \right) \left( \Omega^{-1} \frac{\partial \Omega}{\partial x^{\sigma}} \right) g_{\mu \nu},
\end{multline}
and therefore, by the Weyl-Schouten theorem again, a necessary and sufficient condition for the underlying manifold ${\mathcal{M}}$ to be conformally-flat in dimension ${n = 3}$ is that both the Schouten tensor ${P_{\mu \nu}}$ and its covariant derivatives ${\nabla_{\rho} P_{\mu \nu}}$ be symmetric (i.e. that ${P_{\mu \nu}}$ be a Codazzi tensor). On the other hand, the geometrical significance of the Bach tensor ${B_{\mu \nu}}$ is that (much like the full Weyl tensor ${C_{\rho \sigma \mu \nu}}$ itself) it is trace-free and conformally-invariant in dimension ${n = 4}$, yet is only of rank-2, and is known to be algebraically independent of the Weyl tensor\cite{szekeres}. A representation of the Schouten and Bach tensors for the Reissner-Nordstr\"om metric (representing e.g. a charged, non-rotating black hole of mass $M$ and electric charge $Q$ in Schwarzschild/spherical polar coordinates ${\left( t, r, \theta, \phi \right)}$) using the \texttt{SchoutenTensor} and \texttt{BachTensor} functions is shown in Figure \ref{fig:Figure31}.

\begin{figure}[ht]
\centering
\begin{framed}
\includegraphics[width=0.545\textwidth]{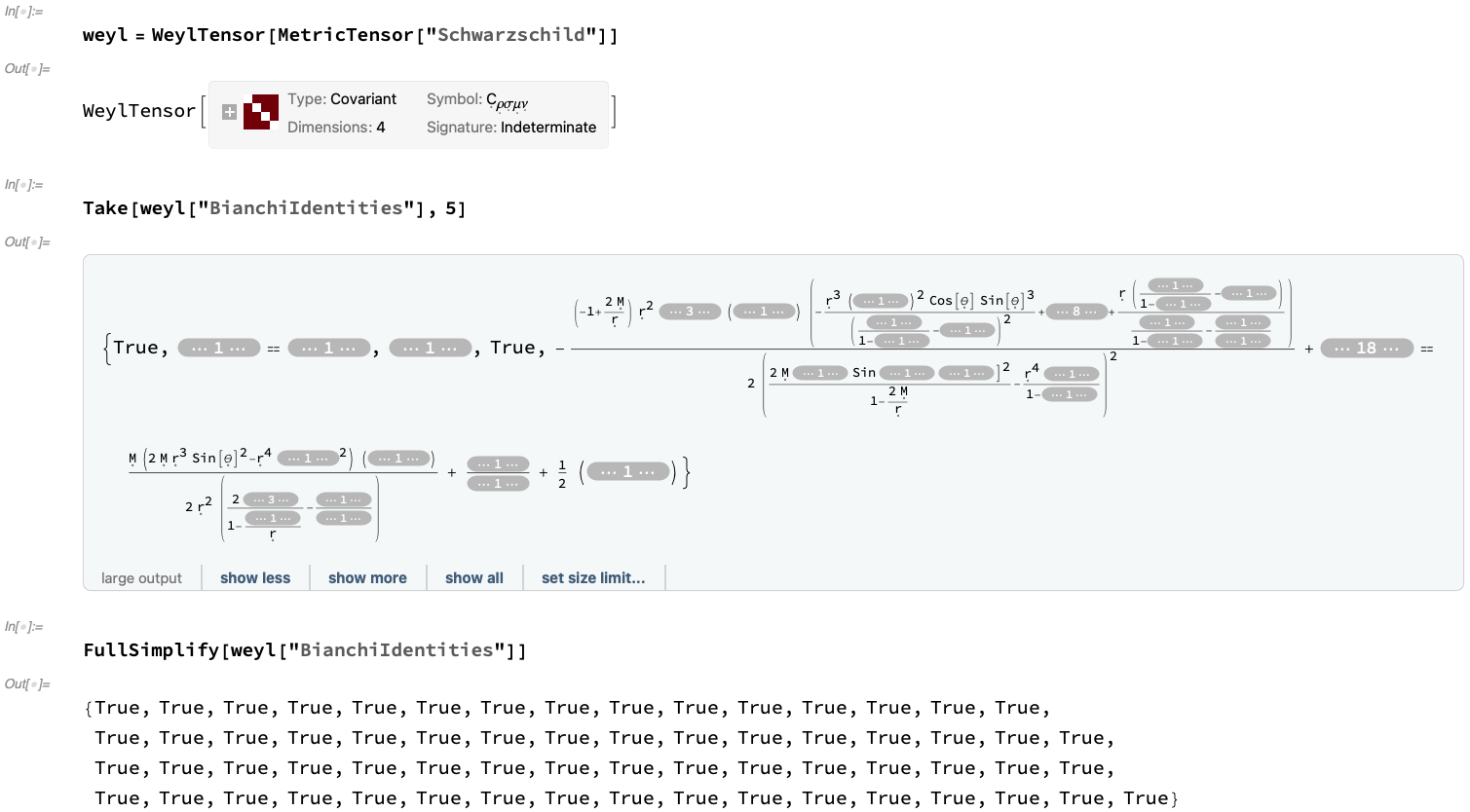}
\vrule
\includegraphics[width=0.445\textwidth]{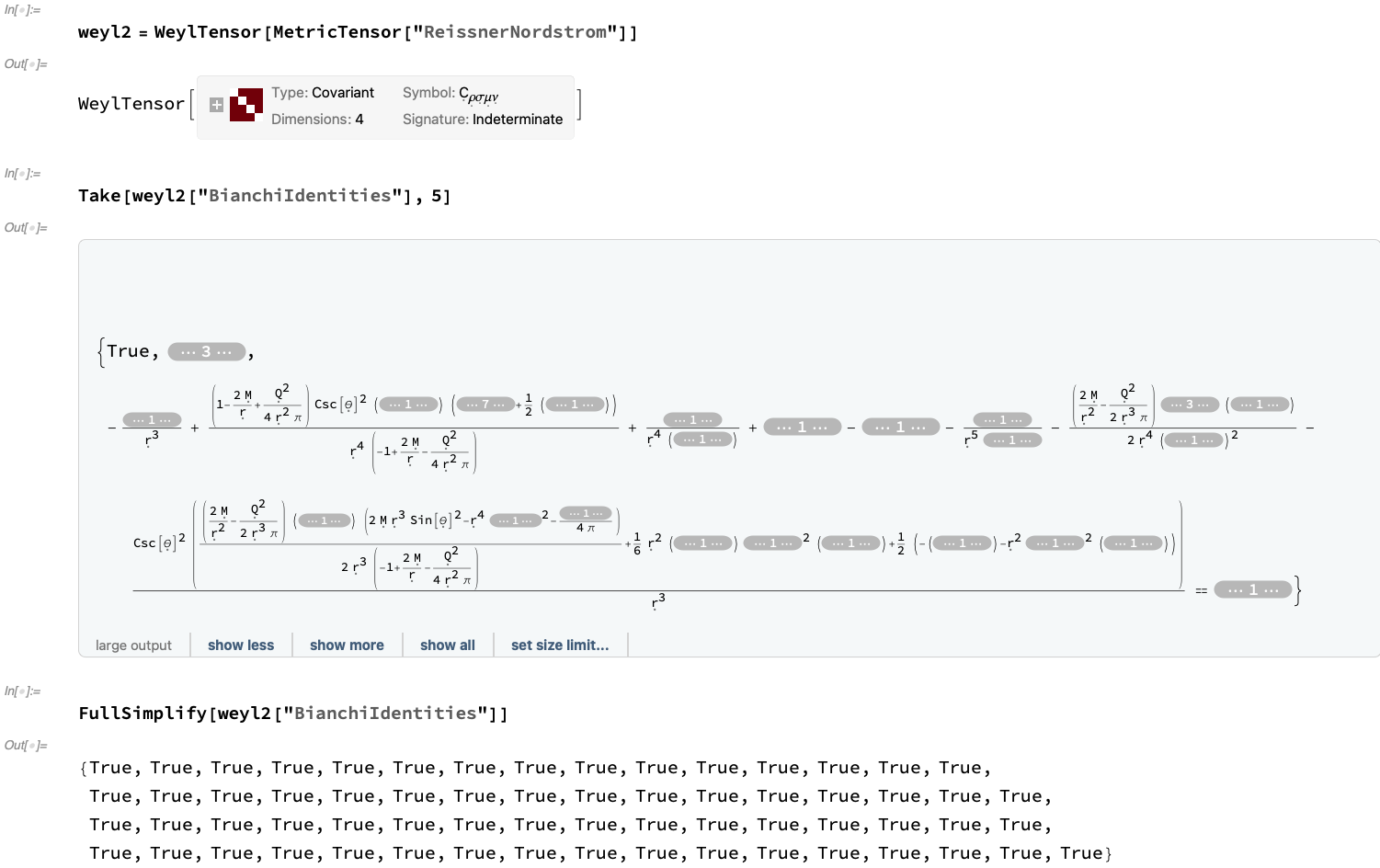}
\end{framed}
\caption{On the left, the list of (partially-contracted) Bianchi identities relating the covariant divergence of the \texttt{WeylTensor} object for a Schwarzschild geometry (representing e.g. an uncharged, non-rotating black hole of mass $M$ in Schwarzschild/spherical polar coordinates ${\left( t, r, \theta, \phi \right)}$) to the covariant derivatives of the corresponding \texttt{SchoutenTensor} object for the same geometry, together with a verification that they all hold identically. On the right, the list of (partially-contracted) Bianchi identities relating the covariant divergence of the \texttt{WeylTensor} object for a Reissner-Nordstr\"om geometry (representing e.g. a charged, non-rotating black hole of mass $M$ and electric charge $Q$ in Schwarzschild/spherical polar coordinates ${\left( t, r, \theta, \phi \right)}$) to the covariant derivatives of the corresponding \texttt{SchoutenTensor} object for the same geometry, together with a verification that they all hold identically.}
\label{fig:Figure32}
\end{figure}

\begin{figure}[ht]
\centering
\begin{framed}
\includegraphics[width=0.495\textwidth]{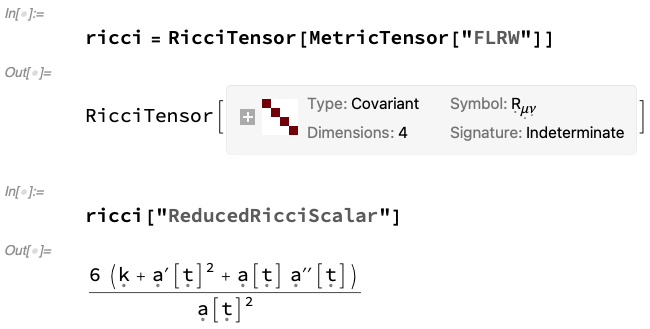}
\vrule
\includegraphics[width=0.495\textwidth]{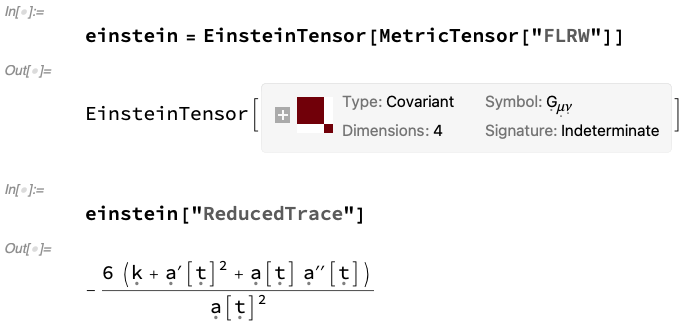}
\end{framed}
\caption{On the left, the Ricci scalar computed using the \texttt{RicciTensor} object for an FLRW geometry (representing e.g. a homogeneous, isotropic and uniformly expanding/contracting universe with global curvature $k$ and scale factor ${a \left( t \right)}$ in spherical polar coordinates ${\left( t, r, \theta, \phi \right)}$). On the right, the Einstein trace computed using the \texttt{EinsteinTensor} object for an FLRW geometry (representing e.g. a homogeneous, isotropic and uniformly expanding/contracting universe with global curvature $k$ and scale factor ${a \left( t \right)}$ in spherical polar coordinates ${\left( t, r, \theta, \phi \right)}$), illustrating that it is indeed equal to the negative trace of the corresponding \texttt{RicciTensor} object in four dimensions.}
\label{fig:Figure33}
\end{figure}

\begin{figure}[ht]
\centering
\begin{framed}
\includegraphics[width=0.395\textwidth]{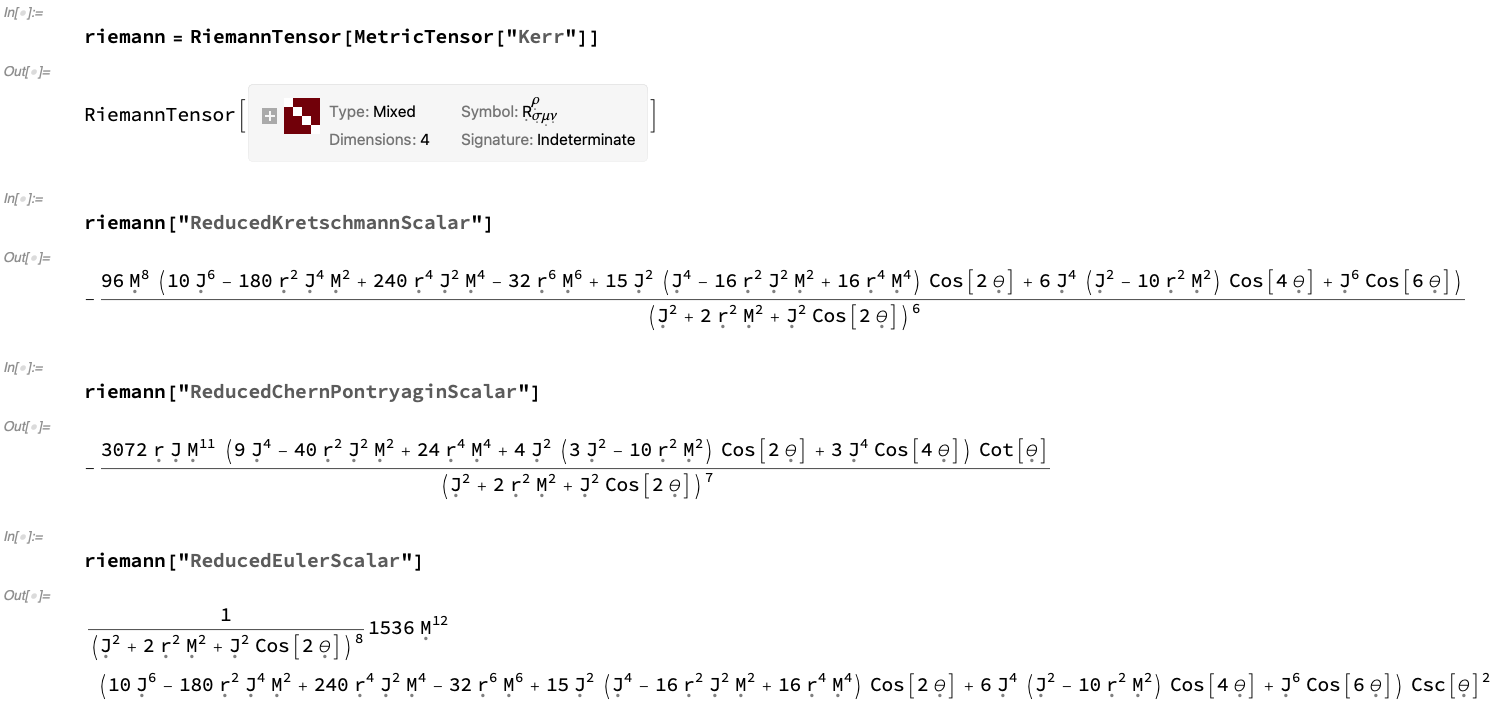}
\vrule
\includegraphics[width=0.595\textwidth]{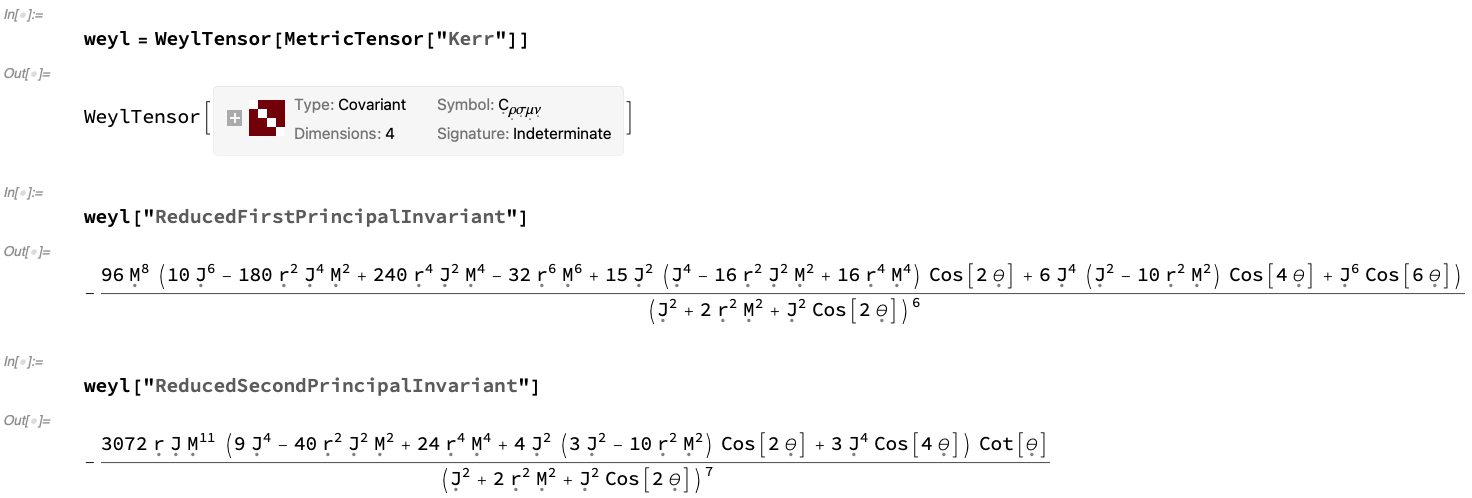}
\end{framed}
\caption{On the left, the Kretschmann scalar, Chern-Pontryagin scalar and Euler scalar computed using the \texttt{RiemannTensor} object for a Kerr geometry (representing e.g. an uncharged, spinning black hole of mass $M$ and angular momentum $J$ in Boyer-Lindquist/oblate spheroidal coordinates ${\left( t, r, \theta, \phi \right)}$). On the right, the first and second principal Weyl tensor invariants computed using the \texttt{WeylTensor} object for a Kerr geometry (representing e.g. an uncharged, spinning black hole of mass $M$ and angular momentum $J$ in Boyer-Lindquist/oblate spheroidal coordinates ${\left( t, r, \theta, \phi \right)}$).}
\label{fig:Figure34}
\end{figure}

By taking appropriate traces of the (differential) Bianchi identities on the Riemann tensor ${R_{\sigma \mu \nu}^{\rho}}$, one can proceed to derive a (partially-contracted) form of the Bianchi identities, asserting a relationship between the covariant divergence of the Weyl tensor ${C_{\rho \sigma \mu \nu}}$ and the covariant derivatives of the corresponding Schouten tensor ${P_{\mu \nu}}$:

\begin{equation}
\nabla_{\rho} C_{\sigma \mu \nu}^{\rho} = \left( n - 3 \right) \left( \nabla_{\mu} P_{\nu \sigma} - \nabla_{\nu} P_{\mu \sigma} \right),
\end{equation}
i.e., in expanded form:

\begin{multline}
\frac{\partial}{\partial x^{\rho}} \left( C_{\sigma \mu \nu}^{\rho} \right) + \Gamma_{\rho \lambda}^{\rho} C_{\sigma \mu \nu}^{\lambda} - \Gamma_{\rho \sigma}^{\lambda} C_{\lambda \mu \nu}^{\rho} - \Gamma_{\rho \mu}^{\lambda} C_{\sigma \lambda \nu}^{\rho} - \Gamma_{\rho \nu}^{\lambda} C_{\sigma \mu \lambda}^{\rho}\\
= \left( n - 3 \right) \left( \left( \frac{\partial}{\partial x^{\mu}} \left( P_{\nu \sigma} \right) - \Gamma_{\mu \nu}^{\lambda} P_{\lambda \sigma} - \Gamma_{\mu \sigma}^{\lambda} P_{\nu \lambda} \right) - \left( \frac{\partial}{\partial x^{\nu}} \left( P_{\mu \sigma} \right) - \Gamma_{\nu \mu}^{\lambda} P_{\lambda \sigma} - \Gamma_{\nu \sigma}^{\lambda} P_{\mu \lambda} \right) \right),
\end{multline}
which is illustrated in Figure \ref{fig:Figure32} for the case of \texttt{WeylTensor} objects obtained from the Schwarzschild and Reissner-Nordstr\"om metrics. Due to their rank-2 nature, the Ricci tensor ${R_{\mu \nu}}$ and the Einstein tensor ${G_{\mu \nu}}$ are associated with only one (linear) scalar curvature invariant each, namely the Ricci scalar ${R = R_{\sigma}^{\sigma}}$\cite{aubin} and the Einstein trace ${G = G_{\sigma}^{\sigma}}$ (which, due to the trace-reversed nature of the relationship between ${R_{\mu \nu}}$ and ${G_{\mu \nu}}$ is simply equal to the Ricci scalar modulo a sign in four dimensions, i.e. ${G = - R}$ in a four-dimensional geometry, and ${G = \left( 1 - \frac{n}{2} \right) R}$ more generally in $n$ dimensions), respectively, as shown in Figure \ref{fig:Figure33} for \texttt{RicciTensor} and \texttt{EinsteinTensor} objects obtained from the FLRW metric (representing e.g. a homogeneous, isotropic and uniformly expanding/contracting universe of global curvature $k$ and scale factor ${a \left( t \right)}$ in spherical polar coordinates ${\left( t, r, \theta, \phi \right)}$). However, the rank-4 Riemann (i.e. ${R_{\sigma \mu \nu}^{\rho}}$) and Weyl (i.e. ${C_{\rho \sigma \mu \nu}}$) tensors each admit multiple (quadratic) scalar invariants\cite{cherubini}\cite{coley}, namely the Kretschmann scalar (typically denoted ${K_1}$), the Chern-Pontryagin scalar (typically denoted ${K_2}$) and the Euler scalar (typically denoted ${K_3}$) in the case of the Riemann tensor:

\begin{equation}
K_1 = R_{\rho \sigma \mu \nu} R^{\rho \sigma \mu \nu}, \qquad K_2 = {}^{\star} R_{\rho \sigma \mu \nu} R^{\rho \sigma \mu \nu}, \qquad \text{ and } \qquad K_3 = {}^{\star} R {}^{\star} {}_{\rho \sigma \mu \nu} R^{\rho \sigma \mu \nu}
\end{equation}
and the first (typically denoted ${I_1}$) and second (typically denoted ${I_2}$) principal invariants in the case of the Weyl tensor:

\begin{equation}
I_1 = C_{\rho \sigma \mu \nu} C^{\rho \sigma \mu \nu}, \qquad \text{ and } \qquad I_2 = {}^{\star} C_{\rho \sigma \mu \nu} C^{\rho \sigma \mu \nu},
\end{equation}
with the third invariant being degenerate:

\begin{equation}
I_3 = {}^{\star} C {}^{\star} {}_{\rho \sigma \mu \nu} C^{\rho \sigma \mu \nu} = - C_{\rho \sigma \mu \nu} C^{\rho \sigma \mu \nu} = - I_1,
\end{equation}
due to the algebraic symmetries of the Weyl tensor. In the above, the ${\star}$ operator is used to denote the (either left-sided or double-sided) Hodge dual of the corresponding tensor within its associated Grassmann algebra, and consequently the Chern-Pontryagin scalar ${K_2}$, Euler scalar ${K_3}$ and second principal Weyl invariant ${I_2}$ are definable only in four dimensions, hence allowing us to rewrite these invariants explicitly in terms of the totally-antisymmetric Levi-Civita symbol ${\varepsilon_{\rho \sigma \mu \nu}}$ as:

\begin{equation}
K_2 = R_{\rho \sigma \mu \nu} \varepsilon^{\rho \sigma \alpha \beta} R_{\alpha \beta}^{\mu \nu},\qquad K_3 = R_{\rho \sigma \mu \nu} R_{\alpha \beta \gamma \delta} \varepsilon^{\rho \sigma \alpha \beta} \varepsilon^{\mu \nu \gamma \delta}, \qquad \text{ and } \qquad I_2 = C_{\rho \sigma \nu \nu} \varepsilon^{\rho \sigma \alpha \beta} C_{\alpha \beta}^{\mu \nu}.
\end{equation}
The Kretschmann scalar ${K_1}$ is commonly used for representing the spacetime curvature within vacuum black hole spacetimes where the Ricci curvature ${R_{\mu \nu}}$ vanishes identically (such as the Schwarzschild and Kerr metrics); the integrals of the Chern-Pontryagin and Euler scalars ${K_2}$ and ${K_3}$ encode more topological information, being directly related to the instanton number and the Euler characteristic of the underlying manifold ${\mathcal{M}}$, respectively. In Figure \ref{fig:Figure34}, the Kretschmann, Chern-Pontyagin and Euler scalars of the \texttt{RiemannTensor} object, and the first and second principal invariants of the \texttt{WeylTensor} object, are computed for the case of the Kerr metric (representing e.g. an uncharged, spinning black hole of mass $M$ and angular momentum $J$ in Boyer-Lindquist/oblate spheroidal coordinates ${\left( t, r, \theta, \phi \right)}$).

\begin{figure}[ht]
\centering
\begin{framed}
\includegraphics[width=0.545\textwidth]{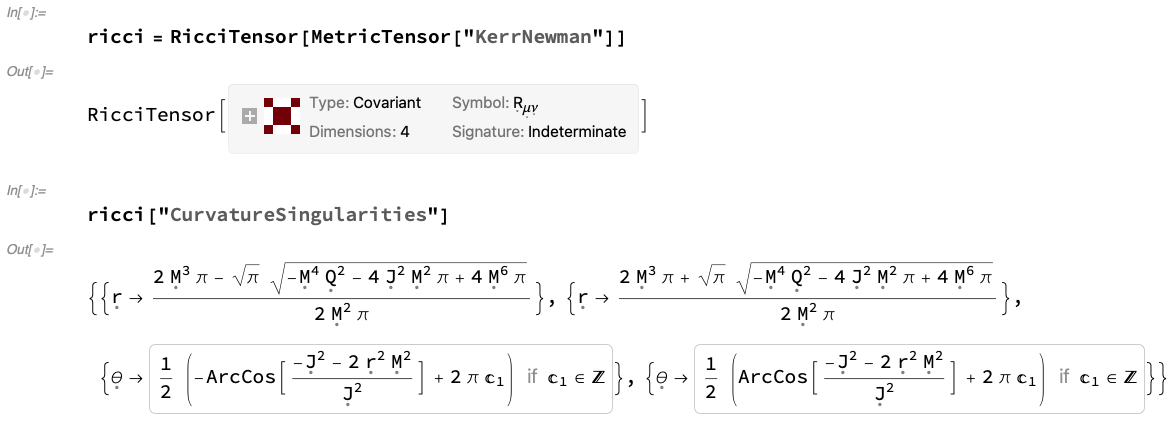}
\vrule
\includegraphics[width=0.445\textwidth]{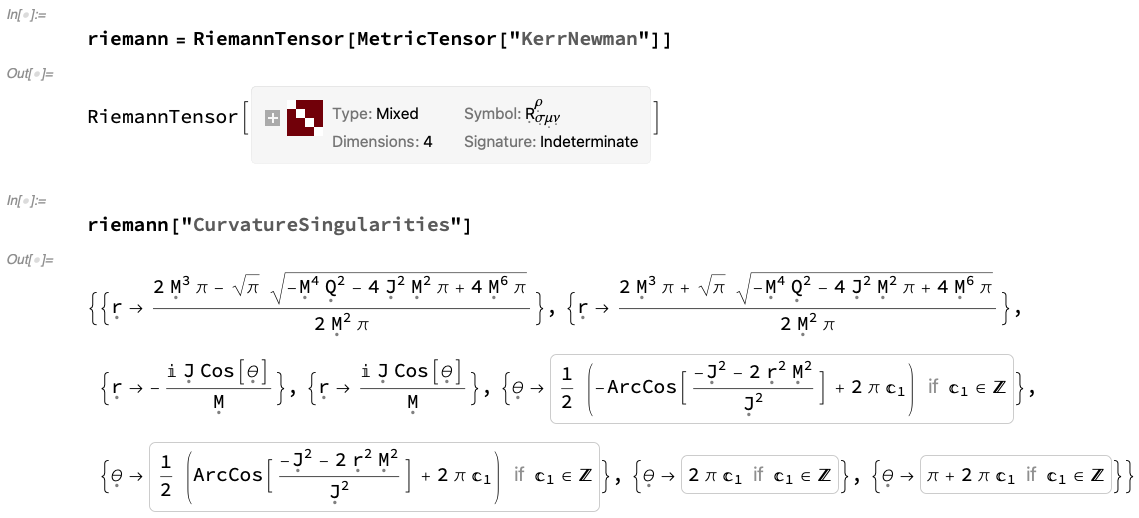}
\end{framed}
\caption{On the left, the list of coordinate values that cause the \texttt{RicciTensor} object for a Kerr-Newman geometry (representing e.g. a charged, spinning black hole of mass $M$, angular momentum $J$ and electric charge $Q$ in Boyer-Lindquist/oblate spheroidal coordinates ${\left( t, r, \theta, \phi \right)}$) to become singular. On the right, the list of coordinate values that cause the \texttt{RiemannTensor} object for a Kerr-Newman geometry (representing e.g.a charged, spinning black hole of mass $M$, angular momentum $J$ and electric charge $Q$ in Boyer-Lindquist/oblate spheroidal coordinates ${\left( t, r, \theta, \phi \right)}$) to become singular.}
\label{fig:Figure73}
\end{figure}

\begin{figure}[ht]
\centering
\begin{framed}
\includegraphics[width=0.495\textwidth]{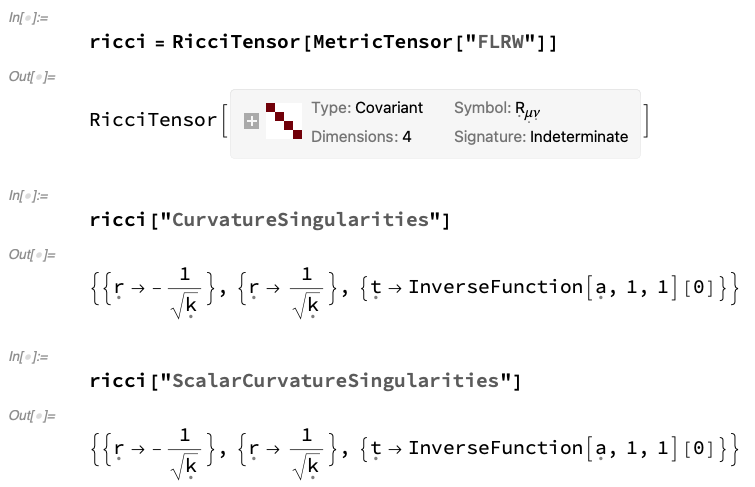}
\vrule
\includegraphics[width=0.495\textwidth]{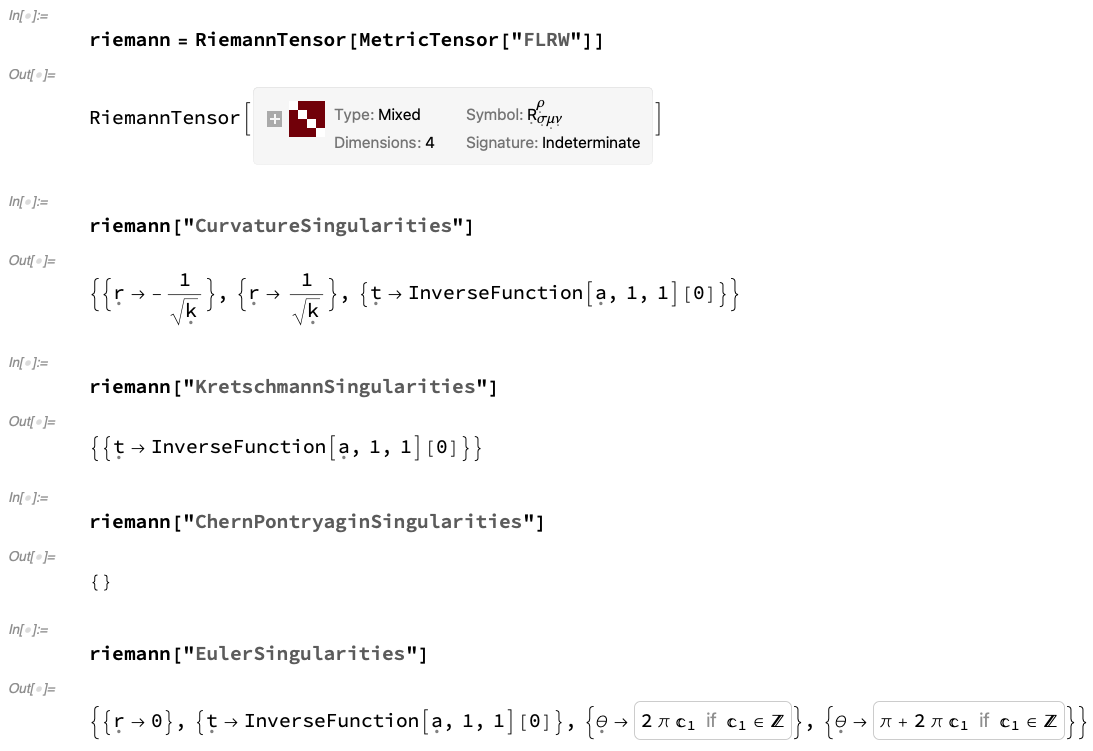}
\end{framed}
\caption{On the left, the lists of coordinate values that cause the \texttt{RicciTensor} object and its trace, i.e. the Ricci scalar, for an FLRW geometry (representing e.g. a homogeneous, isotropic and uniformly expanding/contracting universe with global curvature $k$ and scale factor ${a \left( t \right)}$ in spherical polar coordinates ${\left( t, r, \theta, \phi \right)}$) to become singular. On the right, the lists of coordinate values that cause the \texttt{RiemannTensor} object and its principal invariants, i.e. the Kretschmann, Chern-Pontryagin and Euler scalars, for an FLRW geometry (representing e.g. a homogeneous, isotropic and uniformly expanding/contracting universe with global curvature $k$ and scale factor ${a \left( t \right)}$ in spherical polar coordinates ${\left( t, r, \theta, \phi \right)}$) to become singular.}
\label{fig:Figure74}
\end{figure}

Note that, when performing automatic algebraic simplification of higher-rank tensor expressions (especially those involving arbitrary tensor products of \texttt{RiemannTensor} objects, \texttt{WeylTensor} objects and their covariant derivatives), \textsc{Gravitas} employs a custom implementation of the classic Butler-Portugal algorithm\cite{butler}\cite{manssur} from computational group theory to reduce higher-rank tensors to a canonical form with respect to the various permutation symmetries of their indices, such as the skew and interchange symmetries:

\begin{equation}
R_{\rho \sigma \mu \nu} = - R_{\rho \sigma \nu \mu} = - R_{\sigma \rho \mu \nu} = R_{\mu \nu \rho \sigma},
\end{equation}
in the case of the Riemann tensor ${R_{\sigma \mu \nu}^{\rho}}$. Note, moreover, that, just as with \texttt{MetricTensor} and its ability to detect coordinate singularities in generic (pseudo-)Riemannian manifolds, the curvature singularity structure of any (pseudo-)Riemannian manifold ${\mathcal{M}}$ can also be extracted from the various curvature tensors that \textsc{Gravitas} supports. For instance, the Ricci tensor ${R_{\mu \nu}}$ for the Kerr-Newman metric (representing e.g. a charged, spinning black hole of mass $M$, angular momentum $J$ and electric charge $Q$ in Boyer-Lindquist/oblate spheroidal coordinates ${\left( t, r, \theta, \phi \right)}$) becomes singular at both its interior and exterior horizons ${r_{H}^{\pm}}$:

\begin{equation}
r_{H}^{\pm} = M \pm \sqrt{M^2 - \left( \frac{J}{M} \right)^2 - \frac{Q^2}{4 \pi}},
\end{equation}
and at the angular coordinate values:

\begin{equation}
\forall C \in \mathbb{Z}, \qquad \theta = \pm \frac{1}{2} \left( \arccos \left( \frac{J^2 - 2 r^2 M^2}{J^2} \right) + 2 \pi C \right),
\end{equation}
whereas the full Riemann tensor ${R_{\sigma \mu \nu}^{\rho}}$ for the Kerr-Newman metric also admits additional curvature singularities that are not present in the Ricci tensor ${R_{\mu \nu}}$, specifically at the interior and exterior boundaries of its ergosphere ${r_{E}^{\pm}}$:

\begin{equation}
r_{E}^{\pm} = M \pm \sqrt{M^2 - \left( \frac{J}{M} \right)^2 \cos^2 \left( \theta \right) - \frac{Q^2}{4 \pi}},
\end{equation}
as well as at the angular coordinate values:

\begin{equation}
\forall C \in \mathbb{Z}, \qquad \theta = 2 \pi C, \qquad \text{ and } \qquad \theta = \pi + 2 \pi C,
\end{equation}
the latter of which are notable for not being coordinate singularities of the Kerr-Newman metric itself (in other words they are pure curvature singularities of the Riemann tensor, induced solely by taking partial derivatives of the metric) as shown in Figure \ref{fig:Figure73}, computed directly from the corresponding \texttt{RicciTensor} and \texttt{RiemannTensor} objects. Curvature singularities can also manifest within various scalar curvature invariants on the manifold, as shown in Figure \ref{fig:Figure74}, which shows the singularity structure of the FLRW metric (representing e.g. a homogeneous, isotropic and uniformly expanding/contracting universe with global curvature $k$ and scale factor ${a \left( t \right)}$ in spherical polar coordinates ${\left( t, r, \theta, \phi \right)}$) from the perspective of the Ricci scalar $R$, the Kretschmann scalar ${K_1}$, the Chern-Pontryagin scalar ${K_2}$ and the Euler scalar ${K_3}$, computed again using \texttt{RicciTensor} and \texttt{RiemannTensor}.

\clearpage

\section{Matter Fields and the Einstein Field Equations}
\label{sec:Section3}

\begin{figure}[ht]
\centering
\begin{framed}
\includegraphics[width=0.495\textwidth]{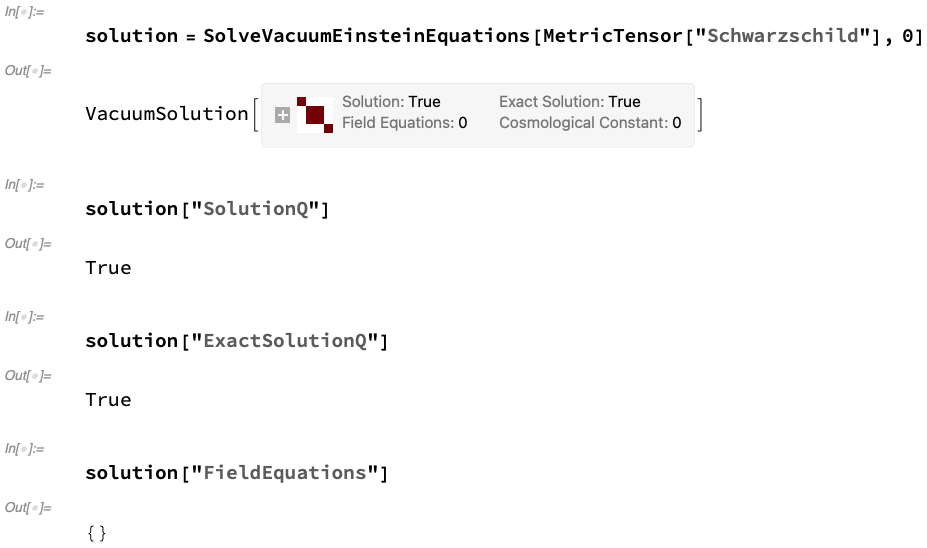}
\vrule
\includegraphics[width=0.495\textwidth]{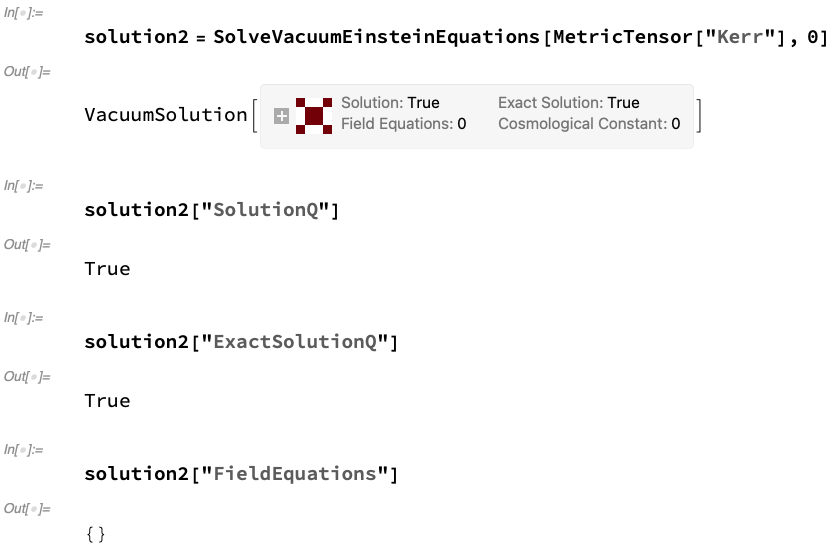}
\end{framed}
\caption{On the left, the \texttt{VacuumSolution} object for a Schwarzschild geometry (representing e.g. an uncharged, non-rotating black hole of mass $M$ in Schwarzschild/spherical polar coordinates ${\left( t, r, \theta, \phi \right)}$) computed using \texttt{SolveVacuumEinsteinEquations}, illustrating that the Schwarzschild metric is an exact solution to the vacuum Einstein field equations, with zero cosmological constant. On the right, the \texttt{VacuumSolution} object for a Kerr geometry (representing e.g. an uncharged, spinning black hole of mass $M$ and angular momentum $J$ in Boyer-Lindquist/oblate spheroidal coordinates ${\left( t, r, \theta, \phi \right)}$) computed using \texttt{SolveVacuumEinsteinEquations}, illustrating that the Kerr metric is an exact solution to the vacuum Einstein field equations, with zero cosmological constant.}
\label{fig:Figure35}
\end{figure}

\begin{figure}[ht]
\centering
\begin{framed}
\includegraphics[width=0.495\textwidth]{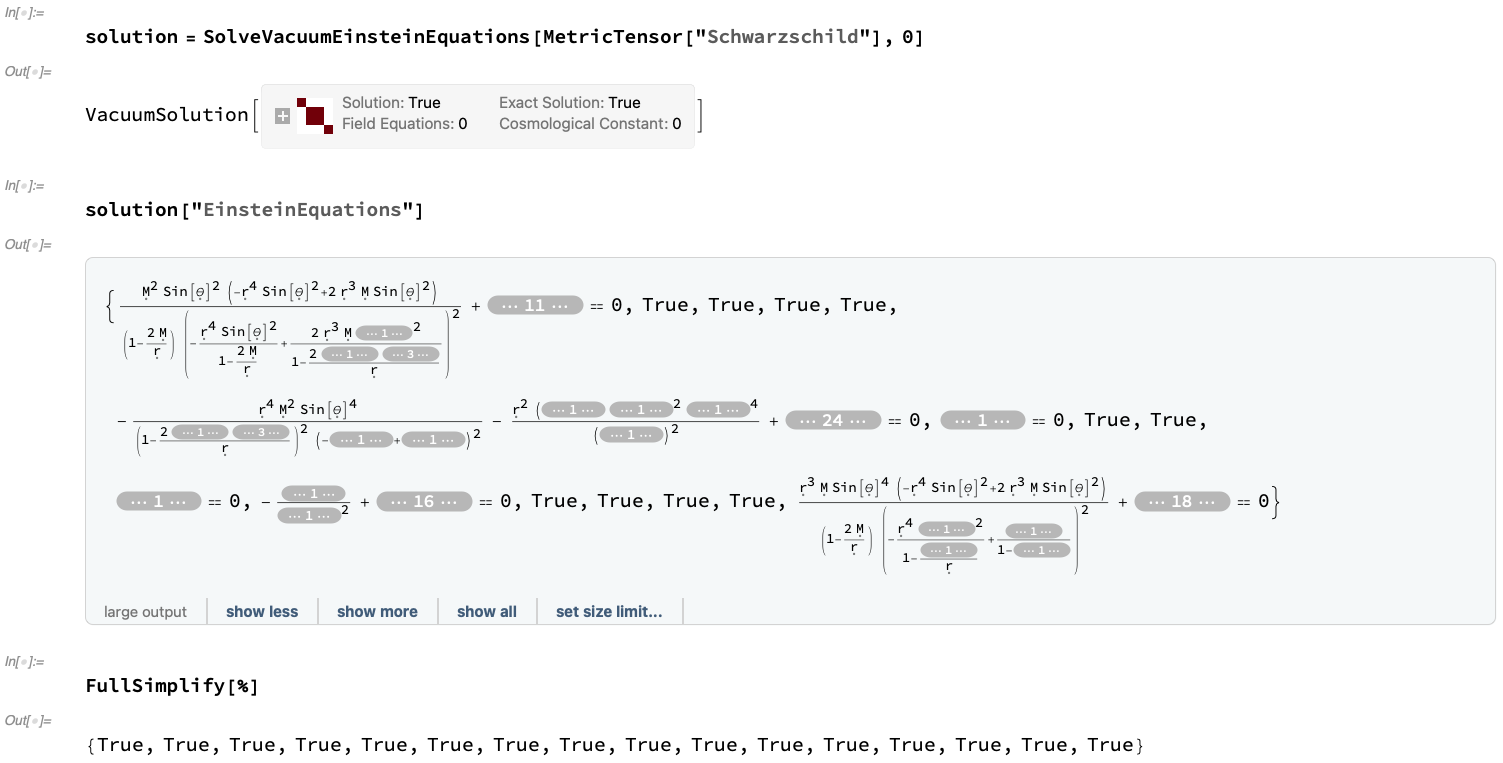}
\vrule
\includegraphics[width=0.495\textwidth]{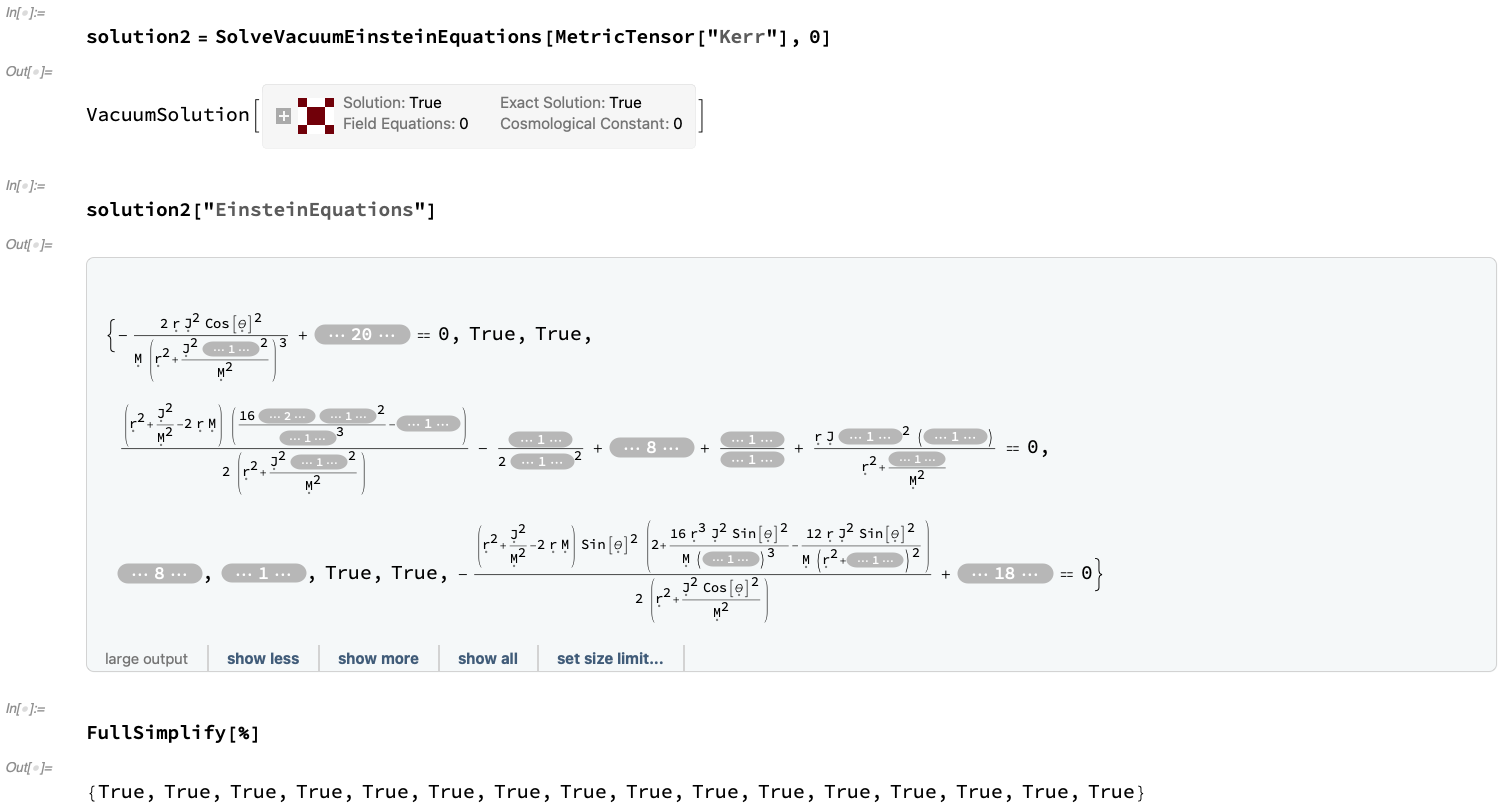}
\end{framed}
\caption{On the left, the list of Einstein field equations asserting that the Einstein tensor, with zero cosmological constant, vanishes, computed using the \texttt{VacuumSolution} object for a Schwarzschild geometry (representing e.g. an uncharged, non-rotating black hole of mass $M$ in Schwarzschild/spherical polar coordinates ${\left( t, r, \theta, \phi \right)}$), together with a verification that they all hold identically. On the right, the list of Einstein field equations asserting that the Einstein tensor, with zero cosmological constant, vanishes, computed using the \texttt{VacuumSolution} object for a Kerr geometry (representing e.g. an uncharged, spinning black hole of mass $M$ and angular momentum $J$ in Boyer-Lindquist/oblate spheroidal coordinates ${\left( t, r, \theta, \phi \right)}$), together with a verification that they all hold identically.}
\label{fig:Figure36}
\end{figure}

\begin{figure}[ht]
\centering
\begin{framed}
\includegraphics[width=0.595\textwidth]{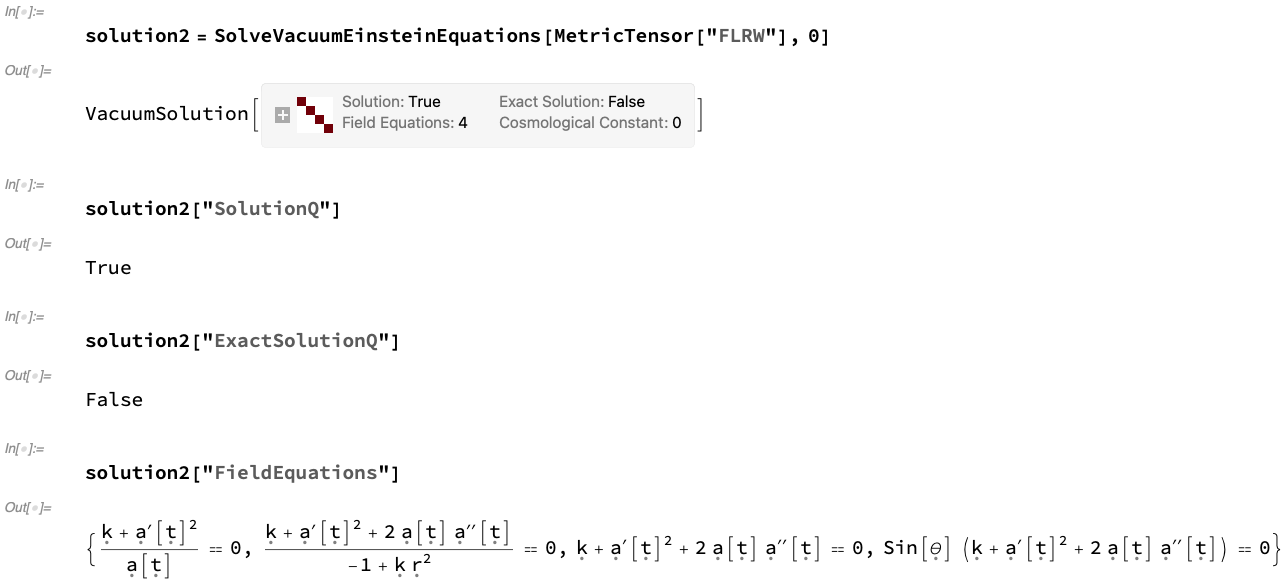}
\vrule
\includegraphics[width=0.395\textwidth]{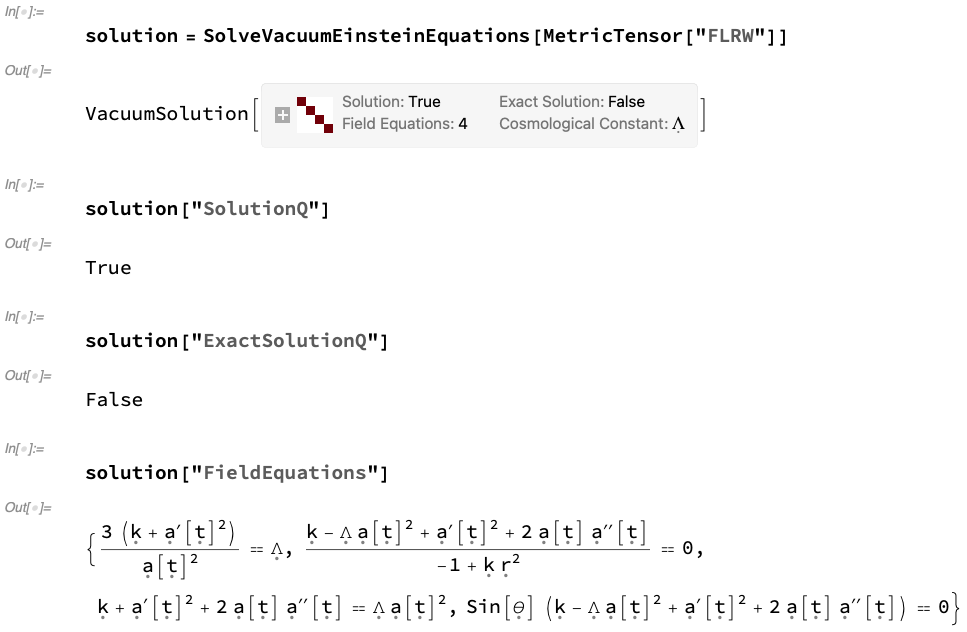}
\end{framed}
\caption{On the left, the \texttt{VacuumSolution} object for an FLRW geometry (representing e.g. a homogeneous, isotropic and uniformly expanding/contracting universe with global curvature $k$ and scale factor ${a \left( t \right)}$ in spherical polar coordinates ${\left( t, r, \theta, \phi \right)}$), with zero cosmological constant, computed using \texttt{SolveVacuumEinsteinEquations}, illustrating that the FLRW metric is a non-exact solution to the vacuum Einstein field equations, with four additional field equations required. On the right, the \texttt{VacuumSolution} object for an FLRW geometry (representing e.g. a homogeneous, isotropic and uniformly expanding/contracting universe with global curvature $k$ and scale factor ${a \left( t \right)}$ in spherical polar coordinates ${\left( t, r, \theta, \phi \right)}$), with non-zero cosmological constant ${\Lambda \neq 0}$, computed using \texttt{SolveVacuumEinsteinEquations}, illustrating that the FLRW metric with non-vanishing cosmological constant is a non-exact solution to the vacuum Einstein field equations, with four additional field equations required.}
\label{fig:Figure37}
\end{figure}

\begin{figure}[ht]
\centering
\begin{framed}
\includegraphics[width=0.495\textwidth]{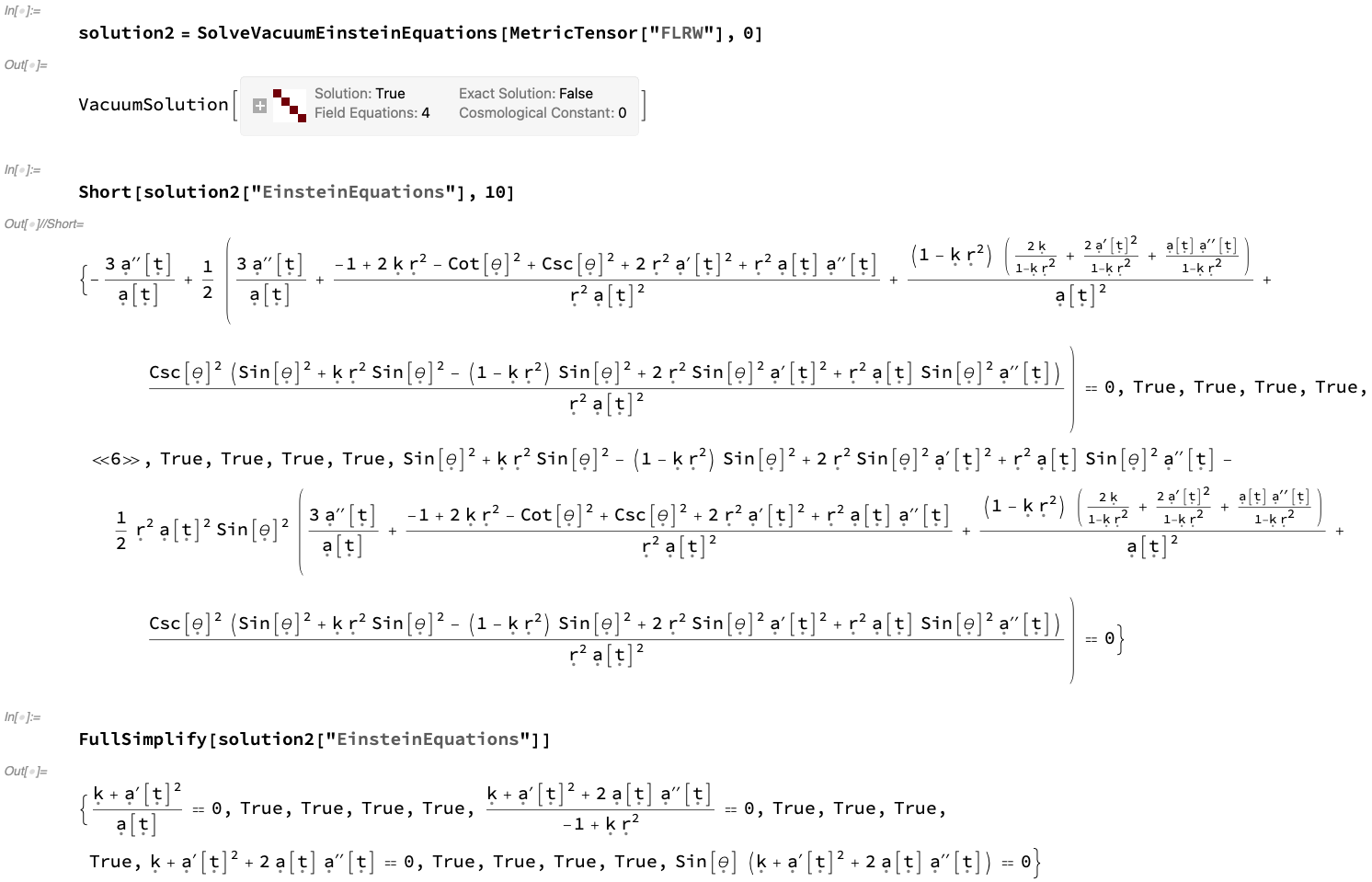}
\vrule
\includegraphics[width=0.495\textwidth]{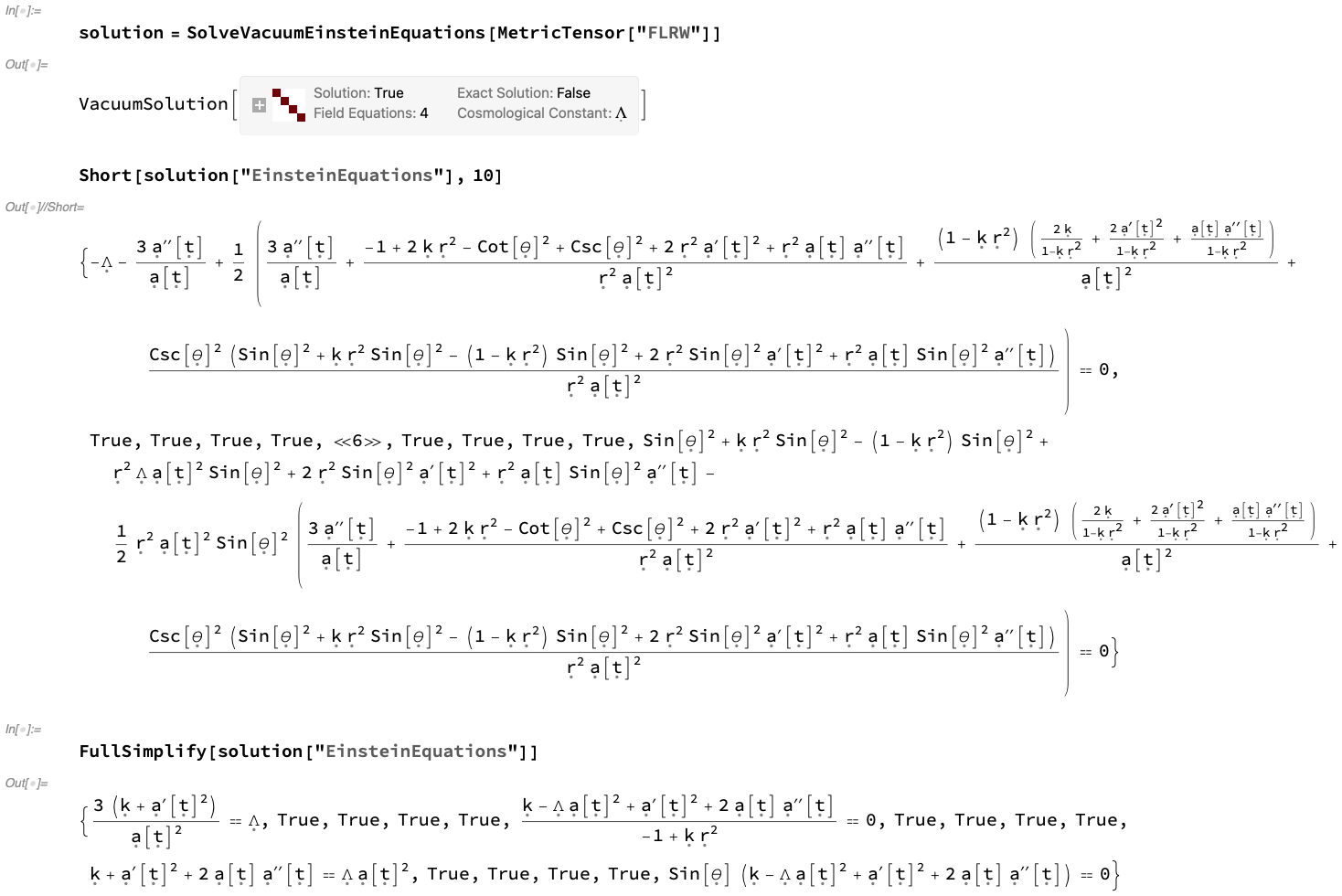}
\end{framed}
\caption{On the left, the list of Einstein field equations asserting that the Einstein tensor, with zero cosmological constant, vanishes, computed using the \texttt{VacuumSolution} object for an FLRW geometry (representing e.g. a homogeneous, isotropic and uniformly expanding/contracting universe with global curvature $k$ and scale factor ${a \left( t \right)}$ in spherical polar coordinates ${\left( t, r, \theta, \phi \right)}$), together with a verification that all but four of them hold identically. On the right, the list of Einstein field equations asserting that the Einstein tensor, plus a non-zero cosmological constant ${\Lambda \neq 0}$, vanishes, computed using the \texttt{VacuumSolution} object for an FLRW geometry (representing e.g. a homogeneous, isotropic and uniformly expanding/contracting universe with global curvature $k$ and scale factor ${a \left( t \right)}$ in spherical polar coordinates ${\left( t, r, \theta, \phi \right)}$), together with a validation that all but four of them hold identically.}
\label{fig:Figure38}
\end{figure}

The simplest non-trivial set of equations of motion that one can impose on a Riemannian or pseudo-Riemannian manifold ${\mathcal{M}}$ are the vacuum Einstein field equations, which assert that the Einstein tensor, plus an optional cosmological constant term ${\Lambda}$ (here playing the role of an arbitrary integration constant), vanishes identically:

\begin{equation}
G_{\mu \nu} + \Lambda g_{\mu \nu} = R_{\mu \nu} - \frac{1}{2} R g_{\mu \nu} + \Lambda g_{\mu \nu} = 0,
\end{equation}
which, due to the aforementioned trace-reversed nature of the relationship between ${R_{\mu \nu}}$ and ${G_{\mu \nu}}$, will occur if and only if the Ricci tensor (plus an optional dimension-dependent function of the cosmological constant term ${\Lambda}$) also vanishes identically:

\begin{equation}
R_{\mu \nu} - \frac{2 \Lambda}{n - 2} g_{\mu \nu} = 0,
\end{equation}
at least in dimensions ${n \neq 2}$\cite{stephani}. Mathematically, solutions to the vacuum Einstein field equations correspond to Einstein manifolds\cite{besse} (i.e. manifolds whose Ricci tensor is proportional to the metric tensor, where the constant of proportionality is related to the cosmological constant ${\Lambda}$ in a dimension-dependent way, as above); in the case of complete four-dimensional manifolds, these correspond to gravitational instanton solutions. Physically, such solutions correspond to the case where the stress-energy tensor ${T_{\mu \nu}}$, and hence the energy-matter content of spacetime, vanishes identically. Representations of the corresponding \texttt{VacuumSolution} objects for the Schwarzschild metric (representing e.g. an uncharged, non-rotating black hole of mass $M$ in Schwarzschild/spherical polar coordinates ${\left( t, r, \theta, \phi \right)}$) and the Kerr metric (representing e.g. an uncharged, spinning black hole of mass $M$ and angular momentum $J$ in Boyer-Lindquist/oblate spheroidal coordinates ${\left( t, r, \theta, \phi \right)}$), computed using the \texttt{SolveVacuumEinsteinEquations} function, are shown in Figure \ref{fig:Figure35}; these examples demonstrate that the Schwarzschild and Kerr metrics are both \textit{exact} solutions of the vacuum Einstein field equations, in the sense that no additional field equations need to be assumed. The complete lists of Einstein field equations for both metrics can be computed directly from the \texttt{VacuumSolution} object, and it can be verified that they all indeed hold identically, as illustrated in Figure \ref{fig:Figure36}. All of these examples thus far have assumed a vanishing cosmological constant (i.e. ${\Lambda = 0}$), and therefore correspond to Ricci-flat solutions in which all components of the Ricci and Einstein tensors ${R_{\mu \nu}}$ and ${G_{\mu \nu}}$ vanish identically. By way of comparison, in Figure \ref{fig:Figure37}, we show representations of the corresponding \texttt{VacuumSolution} objects for the FLRW metric (representing e.g. a homogeneous, isotropic and uniformly expanding/contracting universe with global curvature $k$ and scale factor ${a \left( t \right)}$ in spherical polar coordinates ${\left( t, r, \theta, \phi \right)}$), assuming both zero and non-zero values of the cosmological constant ${\Lambda}$; these examples demonstrate that, in both cases, the FLRW metric is a \textit{non-exact} solution of the vacuum Einstein field equations, in the sense that four additional field equations need to be assumed in each case. Once again, the complete lists of Einstein field equations for the FLRW metric (both with and without a non-zero cosmological constant ${\Lambda}$) can be computed directly from the respective \texttt{VacuumSolution} objects, and it can be verified that all but four of them hold identically, as illustrated in Figure \ref{fig:Figure38}.

\begin{figure}[ht]
\centering
\begin{framed}
\includegraphics[width=0.395\textwidth]{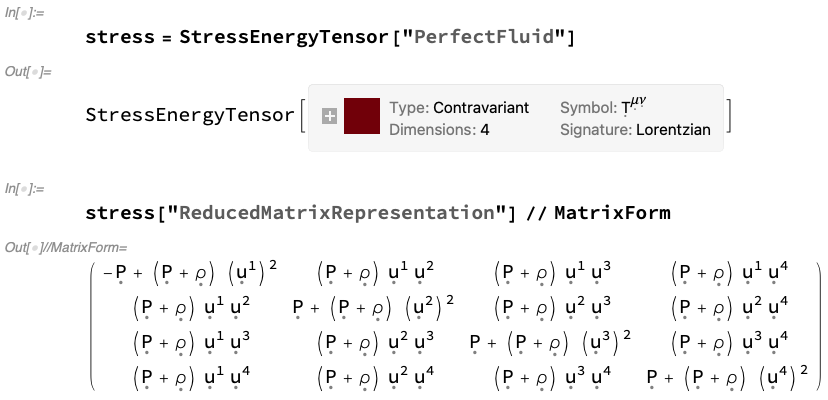}
\vrule
\includegraphics[width=0.595\textwidth]{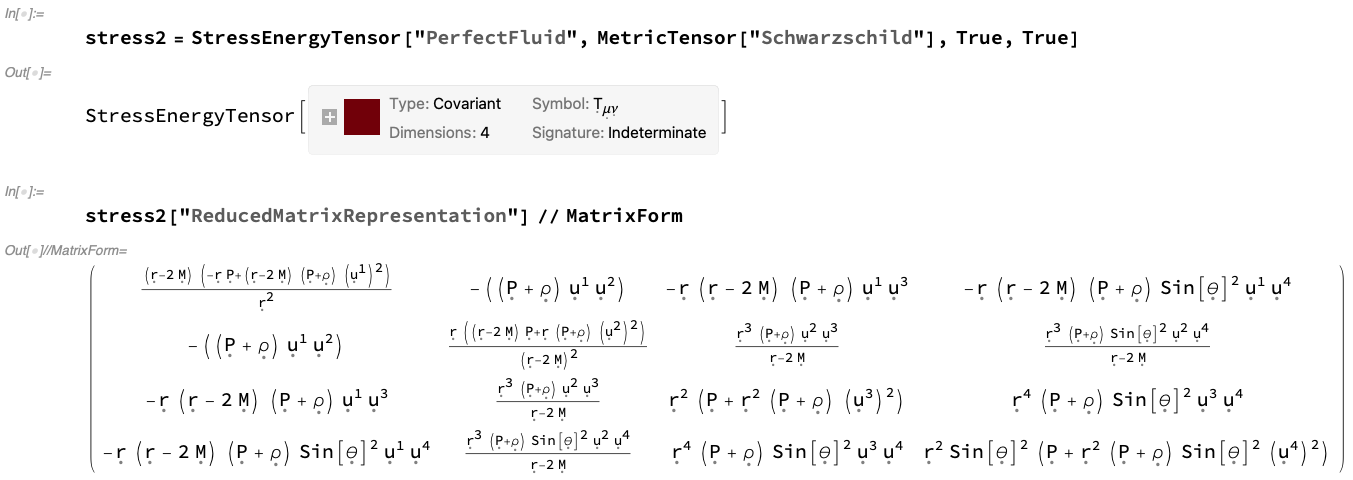}
\end{framed}
\caption{On the left, the \texttt{StressEnergyTensor} object for a perfect relativistic fluid (representing an idealized fluid with mass-energy density ${\rho}$, hydrostatic pressure $P$ and spacetime velocity ${u^{\mu}}$, but with vanishing heat conduction, viscosity and shear stresses) embedded within a Minkowski geometry in explicit contravariant matrix form, with both indices raised/contravariant (default). On the right, the \texttt{StressEnergyTensor} object for a perfect relativistic fluid (representing an idealized fluid with mass-energy density ${\rho}$, hydrostatic pressure $P$ and spacetime velocity ${u^{\mu}}$, but with vanishing heat conduction, viscosity and shear stresses) embedded within a Schwarzschild geometry in explicit covariant matrix form, with both indices lowered/covariant.}
\label{fig:Figure39}
\end{figure}

\begin{figure}[ht]
\centering
\begin{framed}
\includegraphics[width=0.445\textwidth]{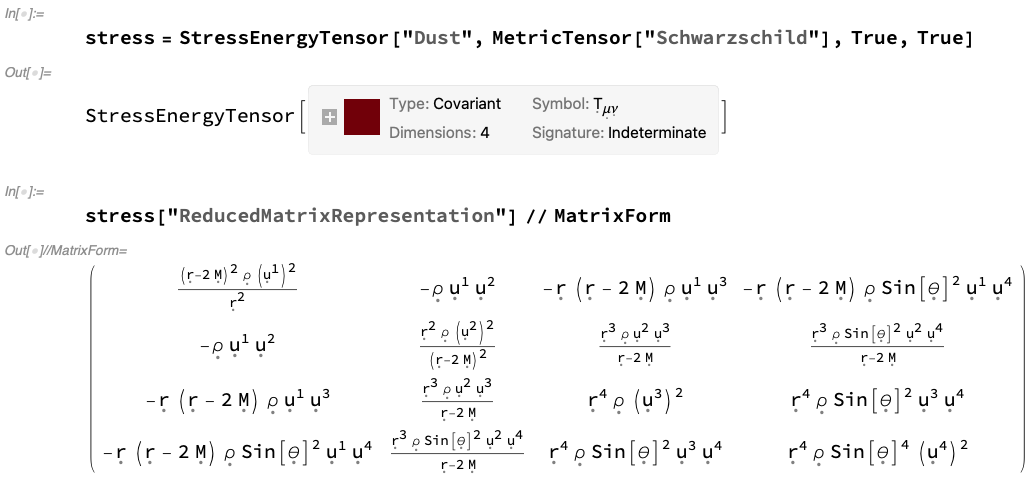}
\vrule
\includegraphics[width=0.545\textwidth]{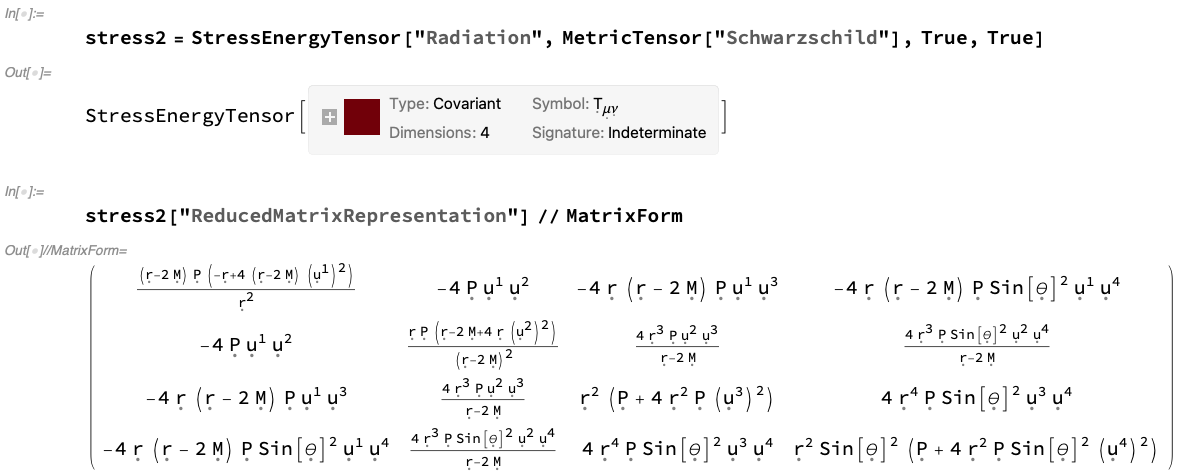}
\end{framed}
\caption{On the left, the \texttt{StressEnergyTensor} object for a perfect relativistic dust (representing an idealized distribution of dust particles with mass-energy density ${\rho}$ and spacetime velocity ${u^{\mu}}$, but with vanishing hydrostatic pressure) embedded within a Schwarzschild geometry in explicit covariant matrix form, with both indices lowered/covariant. On the right, the \texttt{StressEnergyTensor} object for a perfect relativistic radiation distribution (representing an idealized radiation distribution with radiation pressure $P$ and spacetime velocity ${u^{\mu}}$, but whose mass-energy density is equal to the number of spatial dimensions times the radiation pressure, i.e. ${\rho = \left( n - 1 \right) P}$) embedded within a Schwarzschild geometry in explicit covariant matrix form, with both indices lowered/covariant.}
\label{fig:Figure40}
\end{figure}

\begin{figure}[ht]
\centering
\begin{framed}
\includegraphics[width=0.545\textwidth]{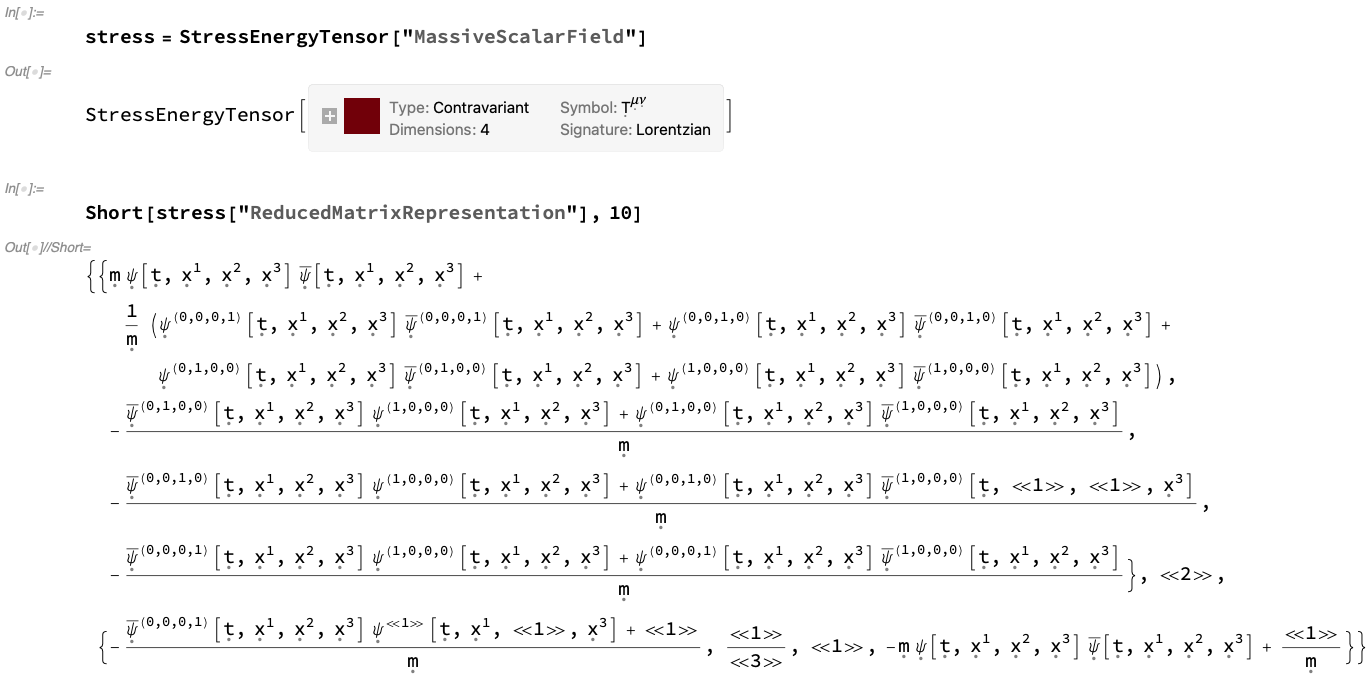}
\vrule
\includegraphics[width=0.445\textwidth]{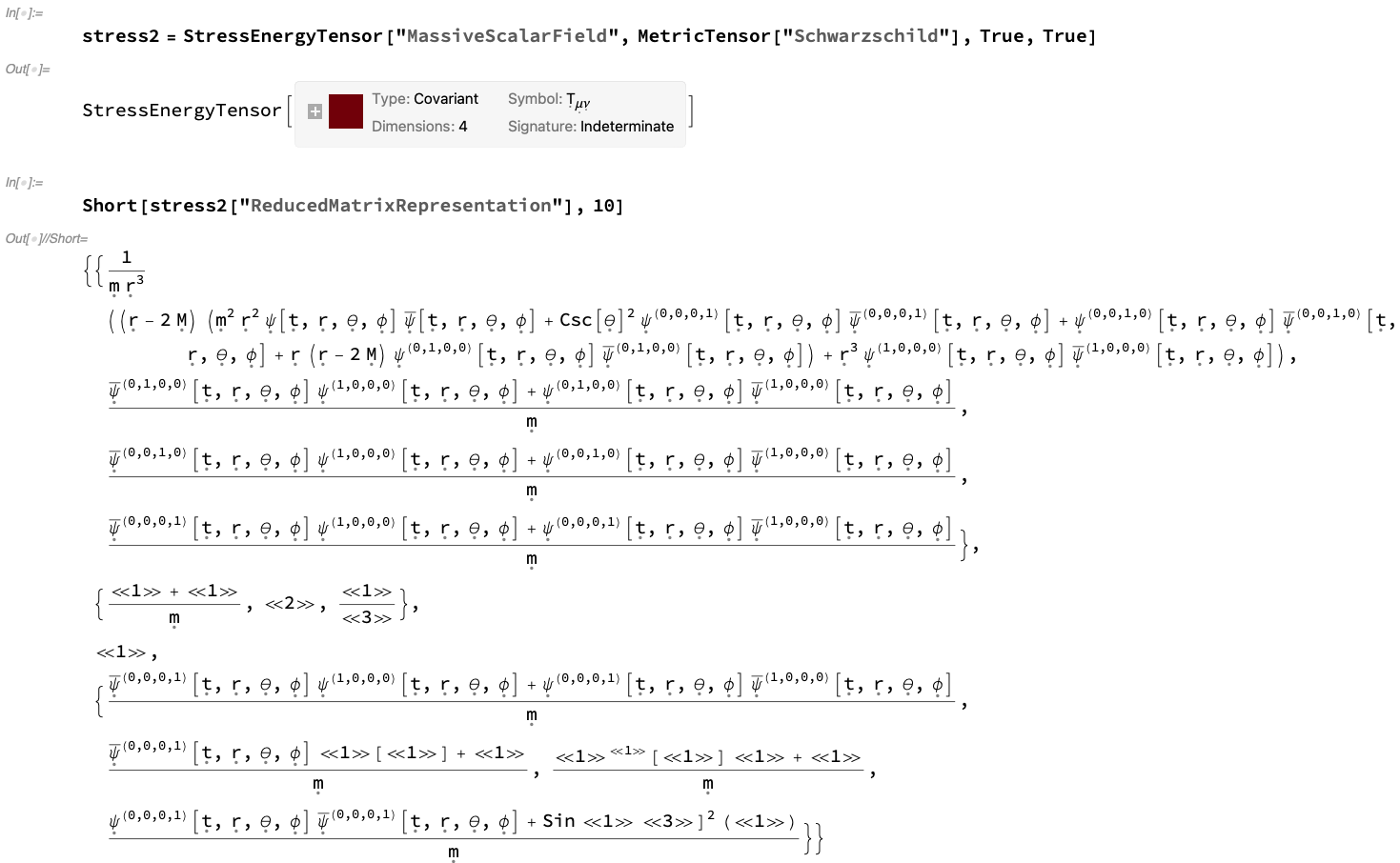}
\end{framed}
\caption{On the left, the \texttt{StressEnergyTensor} object for a massive relativistic scalar field (representing a complex scalar field ${\Psi}$ obeying the massive Klein-Gordon equation with field mass $m$) embedded within a Minkowski geometry in explicit covariant matrix form, with both indices lowered/covariant. On the right, the \texttt{StressEnergyTensor} object for a massive relativistic scalar field (representing a complex scalar field ${\Psi}$ obeying the massive Klein-Gordon equation with field mass $m$) embedded within a Schwarzschild geometry in explicit covariant matrix form, with both indices lowered/covariant.}
\label{fig:Figure41}
\end{figure}

\begin{figure}[ht]
\centering
\begin{framed}
\includegraphics[width=0.495\textwidth]{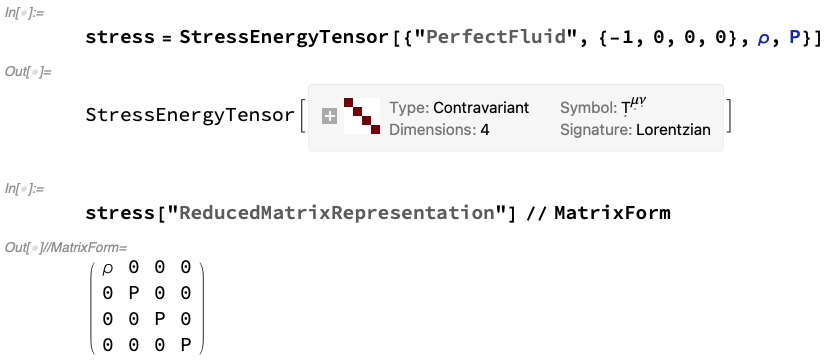}
\vrule
\includegraphics[width=0.495\textwidth]{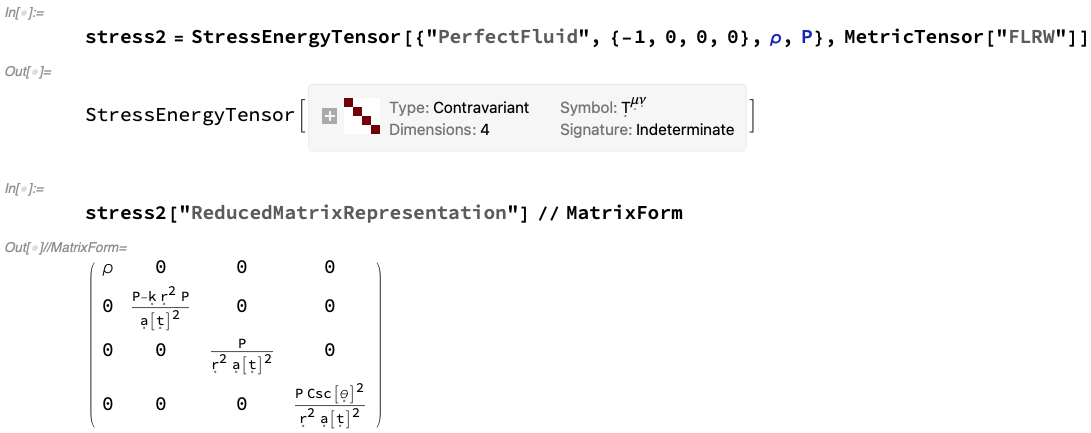}
\end{framed}
\caption{On the left, the \texttt{StressEnergyTensor} object for a modified perfect relativistic fluid (representing an idealized fluid with mass-energy density ${\rho}$, hydrostatic pressure $P$ and spacetime velocity ${\left( -1, 0, 0, 0 \right)}$) embedded within a Schwarzschild geometry in explicit contravariant matrix form (default). On the right, the \texttt{StressEnergyTensor} object for a modified perfect relativistic fluid (representing an idealized fluid with mass-energy density ${\rho}$, hydrostatic pressure $P$ and spacetime velocity ${\left( -1, 0, 0, 0 \right)}$) embedded within an FLRW geometry in explicit contravariant matrix form (default).}
\label{fig:Figure42}
\end{figure}

One can now proceed to introduce a set of non-zero source terms into the Einstein field equations by means of the stress-energy tensor ${T^{\mu \nu}}$, which effectively generalizes the Cauchy stress tensor from continuum mechanics, thus allowing one to quantify the density of energy, momentum and stress (i.e. the flux of energy/momentum) associated with all of the various non-gravitational fields in spacetime\cite{misner}. More precisely, in contravariant/raised-index form ${T^{\mu \nu}}$, the stress-energy tensor represents the flux of spacetime momentum ${P^{\mu}}$ through a codimension-1 hypersurface of constant ${x^{\nu}}$. For instance, representations of the stress-energy tensors for a perfect relativistic fluid (representing e.g. an idealized fluid with mass-energy density ${\rho}$, hydrostatic pressure $P$ and spacetime velocity ${u^{\mu}}$, but with vanishing heat conduction, viscosity and shear stresses)\cite{hawking}:

\begin{equation}
T^{\mu \nu} = \left( \rho + P \right) u^{\mu} u^{\nu} + P g^{\mu \nu},
\end{equation}
embedded within both a Minkowski metric (representing a flat spacetime in Cartesian coordinates ${\left( t, x^1, x^2, x^3 \right)}$) and a Schwarzschild metric (representing e.g. an uncharged, non-rotating black hole of mass $M$ in Schwarzschild/spherical polar coordinates ${\left( t, r, \theta, \phi \right)}$), using the \texttt{StressEnergyTensor} function, are shown in Figure \ref{fig:Figure39}, in the default case where both indices are raised/contravariant, and the case in which both indices are lowered/covariant, respectively. In addition to perfect relativistic fluids, \texttt{StressEnergyTensor} also includes a small library of other in-built relativistic energy-matter distributions (with, much like \texttt{MetricTensor}, many more planned for future inclusion), including, but not limited to: perfect relativistic dust (representing an idealized distribution of dust particles with mass-energy density ${\rho}$ and spacetime velocity ${u^{\mu}}$, but with vanishing hydrostatic pressure) and perfect relativistic radiation (representing an idealized radiation distribution with radiation pressure $P$ and spacetime velocity ${u^{\mu}}$, but whose mass-energy is equal to the number of spatial dimensions times the radiation pressure, i.e. ${\rho = \left( n - 1 \right) P}$), namely:

\begin{equation}
T^{\mu \nu} = \rho u^{\mu} u^{\nu}, \qquad \text{ and } \qquad T^{\mu \nu} = n P u^{\mu} u^{\nu} + P g^{\mu \nu},
\end{equation}
respectively, with both being treated as limiting cases of perfect relativistic fluids, as shown in Figure \ref{fig:Figure40}, both embedded within a Schwarzschild metric and with both indices lowered/covariant in each case; and massive relativistic scalar fields (representing a complex scalar field ${\Psi}$ obeying the massive Klein-Gordon equation with field mass $m$)\cite{fulling}:

\begin{equation}
T^{\mu \nu} = \frac{1}{m} \left( g^{\mu \rho} g^{\nu \sigma} + g^{\mu \sigma} g^{\nu \rho} - g^{\mu \nu} g^{\rho \sigma} \right) \left( \frac{\partial \overline{\Psi}}{\partial x^{\rho}} \right) \left( \frac{\partial \Psi}{\partial x^{\sigma}} \right) - g^{\mu \nu} m \overline{\Psi} \Psi,
\end{equation}
where here we refer to the massive Klein-Gordon equation in curved spacetime, which takes the generic form:

\begin{equation}
\nabla^{\mu} \left( \nabla_{\mu} \Phi \right) - m^2 \Phi = g^{\mu \sigma} \nabla_{\sigma} \left( \nabla_{\mu} \Phi \right) - m^2 \Phi = g^{\mu \sigma} \left( \frac{\partial}{\partial x^{\sigma}} \left( \frac{\partial \Phi}{\partial x^{\mu}} \right) - \Gamma_{\sigma \mu}^{\lambda} \left( \frac{\partial \Phi}{\partial x^{\lambda}} \right) \right) - m^2 \Phi = 0,
\end{equation}
for the case of a real scalar field ${\Phi}$, or, in action integral form:

\begin{multline}
S = \int_{\mathcal{M}} \sqrt{ - \det \left( g_{\mu \nu} \right)} \left( - \frac{1}{2} g^{\mu \nu} \left( \nabla_{\mu} \Phi \right) \left( \nabla_{\nu} \Phi \right) - \frac{1}{2} m^2 \Phi^2 \right) d^n x\\
= \int_{\mathcal{M}} \sqrt{ - \det \left( g_{\mu \nu} \right)} \left( - \frac{1}{2} g^{\mu \nu} \left( \frac{\partial \Phi}{\partial x^{\mu}} \right) \left( \frac{\partial \Phi}{\partial x^{\nu}} \right) - \frac{1}{2} m^2 \Phi^2 \right) d^n x,
\end{multline}
which then generalizes easily to the case of a complex scalar field ${\Psi}$ via complex conjugation:

\begin{multline}
S = \int_{\mathcal{M}} \sqrt{ - \det \left( g_{\mu \nu} \right)} \left( - \frac{1}{2} g^{\mu \nu} \left( \nabla_{\mu} \Psi \right) \left( \nabla_{\nu} \overline{\Psi} \right) - \frac{1}{2} m^2 \Psi \overline{\Psi} \right) d^n x\\
= \int_{\mathcal{M}} \sqrt{ - \det \left( g_{\mu \nu} \right)} \left( - \frac{1}{2} g^{\mu \nu} \left( \frac{\partial \Psi}{\partial x^{\mu}} \right) \left( \frac{\partial \overline{\Psi}}{\partial x^{\nu}} \right) - \frac{1}{2} m^2 \Psi \overline{\Psi} \right) d^n x,
\end{multline}
which is shown in Figure \ref{fig:Figure41} for the illustrative examples of massive (complex) scalar fields embedded within both Minkowski and Schwarzschild metrics, and with both indices lowered/covariant in each case. Relativistic electromagnetic fields are also fully supported within \textsc{Gravitas}, but these require some specialized treatment and will therefore be discussed towards the end of the present section. By default, appropriate formal symbols are assigned to the various parameters of the energy-matter distributions (e.g. mass-energy density ${\rho}$ or spacetime velocity ${u^{\mu}}$), although these defaults can easily be overridden using additional arguments, as shown in Figure \ref{fig:Figure42}.

\begin{figure}[ht]
\centering
\begin{framed}
\includegraphics[width=0.445\textwidth]{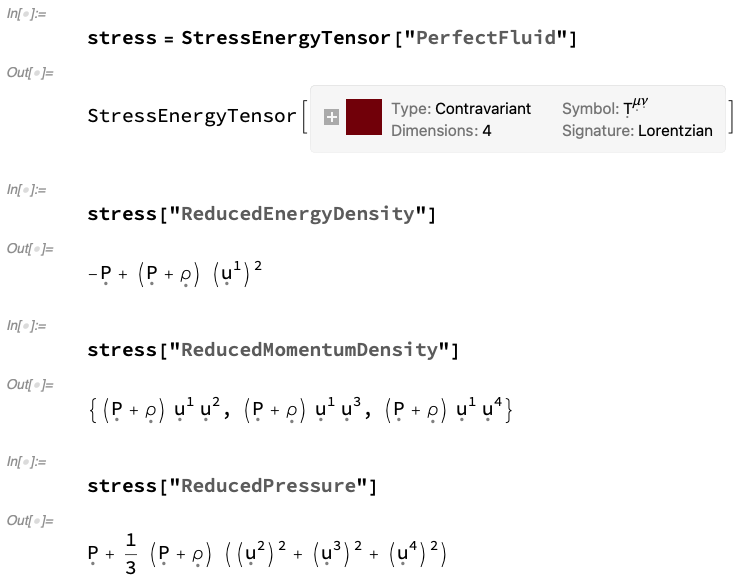}
\vrule
\includegraphics[width=0.545\textwidth]{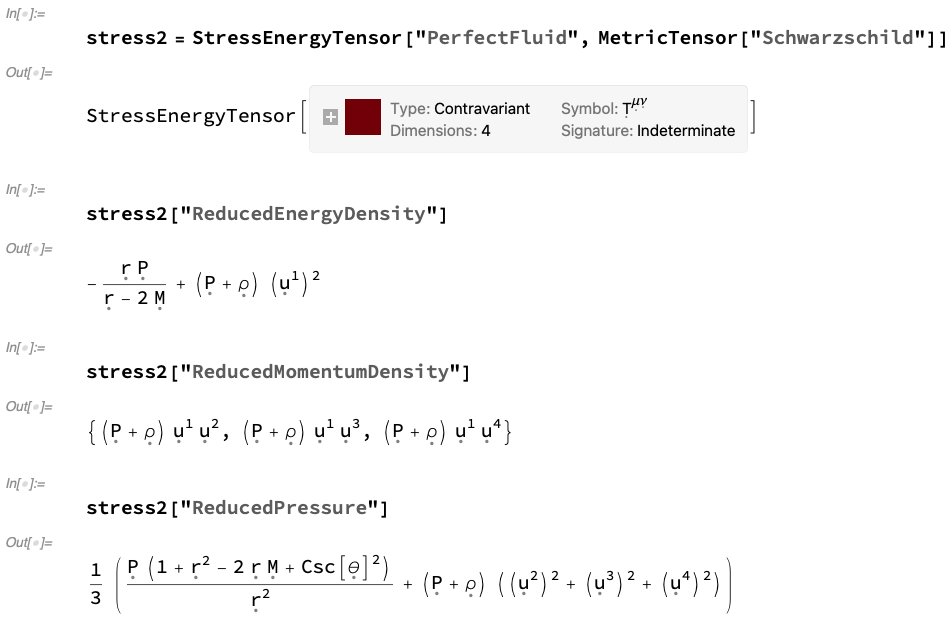}
\end{framed}
\caption{On the left, the relativistic energy density, relativistic momentum density and relativistic pressure of the \texttt{StressEnergyTensor} object for a perfect relativistic fluid (representing an idealized fluid with mass-energy density ${\rho}$, hydrostatic pressure $P$ and spacetime velocity ${u^{\mu}}$, but with vanishing heat conduction, viscosity and shear stresses) embedded within a Minkowski geometry. On the right, the relativistic energy density, relativistic momentum density and relativistic pressure of the \texttt{StressEnergyTensor} object for a perfect relativistic fluid (representing an idealized fluid with mass-energy density ${\rho}$, hydrostatic pressure $P$ and spacetime velocity ${u^{\mu}}$, but with vanishing heat conduction, viscosity and shear stresses) embedded within a Schwarzschild geometry.}
\label{fig:Figure43}
\end{figure}

\begin{figure}[ht]
\centering
\begin{framed}
\includegraphics[width=0.545\textwidth]{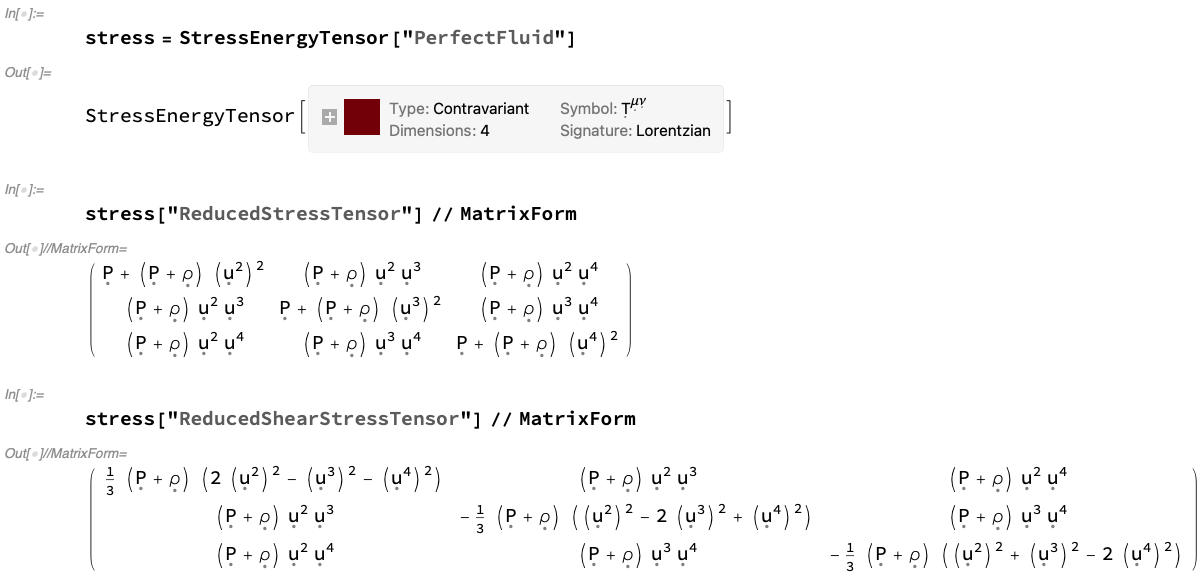}
\vrule
\includegraphics[width=0.445\textwidth]{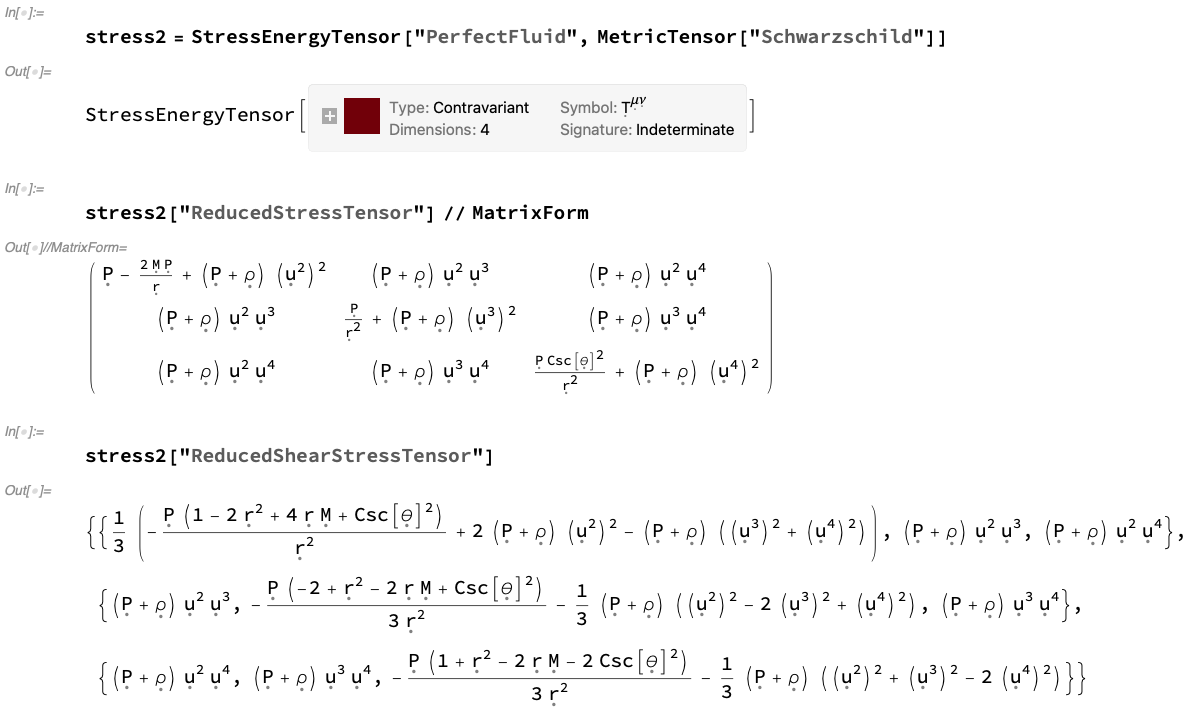}
\end{framed}
\caption{On the left, the relativistic Cauchy stress tensor and the relativistic shear stress tensor of the \texttt{StressEnergyTensor} object for a perfect relativistic fluid (representing an idealized fluid with mass-energy density ${\rho}$, hydrostatic pressure $P$ and spacetime velocity ${u^{\mu}}$, but with vanishing heat conduction, viscosity and shear stresses) embedded within a Minkowski geometry, in explicit matrix form. On the right, the relativistic Cauchy stress tensor and relativistic shear stress tensor of the \texttt{StressEnergyTensor} object for a perfect relativistic fluid (representing an idealized fluid with mass-energy density ${\rho}$, hydrostatic pressure $P$ and spacetime velocity ${u^{\mu}}$, but with vanishing heat conduction, viscosity and shear stresses) embedded within a Schwarzschild geometry, in explicit matrix form.}
\label{fig:Figure44}
\end{figure}

\begin{figure}[ht]
\centering
\begin{framed}
\includegraphics[width=0.595\textwidth]{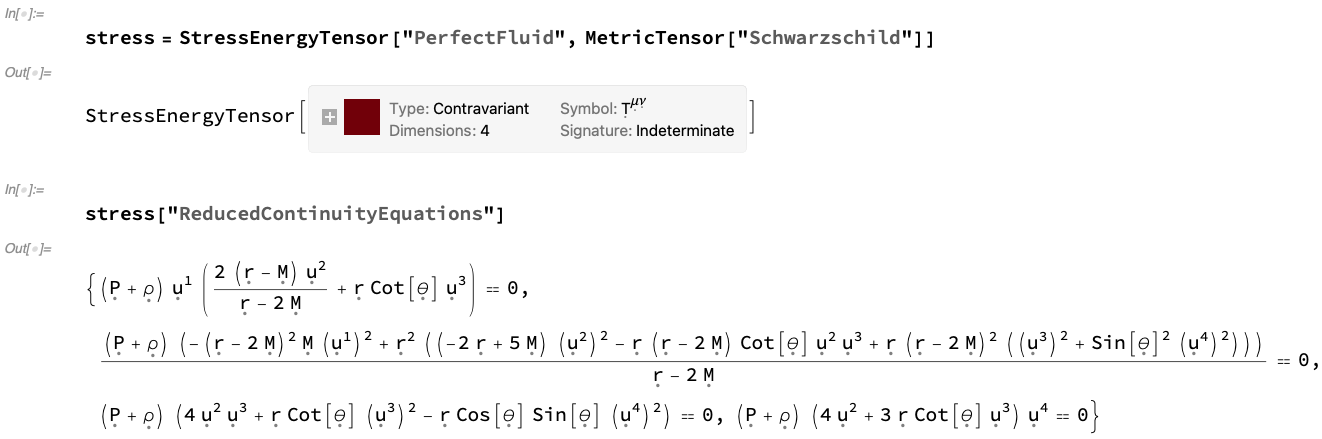}
\vrule
\includegraphics[width=0.395\textwidth]{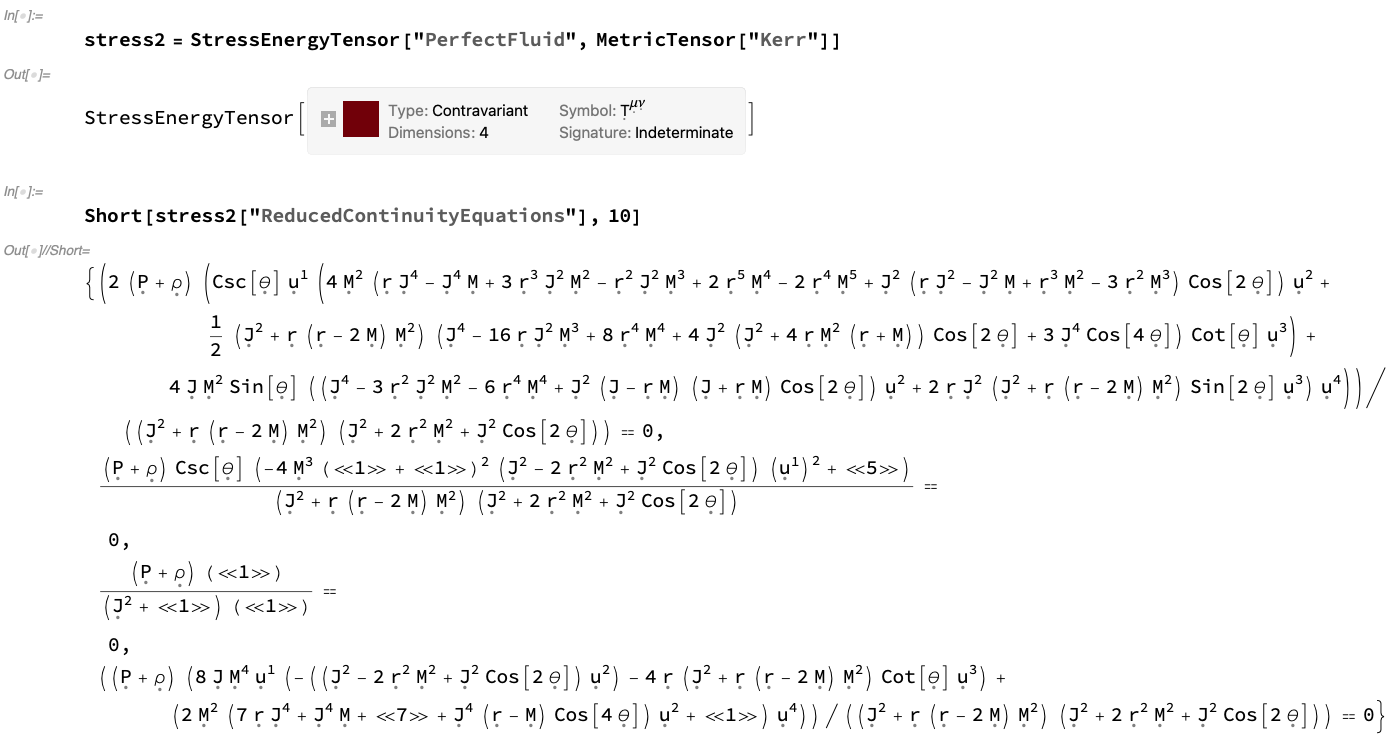}
\end{framed}
\caption{On the left, the list of continuity equations asserting that the covariant divergence of the \texttt{StressEnergyTensor} object for a perfect relativistic fluid (representing an idealized fluid with mass-energy density ${\rho}$, hydrostatic pressure $P$ and spacetime velocity ${u^{\mu}}$, but with vanishing heat conduction, viscosity and shear stresses), embedded within a Schwarzschild geometry, vanishes identically. On the right, the list of continuity equations asserting that the covariant divergence of the \texttt{StressEnergyTensor} object for a perfect relativistic fluid (representing an idealized fluid with mass-energy density ${\rho}$, hydrostatic pressure $P$ and spacetime velocity ${u^{\mu}}$, but with vanishing heat conduction, viscosity and shear stresses), embedded within a Kerr geometry, vanishes identically.}
\label{fig:Figure45}
\end{figure}

\begin{figure}[ht]
\centering
\begin{framed}
\includegraphics[width=0.495\textwidth]{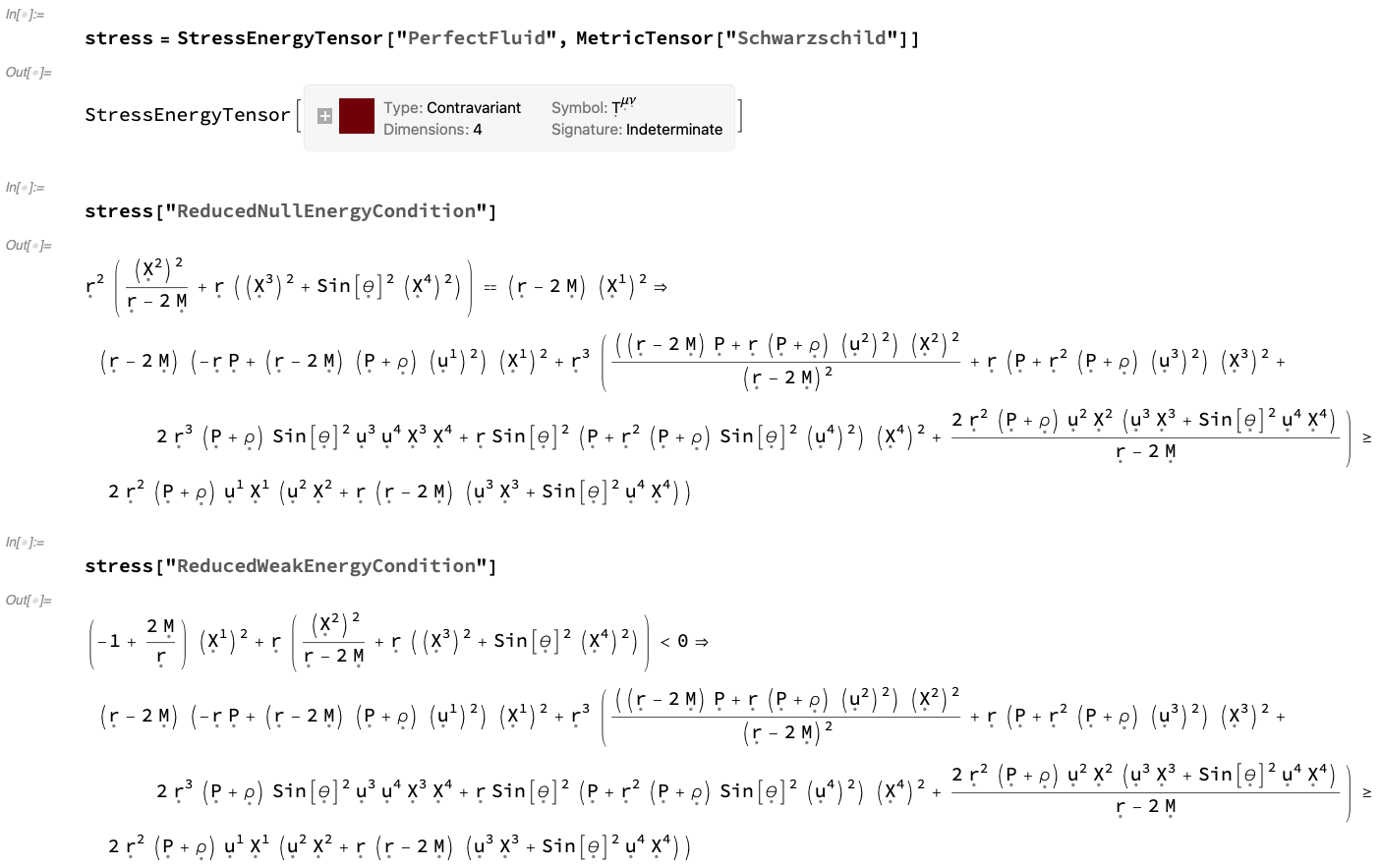}
\vrule
\includegraphics[width=0.495\textwidth]{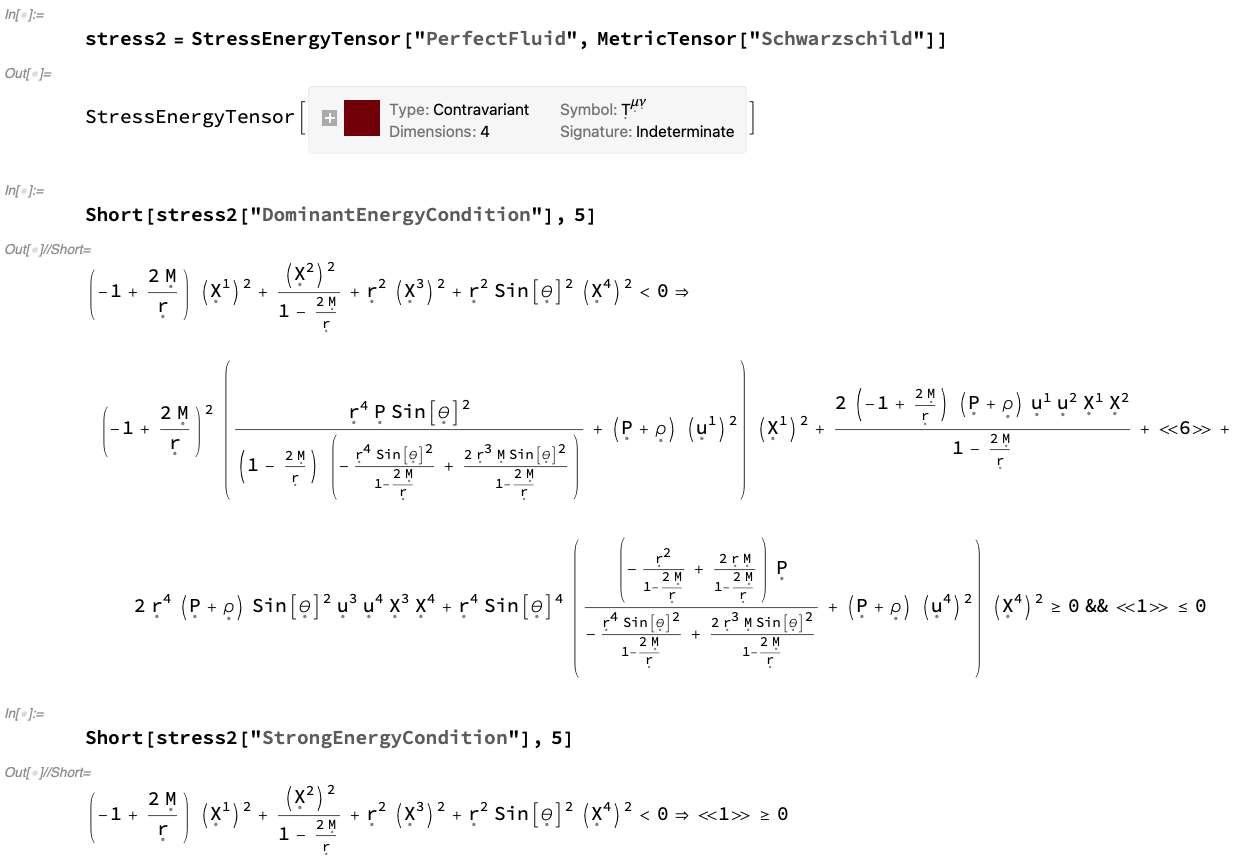}
\end{framed}
\caption{On the left, the null and weak energy conditions on the \texttt{StressEnergyTensor} object for a perfect relativistic fluid (representing an idealized fluid with mass-energy density ${\rho}$, hydrostatic pressure $P$ and spacetime velocity ${u^{\mu}}$, but with vanishing heat conduction, viscosity and shear stresses), embedded within a Schwarzschild geometry. On the right, the dominant and strong energy conditions on the \texttt{StressEnergyTensor} object for a perfect relativistic fluid (representing an idealized fluid with mass-energy density ${\rho}$, hydrostatic pressure $P$ and spacetime velocity ${u^{\mu}}$, but with vanishing heat conduction, viscosity and shear stresses), embedded within a Schwarzschild geometry.}
\label{fig:Figure46}
\end{figure}

A wide variety of physical quantities and properties of the underlying energy-matter distribution, including various energy conditions, can be computed directly from its associated \texttt{StressEnergyTensor} object. For instance, from the definition of the stress-energy tensor ${T^{\mu \nu}}$ in contravariant/raised-index form as a flux of ${P^{\mu}}$ through a hypersurface of constant ${x^{\nu}}$, it is clear that the time-time component ${T^{0 0}}$ represents a relativistic energy density, the time-space and space-time components ${T^{\mu 0}}$ and ${T^{0 \nu}}$ (for ${\mu, \nu \neq 0}$) represent a relativistic momentum density, and the space-space components ${T^{\mu \nu}}$ (for ${\mu, \nu \neq 0}$) represent a relativistic Cauchy/normal stress, which may be further decomposed into a trace part ${\frac{1}{n - 1} T_{\sigma}^{\sigma}}$ (essentially obtained by averaging over all elements on the main diagonal with ${\sigma \neq 0}$), representing a relativistic bulk pressure/isotropic stress, and a trace-free part ${T^{\mu \nu} - \frac{1}{n - 1} T_{\sigma}^{\sigma}}$ (where ${\mu, \nu \neq 0}$, and where one is again averaging over all elements on the main diagonal with ${\sigma \neq 0}$), representing a relativistic shear stress. Figure \ref{fig:Figure43} shows the relativistic energy density ${T^{0 0}}$, relativistic momentum density ${T^{\mu 0}}$ / ${T^{0 \nu}}$ (where ${\mu, \nu \neq 0}$) and relativistic pressure ${\frac{1}{n - 1} T_{\sigma}^{\sigma}}$ (with the sum taken over all elements on the main diagonal with ${\sigma \neq 0}$), computed directly from the \texttt{StressEnergyTensor} object for a perfect relativistic fluid (representing an idealized fluid with mass-energy density ${\rho}$, hydrostatic pressure $P$ and spacetime velocity ${u^{\mu}}$, but with vanishing heat conduction, viscosity and shear stresses) embedded within both a Minkowski metric and a Schwarzschild metric. Likewise, Figure \ref{fig:Figure44} shows the relativistic Cauchy stress tensor ${T^{\mu \nu}}$ (where ${\mu, \nu \neq 0}$) and relativistic shear stress tensor ${T^{\mu \nu} - \frac{1}{n - 1} T_{\sigma}^{\sigma}}$ (where ${\mu, \nu \neq 0}$, and with the sum taken over all elements on the main diagonal with ${\sigma \neq 0}$), both in explicit matrix form, also computed directly from the \texttt{StressEnergyTensor} object for a perfect relativistic fluid, again embedded within both a Minkowski metric and a Schwarzschild metric. As we shall see momentarily, once the full Einstein field equations have been imposed, the contracted Bianchi identities (which assert that the covariant divergence of the Einstein tensor ${G_{\mu \nu}}$ must vanish identically) consequently ensure that the covariant divergence of the stress-energy tensor ${T_{\mu \nu}}$ must also vanish identically:

\begin{multline}
\nabla_{\nu} G^{\mu \nu} = \frac{\partial}{\partial x^{\nu}} \left( G^{\mu \nu} \right) + \Gamma_{\nu \sigma}^{\mu} G^{\sigma \nu} + \Gamma_{\nu \sigma}^{\nu} G^{\mu \sigma} = 0,\\
\implies \qquad \nabla_{\nu} T^{\mu \nu} = \frac{\partial}{\partial x^{\nu}} \left( T^{\mu \nu} \right) + \Gamma_{\nu \sigma}^{\mu} T^{\sigma \nu} + \Gamma_{\nu \sigma}^{\nu} T^{\mu \sigma} = 0,
\end{multline}
which is formally equivalent to the statement that relativistic energy and momentum must be conserved. Figure \ref{fig:Figure45} shows how such continuity equations may be derived directly from the \texttt{StressEnergyTensor} object for a perfect relativistic fluid embedded within both a Schwarzschild metric and a Kerr metric. Furthermore, there exists a hierarchy of standard relativistic \textit{energy conditions} which can assist, at a variety of different levels of generality, in guaranteeing that the mass-energy density of a given region of space is always non-negative\cite{curiel}. For instance, in the weakest case, one has the null energy condition, in which all observers following a future-pointing, lightlike vector field ${\mathbf{X} \in \Gamma \left( \bigsqcup\limits_{\mathbf{x} \in \mathcal{M}} T_{\mathbf{x}} \mathcal{M} \right)}$ must always observe a non-negative mass-energy density ${T_{\mu \nu} X^{\mu} X^{\nu}}$:

\begin{equation}
\forall \mathbf{X} \in \Gamma \left( \bigsqcup\limits_{\mathbf{x} \in \mathcal{M}} T_{\mathbf{x}} \mathcal{M} \right), \qquad g_{\mu \nu} X^{\mu} X^{\nu} = 0, \qquad \implies \qquad T_{\mu \nu} X^{\mu} X^{\nu} \geq 0;
\end{equation}
in a very similar fashion, one can consider the weak energy condition, in which all observers following a \textit{timelike} vector field ${\mathbf{X} \in \Gamma \left( \bigsqcup\limits_{\mathbf{x} \in \mathcal{M}} T_{\mathbf{x}} \mathcal{M} \right)}$ must always observe a non-negative mass-energy density ${T_{\mu \nu} X^{\mu} X^{\nu}}$:

\begin{equation}
\forall \mathbf{X} \in \Gamma \left( \bigsqcup\limits_{\mathbf{x} \in \mathcal{M}} T_{\mathbf{x}} \mathcal{M} \right), \qquad g_{\mu \nu} X^{\mu} X^{\nu} < 0, \qquad \implies \qquad T_{\mu \nu} X^{\mu} X^{\nu} \geq 0.
\end{equation}
The dominant energy condition is a little more subtle to state, since it asserts both that the weak energy condition must hold \textit{and} that all observers following a future-pointing, causal (i.e. lightlike or timelike) vector field ${\mathbf{X} \in \Gamma \left( \bigsqcup\limits_{\mathbf{x} \in \mathcal{M}} T_{\mathbf{x}} \mathcal{M} \right)}$ must always observe mass-energy to be flowing no faster than light, and therefore that the vector field ${- T_{\nu}^{\mu} X^{\nu}}$ must itself be future-pointing and causal:

\begin{multline}
\forall \mathbf{X} \in \Gamma \left( \bigsqcup\limits_{\mathbf{x} \in \mathcal{M}} T_{\mathbf{x}} \mathcal{M} \right), \qquad \left( g_{\mu \nu} X^{\mu} X^{\nu} < 0, \qquad \implies \qquad T_{\mu \nu} X^{\mu} X^{\nu} \geq 0 \right)\\
\wedge \left( g_{\mu \nu} X^{\mu} X^{\nu} \leq 0, \qquad \implies \qquad g_{\mu \nu} \left( - T_{\sigma}^{\mu} X^{\sigma} \right) \left( - T_{\sigma}^{\nu} X^{\sigma} \right) \leq 0 \right).
\end{multline}
Finally, the strong energy condition (which is routinely violated within many standard models of cosmology, such as cosmic inflation\cite{visser}) asserts that all observers following a timelike vector field ${\mathbf{X} \in \Gamma \left( \bigsqcup\limits_{\mathbf{x} \in \mathcal{M}} T_{\mathbf{x}} \mathcal{M} \right)}$ must always observe the tidal tensor/electrogravitic tensor ${R_{\mu \nu} X^{\mu} X^{\nu}}$ to have a non-negative trace:

\begin{equation}
\forall \mathbf{X} \in \Gamma \left( \bigsqcup\limits_{\mathbf{x} \in \mathcal{M}} T_{\mathbf{x}} \mathcal{M} \right), \qquad g_{\mu \nu} X^{\mu} X^{\nu} < 0, \qquad \implies \qquad \left( T_{\mu \nu} - \frac{1}{2} T g_{\mu \nu} \right) X^{\mu} X^{\nu} \geq 0.
\end{equation}
Figure \ref{fig:Figure46} shows how the null, weak, dominant and strong energy conditions may be computed and/or imposed directly from the \texttt{StressEnergyTensor} object for a perfect relativistic fluid embedded within a Schwarzschild metric.

\begin{figure}[ht]
\centering
\begin{framed}
\includegraphics[width=0.495\textwidth]{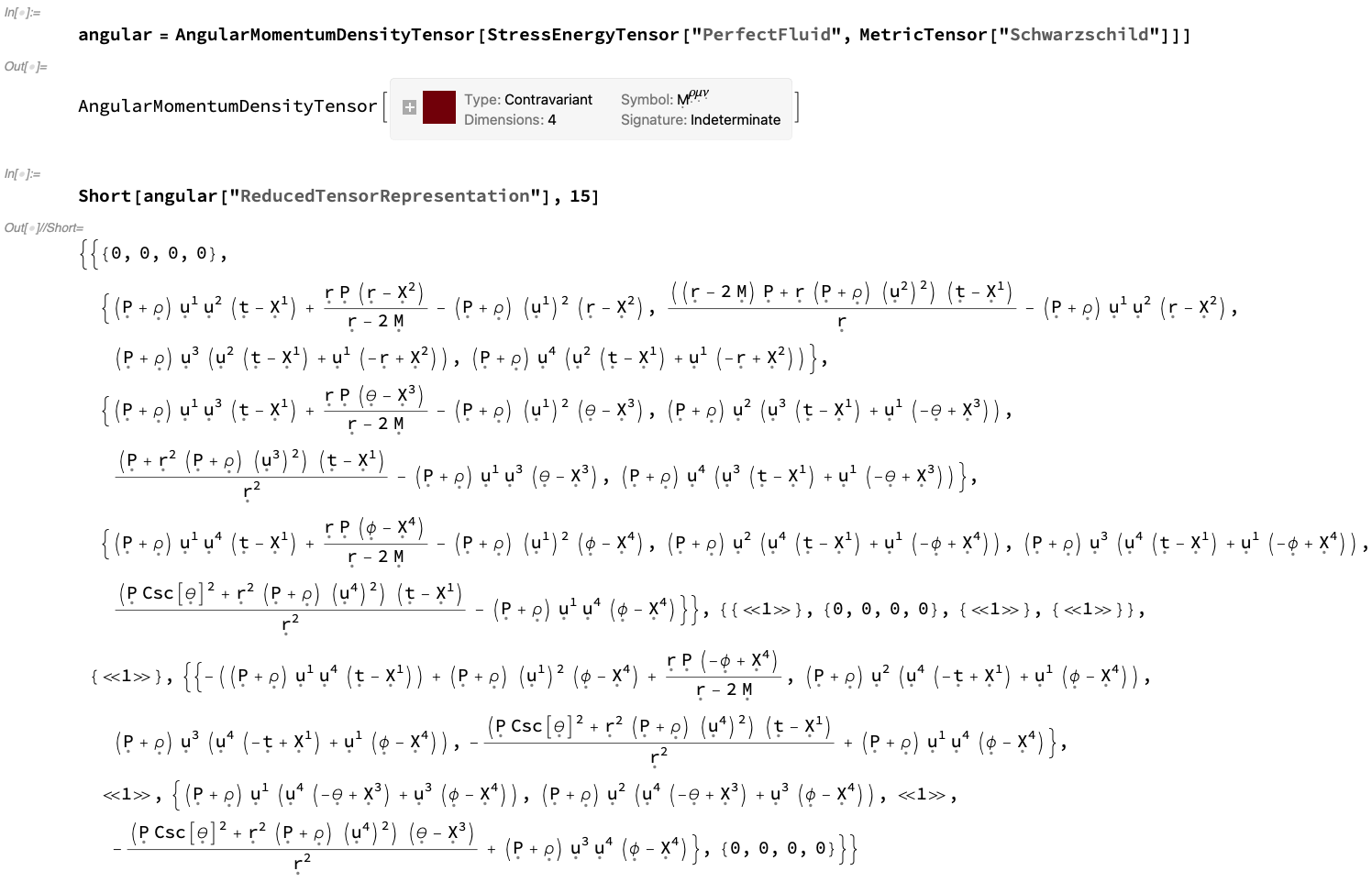}
\vrule
\includegraphics[width=0.495\textwidth]{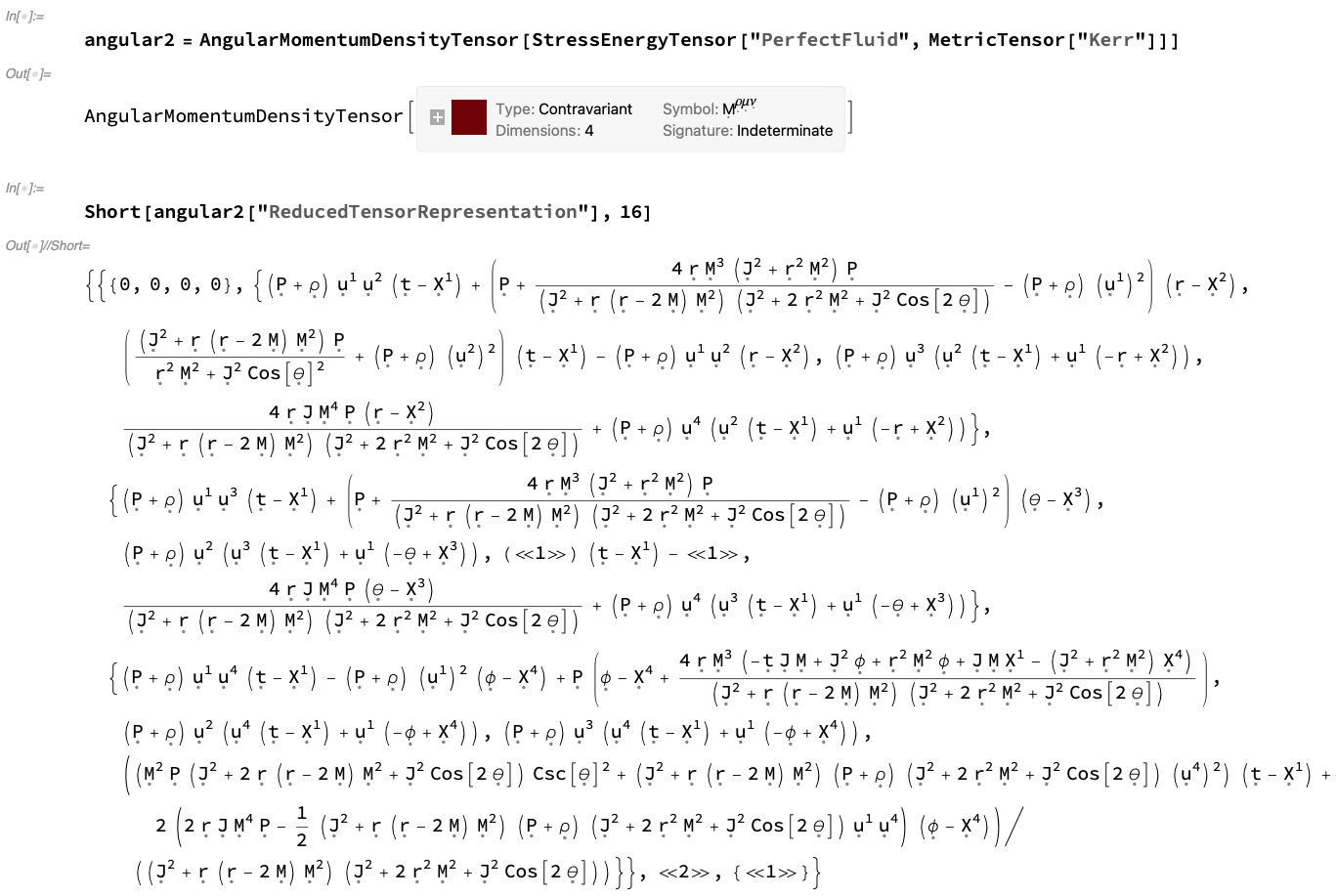}
\end{framed}
\caption{On the left, the \texttt{AngularMomentumDensityTensor} object about the spacetime position vector ${\mathbf{X} = \left( X^1, X^2, X^3, X^4 \right)}$ for a perfect relativistic fluid (representing an idealized fluid with mass-energy density ${\rho}$, hydrostatic pressure $P$ and spacetime velocity ${u^{\mu}}$, but with vanishing heat conduction, viscosity and shear stresses) embedded within a Schwarzschild geometry in explicit contravariant array form, with all indices raised/contravariant. On the right, the \texttt{AngularMomentumDensityTensor} object about the spacetime position vector ${\mathbf{X} = \left( X^1, X^2, X^3, X^4 \right)}$ for a perfect relativistic fluid (representing an idealized fluid with mass-energy density ${\rho}$, hydrostatic pressure $P$ and spacetime velocity ${u^{\mu}}$, but with vanishing heat conduction, viscosity and shear stresses) embedded within a Kerr geometry in explicit contravariant array form, with all indices raised/contravariant.}
\label{fig:Figure47}
\end{figure}

\begin{figure}[ht]
\centering
\begin{framed}
\includegraphics[width=0.545\textwidth]{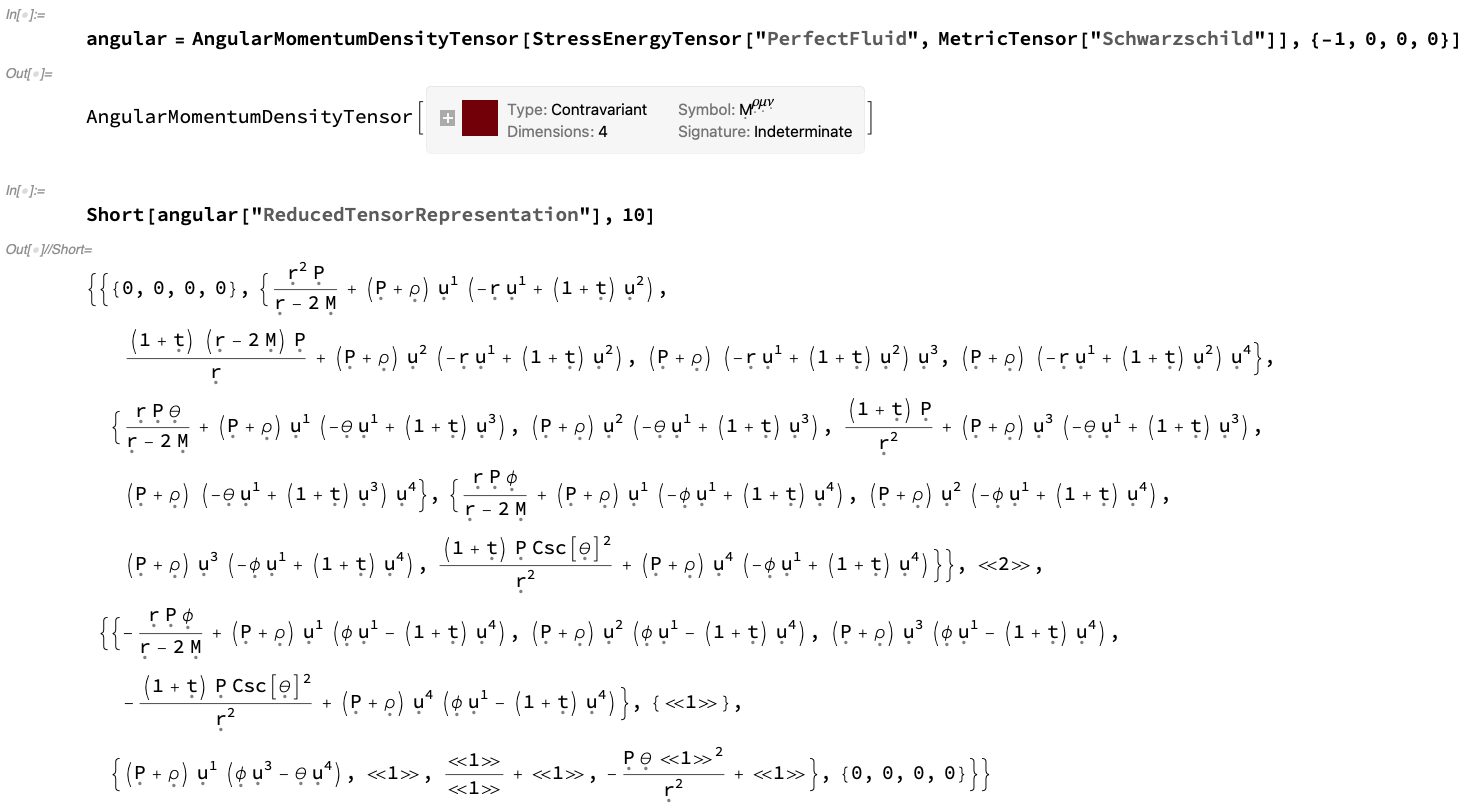}
\vrule
\includegraphics[width=0.445\textwidth]{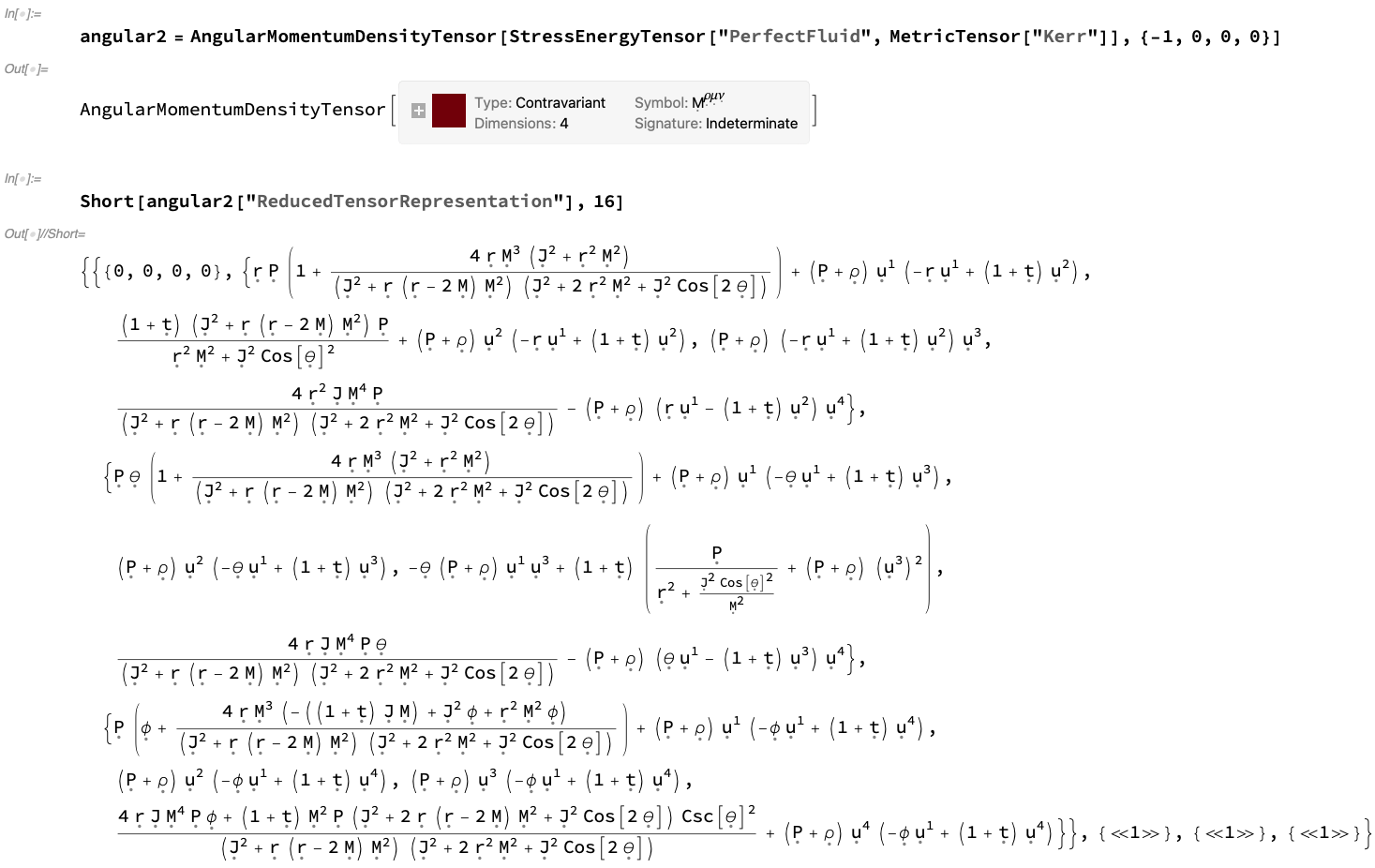}
\end{framed}
\caption{On the left, the \texttt{AngularMomentumDensityTensor} object about the modified spacetime position vector ${\mathbf{X} = \left( -1, 0, 0, 0 \right)}$ for a perfect relativistic fluid (representing an idealized fluid with mass-energy density ${\rho}$, hydrostatic pressure $P$ and spacetime velocity ${u^{\mu}}$, but with vanishing heat conduction, viscosity and shear stresses) embedded within a Schwarzschild geometry in explicit contravariant array form, with all indices raised/contravariant. On the right, the \texttt{AngularMomentumDensityTensor} object about the modified spacetime position vector ${\mathbf{X} = \left( -1, 0, 0, 0 \right)}$ for a perfect relativistic fluid (representing an idealized fluid with mass-energy density ${\rho}$, hydrostatic pressure $P$ and spacetime velocity ${u^{\mu}}$, but with vanishing heat conduction, viscosity and shear stresses) embedded within a Kerr geometry in explicit contravariant array form, with all indices raised/contravariant.}
\label{fig:Figure48}
\end{figure}

\begin{figure}[ht]
\centering
\begin{framed}
\includegraphics[width=0.495\textwidth]{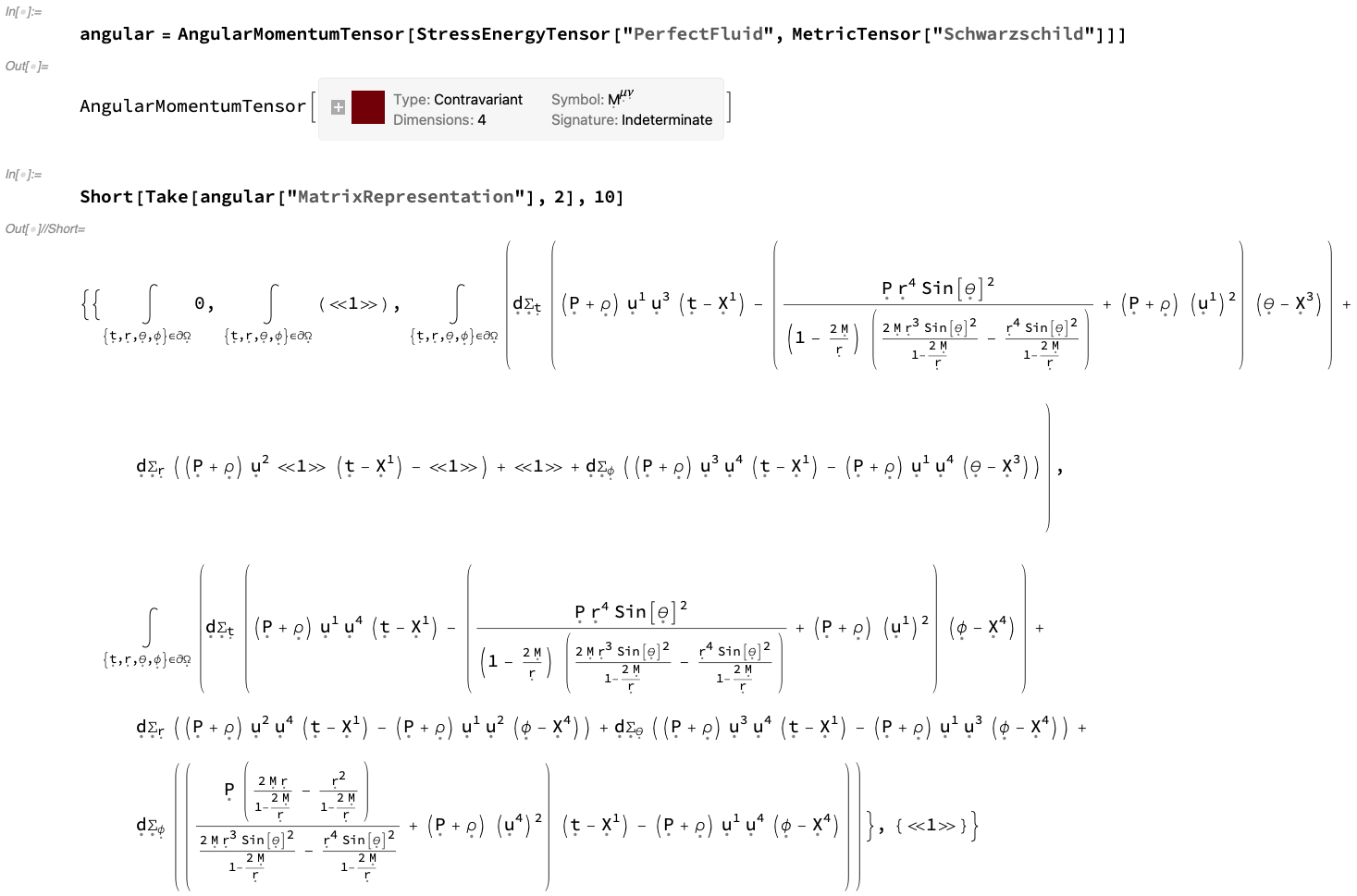}
\vrule
\includegraphics[width=0.495\textwidth]{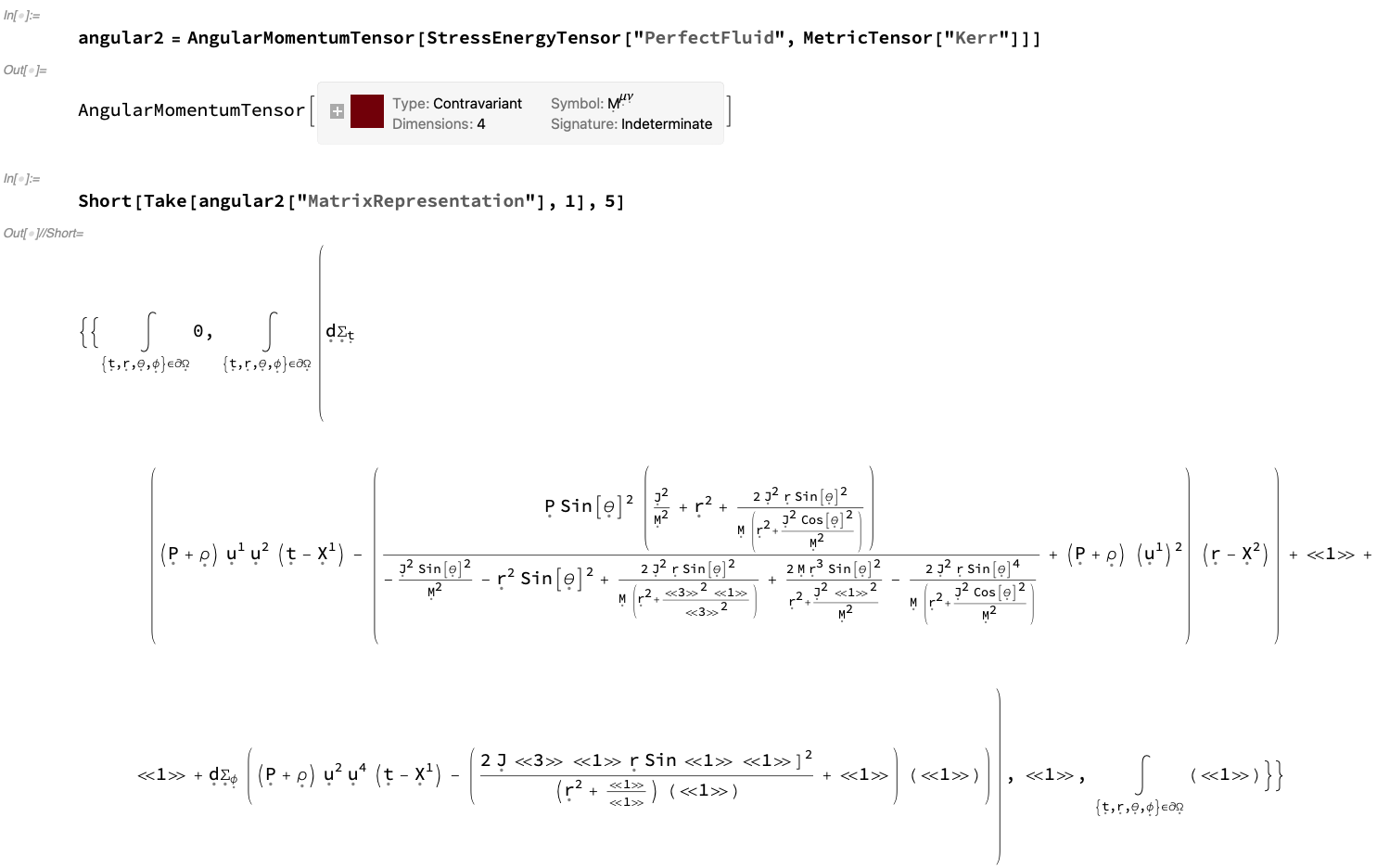}
\end{framed}
\caption{On the left, the \texttt{AngularMomentumTensor} object about the spacetime position vector ${\mathbf{X} = \left( X^1, X^2, X^3, X^4 \right)}$ for a perfect relativistic fluid (representing an idealized fluid with mass-energy density ${\rho}$, hydrostatic pressure $P$ and spacetime velocity ${u^{\mu}}$, but with vanishing heat conduction, viscosity and shear stresses) embedded within a Schwarzschild geometry in explicit contravariant matrix form, with both indices raised/contravariant. On the right, the \texttt{AngularMomentumTensor} object about the spacetime position vector ${\mathbf{X} = \left( X^1, X^2, X^3, X^4 \right)}$ for a perfect relativistic fluid (representing an idealized fluid with mass-energy density ${\rho}$, hydrostatic pressure $P$ and spacetime velocity ${u^{\mu}}$, but with vanishing heat conduction, viscosity and shear stresses) embedded within a Kerr geometry in explicit contravariant matrix form, with both indices raised/contravariant.}
\label{fig:Figure49}
\end{figure}

\begin{figure}[ht]
\centering
\begin{framed}
\includegraphics[width=0.545\textwidth]{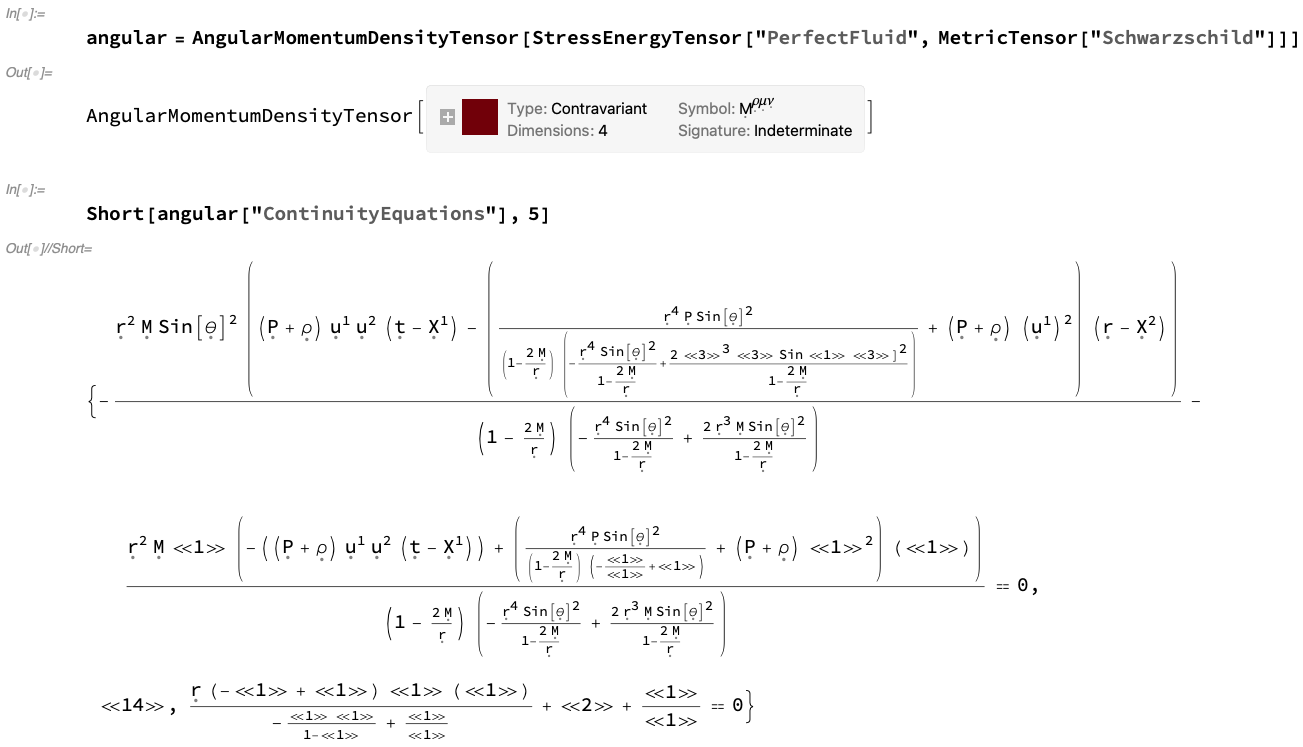}
\vrule
\includegraphics[width=0.445\textwidth]{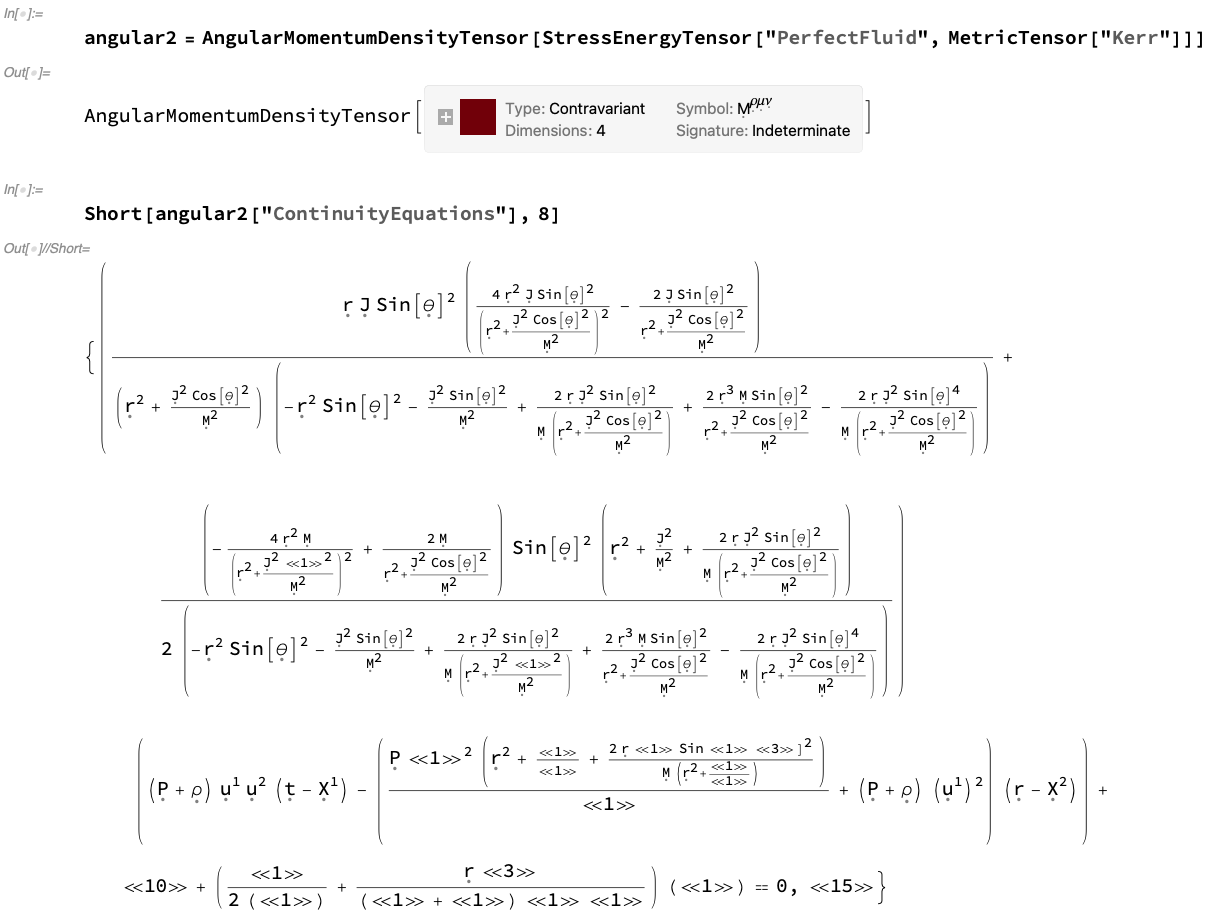}
\end{framed}
\caption{On the left, the list of continuity equations asserting that the covariant divergence of the \texttt{AngularMomentumDensityTensor} object about the spacetime position vector ${\mathbf{X} = \left( X^1, X^2, X^3, X^4 \right)}$ for a perfect relativistic fluid (representing an idealized fluid with mass-energy density ${\rho}$, hydrostatic pressure $P$ and spacetime velocity ${u^{\mu}}$, but with vanishing heat conduction, viscosity and shear stresses), embedded within a Schwarzshild geometry, vanishes identically. On the right, the list of continuity equations asserting that the covariant divergence of the \texttt{AngularMomentumDensityTensor} object about the spacetime position vector ${\mathbf{X} = \left( X^1, X^2, X^3, X^4 \right)}$ for a perfect relativistic fluid (representing an idealized fluid with mass-energy density ${\rho}$, hydrostatic pressure $P$ and spacetime velocity ${u^{\mu}}$, but with vanishing heat conduction, viscosity and shear stresses), embedded within a Kerr geometry, vanishes identically.}
\label{fig:Figure50}
\end{figure}

Once an appropriate relativistic energy-matter distribution over the manifold/spacetime has been selected in the form of a chosen \texttt{StressEnergyTensor} object, one can then proceed to compute other tensorial quantities based upon it, such as its relativistic (orbital) angular momentum. At any given point ${\mathbf{x} \in \mathcal{M}}$ in our manifold/spacetime, we are able to select a distinguished spacetime position vector (i.e. a tangent vector) ${\mathbf{X} \in T_{\mathbf{x}} \mathcal{M}}$, and hence to compute a rank-3 (orbital) angular momentum density tensor\cite{misner} ${M^{\rho \mu \nu}}$ about that position vector:

\begin{equation}
M^{\rho \mu \nu} = \left( x^{\rho} - X^{\rho} \right) T^{\mu \nu} - \left( x^{\mu} - X^{\mu} \right) T^{\rho \nu},
\end{equation}
where, as usual, ${\left\lbrace x^{\mu} \right\rbrace}$ designates our local choice of coordinate basis. Representations of the \texttt{AngularMomentumDensityTensor} objects about the default spacetime position vector ${\mathbf{X} = \left( X^1, X^2, X^3, X^4 \right)}$ for a perfect relativistic fluid (representing an idealized fluid with mass-energy density ${\rho}$, hydrostatic pressure $P$ and spacetime velocity ${u^{\mu}}$, but with vanishing heat conduction, viscosity and shear stresses) embedded within both a Schwarzschild metric and a Kerr metric, are shown in Figure \ref{fig:Figure47}. By default, this generic spacetime position vector ${\mathbf{X}}$ is chosen automatically, although the default can easily be overridden using additional arguments, as shown in Figure \ref{fig:Figure48}. If ${\Omega \subseteq \mathcal{M}}$ denotes a region of our manifold/spacetime with a codimension-1 boundary ${\partial \Omega}$ (typically ${\Omega}$ would represent some compact region of support for our energy-matter fields), then the full rank-2 (orbital) angular momentum tensor ${M^{\mu \nu}}$ about the spacetime position vector ${\mathbf{X} \in T_{\mathbf{x}} \mathcal{M}}$ can be obtained by integrating the rank-3 (orbital) angular momentum density tensor ${M^{\rho \mu \nu}}$ over this codimension-1 hypersurface ${\partial \Omega}$:

\begin{equation}
M^{\mu \nu} = \oint_{\partial \Omega} M^{\mu \nu \rho} d \Sigma_{\rho},
\end{equation}
with respect to the local spacetime coordinates ${\left\lbrace x^{\mu} \right\rbrace}$, with ${d \Sigma_{\rho}}$ being the corresponding volume 1-form for coordinate ${x^{\rho}}$. Representations of the full \texttt{AngularMomentumTensor} objects about the default spacetime position vector ${\mathbf{X} = \left( X^1, X^2, X^3, X^4 \right)}$ for a perfect relativistic fluid embedded within both a Schwarzschild metric and a Kerr metric, with ${\Omega}$ designating a four-dimensional spacetime region and ${d \Omega}$ representing a three-dimensional spacetime hypersurface constituting its boundary, are shown in Figure \ref{fig:Figure49}. The continuity equations on the underlying \texttt{StressEnergyTensor} object, asserting that the covariant divergence ${\nabla_{\nu} T^{\mu \nu}}$ must vanish identically, in turn imply a corresponding set of continuity equations on the \texttt{AngularMomentumDensityTensor} object, asserting that the covariant divergence ${\nabla_{\nu} M^{\rho \mu \nu}}$ must equivalently vanish identically:

\begin{multline}
\nabla_{\nu} T^{\mu \nu} = \frac{\partial}{\partial x^{\nu}} \left( T^{\mu \nu} \right) + \Gamma_{\nu \sigma}^{\mu} T^{\sigma \nu} + \Gamma_{\nu \sigma}^{\nu} T^{\mu \sigma} = 0,\\
\implies \qquad \nabla_{\nu} M^{\rho \mu \nu} = \frac{\partial}{\partial x^{\nu}} \left( M^{\rho \mu \nu} \right) + \Gamma_{\nu \sigma}^{\rho} M^{\sigma \mu \nu} + \Gamma_{\nu \sigma}^{\mu} M^{\rho \sigma \nu} + \Gamma_{\nu \sigma}^{\nu} M^{\rho \mu \sigma} = 0.
\end{multline}
Figure \ref{fig:Figure50} shows how such continuity equations may be derived directly from the \texttt{AngularMomentumDensityTensor} object for a perfect relativistic fluid embedded within both a Schwarzschild metric and a Kerr metric. We emphasize here that we are dealing only with the \textit{orbital} angular momentum tensor ${M^{\mu \nu}}$ since, in full generality, one must perform a spin-orbital decomposition in which the total angular momentum tensor ${J^{\mu \nu}}$ is given as the sum of the orbital angular momentum tensor and the spin tensor ${S^{\mu \nu}}$. However, since we are assuming a Levi-Civita connection on our manifold in which the (Cartan) torsion tensor, and thus also the spin tensor, vanishes identically, such a distinction between ${M^{\mu \nu}}$ and ${J^{\mu \nu}}$ is not necessary for our purposes (support for spin connections, torsion metrics, Riemann-Cartan geometry and Einstein-Cartan field equations within \textsc{Gravitas} is currently planned, but has not yet been fully implemented).

\begin{figure}[ht]
\centering
\begin{framed}
\includegraphics[width=0.445\textwidth]{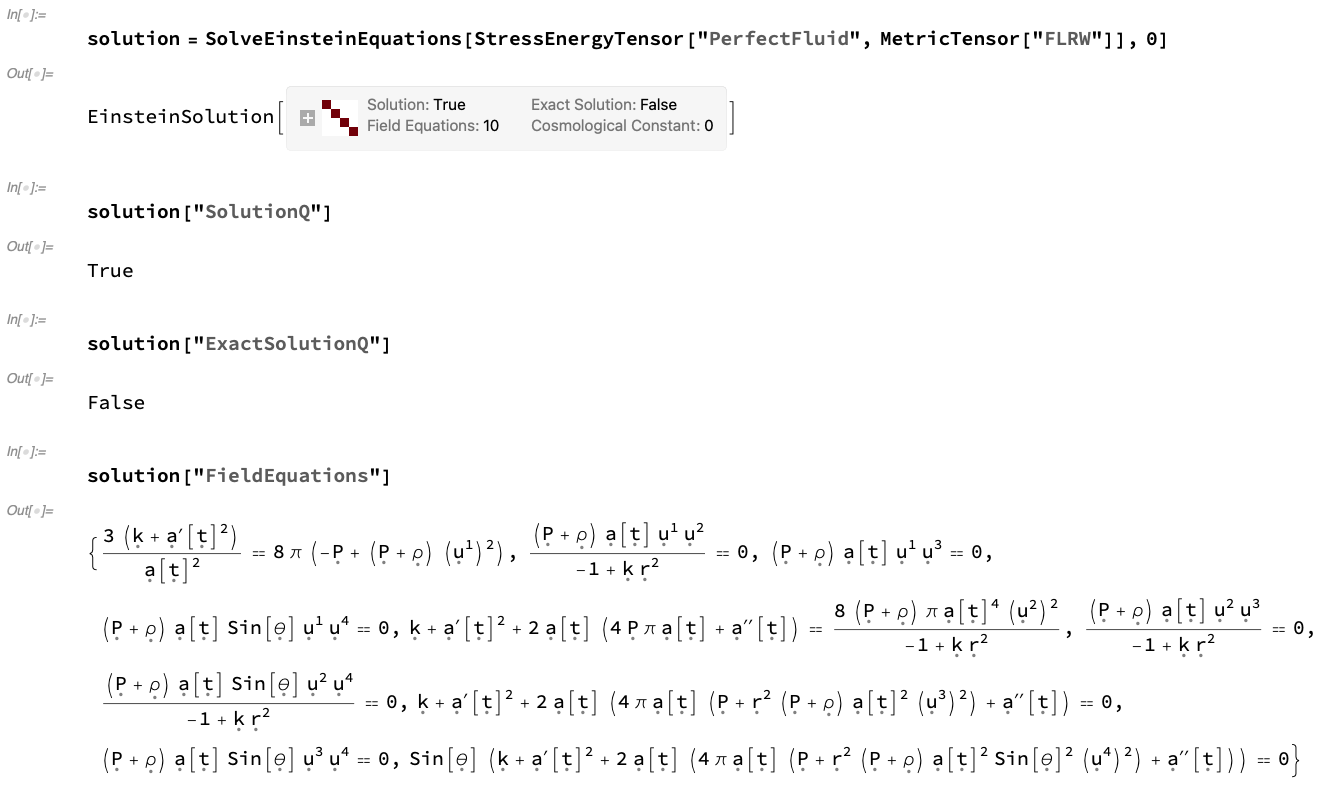}
\vrule
\includegraphics[width=0.545\textwidth]{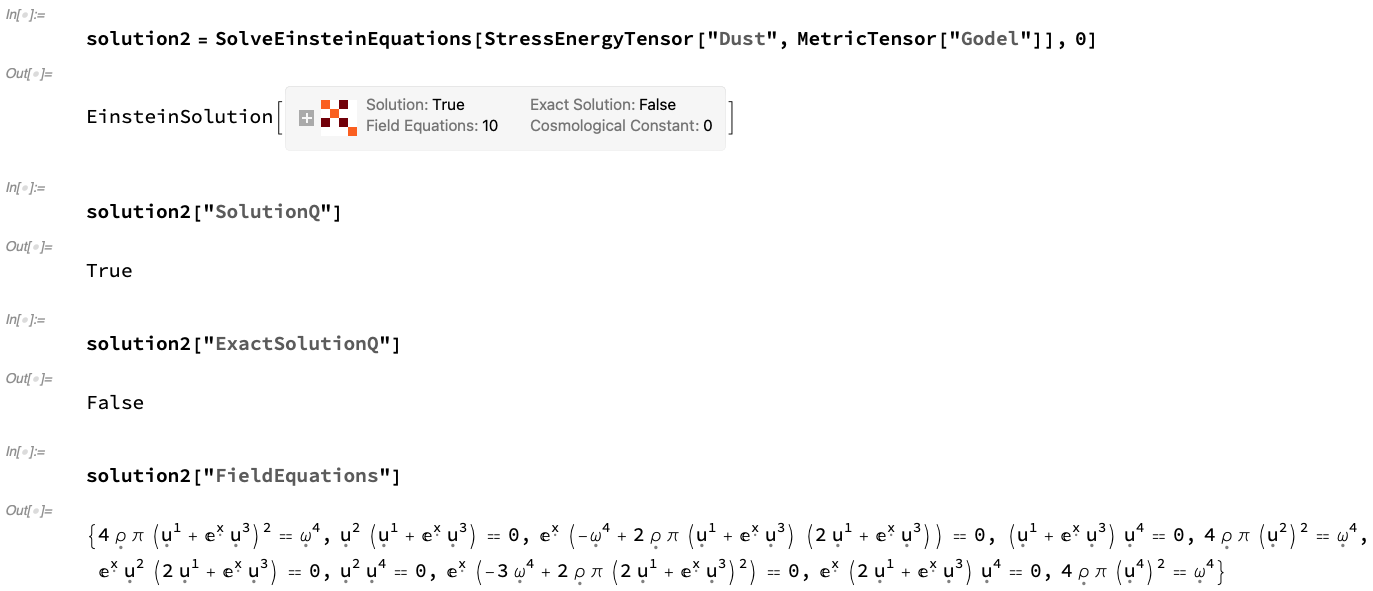}
\end{framed}
\caption{On the left, the \texttt{EinsteinSolution} object for an FLRW geometry equipped with a perfect relativistic fluid (representing a homogeneous, isotropic and uniformly expanding/contracting universe filled with an idealized fluid), with zero cosmological constant, computed using \texttt{SolveEinsteinEquations}, illustrating the the FLRW metric equipped with a perfect relativistic fluid is a non-exact solution to the Einstein field equations, with ten additional field equations required. On the right, the \texttt{EinsteinSolution} object for a G\"odel geometry equipped with a perfect relativistic dust (representing a rotating universe filled with an idealized distribution of dust particles), with zero cosmological constant, computed using \texttt{SolveEinsteinEquations}, illustrating that the G\"odel metric equipped with a perfect relativistic dust is a non-exact solution to the Einstein field equations, with ten additional field equations required.}
\label{fig:Figure51}
\end{figure}

\begin{figure}[ht]
\centering
\begin{framed}
\includegraphics[width=0.445\textwidth]{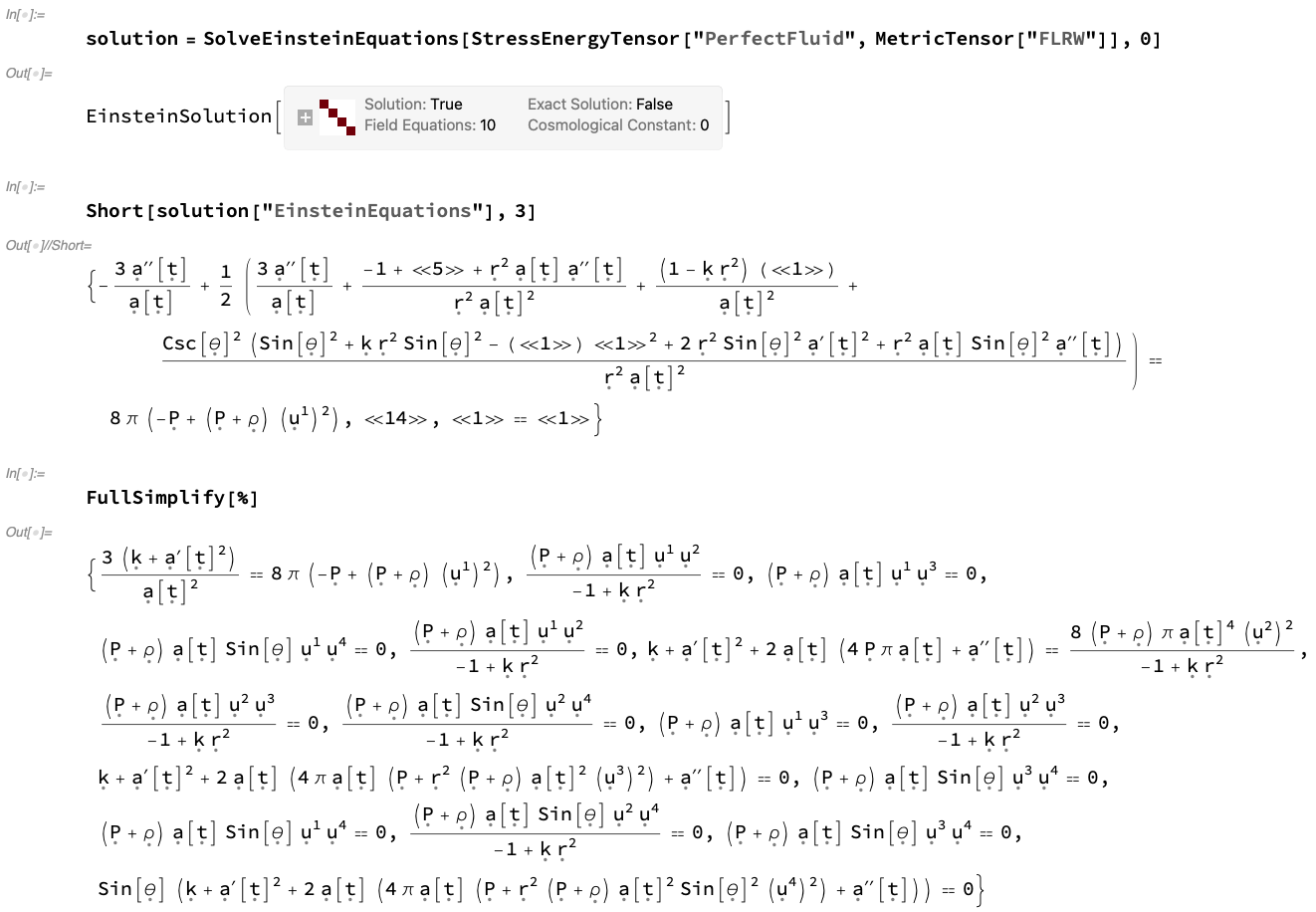}
\vrule
\includegraphics[width=0.545\textwidth]{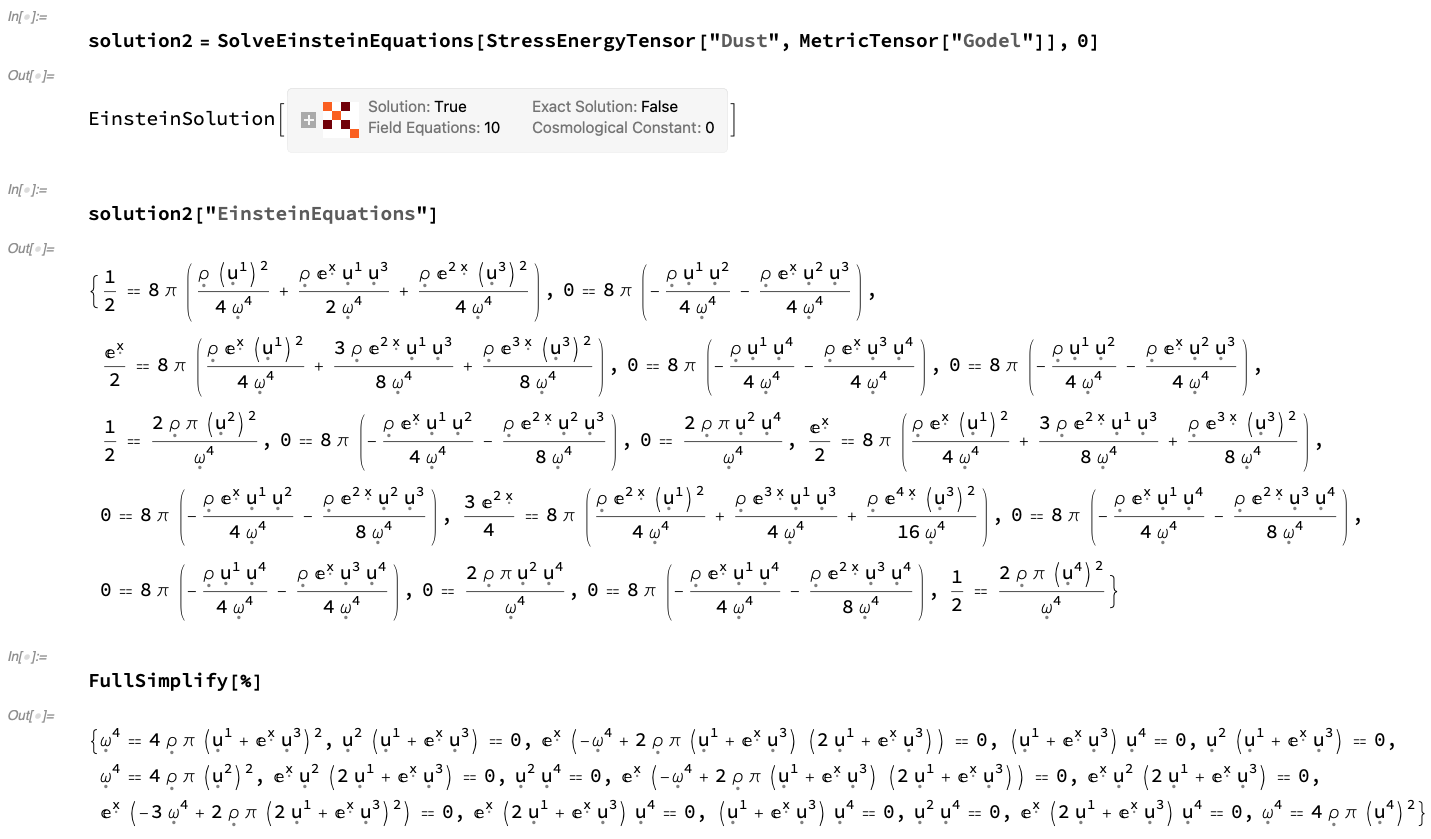}
\end{framed}
\caption{On the left, the list of Einstein field equations asserting that the Einstein tensor, with zero cosmological constant, is equal to ${8 \pi}$ times the stress-energy tensor, computed using the \texttt{EinsteinSolution} object for an FLRW geometry equipped with a perfect relativistic fluid (representing a homogeneous, isotropic and uniformly expanding/contracting universe filled with an idealized fluid), together with a verification that they reduce down to a set of ten canonical field equations. On the right, the list of Einstein field equations asserting that the Einstein tensor, with zero cosmological constant, is equal to ${8 \pi}$ times the stress-energy tensor, computed using the \texttt{EinsteinSolution} object for a G\"odel geometry equipped with a perfect relativistic dust (representing a rotating universe filled with an idealized distribution of dust particles), together with a verification that they reduce down to a set of ten canonical field equations.}
\label{fig:Figure52}
\end{figure}

\begin{figure}[ht]
\centering
\begin{framed}
\includegraphics[width=0.495\textwidth]{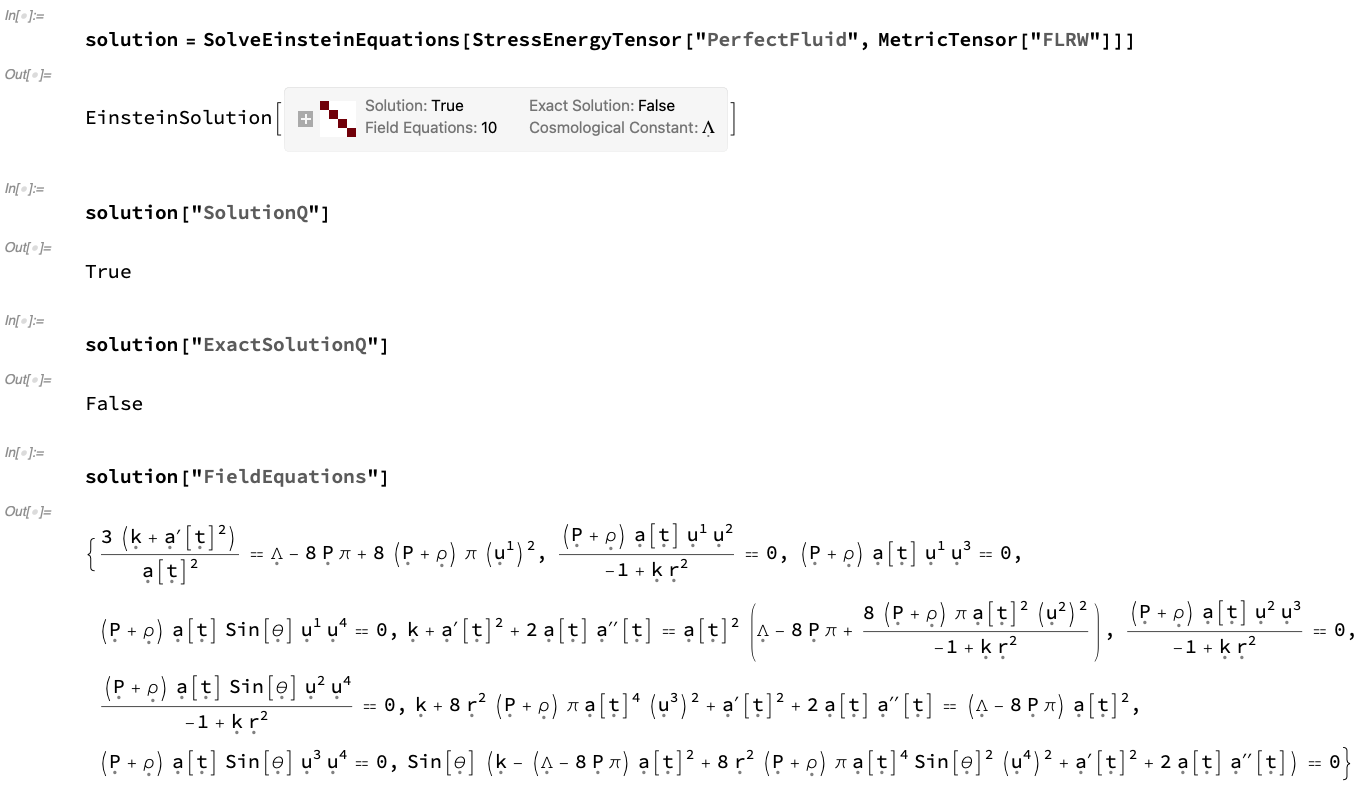}
\vrule
\includegraphics[width=0.495\textwidth]{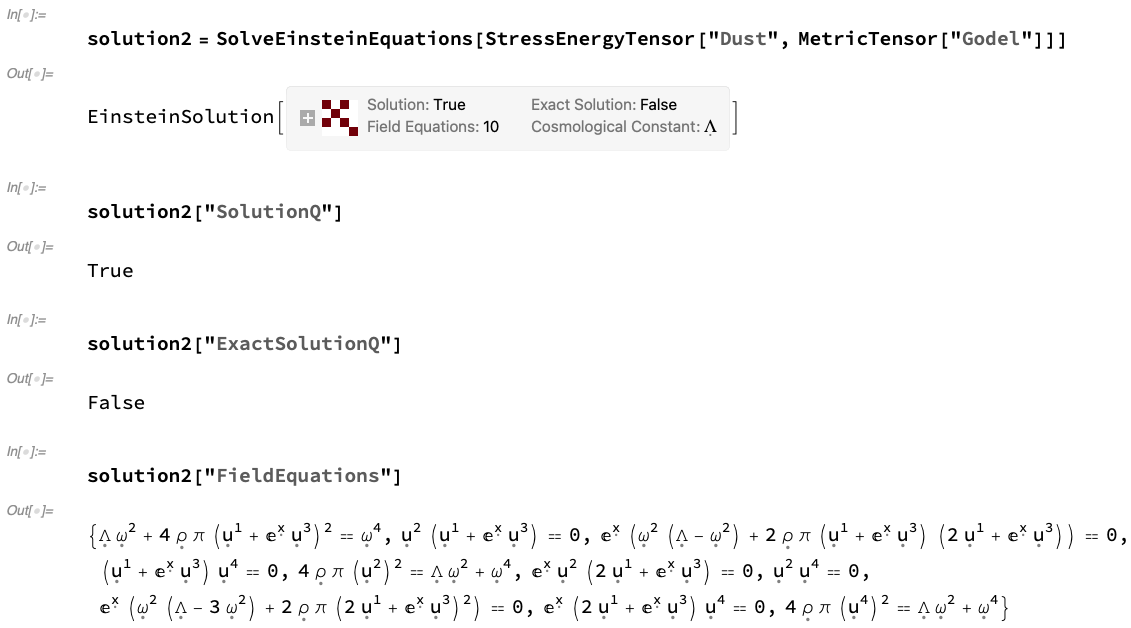}
\end{framed}
\caption{On the left, the \texttt{EinsteinSolution} object for an FLRW geometry equipped with a perfect relativistic fluid (representing a homogeneous, isotropic and uniformly expanding/contracting universe filled with an idealized fluid), with non-zero cosmological constant ${\Lambda \neq 0}$, computed using \texttt{SolveEinsteinEquations}, illustrating that the FLRW metric with non-vanishing cosmological constant and equipped with a perfect relativistic fluid is a non-exact solution to the Einstein field equations, with ten additional field equations required. On the right, the \texttt{EinsteinSolution} object for a G\"odel geometry equipped with a perfect relativistic dust (representing a rotating universe filled with an idealized distribution of dust particles), with non-zero cosmological constant ${\Lambda \neq 0}$, computed using \texttt{SolveEinsteinEquations}, illustrating that the G\"odel metric with non-vanishing cosmological constant and equipped with a perfect relativistic dust is a non-exact solution to the Einstein field equations, with ten additional field equations required.}
\label{fig:Figure53}
\end{figure}

\begin{figure}[ht]
\centering
\begin{framed}
\includegraphics[width=0.445\textwidth]{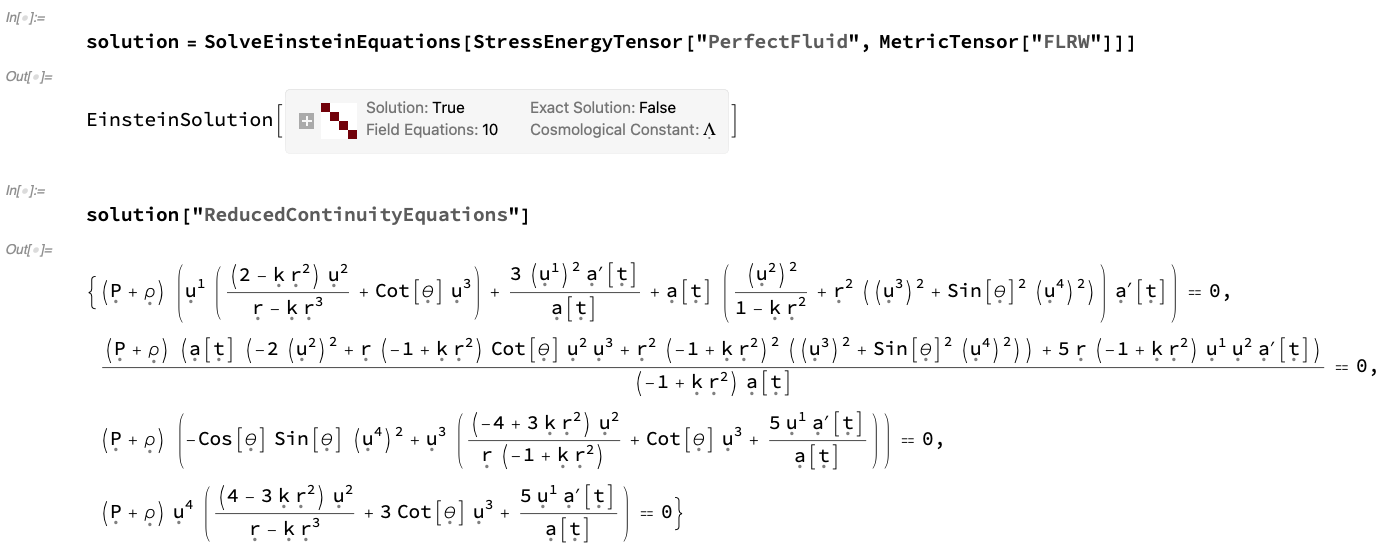}
\vrule
\includegraphics[width=0.545\textwidth]{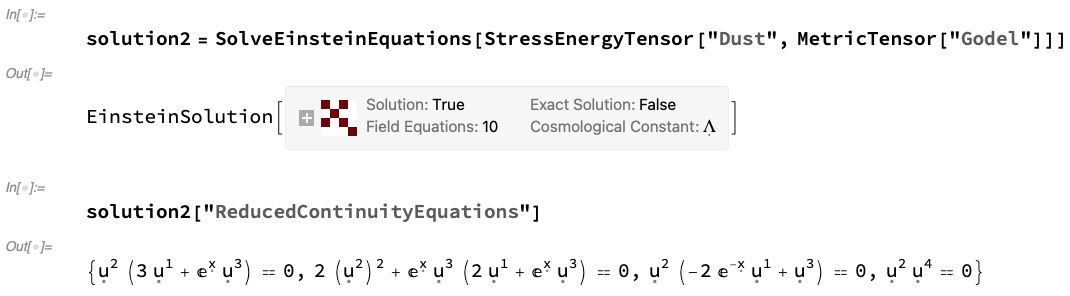}
\end{framed}
\caption{On the left, the list of continuity equations asserting that the covariant divergence of the stress-energy tensor vanishes identically, computed using the \texttt{EinsteinSolution} object for an FLRW geometry equipped with a perfect relativistic fluid (representing a homogeneous, isotropic and uniformly expanding/contracting universe filled with an idealized fluid), with non-zero cosmological constant ${\Lambda \neq 0}$. On the right, the list of continuity equations asserting that the covariant divergence of the stress-energy tensor vanishes identically, computed using the \texttt{EinsteinSolution} object for a G\"odel geometry equipped with a perfect relativistic dust (representing a rotating universe filled with an idealized distribution of dust particles), with non-zero cosmological constant ${\Lambda \neq 0}$.}
\label{fig:Figure54}
\end{figure}

We are now in a position to be able to impose the \textit{full} Einstein field equations on our manifold ${\mathcal{M}}$ (including stress-energy source terms), which assert that the Einstein tensor, plus an optional cosmological constant term ${\Lambda}$, is equal to ${8 \pi}$ times the stress-energy tensor:

\begin{equation}
G_{\mu \nu} + \Lambda g_{\mu \nu} = R_{\mu \nu} - \frac{1}{2} R g_{\mu \nu} + \Lambda g_{\mu \nu} = 8 \pi T_{\mu \nu}.
\end{equation}
Representations of the corresponding \texttt{EinsteinSolution} objects for the FLRW metric (representing e.g. a homogeneous, isotropic and uniformly expanding/contracting universe with global curvature $k$ and scale factor ${a \left( t \right)}$ in spherical polar coordinates ${\left( t, r, \theta, \phi \right)}$) equipped with a perfect relativistic fluid (representing an idealized fluid with mass-energy density ${\rho}$, hydrostatic pressure $P$ and spacetime velocity ${u^{\mu}}$, but with vanishing heat conduction, viscosity and shear stresses), and for the G\"odel metric (representing e.g. a rotating, dust-filled universe with global angular velocity ${\omega}$ in G\"odel's Cartesian-like coordinates ${\left( t, x, y, z \right)}$) equipped with a perfect relativistic dust (representing an idealized distribution of dust particles with mass-energy density ${\rho}$ and spacetime velocity ${u^{\mu}}$, but with vanishing hydrostatic pressure), computed using the \texttt{SolveEinsteinEquations} function, are shown in Figure \ref{fig:Figure51}; these examples demonstrate that the FLRW and G\"odel metrics are both \textit{non-exact} solutions of the full Einstein field equations for these particular energy-matter distributions, in the sense that ten additional field equations need to be assumed in each case. Note that, in the former case, these field equations correspond to a strict generalization of the Friedmann equations in relativistic cosmology\cite{friedmann}, in which the fluid is permitted to be comoving. The complete lists of Einstein field equations for both metric/stress-energy tensor combinations can be computed directly from the \texttt{EinsteinSolution} object, and it can be verified in both cases that they do indeed reduce down to the ten canonical field equations previously mentioned, as illustrated in Figure \ref{fig:Figure52}. Although all of these examples thus far have assumed a vanishing cosmological constant (i.e. ${\Lambda = 0}$), in Figure \ref{fig:Figure53} we show representations of the corresponding \texttt{EinsteinSolution} objects for the FLRW metric equipped with a perfect relativistic fluid and the G\"odel metric equipped with a perfect relativistic dust, assuming in both cases a non-zero value of the cosmological constant ${\Lambda}$; these examples demonstrate that the FLRW and G\"odel metrics are also both \text{non-exact} solutions of the full Einstein field equations for these particular energy-matter distributions when assuming a non-vanishing cosmological constant (i.e. ${\Lambda \neq 0}$) too, in the sense that ten additional field equations still need to be assumed in each case. Figure \ref{fig:Figure54} shows the full lists of continuity equations, asserting that the covariant divergence of the stress-energy tensor ${T_{\mu \nu}}$ must vanish identically, computed directly from the \texttt{EinsteinSolution} objects for a perfect relativistic fluid embedded within an FLRW metric and a perfect relativistic dust embedded within a G\"odel metric, thus ensuring that relativistic energy and momentum are identically conserved.

\begin{figure}[ht]
\centering
\begin{framed}
\includegraphics[width=0.495\textwidth]{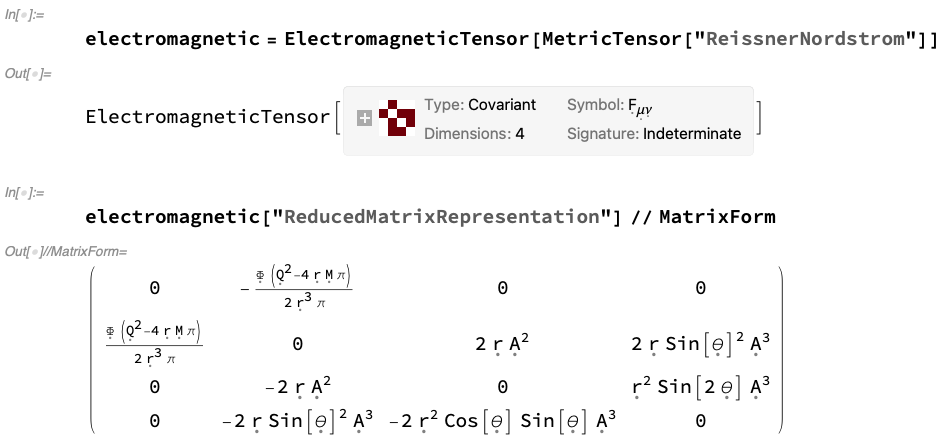}
\vrule
\includegraphics[width=0.495\textwidth]{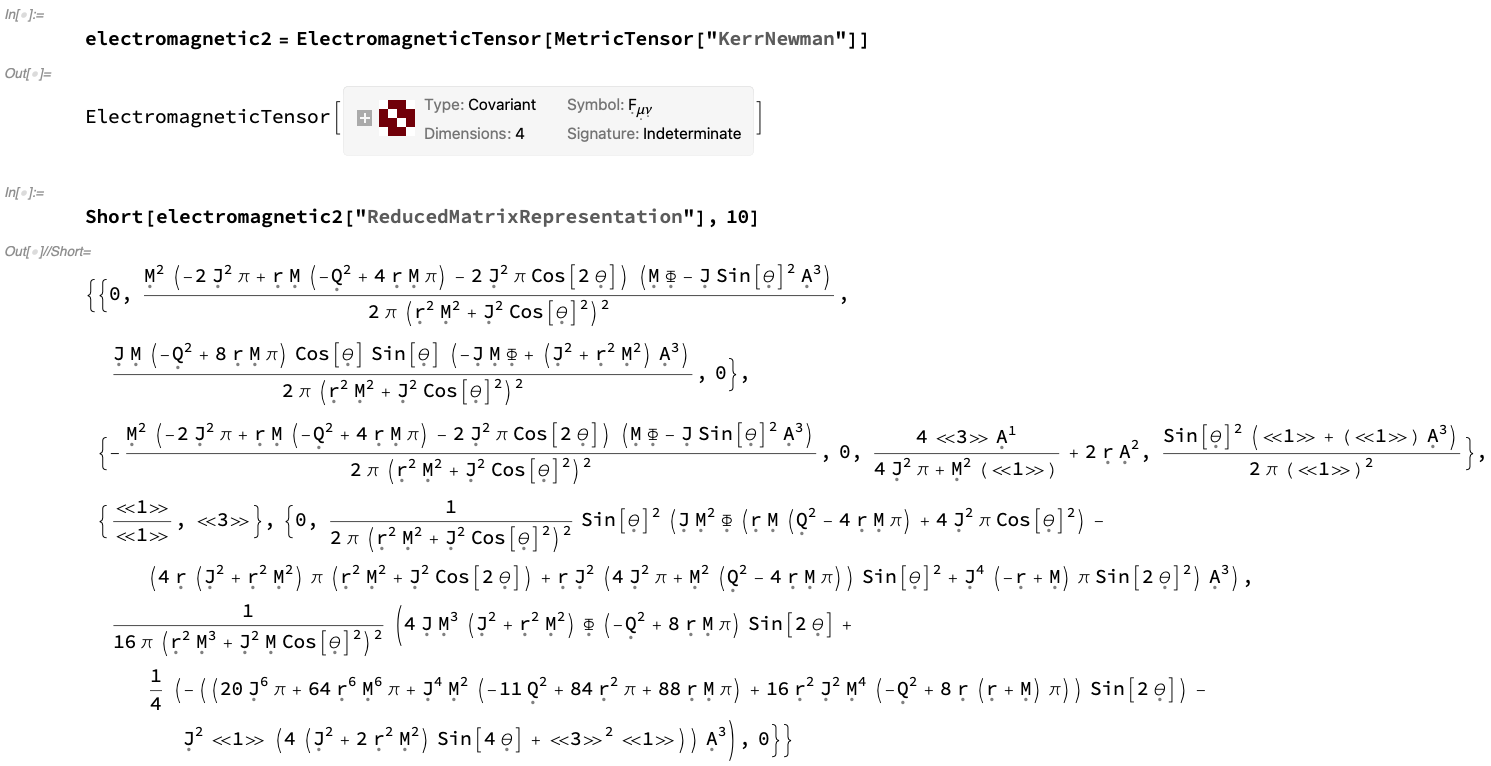}
\end{framed}
\caption{On the left, the \texttt{ElectromagneticTensor} object with spacetime electromagnetic potential ${\mathbf{A} = \left( \Phi, A^1, A^2, A^3 \right)}$ for a Reissner-Nordstr\"om geometry (representing e.g. a charged, non-rotating black hole of mass $M$ and electric charge $Q$ in Schwarzschild/spherical polar coordinates ${\left( t, r, \theta, \phi \right)}$). On the right, the \texttt{ElectromagneticTensor} object with spacetime electromagnetic potential ${\mathbf{A} = \left( \Phi, A^1, A^2, A^3 \right)}$ for a Kerr-Newman geometry (representing e.g. a charged, spinning black hole of mass $M$, angular momentum $J$ and electric charge $Q$ in Boyer-Lindquist/oblate spheroidal coordinates ${\left( t, r, \theta, \phi \right)}$).}
\label{fig:Figure55}
\end{figure}

\begin{figure}[ht]
\centering
\begin{framed}
\includegraphics[width=0.545\textwidth]{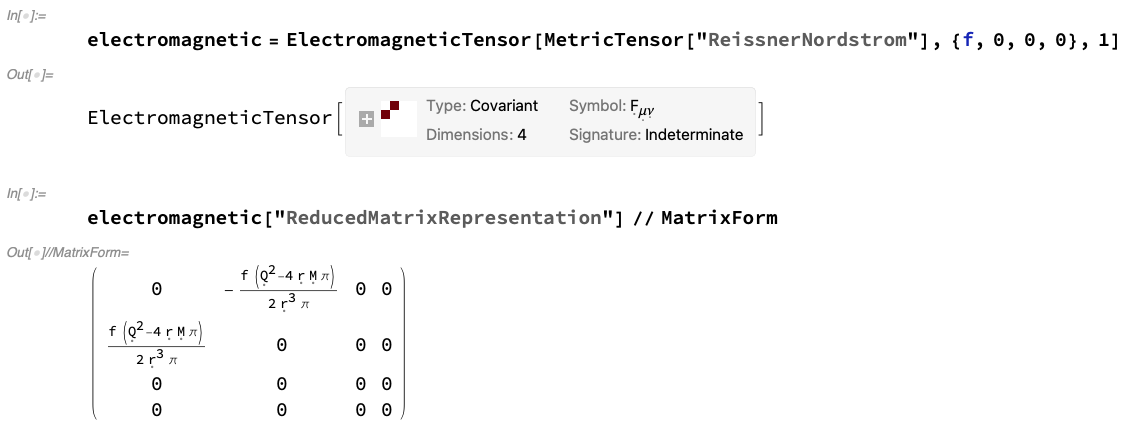}
\vrule
\includegraphics[width=0.445\textwidth]{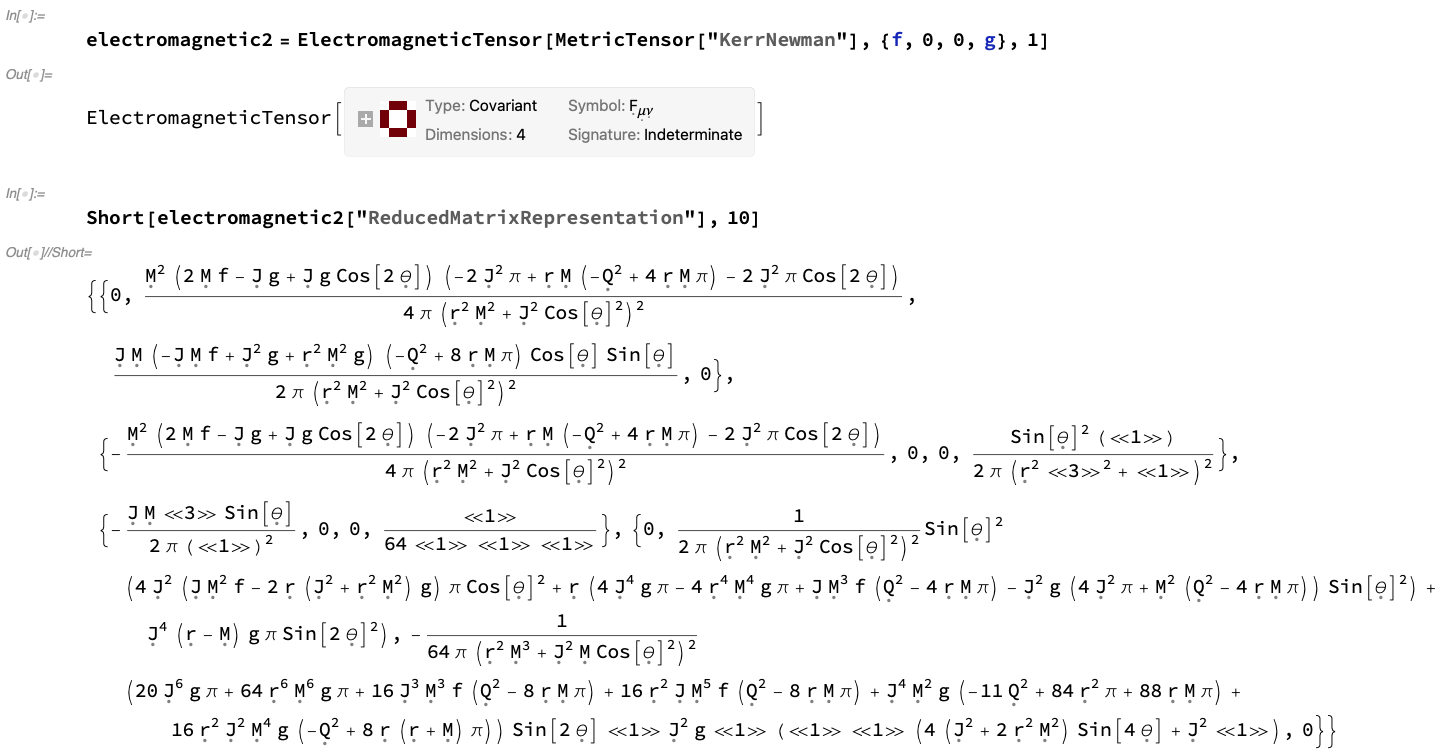}
\end{framed}
\caption{On the left, the \texttt{ElectromagneticTensor} object with modified spacetime electromagnetic potential ${\mathbf{A} = \left( f, 0, 0, 0 \right)}$ for a Reissner-Nordstr\"om geometry (representing e.g. a charged, non-rotating black hole of mass $M$ and electric charge $Q$ in Schwarzschild/spherical polar coordinates ${\left( t, r, \theta, \phi \right)}$). On the right, the \texttt{ElectromagneticTensor} object with modified spacetime electromagnetic potential ${\mathbf{A} = \left( f, 0, 0, g \right)}$ for a Kerr-Newman geometry (representing e.g. a charged, spinning black hole of mass $M$, angular momentum $J$ and electric charge $Q$ in Boyer-Lindquist/oblate spheroidal coordinates ${\left( t, r, \theta, \phi \right)}$).}
\label{fig:Figure56}
\end{figure}

\begin{figure}[ht]
\centering
\begin{framed}
\includegraphics[width=0.545\textwidth]{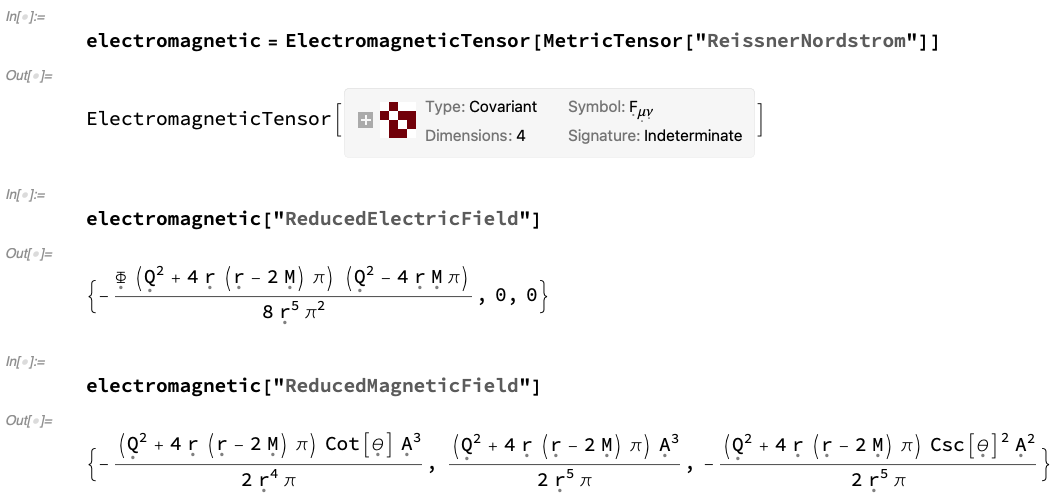}
\vrule
\includegraphics[width=0.445\textwidth]{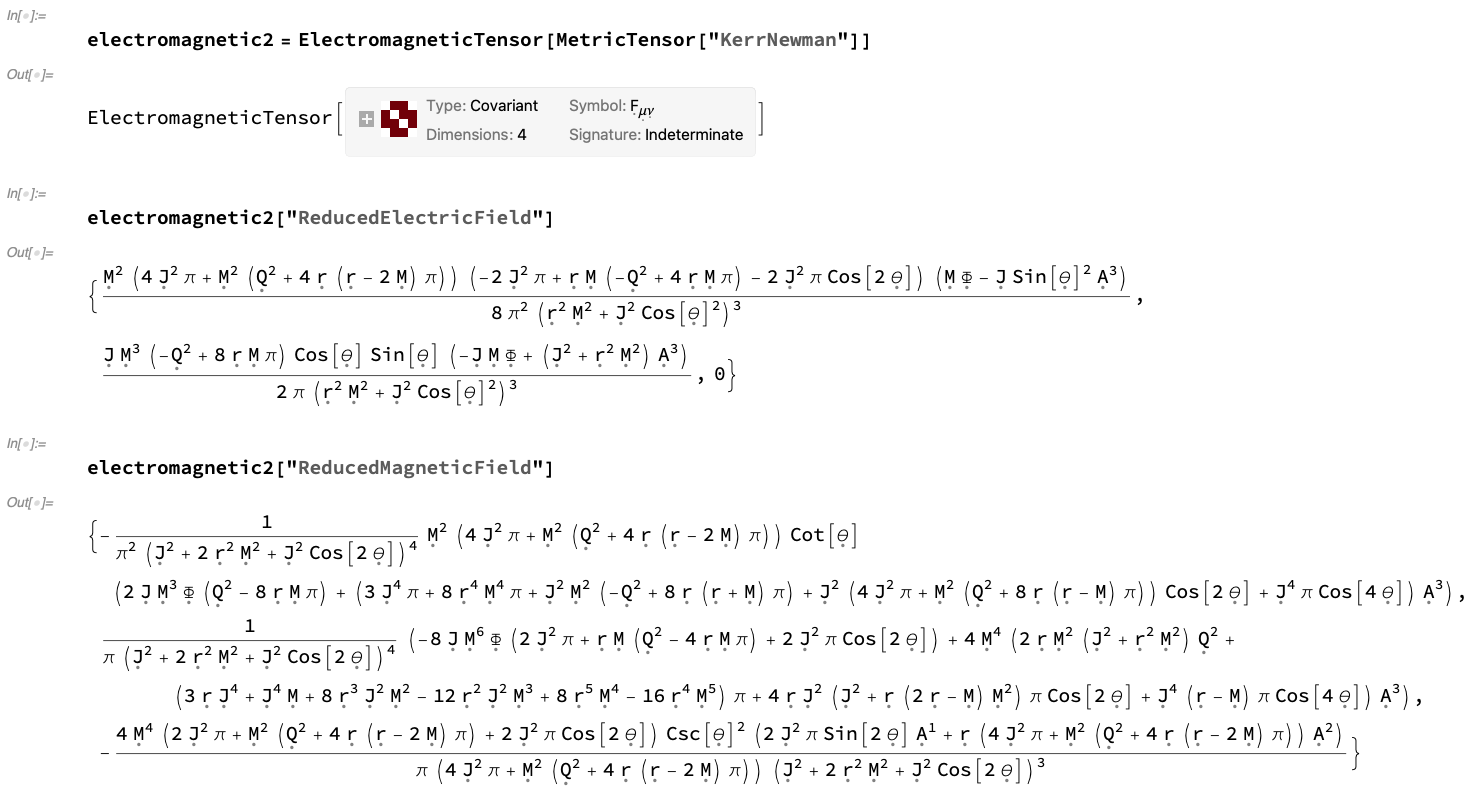}
\end{framed}
\caption{On the left, the relativistic electric and magnetic field vectors of the \texttt{ElectromagneticTensor} object with spacetime electromagnetic potential ${\mathbf{A} = \left( \Phi, A^1, A^2, A^3 \right)}$ for a Reissner-Nordstr\"om geometry (representing e.g. a charged, non-rotating black hole of mass $M$ and electric charge $Q$ in Schwarzschild/spherical polar coordinates ${\left( t, r, \theta, \phi \right)}$). On the right, the relativistic electric and magnetic field vectors of the \texttt{ElectromagneticTensor} object with spacetime electromagnetic potential ${\mathbf{A} = \left( \Phi, A^1, A^2, A^3 \right)}$ for a Kerr-Newman geometry (representing e.g. a charged, spinning black hole of mass $M$, angular momentum $J$ and electric charge $Q$ in Boyer-Lindquist/oblate spheroidal coordinates ${\left( t, r, \theta, \phi \right)}$).}
\label{fig:Figure57}
\end{figure}

\begin{figure}[ht]
\centering
\begin{framed}
\includegraphics[width=0.595\textwidth]{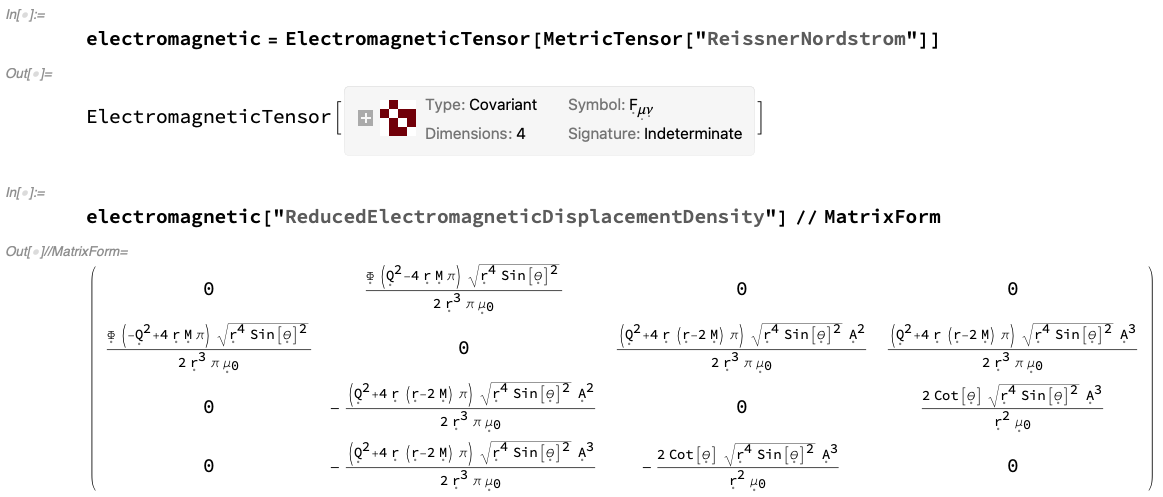}
\vrule
\includegraphics[width=0.395\textwidth]{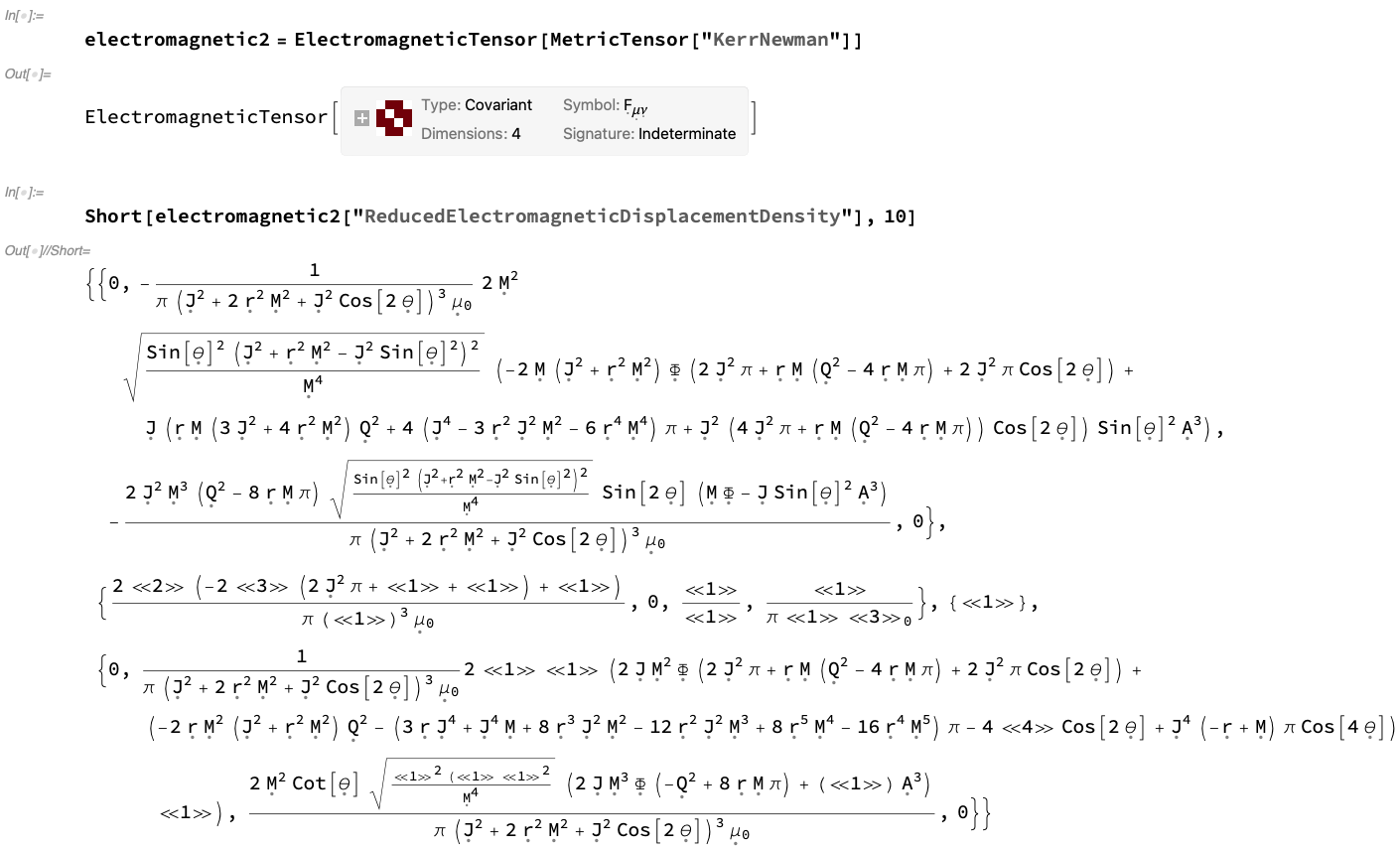}
\end{framed}
\caption{On the left, the relativistic electromagnetic displacement tensor density of the \texttt{ElectromagneticTensor} object with spacetime electromagnetic potential ${\mathbf{A} = \left( \Phi, A^1, A^2, A^3 \right)}$ for a Reissner-Nordstr\"om geometry (representing e.g. a charged, non-rotating black hole of mass $M$ and electric charge $Q$ in Schwarzschild/spherical polar coordinates ${\left( t, r, \theta, \phi \right)}$), in explicit matrix form. On the right, the relativistic electromagnetic displacement tensor density of the \texttt{ElectromagneticTensor} object with spacetime electromagnetic potential ${\mathbf{A} = \left( \Phi, A^1, A^2, A^3 \right)}$ for a Kerr-Newman geometry (representing e.g. a charged, spinning black hole of mass $M$, angular momentum $J$ and electric charge $Q$ in Boyer-Lindquist/oblate spheroidal coordinates ${\left( t, r, \theta, \phi \right)}$), in explicit matrix form.}
\label{fig:Figure58}
\end{figure}

\begin{figure}[ht]
\centering
\begin{framed}
\includegraphics[width=0.595\textwidth]{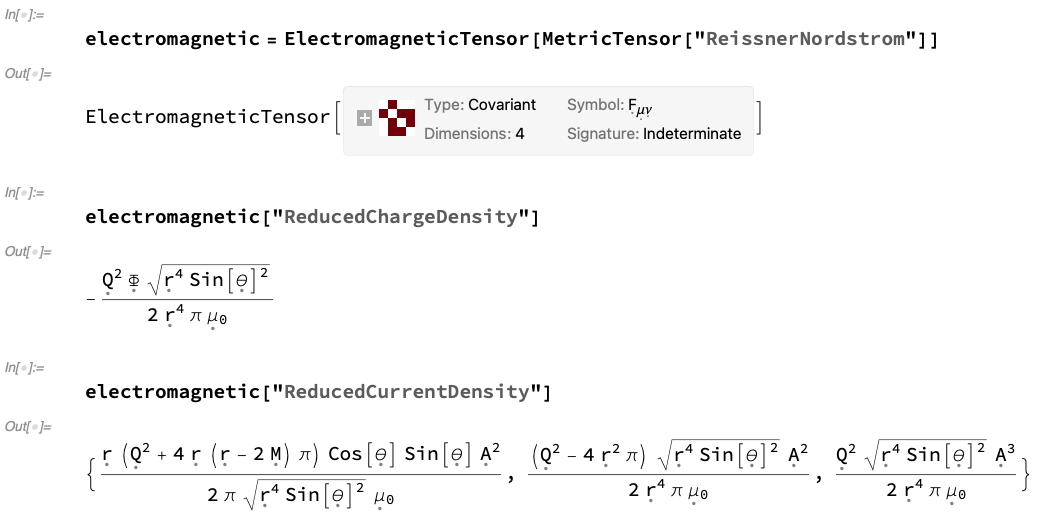}
\vrule
\includegraphics[width=0.395\textwidth]{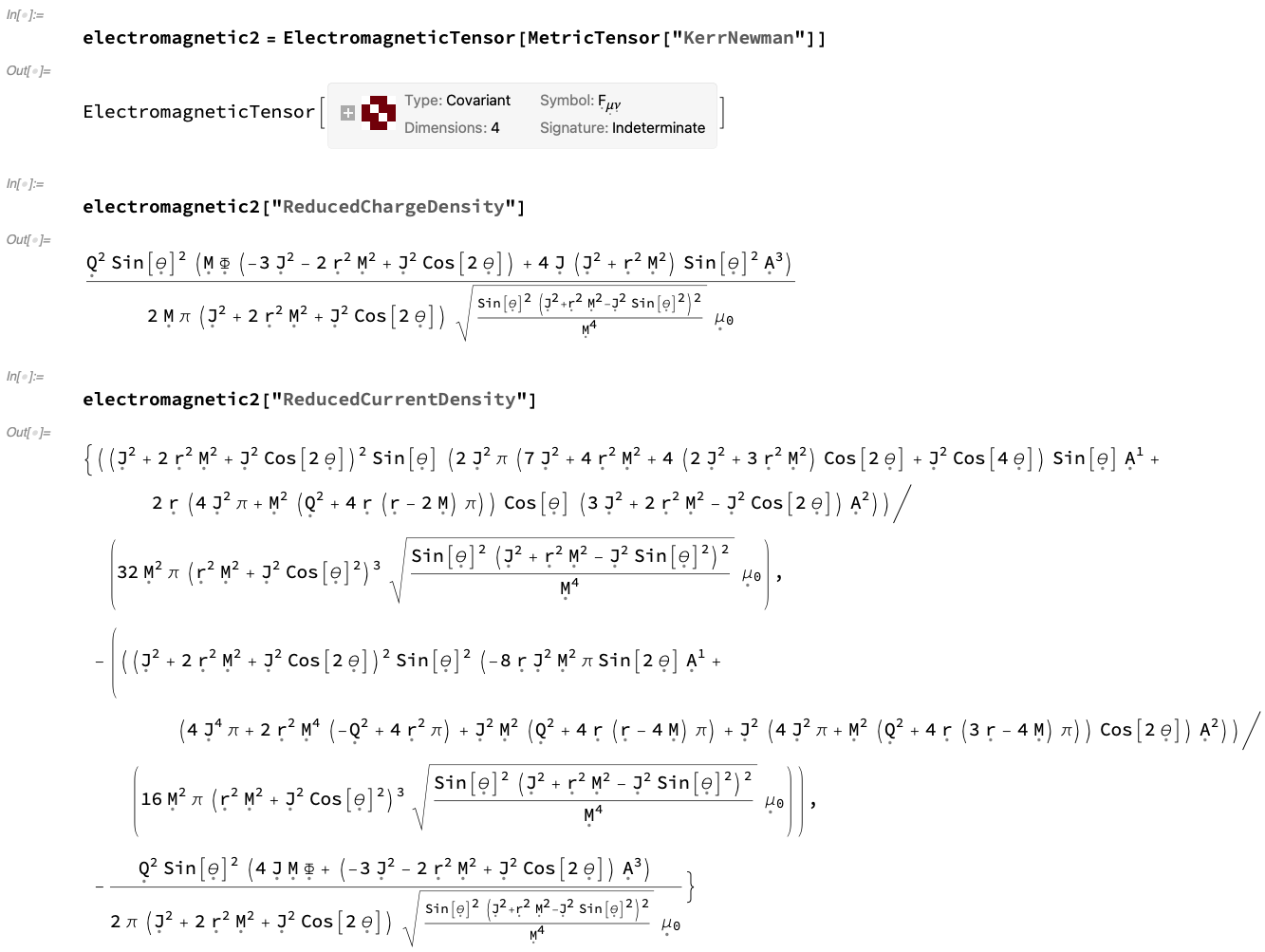}
\end{framed}
\caption{On the left, the relativistic electric charge and electric current densities of the \texttt{ElectromagneticTensor} object with spacetime electromagnetic potential ${\mathbf{A} = \left( \Phi, A^1, A^2, A^3 \right)}$ for a Reissner-Nordstr\"om geometry (representing e.g. a charged, non-rotating black hole of mass $M$ and electric charge $Q$ in Schwarzschild/spherical polar coordinates ${\left( t, r, \theta, \phi \right)}$). On the right, the relativistic electric charge and electric current densities of the \texttt{ElectromagneticTensor} object with spacetime electromagnetic potential ${\mathbf{A} = \left( \Phi, A^1, A^2, A^3 \right)}$ for a Kerr-Newman geometry (representing e.g. a charged, spinning black hole of mass $M$, angular momentum $J$ and electric charge $Q$ in Boyer-Lindquist/oblate spheroidal coordinates ${\left( t, r, \theta, \phi \right)}$).}
\label{fig:Figure59}
\end{figure}

As mentioned previously, the \textsc{Gravitas} framework handles the case of electromagnetic fields somewhat separately from the case of all other non-gravitational fields, specifically since it incorporates specialized functionality for computing and manipulating the electromagnetic field strength tensor/Faraday tensor ${F_{\mu \nu}}$, and for solving the resulting Einstein-Maxwell field equations. We start by considering the spacetime electromagnetic potential ${\mathbf{A}}$, which includes both a timelike component ${A^0}$, which we interpret as an electric scalar potential ${\Phi}$, and spacelike components ${A^{\mu}}$ (for ${\mu \neq 0}$), which we interpret as a magnetic vector potential. By computing the exterior derivative of the 1-form components ${A_{\mu} = g_{\mu \sigma} A^{\sigma}}$ of the spacetime electromagnetic potential ${\mathbf{A}}$, we obtain the electromagnetic tensor\cite{griffiths} ${F_{\mu \nu}}$:

\begin{equation}
F_{\mu \nu} = \frac{\partial}{\partial x^{\mu}} \left( A_{\nu} \right) - \frac{\partial}{\partial x^{\nu}} \left( A_{\mu} \right).
\end{equation}
Representations of the \texttt{ElectromagneticTensor} objects for a Reissner-Nordstr\"om metric (representing e.g. a charged, non-rotating black hole of mass $M$ and electric charge $Q$ in Schwarzschild/spherical polar coordinates ${\left( t, r, \theta, \phi \right)}$) and a Kerr-Newman metric (representing e.g. a charged, spinning black hole of mass $M$, angular momentum $J$ and electric charge $Q$ in Boyer-Lindquist/oblate spheroidal coordinates ${\left( t, r, \theta, \phi \right)}$), assuming the default spacetime electromagnetic potential vector ${\mathbf{A} = \left( \Phi, A^1, A^2, A^3 \right)}$, are shown in Figure \ref{fig:Figure55}. By default, this generic spacetime electromagnetic potential vector ${\mathbf{A}}$ is chosen automatically, although the default can easily be overridden using additional arguments, as shown in Figure \ref{fig:Figure56}. The components of the electric field vector ${E^{\mu}}$ and magnetic field vector ${B^{\mu}}$ can then be recovered straightforwardly from the components of the full electromagnetic tensor ${F_{\mu \nu}}$; these are traditionally represented as covectors/1-forms ${E_{\mu}}$ and ${B_{\mu}}$, but may easily be converted into vector form by means of the inverse metric tensor ${g^{\mu \nu}}$:

\begin{equation}
E_{\mu} = F_{0 \mu}, \qquad \implies \qquad E^{\mu} = g^{\sigma \mu} F_{0 \sigma} = F_{0}^{\mu},
\end{equation}
where ${\mu \neq 0}$ but ${\sigma}$ ranges across all ${\left\lbrace 0, \dots, n - 1 \right\rbrace}$, and:

\begin{equation}
B_{\mu} = - \frac{1}{2} \varepsilon_{\mu \alpha \beta} F^{\alpha \beta}, \qquad \implies \qquad B^{\mu} = - \frac{1}{2} g^{\mu \sigma} \varepsilon_{\sigma \alpha \beta} F^{\alpha \beta} = - \frac{1}{2} \varepsilon_{\alpha \beta}^{\mu} F^{\alpha \beta},
\end{equation}
where ${\mu \neq 0}$ but ${\alpha}$, ${\beta}$ and ${\sigma}$ range across all ${\left\lbrace 0, \dots, n - 1 \right\rbrace}$, and where ${\varepsilon_{\rho \mu \nu}}$ designates, as usual, the totally-antisymmetric Levi-Civita symbol (and hence this duality between electric and magnetic fields is definable only in four dimensions). Figure \ref{fig:Figure57} shows the components of the electric field vector ${E^{\mu}}$ and magnetic field vector ${B^{\mu}}$ (for ${\mu \neq 0}$), computed directly from the \texttt{ElectromagneticTensor} objects for a Reissner-Nordstr\"om metric and a Kerr-Newman metric, and assuming the default spacetime electromagnetic potential vector ${\mathbf{A} = \left( \Phi, A^1, A^2, A^3 \right)}$. Likewise, the components of the rank-2 electromagnetic displacement tensor density ${\mathcal{D}^{\mu \nu}}$ (notable for being the only quantity in the theory of general relativistic electromagnetism that explicitly depends upon the metric tensor ${g_{\mu \nu}}$, or more precisely upon the metric determinant ${\det \left( g_{\mu \nu} \right)}$, thus encoding the fact that electromagnetic fields have the effect of changing the effective speed of light relative to some global coordinate system\cite{landau}) are related to the components of ${F_{\mu \nu}}$ in a highly explicit and elementary way:

\begin{equation}
\mathcal{D}^{\mu \nu} = \frac{\sqrt{ - \det \left( g_{\mu \nu} \right)}}{\mu_0} g^{\mu \alpha} F_{\alpha \beta} g^{\beta \nu} = \frac{\sqrt{ - \det \left( g_{\mu \nu} \right)}}{\mu_0} F^{\mu \nu},
\end{equation}
where ${mu_0}$ denotes the vacuum magnetic permeability constant. Figure \ref{fig:Figure58} shows the electromagnetic displacement tensor density ${\mathcal{D}^{\mu \nu}}$, in explicit matrix form, also computed directly from the \texttt{ElectromagneticTensor} objects for a Reissner-Nordstr\"om metric and a Kerr-Newman metric with spacetime electromagnetic potential ${\mathbf{A} = \left( \Phi, A^1, A^2, A^2 \right)}$. Among other things, the significance of the electromagnetic displacement tensor density ${\mathcal{D}^{\mu \nu}}$ is that its partial divergence (or, equivalently, its covariant divergence) can be decomposed into a timelike component ${\frac{\partial}{\partial x^{\nu}} \left( \mathcal{D}^{0 \nu} \right)}$, representing an electric charge density ${\rho}$, and spacelike components ${\frac{\partial}{\partial x^{\nu}} \left( \mathcal{D}^{\mu \nu} \right)}$ (for ${\mu \neq 0}$), representing an electric current density vector ${J^{\mu}}$, i.e:

\begin{equation}
\rho = \frac{\partial}{\partial x^{\nu}} \left( \mathcal{D}^{0 \nu} \right) = \frac{\partial}{\partial x^{\nu}} \left( \frac{\sqrt{ - \det \left( g_{\mu \nu} \right)}}{\mu_0} g^{0 \alpha} F_{\alpha \beta} g^{\beta \nu} \right) = \frac{\partial}{\partial x^{\nu}} \left( \frac{\sqrt{- \det \left( g_{\mu \nu} \right)}}{\mu_0} F^{0 \nu} \right),
\end{equation}
and:

\begin{equation}
J^{\mu} = \frac{\partial}{\partial x^{\nu}} \left( \mathcal{D}^{\mu \nu} \right) = \frac{\partial}{\partial x^{\nu}} \left( \frac{\sqrt{- \det \left( g_{\mu \nu} \right)}}{\mu_0} g^{\mu \alpha} F_{\alpha \beta} g^{\beta \nu} \right) = \frac{\partial}{\partial x^{\nu}} \left( \frac{\sqrt{- \det \left( g_{\mu \nu} \right)}}{\mu_0} F^{\mu \nu} \right),
\end{equation}
where ${\mu \neq 0}$ but ${\alpha}$, ${\beta}$ and ${\nu}$ range across all ${\left\lbrace 0, \dots, n - 1 \right\rbrace}$. Figure \ref{fig:Figure59} shows how the electric charge density ${\rho}$ and the electric current density vector ${J^{\mu}}$ may be calculated directly from the \texttt{ElectromagneticTensor} objects for the Reissner-Nordstr\"om and Kerr-Newman metrics, assuming spacetime electromagnetic potential ${\mathbf{A} = \left( \Phi, A^1, A^2, A^3 \right)}$.

\begin{figure}[ht]
\centering
\begin{framed}
\includegraphics[width=0.445\textwidth]{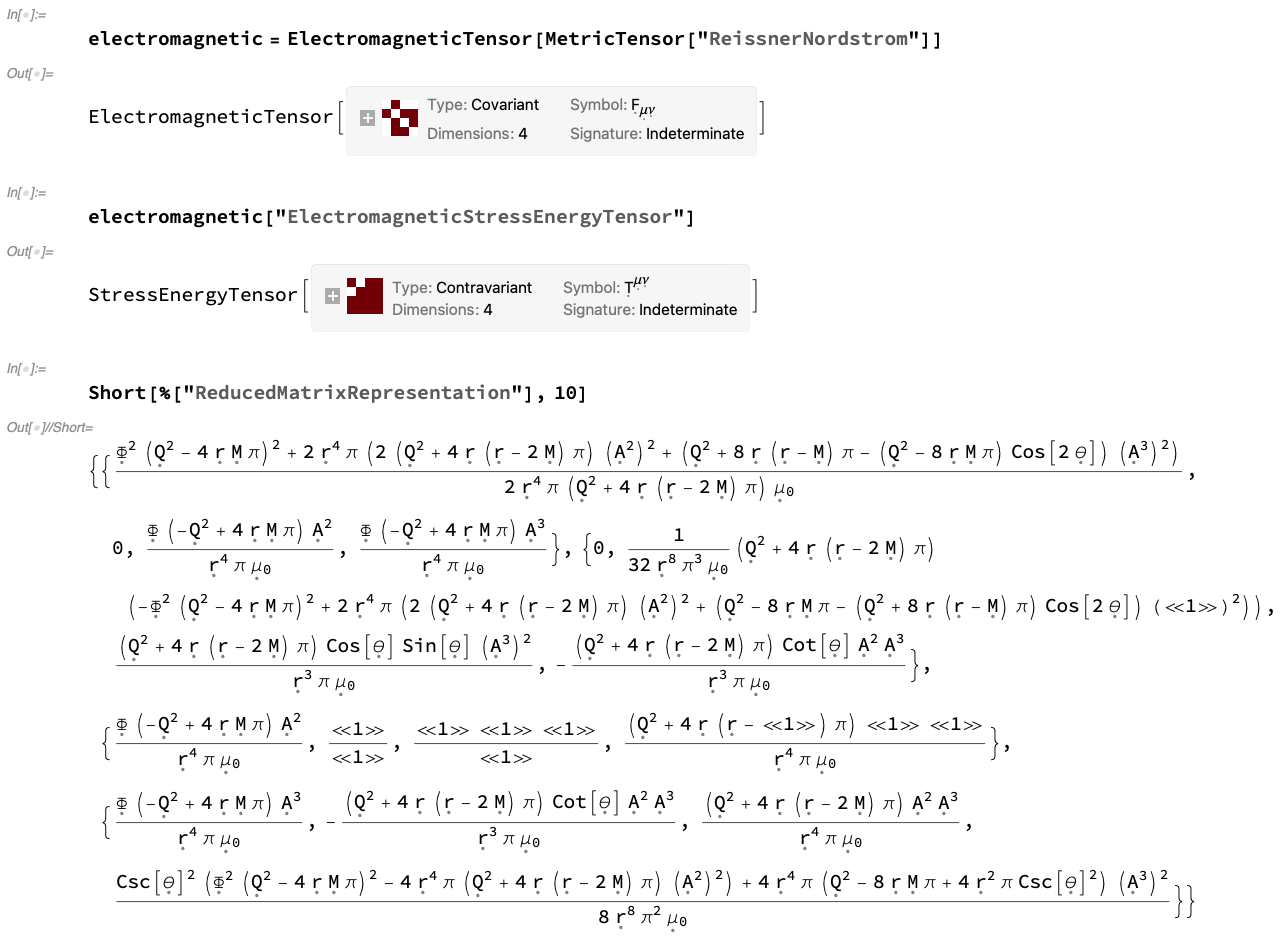}
\vrule
\includegraphics[width=0.545\textwidth]{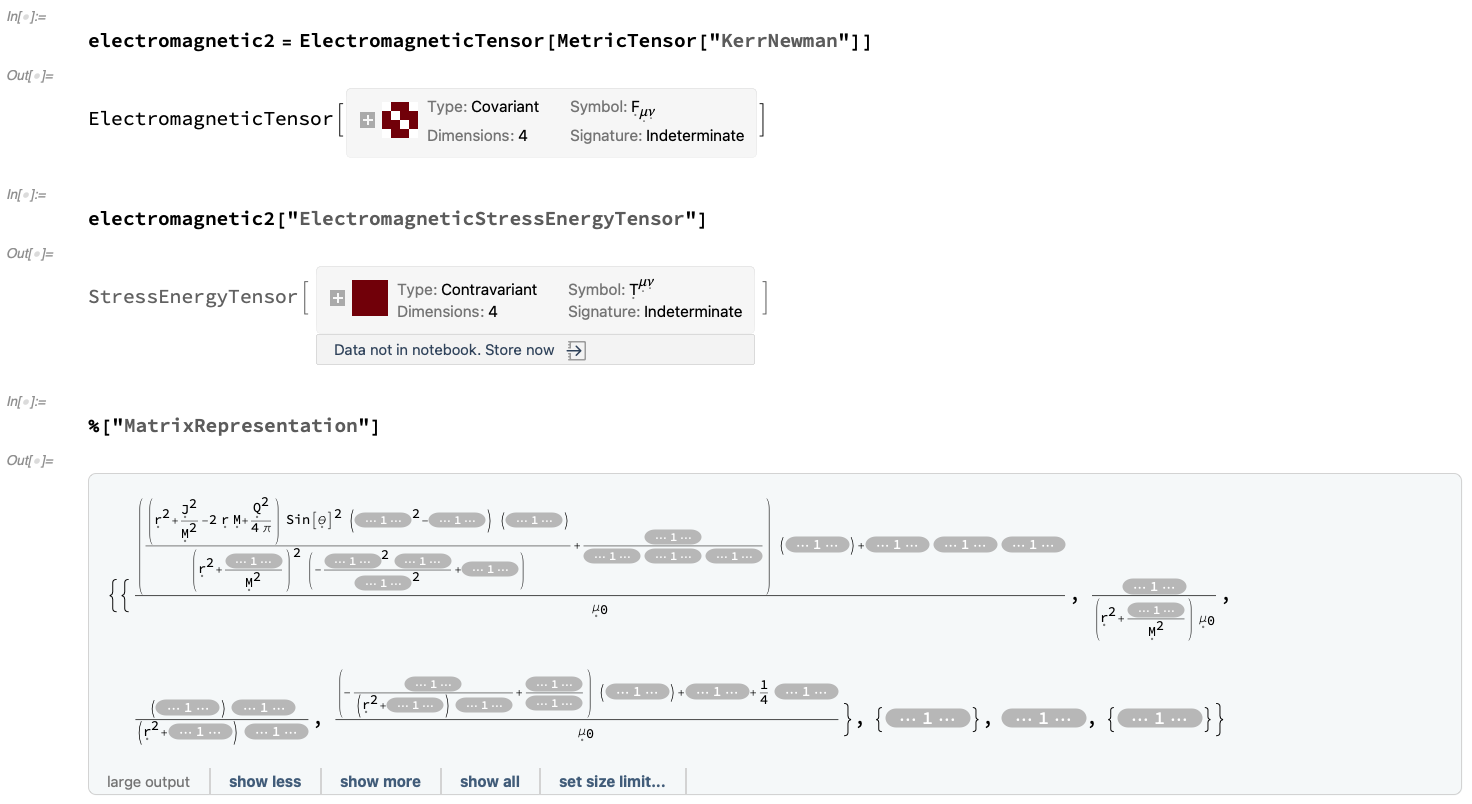}
\end{framed}
\caption{On the left, the electromagnetic \texttt{StressEnergyTensor} object computed from the \texttt{ElectromagneticTensor} object with spacetime electromagnetic potential ${\mathbf{A} = \left( \Phi, A^1, A^2, A^3 \right)}$ for a Reissner-Nordstr\"om geometry (representing e.g. a charged, non-rotating black hole of mass $M$ and electric charge $Q$ in Schwarzschild/spherical polar coordinates ${\left( t, r, \theta, \phi \right)}$). On the right, the electromagnetic \texttt{StressEnergyTensor} object computed from the \texttt{ElectromagneticTensor} object with spacetime electromagnetic potential ${\mathbf{A} = \left( \Phi, A^1, A^2, A^3 \right)}$ for a Kerr-Newman geometry (representing e.g. a charged, spinning black hole of mass $M$, angular momentum $J$ and electric charge $Q$ in Boyer-Lindquist/oblate spheroidal coordinates ${\left( t, r, \theta, \phi \right)}$).}
\label{fig:Figure60}
\end{figure}

\begin{figure}[ht]
\centering
\begin{framed}
\includegraphics[width=0.545\textwidth]{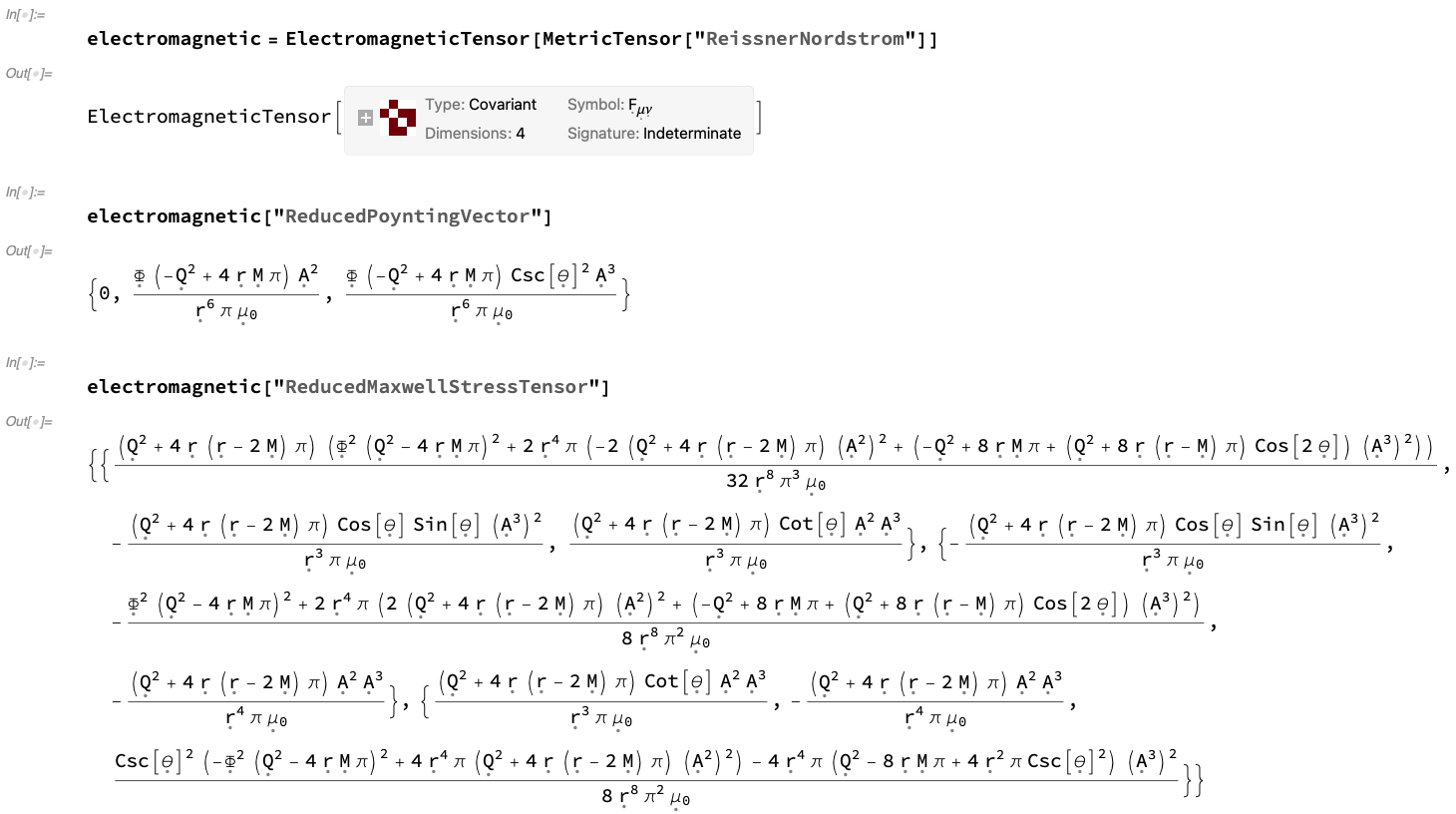}
\vrule
\includegraphics[width=0.445\textwidth]{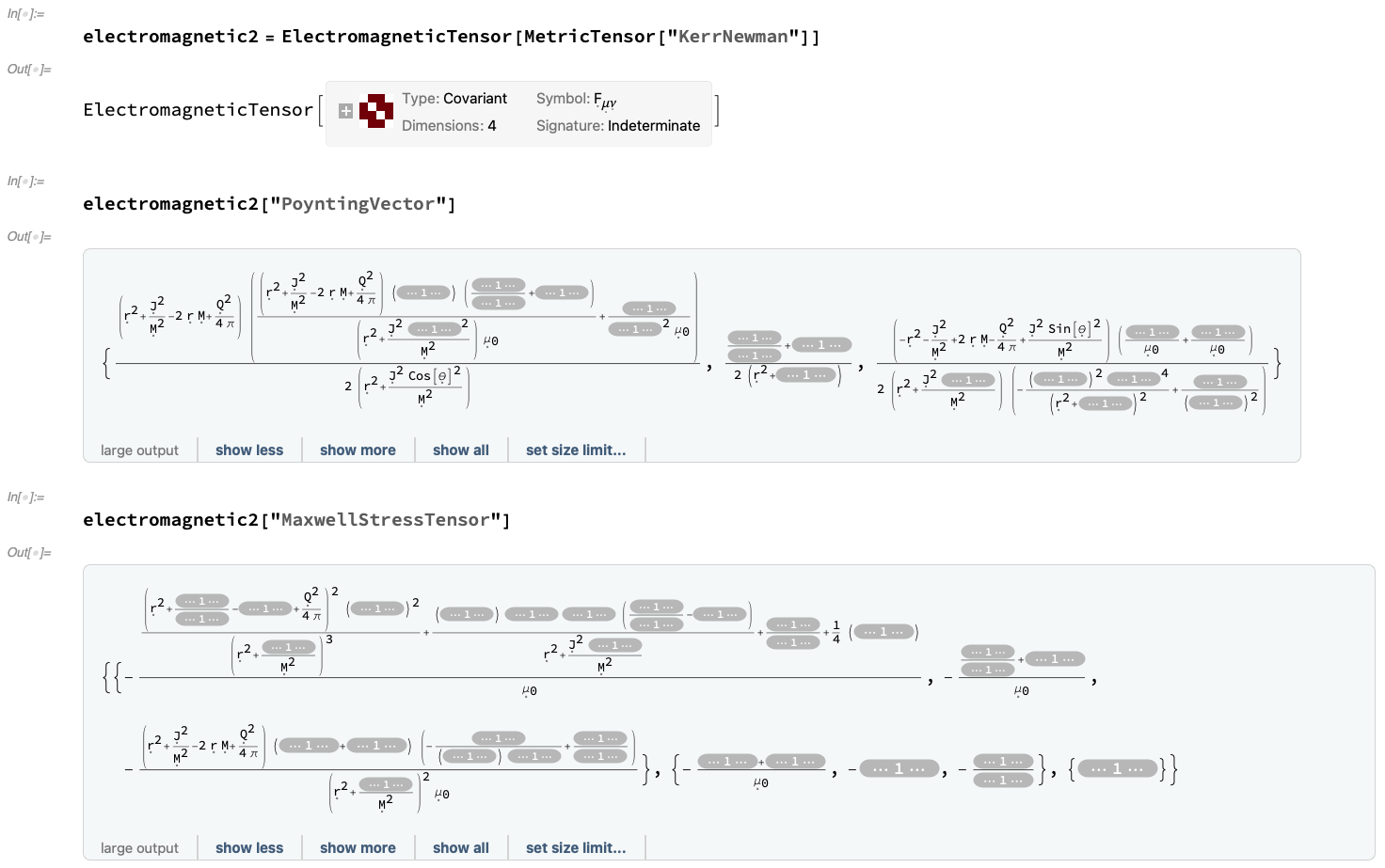}
\end{framed}
\caption{On the left, the relativistic Poynting vector and Maxwell stress tensor of the \texttt{ElectromagneticTensor} object with spacetime electromagnetic potential ${\mathbf{A} = \left( \Phi, A^1, A^2, A^3 \right)}$ for a Reissner-Nordstr\"om geometry (representing e.g. a charged, non-rotating black hole of mass $M$ and electric charge $Q$ in Schwarzschild/spherical polar coordinates ${\left( t, r, \theta, \phi \right)}$), the latter in explicit matrix form. On the right, the relativistic Poynting vector and Maxwell stress tensor of the \texttt{ElectromagneticTensor} object with spacetime electromagnetic potential ${\mathbf{A} = \left( \Phi, A^1, A^2, A^3 \right)}$ for a Kerr-Newman geometry (representing e.g. a charged, spinning black hole of mass $M$, angular momentum $J$ and electric charge $Q$ in Boyer-Lindquist/oblate spheroidal coordinates ${\left( t, r, \theta, \phi \right)}$), the latter in explicit matrix form.}
\label{fig:Figure61}
\end{figure}

\begin{figure}[ht]
\centering
\begin{framed}
\includegraphics[width=0.495\textwidth]{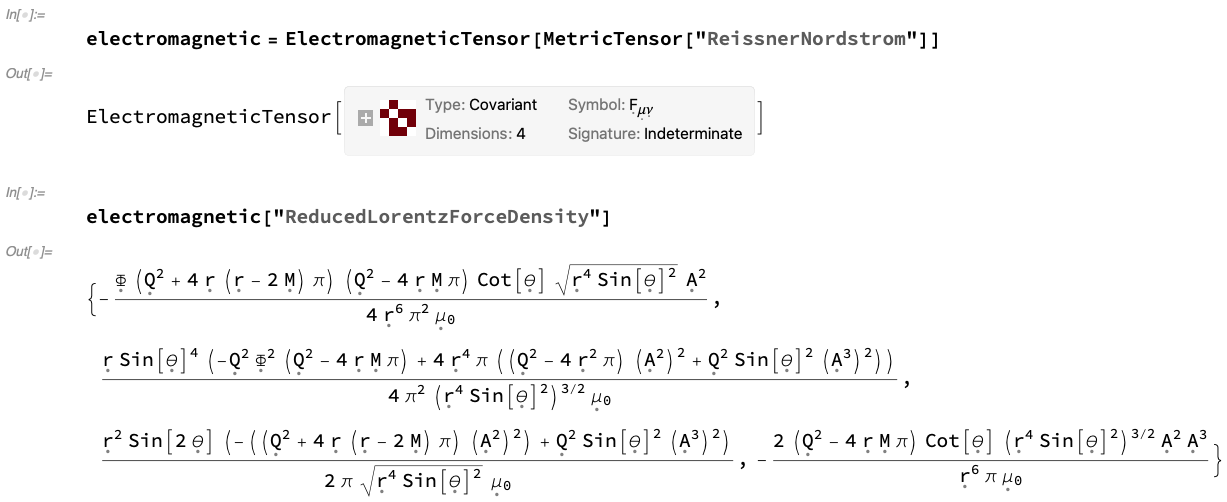}
\vrule
\includegraphics[width=0.495\textwidth]{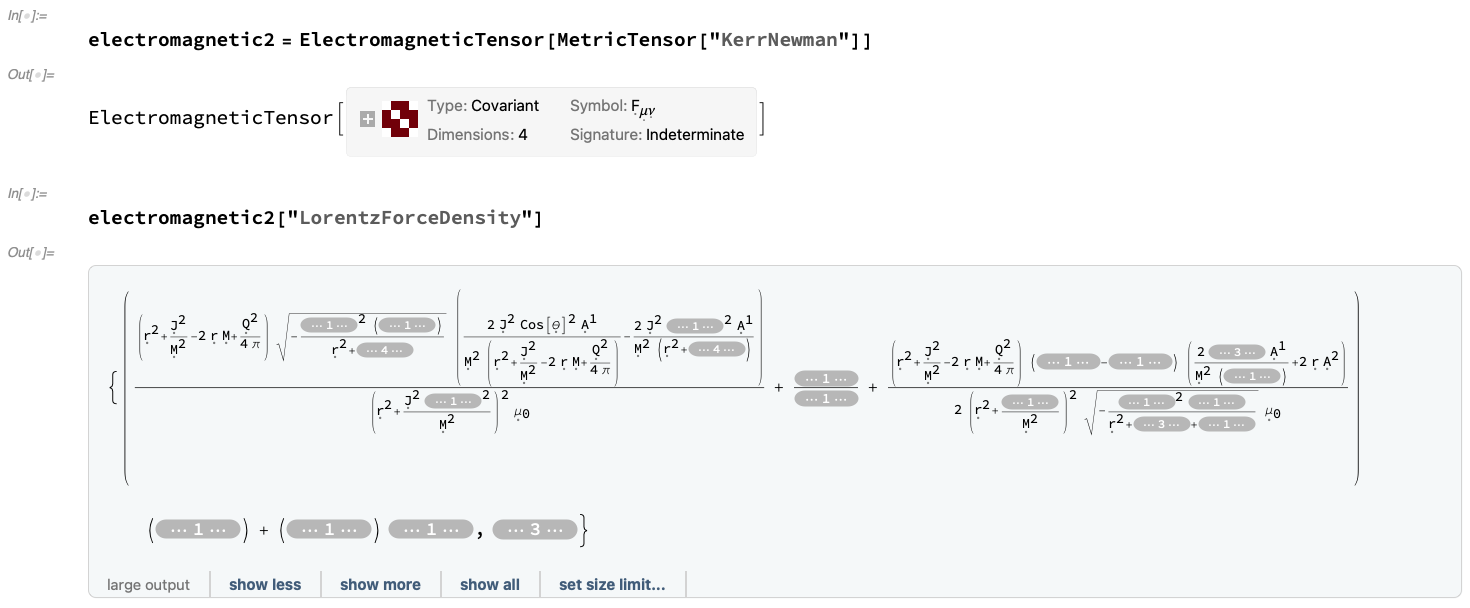}
\end{framed}
\caption{On the left, the covector/1-form density of the Lorentz force for the \texttt{ElectromagneticTensor} object with spacetime electromagnetic potential ${\mathbf{A} = \left( \Phi, A^1, A^2, A^3 \right)}$ for a Reissner-Nordstr\"om geometry (representing e.g. a charged, non-rotating black hole of mass $M$ and electric charge $Q$ in Schwarzschild/spherical polar coordinates ${\left( t, r, \theta, \phi \right)}$). On the right, the covector/1-form density of the Lorentz force for the \texttt{ElectromagneticTensor} object with spacetime electromagnetic potential ${\mathbf{A} = \left( \Phi, A^1, A^2, A^3 \right)}$ for a Kerr-Newman geometry (representing e.g. a charged, spinning black hole of mass $M$, angular momentum $J$ and electric charge $Q$ in Boyer-Lindquist/oblate spheroidal coordinates ${\left( t, r, \theta, \phi \right)}$).}
\label{fig:Figure62}
\end{figure}

\begin{figure}[ht]
\centering
\begin{framed}
\includegraphics[width=0.595\textwidth]{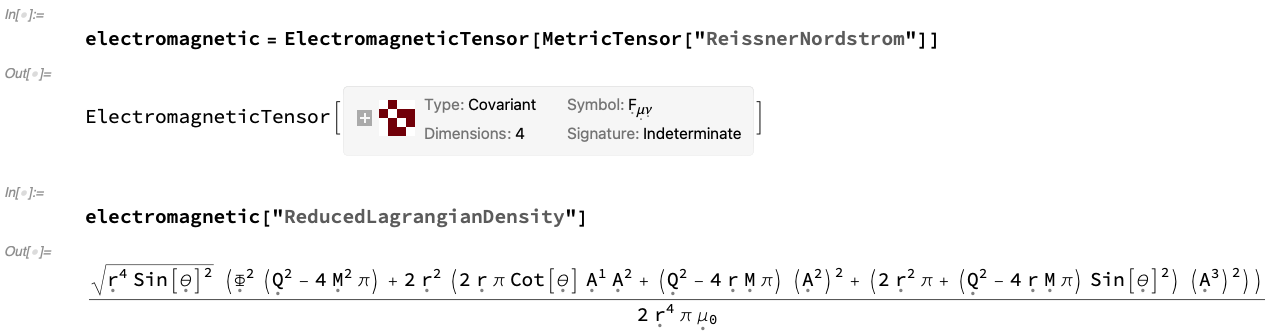}
\vrule
\includegraphics[width=0.395\textwidth]{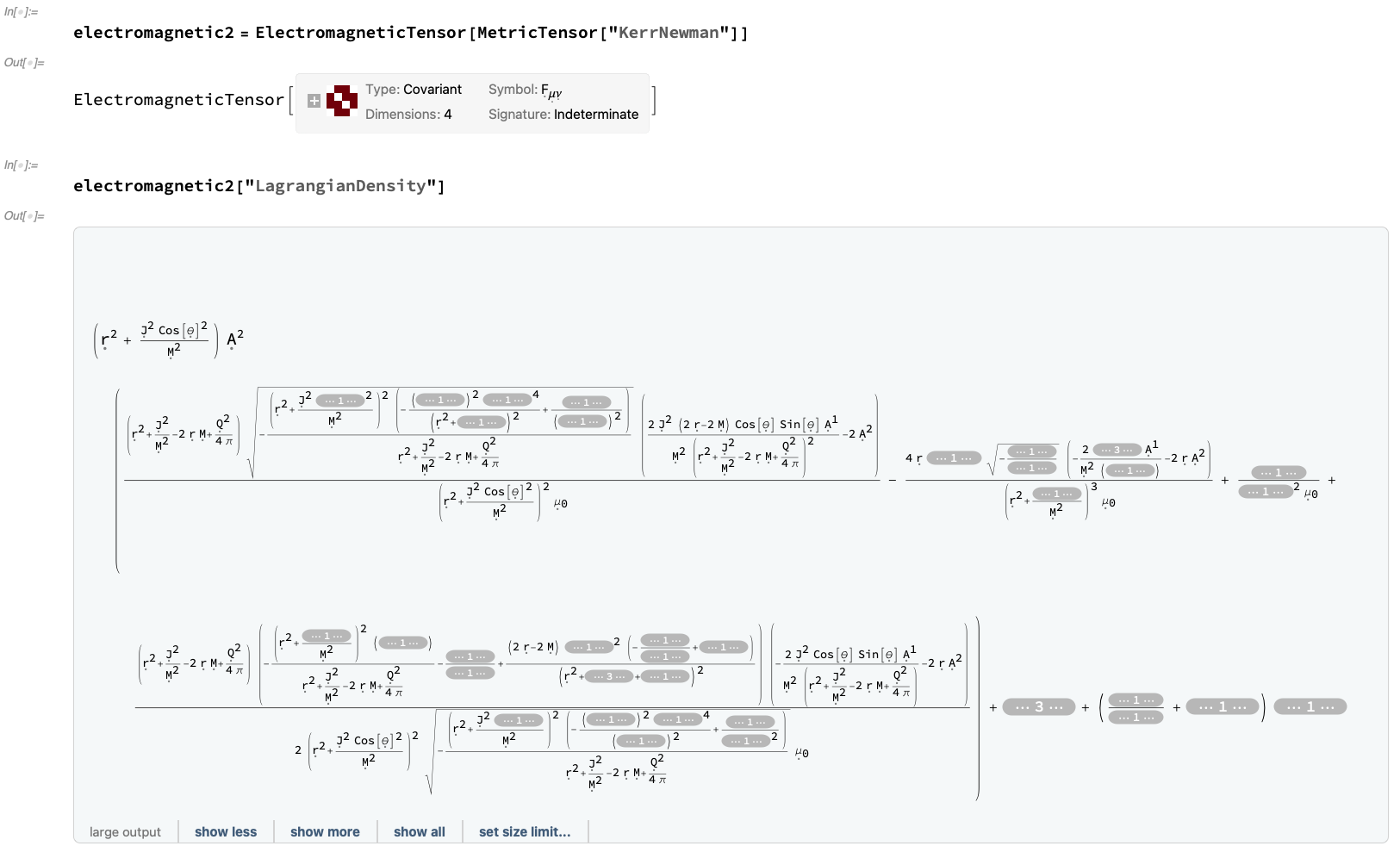}
\end{framed}
\caption{On the left, the relativistic Lagrangian density of the \texttt{ElectromagneticTensor} object with spacetime electromagnetic potential ${\mathbf{A} = \left( \Phi, A^1, A^2, A^3 \right)}$ for a Reissner-Nordstr\"om geometry (representing e.g. a charged, non-rotating black hole of mass $M$ and electric charge $Q$ in Schwarzschild/spherical polar coordinates ${\left( t, r, \theta, \phi \right)}$). On the right, the relativistic Lagrangian density of the \texttt{ElectromagneticTensor} object with spacetime electromagnetic potential ${\mathbf{A} = \left( \Phi, A^1, A^2, A^3 \right)}$ for a Kerr-Newman geometry (representing e.g. a charged, spinning black hole of mass $M$, angular momentum $J$ and electric charge $Q$ in Boyer-Lindquist/oblate spheroidal coordinates ${\left( t, r, \theta, \phi \right)}$).}
\label{fig:Figure63}
\end{figure}

The total contribution to the stress-energy tensor ${T^{\mu \nu}}$ due to the presence of such an electromagnetic field may be defined purely in terms of the electromagnetic tensor ${F_{\mu \nu}}$ as:

\begin{equation}
T^{\mu \nu} = \frac{1}{\mu_0} \left( F^{\mu \alpha} F_{\alpha}^{\nu} - \frac{1}{4} g^{\mu \nu} F_{\alpha \beta} F^{\alpha \beta} \right)
\end{equation}
as shown in Figure \ref{fig:Figure60}, in which a \texttt{StressEnergyTensor} object is computed directly from an \texttt{ElectromagneticTensor} object for the case of both a Reissner-Nordstr\"om metric and a Kerr-Newman metric, assuming default spacetime electromagnetic potential ${\mathbf{A} = \left( \Phi, A^1, A^2, A^3 \right)}$. For this particular, purely electromagnetic, form of the \texttt{StressEnergyTensor} object, the analog of the relativistic momentum density components, namely the time-space and space-time components ${T^{\mu 0}}$ and ${T^{0 \nu}}$ (for ${\mu, \nu \neq 0}$), represent instead the components of the relativistic Poynting vector ${\mathbf{S}}$ (or, more precisely, the components of its covector/1-form representation ${S_{\mu}}$); explicitly, one has:

\begin{equation}
\mathbf{S} = \frac{1}{\mu_0} \mathbf{E} \times \mathbf{B}, \qquad \text{ i.e. } \qquad S_{\mu} = \frac{1}{\mu_0} \left( \epsilon_{\alpha \beta \mu} E^{\alpha} B^{\beta} \right),
\end{equation}
from which the components of the true (contravariant) Poynting vector ${S^{\mu}}$ may be recovered by raising indices as appropriate. Moreover, the analog of the relativistic Cauchy/normal stress components, namely the space-space components ${T^{\mu \nu}}$ (for ${\mu, \nu \neq 0}$), represent instead the components of the relativistic Maxwell stress tensor ${\sigma_{\mu \nu}}$; explicitly, one has:

\begin{equation}
\sigma_{\mu \nu} = E_{\mu} E_{\nu} + \frac{1}{\mu_0} B_{\mu} B_{\nu} - \frac{1}{2} \left( E_{\sigma} E^{\sigma} + \frac{1}{\mu_0} B_{\sigma} B^{\sigma} \right) \delta_{\mu \nu},
\end{equation}
where ${\delta_{\mu \nu}}$ designates, as usual, the identity tensor/Kronecker delta function. Figure \ref{fig:Figure61} shows the relativistic (contravariant) Poynting vector ${S^{\mu}}$ (where ${\mu \neq 0}$) and the relativistic Maxwell stress tensor ${\sigma_{\mu \nu}}$ (where ${\mu, 
\nu \neq 0}$), with the latter represented in explicit matrix form, computed directly from the \texttt{ElectromagneticTensor} objects for the Reissner-Nordstr\"om and Kerr-Newman metrics, assuming spacetime electromagnetic potential ${\mathbf{A} = \left( \Phi, A^1, A^2, A^3 \right)}$. The density of the Lorentz force (i.e. the combination of electric and magnetic forces exerted on a point charge) is typically represented as a covariant rank-1 tensor density, i.e. a covector/1-form density, denoted ${f_{\mu}}$, and calculated in terms of a projection of the electromagnetic tensor ${F_{\mu \nu}}$ onto the spacetime current density (i.e. the partial divergence of the electromagnetic displacement tensor density ${\mathcal{D}^{\mu \nu}}$):

\begin{equation}
f_{\mu} = F_{\mu \nu} \frac{\partial}{\partial x^{\sigma}} \left( D^{\nu \sigma} \right) = F_{\mu \nu} \frac{\partial}{\partial x^{\sigma}} \left( \frac{\sqrt{- \det \left( g_{\mu \nu} \right)}}{\mu_0} g^{\nu \alpha} F_{\alpha \beta} g^{\beta \sigma} \right) = F_{\mu \nu} \left( \frac{\sqrt{- \det \left( g_{\mu \nu} \right)}}{\mu_0} F^{\nu \sigma} \right),
\end{equation}
and its calculation from the \texttt{ElectromagneticTensor} objects for the Reissner-Nordstr\"om and Kerr-Newman metrics with spacetime electromagnetic potential ${\mathbf{A} = \left( \Phi, A^1, A^2, A^3 \right)}$ is shown in Figure \ref{fig:Figure62}. Finally, the (scalar) relativistic Lagrangian density ${\mathcal{L}}$ for the electromagnetic field may be computed as:

\begin{multline}
\mathcal{L} = - \frac{\sqrt{- \det \left( g_{\mu \nu} \right)}}{4 \mu_0} F_{\alpha \beta} F^{\alpha \beta} + A_{\alpha} \frac{\partial}{\partial x^{\sigma}} \left( \mathcal{D}^{\alpha \sigma} \right)\\
= - \frac{\sqrt{- \det \left( g_{\mu \nu} \right)}}{4 \mu_0} F_{\alpha \beta} F^{\alpha \beta} + A_{\alpha} \frac{\partial}{\partial x^{\sigma}} \left( \frac{\sqrt{- \det \left( g_{\mu \nu} \right)}}{\mu_0} g^{\alpha \mu} F_{\mu \nu} g^{\nu \sigma} \right)\\
= - \frac{\sqrt{- \det \left( g_{\mu \nu} \right)}}{4 \mu_0} F_{\alpha \beta} F^{\alpha \beta} + A_{\alpha} \frac{\partial}{\partial x^{\sigma}} \left( \frac{\sqrt{- \det \left( g_{\mu \nu} \right)}}{\mu_0} F^{\alpha \sigma} \right),
\end{multline}
as shown in Figure \ref{fig:Figure63} for the case of \texttt{ElectromagneticTensor} objects for the Reissner-Nordstr\"om and Kerr-Newman metrics, again with spacetime electromagnetic potential ${\mathbf{A} = \left( \Phi, A^1, A^2, A^3 \right)}$.

\begin{figure}[ht]
\centering
\begin{framed}
\includegraphics[width=0.595\textwidth]{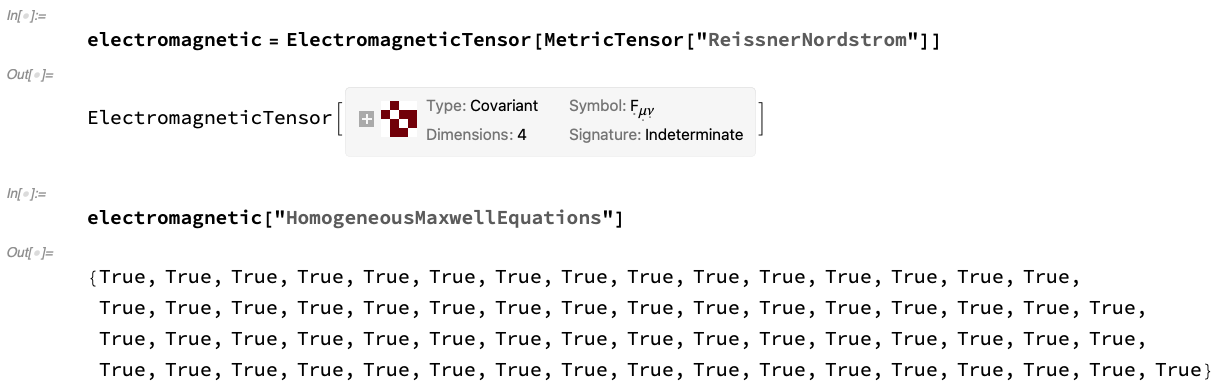}
\vrule
\includegraphics[width=0.395\textwidth]{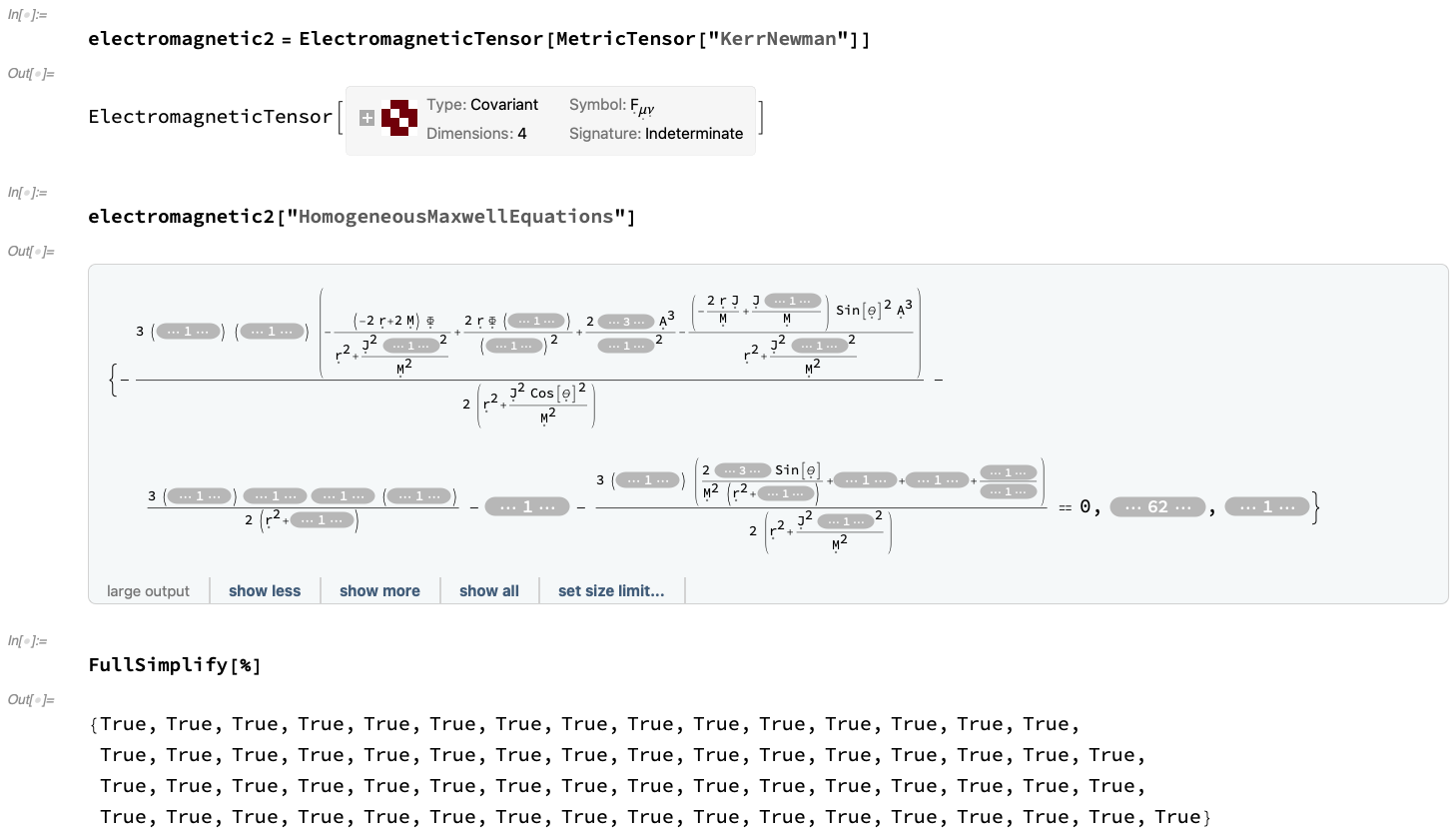}
\end{framed}
\caption{On the left, the list of homogeneous Maxwell equations asserting the symmetries of the covariant derivatives of the \texttt{ElectromagneticTensor} object with spacetime electromagnetic potential ${\mathbf{A} = \left( \Phi, A^1, A^2, A^3 \right)}$ for a Reissner-Nordstr\"om geometry (representing e.g. a charged, non-rotating black hole of mass $M$ and electric charge $Q$ in Schwarzschild/spherical polar coordinates ${\left( t, r, \theta, \phi \right)}$), together with a verification that they all hold identically. On the right, the list of homogeneous Maxwell equations asserting the symmetries of the covariant derivatives of the \texttt{ElectromagneticTensor} object with spacetime electromagnetic potential ${\mathbf{A} = \left( \Phi, A^1, A^2, A^3 \right)}$ for a Kerr-Newman geometry (representing e.g. a charged, spinning black hole of mass $M$, angular momentum $J$ and electric charge $Q$ in Boyer-Lindquist/oblate spheroidal coordinates ${\left( t, r, \theta, \phi \right)}$), together with a verification that they all hold identically.}
\label{fig:Figure64}
\end{figure}

\begin{figure}[ht]
\centering
\begin{framed}
\includegraphics[width=0.595\textwidth]{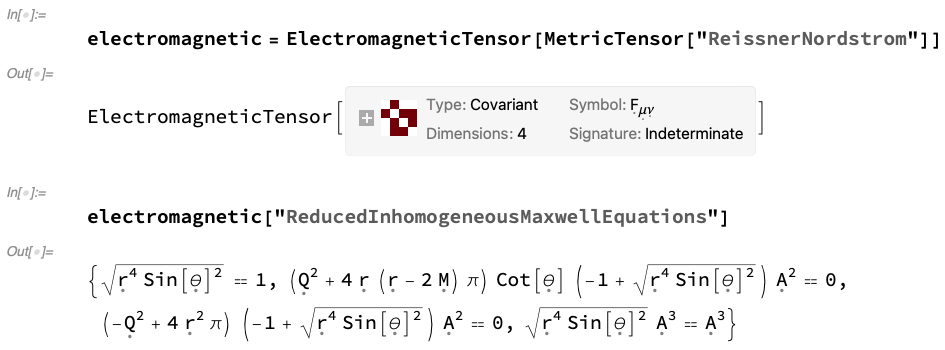}
\vrule
\includegraphics[width=0.395\textwidth]{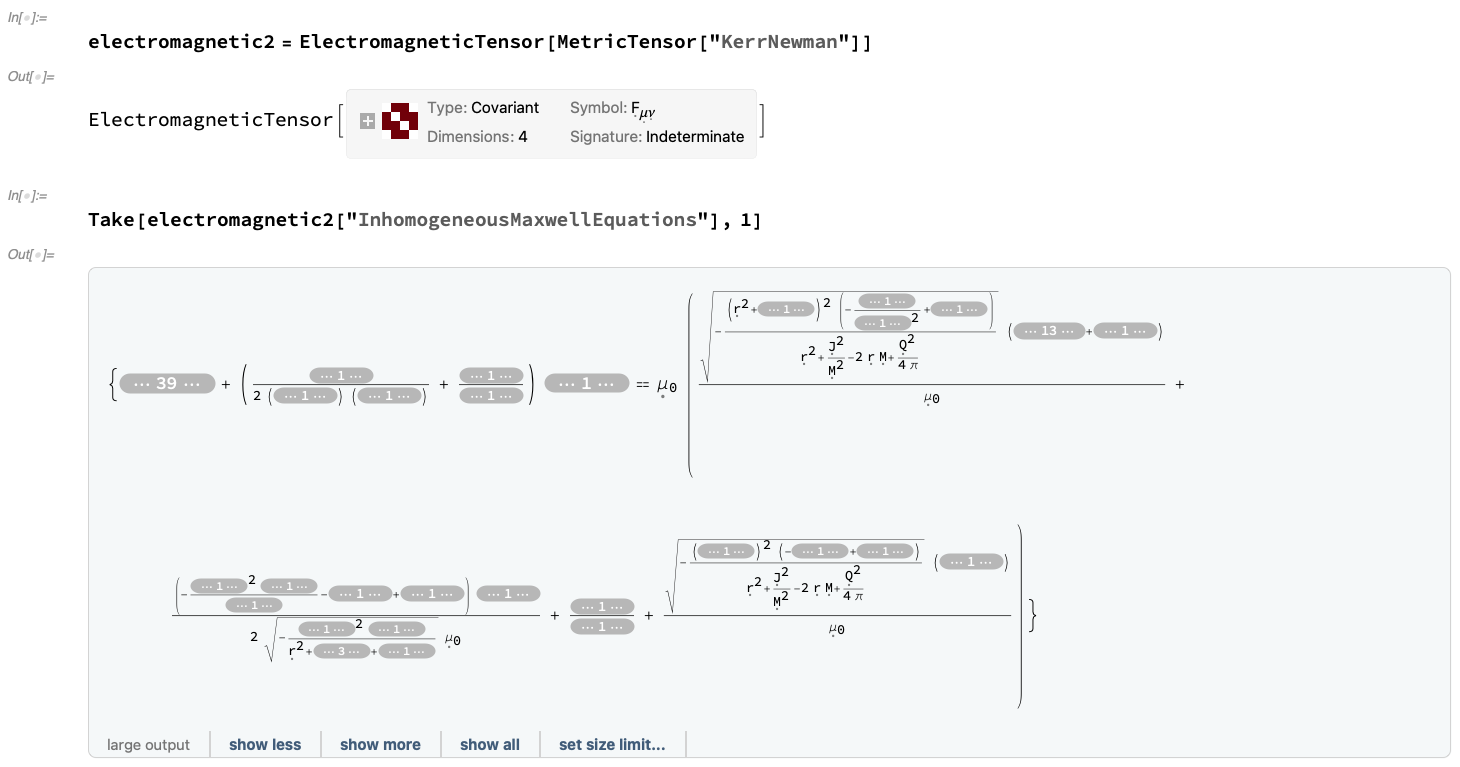}
\end{framed}
\caption{On the left, the list of inhomogeneous Maxwell equations relating the covariant divergence of the \texttt{ElectromagneticTensor} object with spacetime electromagnetic potential ${\mathbf{A} = \left( \Phi, A^1, A^2, A^3 \right)}$ for a Reissner-Nordstr\"om geometry (representing e.g. a charged, non-rotating black hole of mass $M$ and electric charge $Q$ in Schwarzschild/spherical polar coordinates ${\left( t, r, \theta, \phi \right)}$) and the partial divergence of the electromagnetic displacement tensor density. On the right, the list of inhomogeneous Maxwell equations relating the covariant divergence of the \texttt{ElectromagneticTensor} object with spacetime electromagnetic potential ${\mathbf{A} = \left( \Phi, A^1, A^2, A^3 \right)}$ for a Kerr-Newman geometry (representing e.g. a charged, spinning black hole of mass $M$, angular momentum $J$ and electric charge $Q$ in Boyer-Lindquist/oblate spheroidal coordinates ${\left( t, r, \theta, \phi \right)}$) and the partial divergence of the electromagnetic displacement tensor density.}
\label{fig:Figure65}
\end{figure}

\begin{figure}[ht]
\centering
\begin{framed}
\includegraphics[width=0.545\textwidth]{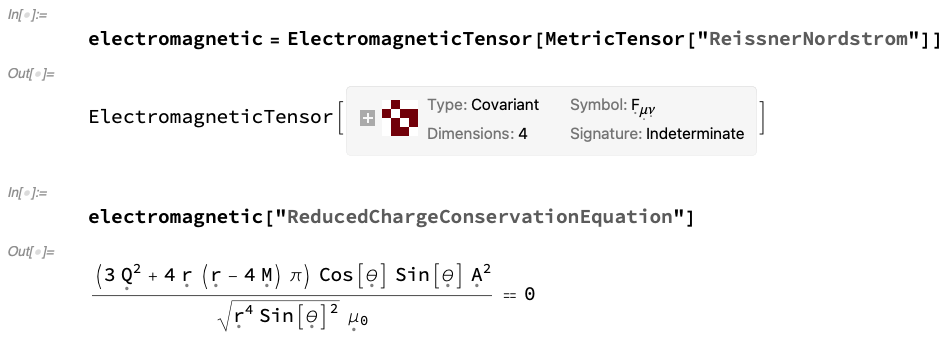}
\vrule
\includegraphics[width=0.445\textwidth]{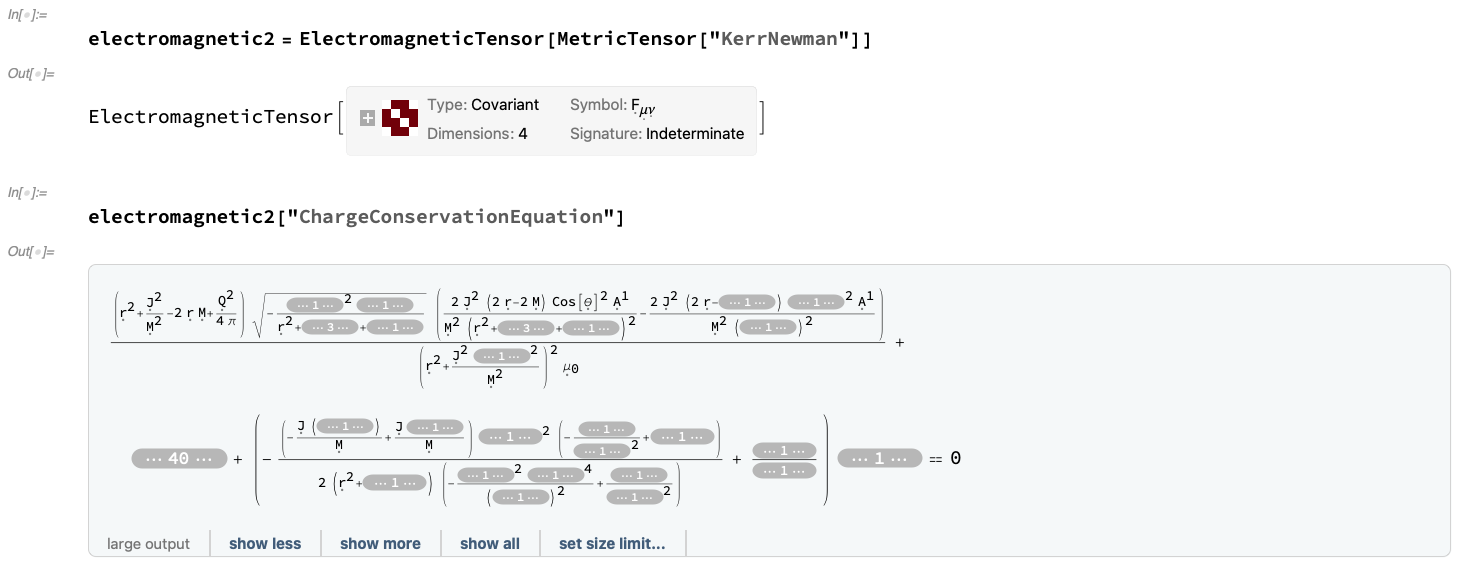}
\end{framed}
\caption{On the left, the charge conservation equation asserting that the covariant divergence of the partial divergence of the electromagnetic displacement tensor density of the \texttt{ElectromagneticTensor} object with spacetime electromagnetic potential ${\mathbf{A} = \left( \Phi, A^1, A^2, A^3 \right)}$ for a Reissner-Nordstr\"om geometry (representing e.g. a charged, non-rotating black hole of mass $M$ and electric charge $Q$ in Schwarzschild/spherical polar coordinates ${\left( t, r, \theta, \phi \right)}$) vanishes identically. On the right, the charge conservation equation asserting that the covariant divergence of the partial divergence of the electromagnetic displacement tensor density of the \texttt{ElectromagneticTensor} object with spacetime electromagnetic potential ${\mathbf{A} = \left( \Phi, A^1, A^2, A^3 \right)}$ for a Kerr-Newman geometry (representing e.g. a charged, spinning black hole of mass $M$, angular momentum $J$ and electric charge $Q$ in Boyer-Lindquist/oblate spheroidal coordinates ${\left( t, r, \theta, \phi \right)}$) vanishes identically.}
\label{fig:Figure66}
\end{figure}

Gauss's law for magnetism (stating that the magnetic field ${\mathbf{B}}$ is divergence-free, and therefore that there cannot exist any magnetic monopoles) and the Maxwell-Faraday equation (stating that spatially-varying electric fields ${\mathbf{E}}$ are always accompanied by corresponding time-varying magnetic fields ${\mathbf{B}}$) now both emerge as consequences of the (differential) Bianchi identities asserting the symmetries of the covariant derivatives of the electromagnetic tensor ${F_{\mu \nu}}$, namely:

\begin{multline}
\nabla_{\rho} F_{\mu \nu} + \nabla_{\mu} F_{\nu \rho} + \nabla_{\nu} F_{\rho \mu} = \left( \frac{\partial}{\partial x^{\rho}} \left( F_{\mu \nu} \right) - \Gamma_{\rho \mu}^{\sigma} F_{\sigma \nu} - \Gamma_{\rho \nu}^{\sigma} F_{\mu \sigma} \right)\\
+ \left( \frac{\partial}{\partial x^{\mu}} \left( F_{\nu \rho} \right) - \Gamma_{\mu \nu}^{\sigma} F_{\sigma \rho} - \Gamma_{\mu \rho}^{\sigma} F_{\nu \sigma} \right) + \left( \frac{\partial}{\partial x^{\nu}} \left( F_{\rho \mu} \right) - \Gamma_{\nu \rho}^{\sigma} F_{\sigma \mu} - \Gamma_{\nu \mu}^{\sigma} F_{\rho \sigma} \right) = 0,
\end{multline}
otherwise known as the homogeneous Maxwell equations, which hold identically for any electromagnetic tensor ${F_{\mu \nu}}$ derived from a spacetime electromagnetic potential ${\mathbf{A}}$ in the usual way, as illustrated in Figure \ref{fig:Figure64} for the case of \texttt{ElectromagneticTensor} objects obtained from the Reissner-Nordstr\"om and Kerr-Newman metrics with spacetime electromagnetic potential ${\mathbf{A} = \left( \Phi, A^1, A^2, A^3 \right)}$. On the other hand, Gauss's flux theorem (relating the distribution of electric charge ${\rho}$ to the configuration of the electric field ${\mathbf{E}}$) and Amp\`ere's circuital law (relating the electric current ${\mathbf{J}}$ passing through a closed loop to the circulation of the magnetic field ${\mathbf{B}}$ around that loop) both emerge as consequences of the \textit{inhomogeneous} Maxwell equations, asserting that the covariant divergence of the electromagnetic tensor ${F_{\mu \nu}}$ is related to the spacetime current density, otherwise known as the partial divergence of the electromagnetic displacement tensor density ${\mathcal{D}^{\mu \nu}}$:

\begin{multline}
\nabla_{\nu} F^{\mu \nu} = \frac{\partial}{\partial x^{\nu}} \left( F^{\mu \nu} \right) + \Gamma_{\nu \sigma}^{\mu} F^{\sigma \nu} + \Gamma_{\nu \sigma}^{\nu} F^{\mu \sigma} = \mu_0 \frac{\partial}{\partial x^{\sigma}} \left( \mathcal{D}^{\mu \sigma} \right)\\
= \mu_0 \frac{\partial}{\partial x^{\sigma}} \left( \frac{\sqrt{- \det \left( g_{\mu \nu} \right)}}{\mu_0} g^{\mu \alpha} F_{\alpha \beta} g^{\beta \sigma} \right) = \mu_0 \frac{\partial}{\partial x^{\sigma}} \left( \frac{\sqrt{- \det \left( g_{\mu \nu} \right)}}{\mu_0} F^{\mu \sigma} \right),
\end{multline}
as illustrated in Figure \ref{fig:Figure65}, again for the case of \texttt{ElectromagneticTensor} objects obtained from the Reissner-Nordstr\"om and Kerr-Newman metrics with spacetime electromagnetic potential ${\mathbf{A} = \left( \Phi, A^1, A^2, A^3 \right)}$. Unlike the homogeneous Maxwell equations, the inhomogeneous equations are not guaranteed to hold identically for a generic electromagnetic tensor ${F_{\mu \nu}}$ obtained from an arbitrary spacetime electromagnetic potential ${\mathbf{A}}$, and instead must be imposed as a constraint. Finally, we can see that the inhomogeneous Maxwell equations imply a set of continuity equations on the spacetime current density, i.e. a statement that the covariant divergence of the partial divergence of the electromagnetic displacement tensor density ${\mathcal{D}^{\mu \nu}}$ vanishes identically:

\begin{multline}
\nabla_{\mu} \left( \frac{\partial}{\partial x^{\sigma}} \left( \mathcal{D}^{\mu \sigma} \right) \right) = \frac{\partial}{\partial x^{\mu}} \left( \frac{\partial}{\partial x^{\sigma}} \left( \mathcal{D}^{\mu \sigma} \right) \right) + \Gamma_{\mu \lambda}^{\mu} \left( \frac{\partial}{\partial x^{\sigma}} \left( \mathcal{D}^{\lambda \sigma} \right) \right)\\
= \frac{\partial}{\partial x^{\mu}} \left( \frac{\partial}{\partial x^{\sigma}} \left( \frac{\sqrt{- \det \left( g_{\mu \nu} \right)}}{\mu_0} g^{\mu \alpha} F_{\alpha \beta} g^{\beta \sigma} \right) \right) + \Gamma_{\mu \lambda}^{\mu} \left( \frac{\partial}{\partial x^{\sigma}} \left( \frac{\sqrt{- \det \left( g_{\mu \nu} \right)}}{\mu_0} g^{\lambda \alpha} F_{\alpha \beta} g^{\beta \sigma} \right) \right)\\
= \frac{\partial}{\partial x^{\mu}} \left( \frac{\partial}{\partial x^{\sigma}} \left( \frac{\sqrt{- \det \left( g_{\mu \nu} \right)}}{\mu_0} F^{\mu \sigma} \right) \right) + \Gamma_{\mu \lambda}^{\mu} \left( \frac{\partial}{\partial x^{\sigma}} \left( \frac{\sqrt{- \det \left( g_{\mu \nu} \right)}}{\mu_0} F^{\lambda \sigma} \right) \right) = 0,
\end{multline}
fomrally expressing the conservation of total electric charge, as demonstrated in Figure \ref{fig:Figure66}, once again for the case of \texttt{ElectromagneticTensor} objects obtained from the Reissner-Nordstr\"om and Kerr-Newman metrics with spacetime electromagnetic potential ${\mathbf{A} = \left( \Phi, A^1, A^2, A^3 \right)}$.

\begin{figure}[ht]
\centering
\begin{framed}
\includegraphics[width=0.495\textwidth]{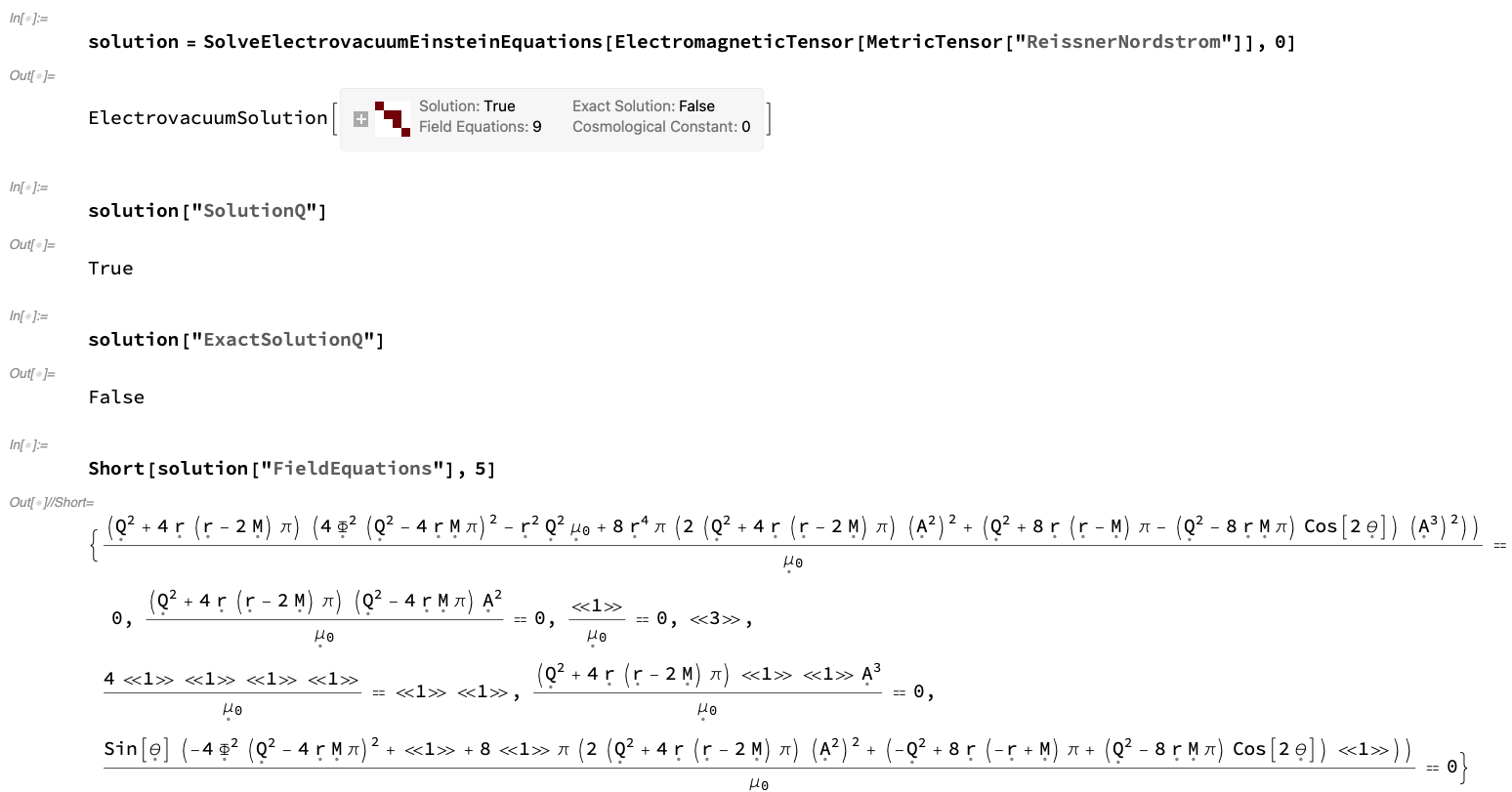}
\vrule
\includegraphics[width=0.495\textwidth]{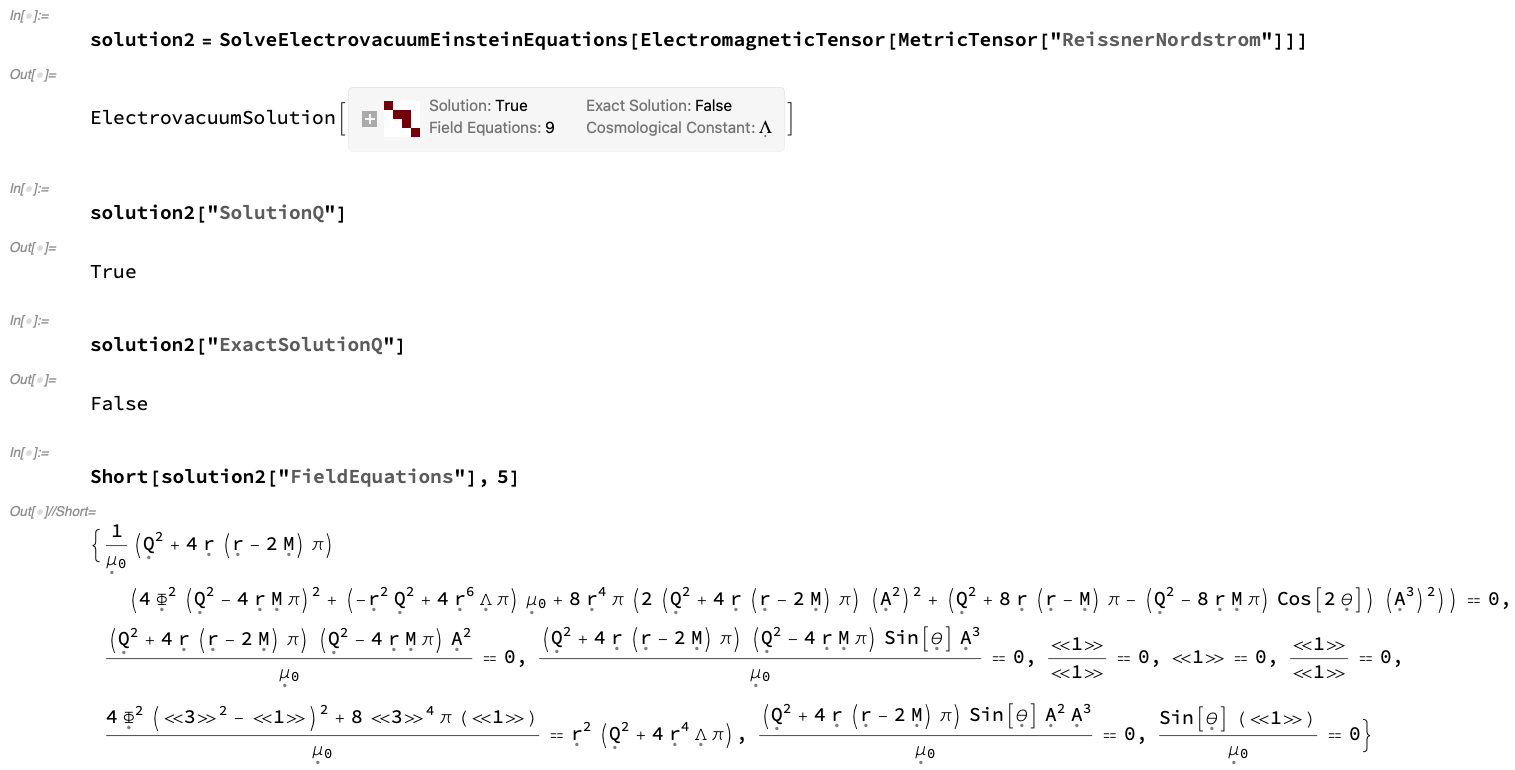}
\end{framed}
\caption{On the left, the \texttt{ElectrovacuumSolution} object for a Reissner-Nordstr\"om geometry (representing e.g. a charged, non-rotating black hole of mass $M$ and electric charge $Q$ in Schwarzschild/spherical polar coordinates ${\left( t, r, \theta, \phi \right)}$) equipped with a spacetime electromagnetic potential ${\mathbf{A} = \left( \Phi, A^1, A^2, A^3 \right)}$, with zero cosmological constant, computed using \texttt{SolveElectrovacuumEinsteinEquations}, illustrating that it corresponds to a non-exact solution to the vacuum Einstein-Maxwell equations, with nine additional field equations required. On the right, the \texttt{ElectrovacuumSolution} object for a Reissner-Nordstr\"om geometry (representing e.g. a charged, non-rotating black hole of mass $M$ and electric charge $Q$ in Schwarzschild/spherical polar coordinates ${\left( t, r, \theta, \phi \right)}$) equipped with a spacetime electromagnetic potential ${\mathbf{A} = \left( \Phi, A^1, A^2, A^3 \right)}$, with non-zero cosmological constant ${\Lambda \neq 0}$, computed using \texttt{SolveElectrovacuumEinsteinEquations}, illustrating that it corresponds to a non-exact solution to the vacuum Einstein-Maxwell equations, with nine additional field equations required.}
\label{fig:Figure67}
\end{figure}

\begin{figure}[ht]
\centering
\begin{framed}
\includegraphics[width=0.445\textwidth]{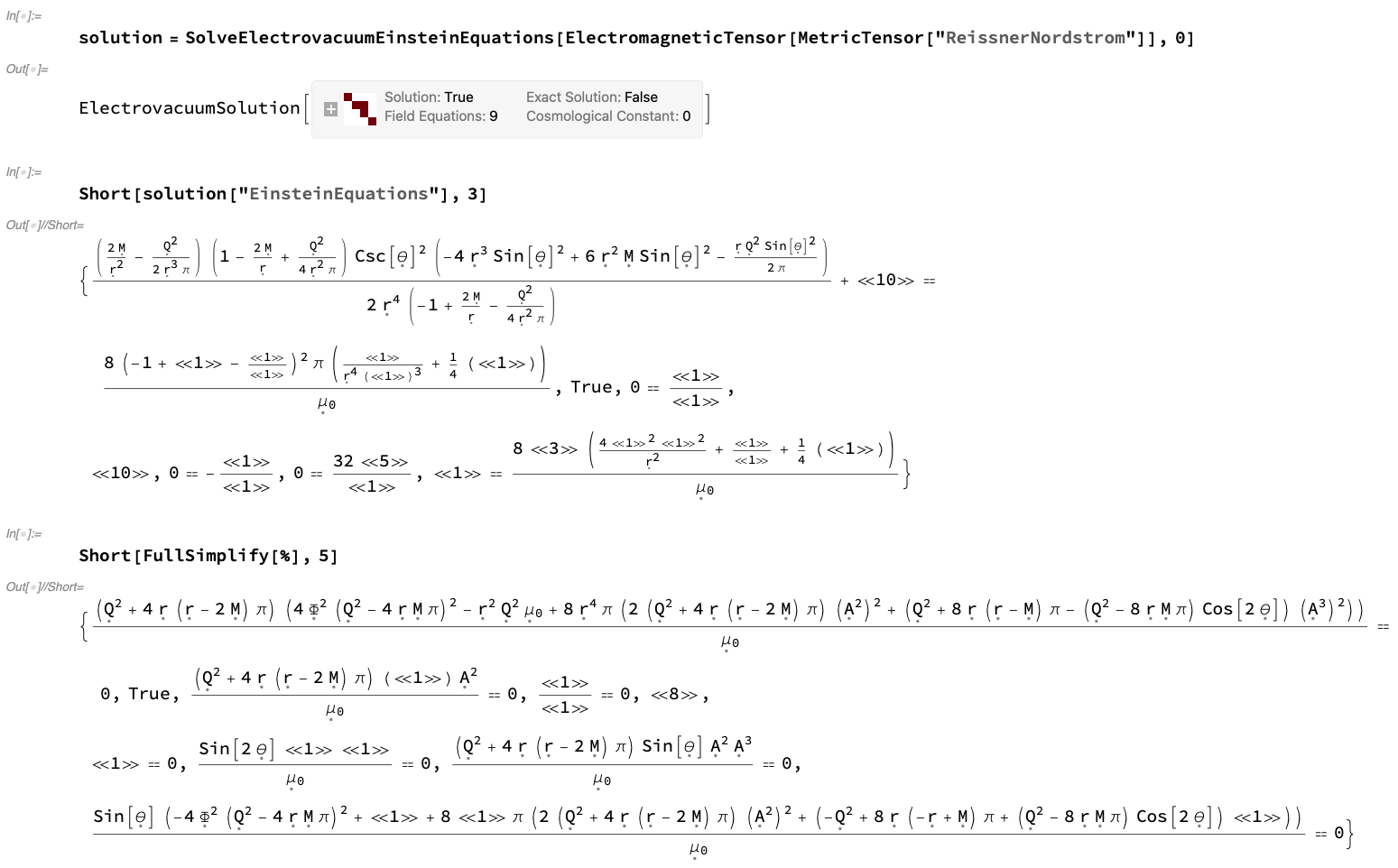}
\vrule
\includegraphics[width=0.545\textwidth]{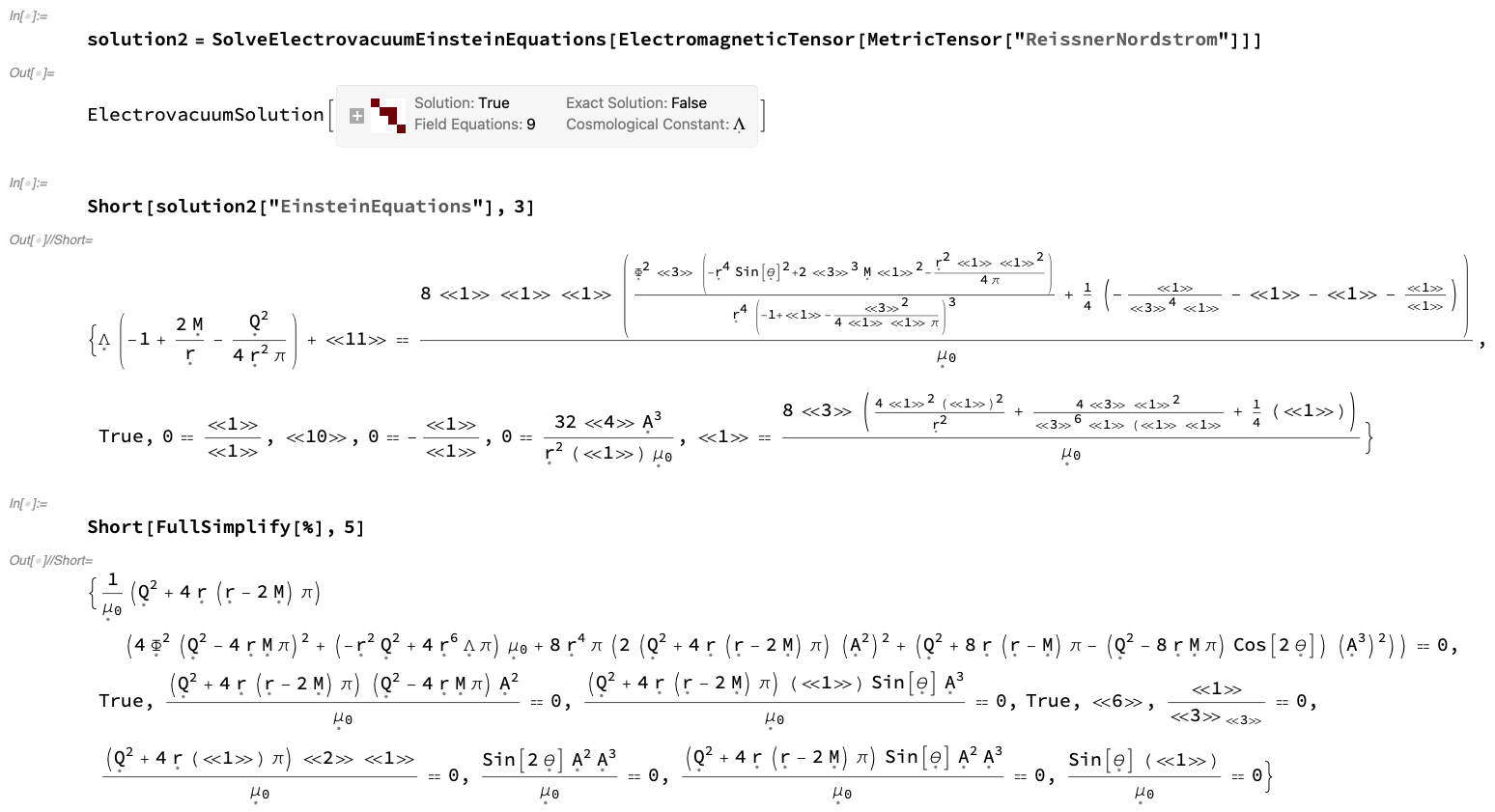}
\end{framed}
\caption{On the left, the list of Einstein-Maxwell field equations asserting that the Einstein tensor, with zero cosmological constant, is equal to ${8 \pi}$ times the electromagnetic stress-energy tensor, computed using the \texttt{ElectrovacuumSolution} object for a Reissner-Nordstr\"om geometry (representing e.g. a charged, non-rotating black hole of mass $M$ and electric charge $Q$ in Schwarzschild/spherical polar coordinates ${\left( t, r, \theta, \phi \right)}$) equipped with spacetime electromagnetic potential ${\mathbf{A} = \left( \Phi, A^1, A^2, A^3 \right)}$, together with a verification that they reduce down to a set of nine canonical field equations. On the right, the list of Einstein-Maxwell field equations asserting that the Einstein tensor, with non-zero cosmological constant ${\Lambda \neq 0}$, is equal to ${8 \pi}$ times the electromagnetic stress-energy tensor, computed using the \texttt{ElectrovacuumSolution} object for a Reissner-Nordstr\"om geometry (representing e.g. a charged, non-rotating black hole of mass $M$ and electric charge $Q$ in Schwarzschild/spherical polar coordinates ${\left( t, r, \theta, \phi \right)}$) equipped with spacetime electromagnetic potential ${\mathbf{A} = \left( \Phi, A^1, A^2, A^3 \right)}$, together with a verification that they reduce down to a set of nine canonical field equations.}
\label{fig:Figure68}
\end{figure}

\begin{figure}[ht]
\centering
\begin{framed}
\includegraphics[width=0.445\textwidth]{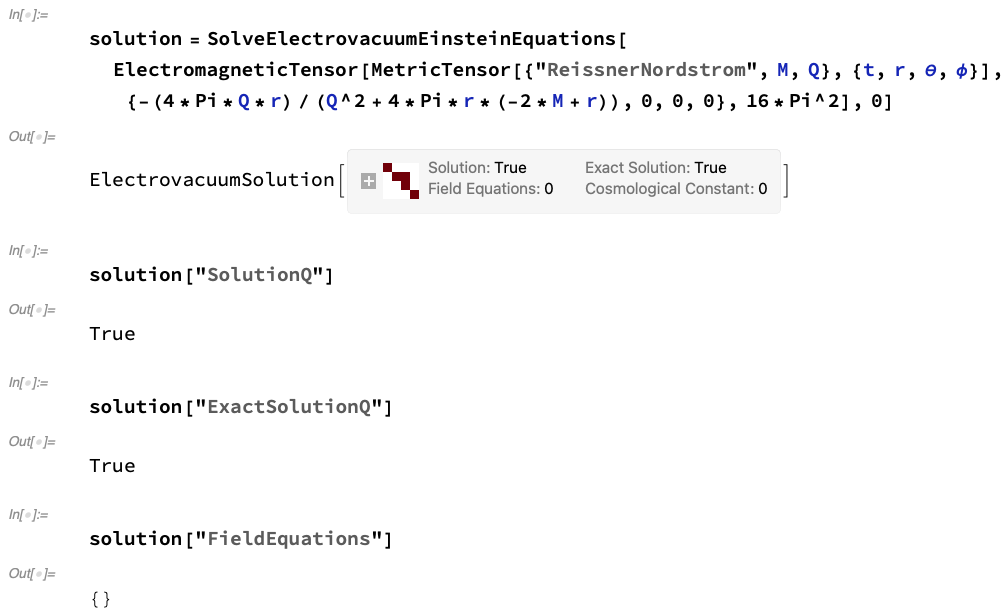}
\vrule
\includegraphics[width=0.545\textwidth]{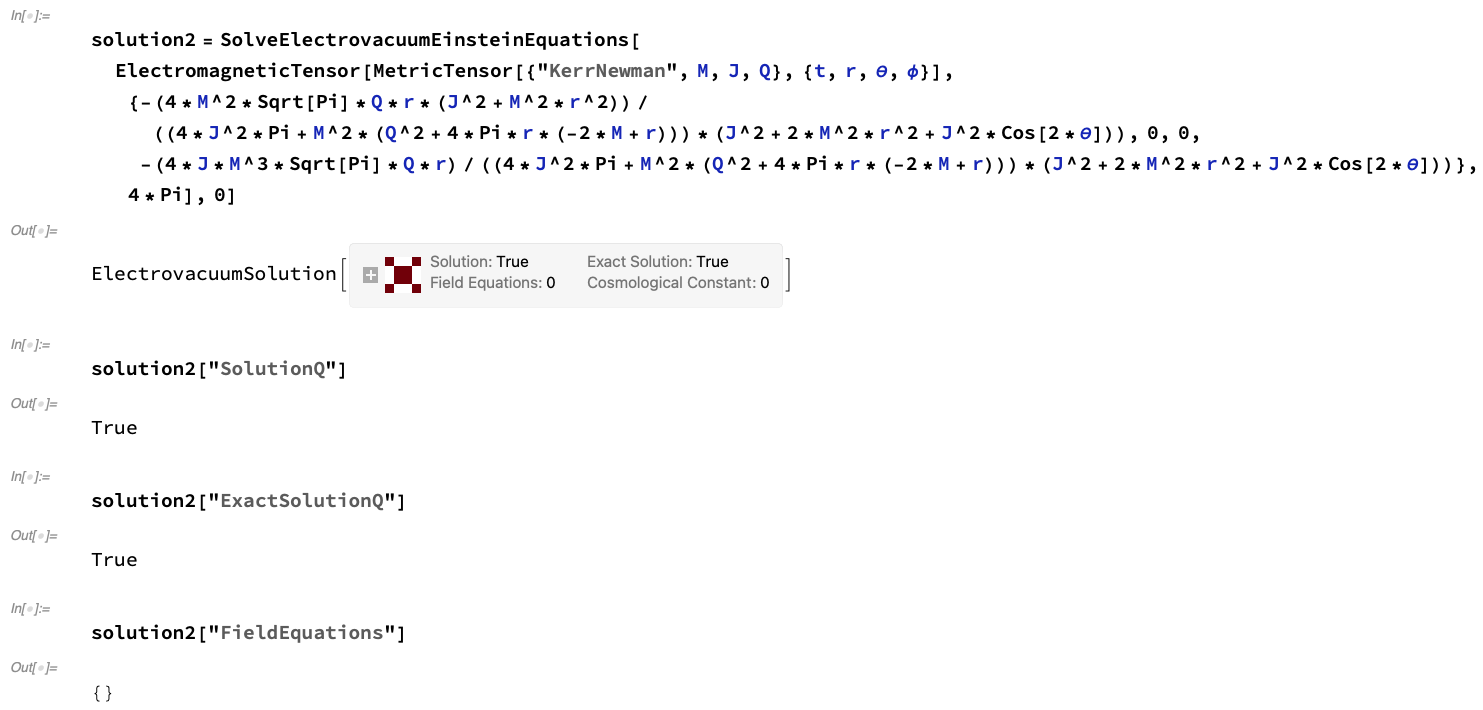}
\end{framed}
\caption{On the left, the \texttt{ElectrovacuumSolution} object for a Reissner-Nordstr\"om geometry (representing e.g. a charged, non-rotating black hole of mass $M$ and electric charge $Q$ in Schwarzschild/spherical polar coordinates ${\left( t, r, \theta, \phi \right)}$) equipped with modified spacetime electromagnetic potential ${\mathbf{A}}$, computed using \texttt{SolveElectrovacuumEinsteinEquations}, illustrating that it corresponds to an exact solution to the vacuum Einstein-Maxwell equations, with zero cosmological constant. On the right, the \texttt{VacuumSolution} object for a Kerr-Newman geometry (representing e.g. a charged, spinning black hole of mass $M$, angular momentum $J$ and electric charge $Q$ in Boyer-Lindquist/oblate spheroidal coordinates ${\left( t, r, \theta, \phi \right)}$) equipped with modified spacetime electromagnetic potential ${\mathbf{A}}$, computed using \texttt{SolveElectrovacuumEinsteinEquations}, illustrating that it corresponds to an exact solution to the vacuum Einstein-Maxwell equations, with zero cosmological constant.}
\label{fig:Figure69}
\end{figure}

\begin{figure}[ht]
\centering
\begin{framed}
\includegraphics[width=0.495\textwidth]{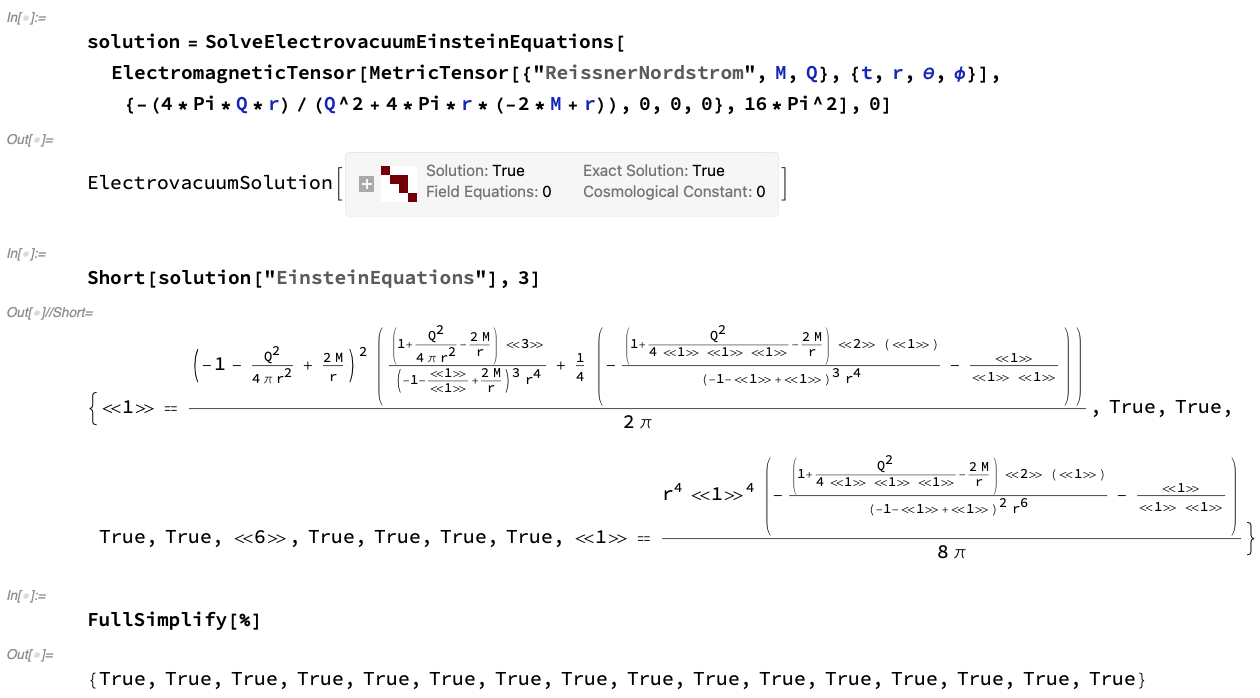}
\vrule
\includegraphics[width=0.495\textwidth]{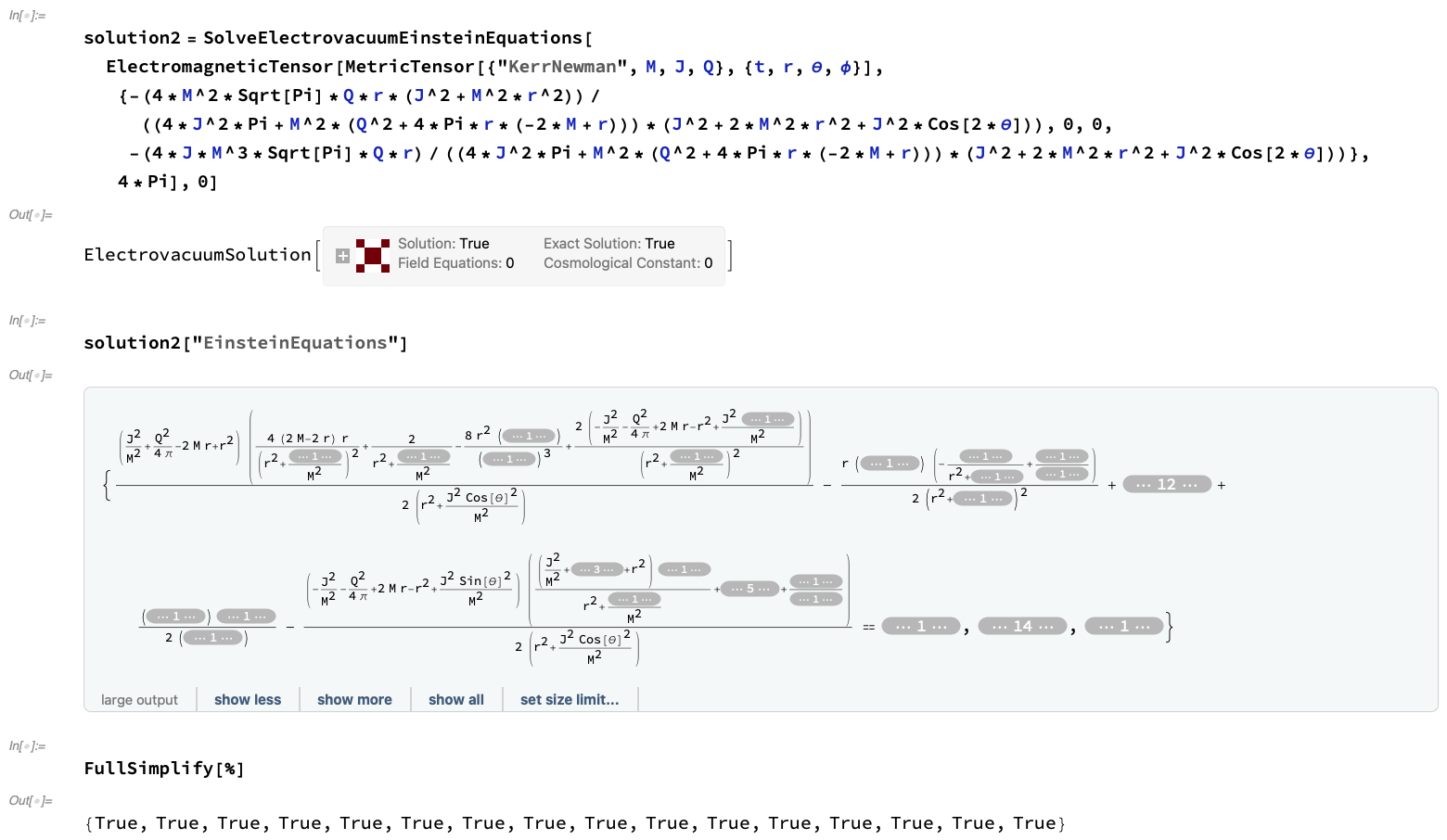}
\end{framed}
\caption{On the left, the list of Einstein-Maxwell field equations asserting that the Einstein tensor, with zero cosmological constant, is equal to ${8 \pi}$ times the electromagnetic stress-energy tensor, computed using the \texttt{ElectrovacuumSolution} object for a Reissner-Nordstr\"om geometry (representing e.g. a charged, non-rotating black hole of mass $M$ and electric charge $Q$ in Schwarzschild/spherical polar coordinates ${\left( t, r, \theta, \phi \right)}$) equipped with modified spacetime electromagnetic potential ${\mathbf{A}}$, together with a verification that they all hold identically. On the right, the list of Einstein-Maxwell field equations asserting that the Einstein tensor, with zero cosmological constant, is equal to ${8 \pi}$ times the electromagnetic stress-energy tensor, computed using the \texttt{ElectrovacuumSolution} object for Kerr-Newman geometry (representing e.g. a charged, spinning black hole of mass $M$, angular momentum $J$ and electric charge $Q$ in Boyer-Lindquist/oblate spheroidal coordinates ${\left( t, r, \theta, \phi \right)}$) equipped with modified spacetime electromagnetic potential ${\mathbf{A}}$, together with a verification that they all hold identically.}
\label{fig:Figure70}
\end{figure}

\begin{figure}[ht]
\centering
\begin{framed}
\includegraphics[width=0.495\textwidth]{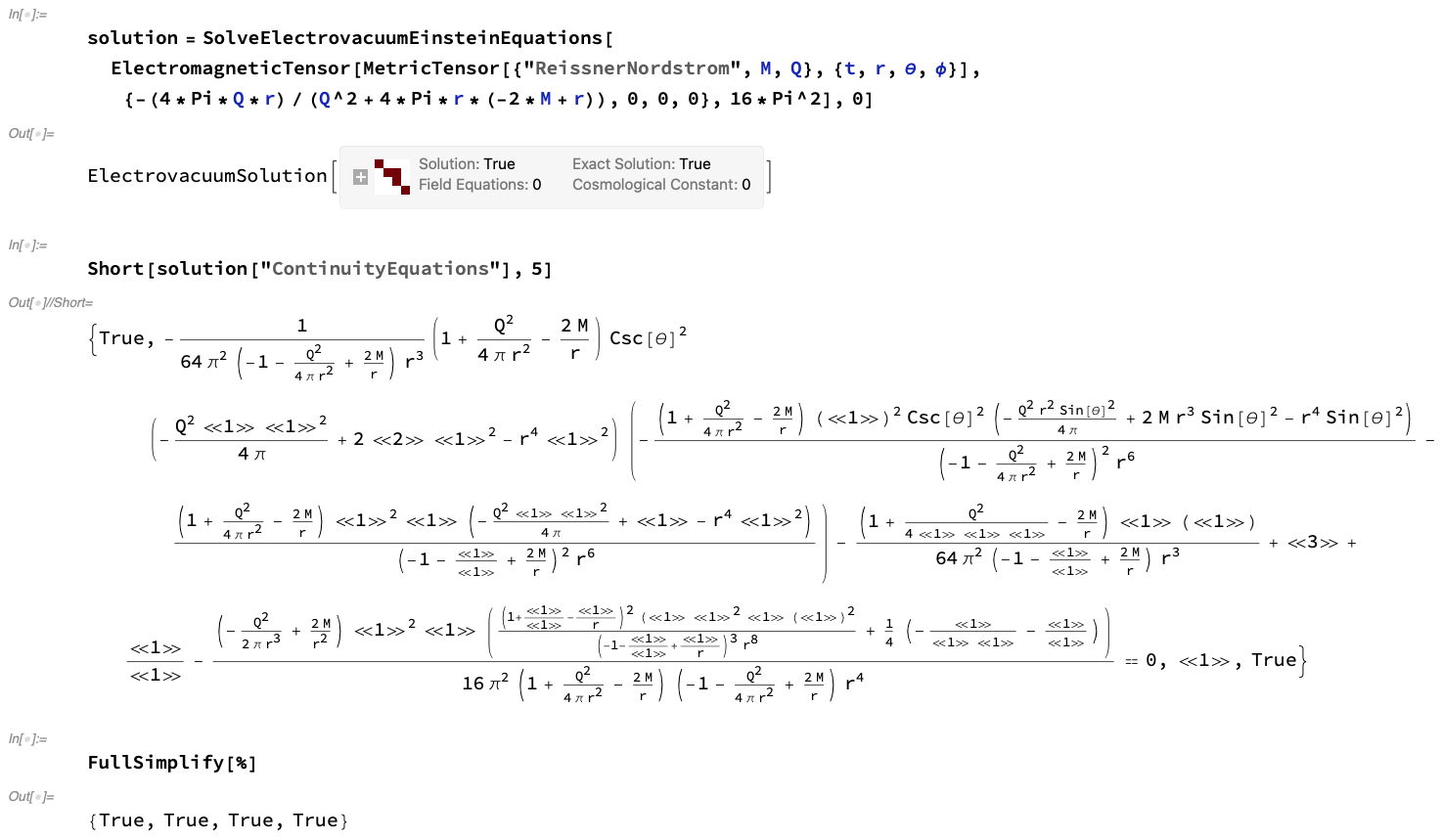}
\vrule
\includegraphics[width=0.495\textwidth]{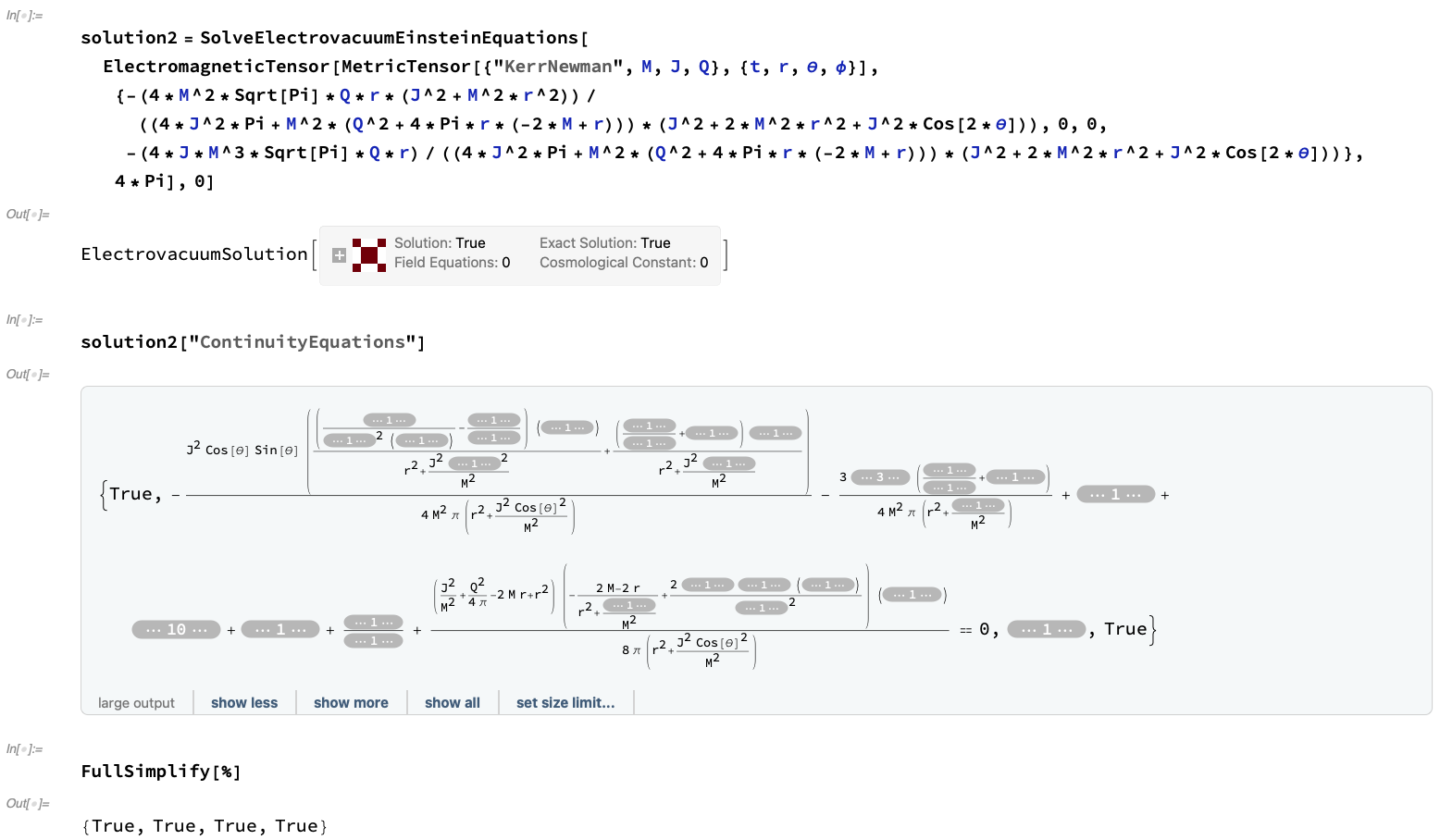}
\end{framed}
\caption{On the left, the list of continuity equations asserting that the covariant divergence of the electromagnetic stress-energy tensor vanishes, computed using the \texttt{ElectrovacuumSolution} object for a Reissner-Nordstr\"om geometry (representing e.g. a charged, non-rotating black hole of mass $M$ and electric charge $Q$ in Schwarzschild/spherical polar coordinates ${\left( t, r, \theta, \phi \right)}$) equipped with modified spacetime electromagnetic potential ${\mathbf{A}}$, with zero cosmological constant, together with a verification that they all hold identically. On the right, the list of continuity equations asserting that the covariant divergence of the electromagnetic stress-energy tensor vanishes, computed using the \texttt{ElectrovacuumSolution} object for a Kerr-Newman geometry (representing e.g. a charged, spinning black hole of mass $M$, angular momentum $J$ and electric charge $Q$ in Boyer-Lindquist/oblate spheroidal coordinates ${\left( t, r, \theta, \phi \right)}$) equipped with modified spacetime electromagnetic potential ${\mathbf{A}}$, with zero cosmological constant, together with a verification that they all hold identically.}
\label{fig:Figure71}
\end{figure}

\begin{figure}[ht]
\centering
\begin{framed}
\includegraphics[width=0.495\textwidth]{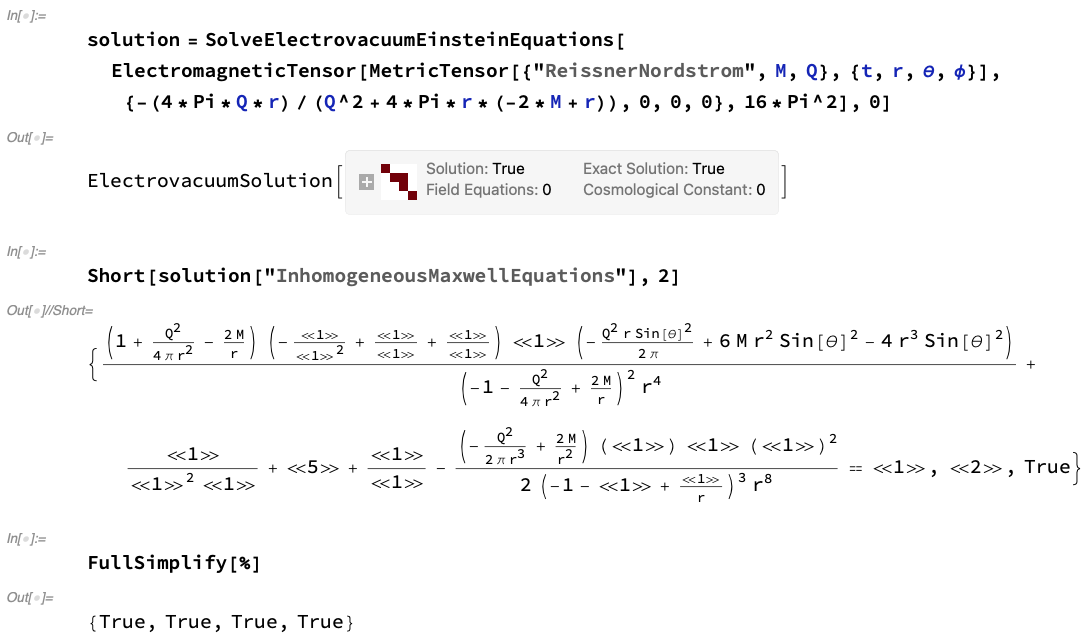}
\vrule
\includegraphics[width=0.495\textwidth]{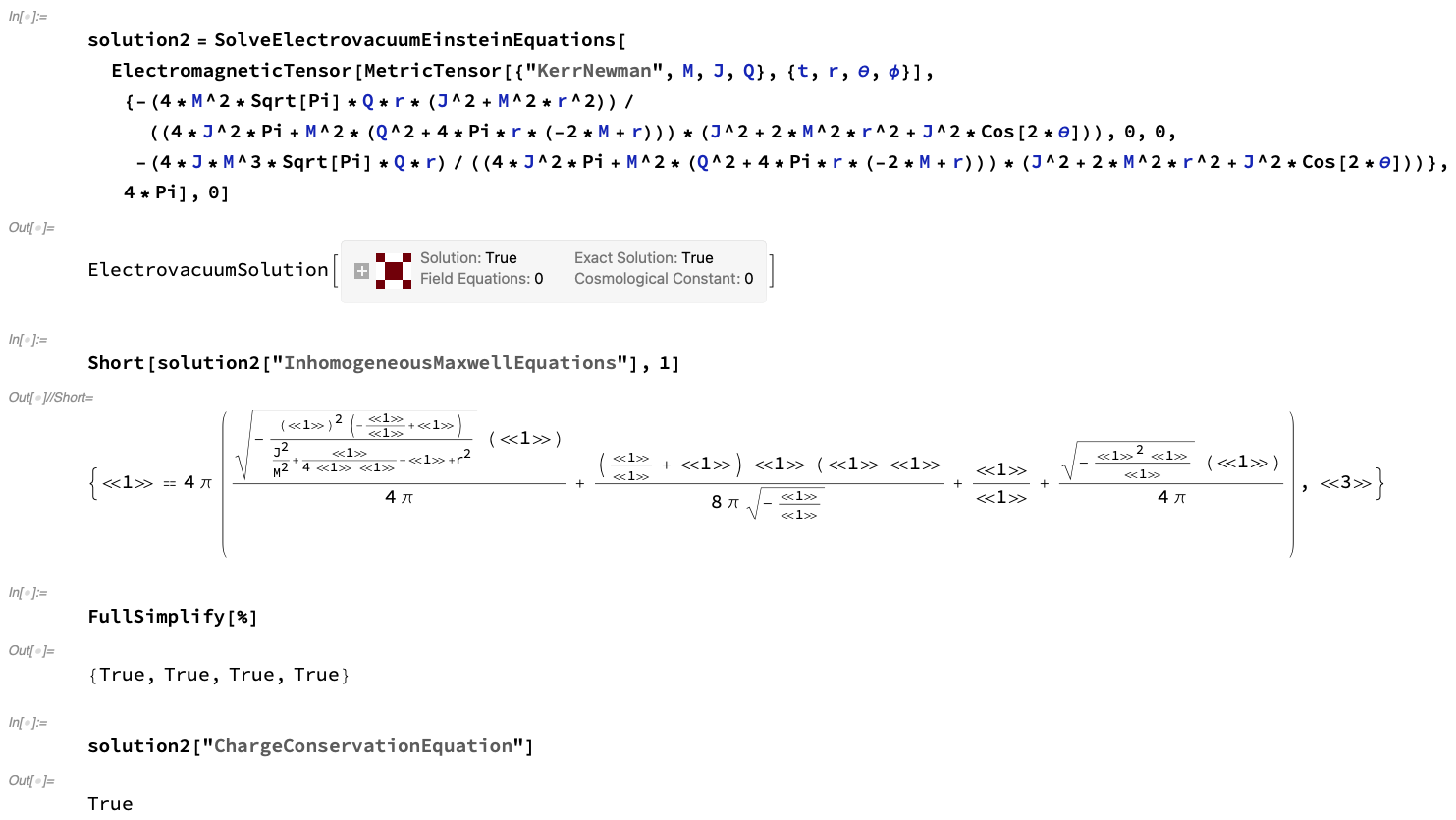}
\end{framed}
\caption{On the left, the list of inhomogeneous Maxwell equations relating the covariant divergence of the electromagnetic tensor and the partial divergence of the electromagnetic displacement tensor density, computed using the \texttt{ElectrovacuumSolution} object for a Reissner-Nordstr\"om geometry (representing e.g. a charged, non-rotating black hole of mass $M$ and electric charge $Q$ in Schwarzschild/spherical polar coordinates ${\left( t, r, \theta, \phi \right)}$) equipped with modified spacetime electromagnetic potential ${\mathbf{A}}$, with zero cosmological constant, together with a verification that they all hold identically. On the right, the list of inhomogeneous Maxwell equations relating the covariant divergence of the electromagnetic tensor and the partial divergence of the electromagnetic displacement tensor density, computed using the \texttt{ElectrovacuumSolution} object for a Kerr-Newman geometry (representing e.g. a charged, spinning black hole of mass $M$, angular momentum $J$ and electric charge $Q$ in Boyer-Lindquist/oblate spheroidal coordinates ${\left( t, r, \theta, \phi \right)}$) equipped with modified spacetime electromagnetic potential ${\mathbf{A}}$, with zero cosmological constant, together with a verification that they all hold identically.}
\label{fig:Figure72}
\end{figure}

We can now proceed to impose the (vacuum) \textit{Einstein-Maxwell} field equations on our manifold ${\mathcal{M}}$, which assert that the Einstein tensor, plus an optional cosmological constant term ${\Lambda}$, is equal to ${8 \pi}$ times the electromagnetic stress-energy tensor alone:

\begin{equation}
G_{\mu \nu} + \Lambda g_{\mu \nu} = R_{\mu \nu} - \frac{1}{2} R g_{\mu \nu} + \Lambda g_{\mu \nu} = \frac{8 \pi}{\mu_0} \left( F_{\mu}^{\alpha} F_{\alpha \nu} - \frac{1}{4} g_{\mu \nu} F_{\alpha \beta} F^{\alpha \beta} \right),
\end{equation}
and which consequently characterize the class of \textit{electrovacuum} solutions to the Einstein field equations, i.e. solutions in which electromagnetic fields constitute the only non-gravitational fields present within the spacetime. Representations of the corresponding \texttt{ElectrovacuumSolution} objects for the Reissner-Nordstr\"om metric (representing e.g. a charged, non-rotating black hole of mass $M$ and electric charge $Q$ in Schwarzschild/spherical polar coordinates ${\left( t, r, \theta, \phi \right)}$), assuming the default spacetime electromagnetic potential vector ${\mathbf{A} = \left( \Phi, A^1, A^2, A^3 \right)}$, and assuming both a vanishing cosmological constant ${\Lambda = 0}$ and a non-vanishing cosmological constant ${\Lambda \neq 0}$, computed using the \texttt{SolveElectrovacuumEinsteinEquations} function, are shown in Figure \ref{fig:Figure67}; these examples demonstrate that the Reissner-Nordstr\"om metric, both with and without cosmological constant, represents a \textit{non-exact} solution of the (vacuum) Einstein-Maxwell field equations if one adopts a fully generic form of the spacetime electromagnetic potential ${\mathbf{A} = \left( \Phi, A^1, A^2, A^3 \right)}$, in the sense that nine additional field equations need to be assumed in each case. The complete lists of (vacuum) Einstein-Maxwell field equations for the Reissner-Nordstr\"om metric, both with and without cosmological constant, can be computed directly from the \texttt{ElectrovacuumSolution} object, and it can be verified in both cases that they do indeed reduce down to the nine canonical field equations previously mentioned, as illustrated in Figure \ref{fig:Figure68}. We can force the electrovacuum solution to be exact by assuming a more restricted form of the spacetime electromagnetic potential, given now by the covector/1-form ${A_{\mu} = \left( \frac{Q}{r}, 0, 0, 0 \right)}$, or, in (contravariant) vector form:

\begin{equation}
A^{\mu} = \left( - \frac{4 \pi Q r}{Q^2 + 4 \pi r \left( r - 2 M \right)}, 0, 0, 0 \right),
\end{equation}
with vacuum magnetic permeability set to ${\mu_0 = 16 \pi^2}$; we can, analogously, force the solution to the (vacuum) Einstein-Maxwell equations for the Kerr-Newman metric (representing e.g. a charged, spinning black hole of mass $M$, angular momentum $J$ and electric charge $Q$ in Boyer-Lindquist/oblate spheroidal coordinates ${\left( t, r, \theta, \phi \right)}$) to be exact by assuming a spacetime electromagnetic potential given by the covector/1-form\cite{carter}:

\begin{equation}
A_{\mu} = \left( \frac{r Q}{2 \sqrt{\pi} \left( r^2 + \left( \frac{J}{M} \right)^2 \cos^2 \left( \theta \right) \right)}, 0, 0, - \frac{\left( \frac{J}{M} \right) r Q \sin^2 \left( \theta \right)}{2 \sqrt{\pi} \left( r^2 + \left( \frac{J}{M} \right)^2 \cos^2 \left( \theta \right) \right)} \right),
\end{equation}
or, in (contravariant) vector form:

\begin{multline}
A^{\mu} = \left( - \frac{4 M^2 \sqrt{\pi} Q r \left( J^2 + M^2 r^2 \right)}{\left( 4 J^2 \pi + M^2 \left( Q^2 + 4 \pi r \left( r - 2 M \right) \right) \right) \left( J^2 + 2 M^2 r^2 + J^2 \cos \left( 2 \theta \right) \right)}, 0, 0, \right.\\
\left. - \frac{4 J M^3 \sqrt{\pi} Q r}{\left( 4 J^2 \pi + M^2 \left( Q^2 + 4 \pi r \left( r - 2 M \right) \right) \right) \left( J^2 + 2 M^2 r^2 + J^2 \cos \left( 2 \theta \right) \right)} \right),
\end{multline}
with vacuum magnetic permeability set to ${\mu_0 = 4 \pi}$. Representations of the corresponding \texttt{ElectrovacuumSolution} objects for the Reissner-Nordstr\"om and Kerr-Newman metrics, assuming these modified forms of the spacetime electromagnetic potential vector ${\mathbf{A}}$, and with vanishing cosmological constant, are shown in Figure \ref{fig:Figure69}; these examples demonstrate that both the Reissner-Nordstr\"om and Kerr-Newman metrics represent \textit{exact} solutions to the (vacuum) Einstein-Maxwell field equations if one adopts these restricted forms of the spacetime electromagnetic potential ${\mathbf{A}}$, in the sense that no additional field equations need to be assumed. The complete lists of (vacuum) Einstein-Maxwell field equations for the Reissner-Nordstr\"om and Kerr-Newman metrics, with vanishing cosmological constant, can be computed directly from the \texttt{ElectrovacuumSolution} object, and it can be verified in both cases that they all indeed hold identically, as illustrated in Figure \ref{fig:Figure70}. Figure \ref{fig:Figure71} shows the full lists of continuity equations, asserting that the covariant divergence of the electromagnetic stress-energy tensor ${T_{\mu \nu}}$ must vanish, computed directly from the \texttt{ElectrovacuumSolution} objects for both a Reissner-Nordstr\"om metric and a Kerr-Newman metric, demonstrating that they all hold identically. Finally, Figure \ref{fig:Figure72} shows the full lists of inhomogeneous Maxwell equations, asserting that the covariant divergence of the electromagnetic tensor ${F_{\mu \nu}}$ is related to the spacetime current density (i.e. the partial divergence of the electromagnetic displacement tensor density ${\mathcal{D}^{\mu \nu}}$), again computed directly from the \texttt{ElectrovacuumSolution} objects for both a Reissner-Nordstr\"om and a Kerr-Newman metric, and again demonstrating that they all hold identically in these cases.

\clearpage

\section{Concluding Remarks}
\label{sec:Section4}

This article has sought to provide a reasonably comprehensive overview of the main capabilities of \textsc{Gravitas}'s powerful symbolic subsystem for representing and manipulating general analytic/continuous representations of Riemannian and pseudo-Riemannian metrics and spacetime geometries in arbitrary coordinate systems, as well as of its purpose-built tensor calculus and differential geometry functionality, before finally concluding with a demonstration of \textsc{Gravitas}'s sophisticated analytical and numerical algorithms for solving the Einstein (and Einstein-Maxwell) field equations with a high level of generality and in the presence of arbitrary energy-matter configurations and/or electromagnetic fields. In the forthcoming second article within this series introducing the framework, we aim to focus more on \textsc{Gravitas}'s \textit{numerical} relativity capabilities, highlighting the procedures for representing and configuring Cauchy-type initial data, gauge conditions and (hyperbolic) evolution equations for numerical relativity simulations with arbitrary coordinate structure, including different methods for handling metric decomposition and the enforcement of the ADM Hamiltonian and momentum constraint equations. We also intend to outline how the various hypergraph-based adaptive refinement algorithms and visualization tools work (and why they allow for the representation of a much larger and more general class of coordinate systems and spacetime topologies than traditional numerical relativity techniques), as well as to demonstrate some exemplar applications of \textsc{Gravitas} to a handful of perennial problems in numerical relativity, including simulated mergers of binary black hole systems (and the extraction of the corresponding gravitational wave signatures), and idealized gravitational collapse simulations via full general relativistic hydrodynamics. Part of the objective here will be to illustrate how these advanced numerical relativity capabilities integrate and interoperate seamlessly with the highly general analytic and symbolic functionality introduced within the present article.

In addition to the obvious planned expansions to \textsc{Gravitas}'s in-built libraries of metric tensors, geometries, coordinate systems and stress-energy models, there exist many exciting potential extensions to \textsc{Gravitas}'s core symbolic subsystem itself, some of which are merely planned but many of which are currently under active research and development. Forthcoming domains of symbolic and analytic functionality that are under either active development or planning include: support for the tetrad formalism and for frame fields  in general (including the Newman-Penrose formalism and spin coefficients as special cases); support for connections defined on spinor bundles, in addition to those on vector bundles (and, along with it, support for both Levi-Civita and affine spin connections, torsion and contorsion tensors, spin tensors, Riemann-Cartan geometries and Einstein-Cartan field equations); support for scalar-tensor gravity theories (such as Brans-Dicke theory and its relatives); and support for gravitoelectromagnetic formalism and the Bel decomposition of the Riemann tensor into electrogravitic, magnetogravitic and topogravitic components. Some ongoing and planned research applications of the \textsc{Gravitas} framework include the investigation of black hole thermodynamics and the ER=EPR conjecture within generic discrete spacetime settings\cite{shah} (and the relationship between computational complexity, entropy and discrete spacetime geometry more generally\cite{gorard11}), the study of certain global topological and homotopical features of discrete spacetimes\cite{arsiwalla}\cite{arsiwalla2}, the simulation of relativistic Bondi-Hoyle fluid accretion onto (potentially spinning, potentially electrically charged) black holes\cite{papadopoulos}\cite{font}, and the simulation of binary neutron star mergers\cite{baiotti}\cite{bernuzzi}. The rationale for pursuing the latter two research directions in particular is as part of an ongoing effort to ascertain potential astrophysical and/or cosmological probes of spacetime discreteness in the form of measurable deviations from the predictions of classical general relativity (for instance in either the electromagnetic, gravitational or neutrino radiation signatures emitted from high-energy astrophysical events) that may remain detectable even at high redshift.

\section*{Acknowledgments}

The author would like to acknowledge Nikola Bukowiecka, Federico Semenzato, Jacopo Uggeri and several others for their early ``battle-testing'' of the \textsc{Gravitas} framework, and for discovering various bugs/limitations and offering numerous helpful suggestions (additional thanks are also due to Nikola Bukowiecka for her diligent proofreading of the present manuscript, and for many valuable discussions). The author is also grateful to the DAMTP GR journal club at the University of Cambridge for allowing him to demonstrate a very early prototype of \textsc{Gravitas} and receive highly constructive feedback on its design. Finally, the author would like to thank Jos\'e Martin-Garc\'ia and Stephen Wolfram for innumerable and invaluable design discussions and feature suggestions, and for their consistent encouragement and enthusiasm throughout all stages of \textsc{Gravitas}'s (ongoing) development.

\end{document}